\documentclass[12pt,openany]{book}
\usepackage[Lenny]{fncychap}

\pdfoutput=1
\usepackage{graphicx}
\usepackage{youngtab}
\usepackage[centertags]{amsmath}
\usepackage{amssymb,amsthm,amsfonts,mathrsfs,bm,bbm}
\usepackage{ccaption}
\usepackage[usenames]{color}
\usepackage{tikz}
\usetikzlibrary{decorations.pathmorphing}
\usepackage[mathscr]{eucal}

\usepackage{tocloft}
\usepackage[
      colorlinks=true,
      linkcolor=blue,
      urlcolor=blue,
      filecolor=black,
      citecolor=red,
      pdfstartview=FitV,
      pdftitle={Holographic entanglement entropy},
        pdfauthor={Mukund Rangamani, Tadashi Takayanagi}
      ]{hyperref}

\usepackage{rotating}
\makeatletter
\@addtoreset{equation}{section}

\makeatletter
\renewcommand\section{\@startsection {section}{1}{\z@}%
                                   {-3.5ex \@plus -1ex \@minus -.2ex}%nn
                                   {2.3ex \@plus.2ex}%
                                   {\normalfont\large\bfseries}}
\renewcommand\subsection{\@startsection{subsection}{2}{\z@}%
                                     {-3.25ex\@plus -1ex \@minus -.2ex}%
                                     {1.5ex \@plus .2ex}%
                                     {\normalfont\bfseries}}

  \captionnamefont{\bfseries}
  \captiontitlefont{\small\sffamily}
  \captiondelim{: }
  \hangcaption

\def\vev#1{\langle\, #1 \, \rangle}
\def\ket#1{\mid  \! #1  \rangle}
\def\bra#1{\langle  #1 \! \mid}

\def\Tr#1#2{{\rm Tr}_{#1}\left(#2\right)}
\def\tr{\rm Tr}

\def\AdS#1{AdS$_{#1}$}
\def\SAdS#1{Schwarzschild-AdS$_{#1}$}

\def\lads{\ell_\text{AdS}}
\def\GN{G_N^{(d+1)}}
\def\ceff{c_\text{eff}}

\def\skL{\text{\sf{L}}}
\def\skR{\text{\sf{R}}}

\def\bulkJ{{\tilde J}}

\def\AdS#1{\AdS$_{#1}$}

\def\tmi{{\sf I_3}}

\def\bulk{{\cal M}} 
\def\bdy{{\cal B}}

\def\regA{{\cal A}} 
\def\regAc{{\cal A}^c} 
\def\rhoA{{\rho_{\regA}}} 
\def\modA{{\cal K}_\regA}
\def\rhoAc{{\rho_{\regAc}}} 
\def\entsurf{
\partial \regA} 
\def\domdA{D[\regA]} 
\def\domdAc{D[\regAc]} 
\def\domd#1{D[#1]}
\def\extrA{{\cal E}_\regA} 
\def\extr#1{{\cal E}_{#1}} 
\def\CIS{{\Xi_\regA}} 
\def\homsurfA{{\cal R}_{\regA}} 
\def\homsurf#1{{\cal R}_{#1}} 
\def\homsurfAc{{\cal R}_{{\regA}^c}} 
\def\CWA{{\cal W}_{\cal C}[\regA]} 
 
\def\CW#1{{\cal W}_{\cal C}[{#1}]}
\def\EW#1{{\cal W}_{\cal E}[{#1}]}
\def\EWA{{\cal W}_{\cal E}[\regA]} 
\def\EWAc{{\cal W}_{\cal E}[\regAc]}

\newcommand{\ball}{
\begin{tikzpicture}
     \draw (0,0) arc (180:360:1mm and 0.5mm) -- (0.1,-0.2) -- cycle;
    \shade[left color=red!15!white,right color=blue!40!white,opacity=0.3] (0,0) arc (180:360:1mm and 0.5mm) -- (0.1,-0.2) -- cycle;
      \draw (0,0) arc (180:0:1mm and 0.5mm) -- (0.1,0.2) -- cycle;
    \shade[left color=blue!15!white,right color=blue!40!white,opacity=0.3] (0,0) arc (180:0:1mm and 0.5mm) -- (0.1,0.2) -- cycle;
        \draw[ color=red!50, fill=orange!50] (0,0) arc (180:540:1mm and 0.5mm);
\end{tikzpicture}
}
\newcommand{\disc}{
\begin{tikzpicture}
        \draw[ color=black, fill=orange!50] circle (0.5ex););
\end{tikzpicture}
}

\def\rhoB{\rho_{\ball}}
\def\regAB{{\cal A}_{_{\disc}}}
\def\homsurfAB{\homsurf{{\disc}}}

\def\OH{{\cal O}_{\scriptstyle{H}}}
\def\OL{{\cal O}_{\scriptstyle{L}}}
\def\DH{\Delta_{\scriptscriptstyle{H}}}
\def\DL{\Delta_{\scriptscriptstyle{L}}}

\def\aH{\alpha_{\scriptscriptstyle{H}}}
\def\ah{\alpha_{\scriptscriptstyle{h}}}
\def\aO{\alpha_{\scriptscriptstyle{{\cal O}}}}

\def\AdS#1{AdS$_{#1}$}
\def\SAdS#1{Schwarzschild-AdS$_{#1}$}

\def\vev#1{\langle\, #1 \, \rangle} 
\def\ket#1{\mid \! #1\rangle} 
\def\bra#1{\langle \, #1 \! \mid\! \ }

\def\rh{r_+}
\def\Hq{{\cal H}_\text{qubit}}

\def\phA{\varphi_\regA}

\newcommand{\be}{\begin{equation}}
\newcommand{\ba}{\begin{eqnarray}}
\newcommand{\ea}{\end{eqnarray}}
\newcommand{\ee}{\end{equation}}

\newcommand{\ap}{\alpha}

\newcommand{\lb}{\rangle}
\newcommand{\bea}{\begin{eqnarray}}
\newcommand{\eea}{\end{eqnarray}}
\newcommand{\bes}{\begin{equation*}}
\newcommand{\beas}{\begin{eqnarray*}}
\newcommand{\eeas}{\end{eqnarray*}}
\newcommand{\bas}{\begin{array*}}
\newcommand{\eas}{\end{array*}}
\newcommand{\ees}{\end{equation*}}

\newcommand{\ep}{\epsilon}

\definecolor{rust}{rgb}{0.8,0.2,0.2} 
\definecolor{purple}{rgb}{0.8,0.1,0.9} 
\definecolor{olivegreen}{rgb}{0,0.52,0.17}

\def\tE{t_\text{\tiny E}}
\newcommand{\fixM}{\mathbf{e}}

\newcommand{\CSA}{\Sigma_{_t}}
\newcommand{\bulkCS}{{\tilde \Sigma}_{_t}}

\title{{\bf \Huge Holographic Entanglement Entropy}}

\author{\normalsize
Mukund Rangamani$^a$ \& \  Tadashi Takayanagi$^{\,b}$\\
\small\sl mukund@physics.ucdavis.edu, takayana@yukawa.kyoto-u.ac.jp \\
\small \sl $^a$  Center for Quantum Mathematics and Physics (QMAP), \\
\small \sl Department of Physics, University of California, Davis, CA 95616 USA.\\
\small \sl $^b$ Yukawa Institute for Theoretical Physics (YITP),
\\[-1.5mm]
\small \sl  Kyoto University, Kyoto 606-8502, Japan. \\
}

\begin{document}

\setlength{\baselineskip}{16pt}
\begin{titlepage}
\maketitle
% \vspace{-36pt}

%Abstract
% \begin{abstract}
%  Synopsis for the book on holographic entanglement entropy.
%  \end{abstract}
\thispagestyle{empty}
\setcounter{page}{0}
\end{titlepage}

\renewcommand{\thefootnote}{\arabic{footnote}}
%______________________________________

%%%%%%%%%%%%%%%%%%%%%%%%%%%%%%%%%%%%%%%%%%%%
\frontmatter

\chapter{Abstract}

\begin{picture}(0,0)(0,0)
\put(380,150){YITP-16-106} 
\put(380,130){IPMU 16-0135}
\end{picture}

We review the developments in the past decade on holographic entanglement entropy, a subject that has garnered much attention owing to its potential to teach us about the emergence of spacetime in holography. 
We provide  an introduction to the concept of entanglement entropy in quantum field theories, review the holographic proposals for computing the same, providing some justification for where these proposals arise from in the first two parts. The final part addresses recent developments linking entanglement and geometry. We provide an overview of the various arguments and technical developments that teach us how to use field theory entanglement to detect geometry. Our discussion is by design eclectic; we have chosen to focus on developments that appear to us most promising for further insights into the holographic map. 

This is a preliminary draft of a few chapters of a book which will appear sometime in the near future, to be published by Springer, as part of their Lecture Notes in Physics series. The book in addition contains a discussion of application of holographic ideas to computation of entanglement entropy in strongly coupled field theories, and discussion of tensor networks and holography, which we have chosen to exclude from the current manuscript.

%%%%%%%%%%%%%%%%%%%%%%%%%%%%%%%%%%%%%%%%%%%%

\chapter{Acknowledgments}

We have been extremely fortunate to benefit from the wisdom and deep physical intuition of our wonderful collaborators  Veronika Hubeny and Shinsei Ryu who played a pivotal role in helping us develop the basic picture relating quantum entanglement and holography. The importance of their role in shaping the story of holography entanglement entropy cannot be overstated. 

We have also enjoyed many excellent collaborations in our explorations over the past decade on this subject: thanks to Tatsuo  Azeyanagi, Jyotirmoy Bhattacharya, Pawel Caputa, Sumit Das, Xi Dong, Mitsutoshi Fujita, Simon Gentle, Kanato Goto, Thomas Hartman, Felix Haehl, Song He,  Matthew Headrick, Andreas Karch, Albion Lawrence, Aitor Lewkowycz, Wei Li, Don Marolf, Masamichi Miyaji, Ali Mollabashi, K. Narayan, Tatsuma Nishioka, Masahiro Nozaki, Tokiro  Numasawa, Noriaki  Ogawa, Eric Perlmutter, Max Rota, Noburo Shiba, Joan Simon, Andrius  Stikonas, Moshe Rozali, Sandip Trivedi, Erik Tonni, Henry Maxfield, Tomonori Ugajin, Alexandre Vincart-Emard, Kento Watanabe, Xueda Wen, and Anson Wong for many fun discussions and for helping us understand various aspects of the story we are about to relate. An especial thanks to Max Rota for his useful comments on a draft of this manuscript.

We would especially like to single out the influence of Horacio Casini, Matt Headrick, Don Marolf, Rob Myers, and Mark Van Raamsdonk whose perspicacious insights have contributed immensely to our understanding of entanglement and holography.

%%%%%%%%%%%%%%%%%%%%%%%%%%%%%%%%%%%%%%%%%%%%
\newpage 
\tableofcontents
\cleardoublepage

\mainmatter

%~~~~~~~~~~~~~~~~~~~~~~~~~~~~~~~~~~~~~~~~~~~~~~~
\part*{Introduction}
\label{part:intro}
\chapter{Introduction}
\label{sec:intro}
%~~~~~~~~~~~~~~~~~~~~~~~~~~~~~~~~~~~~~~~~~~~~~~

Quantum mechanics distinguishes itself from classical physics via the presence of entanglement. Classically, one is conditioned to imagine situations wherein components of a single system  may be separated into non-interacting parts, which we can separately examine, and then put back together to reconstruct the full system. This intuition fails  spectacularly in quantum mechanics, since the separate pieces, whilst non-interacting, could nevertheless be entangled. As Schr\"odinger put it quite clearly \cite{Schrodinger:1935zz}:
\begin{quote}
{\em The best possible knowledge of a whole does not necessarily include the best possible knowledge of all its parts, even though they may be entirely separate and therefore virtually capable of being `best possibly known', i.e., of possessing, each of them, a representative of its own. The lack of knowledge is by no means due to the interaction being insufficiently known — at least not in the way that it could possibly be known more completely — it is due to the interaction itself.}
\end{quote}

This quintessential feature of the quantum world has been a source of great theoretical interest over the intervening decades. The initial debate about ``spooky action at a distance'' consequent of the Einstein-Podolsky-Rosen (EPR) \cite{Einstein:1935rr} gedanken experiments involving  entangled spins laid the foundations for more detailed investigation in later years. With the passage of time, the advent of John Bell's seminal understanding of quantum correlations \cite{Bell:1966aa},  and our improved understanding of the physical implications, we now view the presence of entanglement as a fungible resource in quantum systems, which can be exploited for several tasks. This perspective has been immensely bolstered by the rapid development of the subject of quantum information over the past few decades. An excellent resource for getting acquainted with the subject is the classic textbook  by Nielsen and Chuang \cite{Nielsen:2010aa}.

The presence of quantum entanglement is cleanly exhibited by the simplest of systems: two qubits. We have the total Hilbert space, which is a tensor product of single qubit Hilbert spaces ${\cal H} = \Hq \otimes \Hq = \text{Span}\{ \ket{00}\, \ket{01}, \ket{10}, \ket{11}\}$. The basis states are clearly {\em separable}, in that we can isolate each qubit individually whilst leaving the other unaffected. On the other hand, the  EPR/Bell/cat state
\begin{equation}
\frac{1}{\sqrt{2}}\, \left(\ket{00} + \ket{11} \right) ,
\label{}
\end{equation}
doesn't admit a separation into individual components that may be identified as elements of one qubit or the other. More pertinently, the state of the two qubits is correlated; knowing that the first qubit is in a particular state determines the state of the second. Such non-separable states are said to be {\em entangled}.

While these ideas involving quantum entanglement are easy to illustrate and intuit in simple systems involving a few qubits, it should be apparent that the essential concepts continue to hold in continuum systems. In recent years, we have come to appreciate that the entanglement structure encoded in a many-body wavefunction provides important insight into the structure of the quantum state under consideration. The simplest illustrative example is the notion of topological entanglement entropy in $(2+1)$-dimensional topological field theories. These are systems with no dynamical degrees of freedom, which nevertheless exhibit interesting phase structure. It was realized in \cite{Kitaev:2005dm,Levin:2006aa} that the entanglement entropy provides a useful order parameter for characterizing the distinct phases in such systems. More generally perhaps, one can view modern efforts to characterize the ground states of interacting many-body systems in terms of understanding the potential entanglement structure of the wavefunctions (see, e.g., \cite{Vishwanath:2014aa}). In the continuum limit, when we focus on quantum field theories (QFTs), it is efficacious to pass over from the many-body wavefunction to the wavefunctional localized onto some spatial domain.

While the application of  quantum entanglement to distinguish phases of many-body dynamics would have been fascinating in its own right, a powerful connection between gravitational dynamics and entanglement, which has emerged in the context of holography, provides further reason to delve deeper into the subject. The notion of holography in high energy physics connotes an important duality between two disparately presented physical systems. On the one hand, one has a quantum mechanical system of a familiar kind and on the other one has a theory of quantum gravity, one in which the geometry itself fluctuates quantum mechanically. While one would a-priori assume that these two situations are unrelated, the remarkable gauge/gravity or AdS/CFT correspondence put forth by Maldacena nearly two decades ago \cite{Maldacena:1997re} suggests that they are two representations of the same physical system in certain situations. This statement is usually codified by the statement that ``Quantum gravity in an asymptotically Anti de Sitter (AdS) spacetime is dual, i.e., physically equivalent to a standard quantum field theory.'' One heuristically thinks of the QFT as living on the boundary of the AdS spacetime as a useful mnemonic. Whilst this is appropriate for intuition building and setting up some of the basic elements of the correspondence, it should be emphasized that the QFT is a separate entity. Importantly, spacetime in which gravity operates is emergent from the collective dynamics of the quantum fields;  the latter by themselves reside on a rigid spacetime sans gravity.

The question which has been actively researched since the early days of the AdS/CFT correspondence is how does the geometric picture emerge from the QFT dynamics? What are the building blocks of the gravitational spacetime?  Surprisingly, the answer to these questions seems intricately tied to the entanglement structure of the states of the QFT. The genesis of these ideas dates back to the observation of Ryu-Takayanagi (RT) \cite{Ryu:2006bv,Ryu:2006ef},  who proposed that the entanglement entropy associated with a spatial region in a holographic QFT is given by the area of a particular minimal area surface in the dual geometry. Inspired by this claim, and its generalization by Hubeny-Rangamani-Takayanagi (HRT) \cite{Hubeny:2007xt} to time-dependent states, Swingle \cite{Swingle:2009bg} and Van Raamsdonk \cite{VanRaamsdonk:2009ar,VanRaamsdonk:2010pw} argued that the essential building block of the spacetime geometry should somehow be related to the entanglement structure of the quantum state in the QFT. This philosophy has since  been codified by Maldacena and Susskind   \cite{Maldacena:2013xja} into the pithy epigram ``ER = EPR'', which refers to a geometric construct, the Einstein-Rosen bridge (ER), being related to the entanglement structure suggested by the Einstein-Podolsky-Rosen  (EPR) gendanken experiment.

Our aim in this book is to provide a sampler of the developments in the subject over the past decade, taking the reader on a tour through the quantum entanglement landscape. It is to some extent remarkable that we have come to appreciate (or perhaps re-appreciate) the central role played by this concept in the context of quantum field theories. Our discussion will necessarily be eclectic -- we shall summarize salient developments in the subject which has played a role in shaping our understanding of holography, but will elide over some of the discussion of computing entanglement entropy in QFTs and its application to many-body systems.

\paragraph{Synopsis of the book:} The book is divided into four thematically distinct parts.
\begin{enumerate}
\item Part \ref{part:ee} describes how to think about entanglement in quantum mechanics and QFTs and lays out the basic formalism we will need  for our discussion. We review the construction of density matrix elements and the computation of R\'enyi and Von Neumann entropies and illustrate the general discussion with examples from two-dimensional conformal field theories (CFTs).
\item Part \ref{part:hee} then turns to holography and describes the various ideas for computing entanglement entropy in field theories which are amenable to such a holographic description. To keep the discussion brief, we only give a rather quick review of the holographic map between QFTs and their gravitational avatars. Later in the book,  we will describe in some detail under what conditions we expect the holographic dual of a QFT to be given by classical gravitational dynamics. Our primary goal here will be to give a working knowledge of the holographic entanglement entropy proposals and therefore will defray the conceptual questions for subsequent discussion.
\item Part \ref{part:qg} then turns to the developments which provide a direct link between geometry and entanglement. This is a rapidly evolving area at the forefront of current research focused on how we can use quantum entanglement in field theories as underlying the holographic map; in the colloquial phrasing, ``entanglement builds geometry''. We will review at a heuristic level many of the ideas that have been developed in the past few years, and also take the opportunity to comment on some of the open issues. Given the rapid flux of ideas, we will try to focus on those that we feel hold the most promise for future investigation, and thus will not attempt to give a comprehensive survey. 

In the published version, we also take the opportunity to explain concepts from tensor networks which have been suggested as useful toy models for understanding the holographic dictionary. An additional part will focus on recent studies of entanglement as a diagnostic of quantum dynamics.  We describe, primarily, applications which are easily amenable to holographic analysis such as quench dynamics and entanglement in the presence of Fermi surfaces. The examples we have chosen illustrate the general lessons we can learn about entanglement propagation in interacting quantum systems. 
\end{enumerate}

\paragraph{Other resources:} We list a series of references that the reader may wish to consult for various concepts that will come up during the course of our discussion.
\begin{itemize}
\item {\bf Quantum Entanglement:} A good review of the developments in quantum entanglement  from a foundational and operational perspective can be found in \cite{Horodecki:2009aa}, while \cite{Nielsen:2010aa}  provides a good introduction to the concepts from a quantum computational perspective.

\item {\bf Entanglement in QFTs:} A good introductory discussion about entanglement entropy in quantum field theories can be found in \cite{Calabrese:2005zw}. Computational techniques and results for two-dimensional conformal field theories are reviewed in \cite{Calabrese:2009qy}.

\item {\bf Holography:} The original papers \cite{Maldacena:1997re} which obtained the statement of the correspondence, as well as the formulation of the map between the bulk and boundary theories developed in
\cite{Gubser:1998bc,Witten:1998qj}, are mandatory reading for any serious student of the subject.
As such, this a vast subject which is hard to review in short order, but \cite{Aharony:1999ti} does a great job of laying out the essentials despite dating back to the genesis of the subject. Other reviews such as \cite{DHoker:2002aw} provide a useful complementary perspective.  There is a recently published book \cite{Ammon:2015wua} which could be a valuable resource.

\item {\bf Holographic entanglement entropy:} The original papers \cite{Ryu:2006bv} and \cite{Hubeny:2007xt} which developed the holographic methods contain many examples and review other salient features.  A review of developments in holographic entanglement entropy which includes most of the early developments is \cite{Nishioka:2009un}. A more recent review emphasizing the connection to gravity is \cite{VanRaamsdonk:2016exw}.
\end{itemize}

%~~~~~~~~~~~~~~~~~~~~~~~~~~~~~~~~~~~~~~~~~~~~~~~
\part{Quantum Entanglement}
\label{part:ee}
%~~~~~~~~~~~~~~~~~~~~~~~~~~~~~~~~~~~~~~~~~~~~~~

%~~~~~~~~~~~~~~~~~~~~~~~~~~~~~~~~~~~~~~~~~~~~~~~
\chapter{Entanglement in QFT}
\label{sec:qft}
%~~~~~~~~~~~~~~~~~~~~~~~~~~~~~~~~~~~~~~~~~~~~~~

As presaged in \S\ref{sec:intro}, we will primarily be interested in understanding entanglement in holographic
field theories. But before we get to this particular set of quantum systems, it is useful to build some intuition in a more familiar setting. In this and the next section, we will therefore focus our attention on getting some insight into the concept of entanglement and learn some of the techniques which are used to characterize it. The discussion here will also serve to build some technical machinery which will be useful in the holographic context.

%~~~~~~~~~~~~~~~~~~~~~~~~~~~~~~~~~~~~~~~~~~~~~~~
\section{Entanglement in lattice systems}
\label{sec:eelat}
%~~~~~~~~~~~~~~~~~~~~~~~~~~~~~~~~~~~~~~~~~~~~~~

% Figure
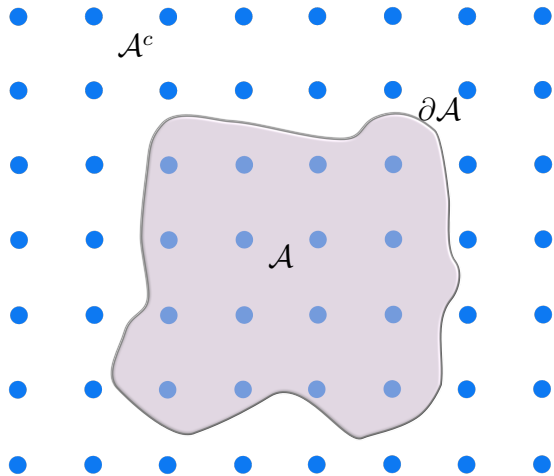
\begin{figure}[htbp]
\begin{center}
\includegraphics[width=3in]{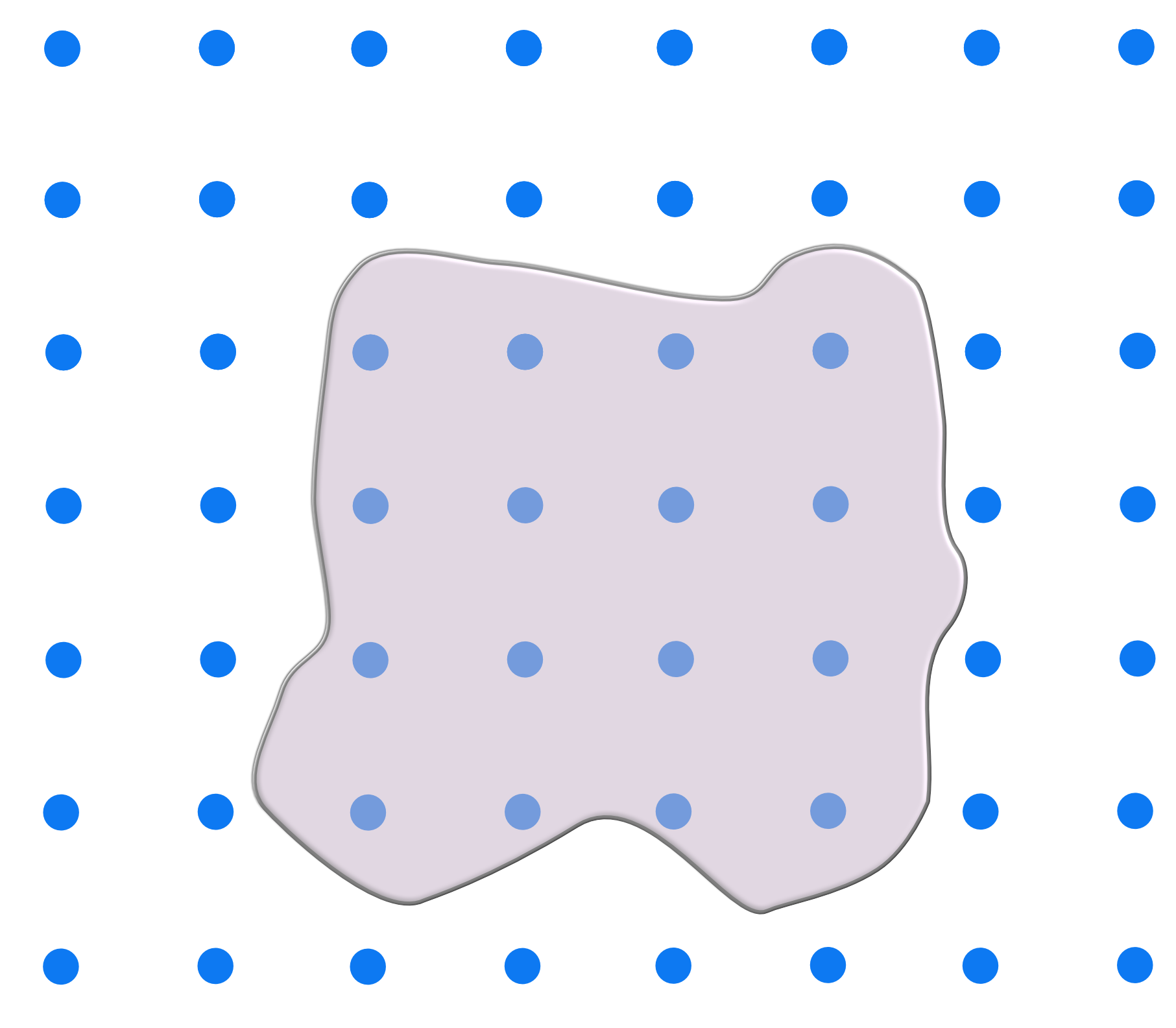}
 \begin{picture}(0,0)(0,0)
 \put(-110,90){\makebox(0,0){$\regA$}}
 \put(-165,170){\makebox(0,0){$\regAc$}}
 \put(-50,146){\makebox(0,0){$\entsurf$}}
 % \put(-115,10){\makebox(0,0){$\leftrightarrow$}}
 \end{picture}
 \caption{A discrete latticized quantum system with a Hilbert space ${\cal H}_\alpha$ at every site. We have indicated the region $\regA$ by shading the enclosed sites while the unshaded area indicates $\regAc$. We take the lattice spacing to be $\ep$.}
\label{f:lattice}
\end{center}
\end{figure}

Let us begin our discussion of entanglement entropy in QFTs by first considering a discrete problem. Imagine that we are given a lattice model, with degrees of freedom localized on the lattice sites, cf., Fig.~\ref{f:lattice}. The lattice spacing will be taken to be $\ep$. For the present, we will assume that  at each site we have a finite-dimensional Hilbert space ${\cal H}_{\alpha}$ with $\alpha$ indexing the sites.  For instance, we can consider a single qubit per site, so ${\cal H}_\alpha \cong \Hq$ for each value of
$\alpha$. A pure quantum state of the system then is an element of the tensor product Hilbert space:
\begin{equation}
\ket{\Psi}  \in \otimes_\alpha\, {\cal H}_\alpha\,.
\end{equation}

We want to understand how a subset of the lattice degrees of freedom are entangled with the rest in a given state of the above kind. Since we have the spatial information of the lattice, we can do the following: we demarcate the lattice sites into two sets by drawing a fiducial boundary across the lattice. We will label the region within the boundary as $\regA$ and the region outside as $\regAc$ and call the artificial boundary, the {\em entangling surface}, $\entsurf$. We have ensured by this spatial decomposition a particular {\em bipartitioning} of the lattice Hilbert space:
\begin{align}
\otimes_\alpha{\cal H}_\alpha \cong {\cal H}_\regA \otimes {\cal H}_{\regAc}
\end{align}

Now, given a bipartitioning of a Hilbert space into two separate tensor factors, we can construct an operator that acts on one of the factors, say ${\cal H}_\regA$, by tracing out the other (in this case ${\cal H}_\regAc$).  This operator is the {\em  reduced density matrix}  $\rhoA$ and our definition can be formalized as
\begin{equation}
\rhoA = \Tr{\regAc}{ \ket{\Psi} \,\bra{\Psi}}
\label{eq:rhoAdef}
\end{equation}
The definition here is in accord with the intuition  of capturing the state of the degrees of freedom in $\regA$  assuming complete ignorance of what happens in $\regAc$. If the wavefunction for the state $\ket{\Psi}$ in question is factorized, then clearly one would have a pure state in ${\cal H}_\regA$. The presence of quantum entanglement however leaves open the possibility that the price we pay for our ignorance is that we end up with a density matrix, i.e., a list of probabilities for the occurrence of various states in ${\cal H}_\regA$.

As presaged, we are interested in quantifying the amount of entanglement that exists in $\ket{\Psi}$ partitioned  as dictated by the spatial decomposition described above. This can be gleaned from the von Neumann entropy of the reduced density matrix, which is often referred to as the \emph{entanglement entropy}. To wit,
\begin{equation}
S_\regA = -\Tr{\regA}{\; \rhoA \,\log \rhoA}
\label{eq:Sadef}
\end{equation}
The definition calls for taking the logarithm of the operator. In a finite system, one can imagine explicitly diagonalizing the operator $\rhoA$ and obtaining its eigenvalues $\lambda_i$. These are sometimes referred to as comprising the entanglement spectrum. In terms of these, we have simply
\begin{equation}
S_\regA = - \sum_i \, \lambda_i \,\log \lambda_i
\label{}
\end{equation}

It is also convenient to define another set of `entropies' called the R\'enyi entropies
\cite{Renyi:1960fk}, which are simply defined in terms of the moments of the reduced density matrix:
\begin{align}
S^{(q)}_\regA = \frac{1}{1-q} \, \log \Tr{\regA}{\rhoA^q} = \frac{1}{1-q}\, \log \left(\sum_i \, \lambda_i^q\right)
\label{eq:Rendef}
\end{align}
The canonical definition here requires that $q\in {\mathbb Z}_+$, but we will see that oftentimes it is efficacious to analytically continue the definition to $q \in {\mathbb R}_+$.
The rationale for defining the R\'enyi entropies will become apparent soon when we discuss the replica trick for computing entanglement entropy. The key point to note is the fact that if
we consider $q \to 1$, then the R\'enyi entropies converge to $S_\regA$, i.e.,
\begin{align}
S_\regA  = \lim_{q \to 1} S_\regA^{(q)} \,.
\label{eq:ree}
\end{align}

As one might appreciate from their definition, the R\'enyi entropies capture the moments of the reduced density matrix, and turn out to be very useful for probing the
\emph{purity} of the system. Recall that a pure state density matrix is nothing but a projection operator
$\rho_\psi = \ket{\psi}\bra{\psi}$. If  it is appropriately normalized $\tr(\rho_\psi) =1$, then one notes simply that $\tr(\rho_\psi^2)=1$ again. However, if $\rho$ is a mixed state, then we expect that $\tr(\rho^2) <1$, and thus the R\'enyi entropies provide a good measure of quantum purity (despite being non-linear).

Since \eqref{eq:ree} requires us to have already explored the behaviour of R\'enyi entropies away from positive integral values of the R\'enyi index $q$, it is convenient to introduce another quantity, which we will call the \emph{modular entropy}
\begin{align}
\tilde{S}^{(q)}_\regA = \frac{1}{q^2} \, \partial_q \left(\frac{q-1}{q}\, S^{(q)}_\regA\right) \,.
\label{eq:modent}
\end{align}
This object was introduced in the holographic context in \cite{Dong:2016fnf} motivated by the intuition that it is closer to an entropy than the R\'enyi entropy $S^{(q)}_\regA$.

An analogy with classical thermodynamics is useful in understanding the modular entropy $\tilde{S}^{(q)}_\regA $. First note that the R\'enyi entropies have a close analogy to the thermodynamic free energies at a temperature $\frac{1}{q}$. This is best seen by  defining the \emph{modular Hamiltonian}
\begin{equation}
\modA = -\log \rhoA \,.
\label{eq:modHam}
\end{equation}
Whilst formal in its definition, owing to the non-linearity of the map from the reduced density matrix to the modular Hamiltonian, we will have much use for this concept. Now clearly, modulo the normalizing prefactor o
of $1-q$, we can view the R\'enyi entropy as the modular free energy, since
\begin{equation}
S^{(q)}_\regA  = \frac{1}{1-q}\, \log \Tr{\regA}{e^{-q\,\modA}}\,,
\label{}
\end{equation}
involves modular evolution by an amount $q$, the inverse temperature. The modular entropy can then  easily be seen to be the derivative of the logarithm of the generating function with respect to the inverse temperature.
\begin{equation}
\tilde{S}^{(q)}_\regA =-\frac{1}{q^2}\,\partial_q \left(\frac{1}{q} \log \Tr{\regA}{e^{-q\,\modA}} \right) \
\end{equation}
The thermodynamic versions of these statements are the usual definitions:
\begin{equation}
\begin{split}
F &= - T\, \log \mathscr{Z} = -\frac{1}{\beta}\, \log \Tr{}{e^{-\beta H}} \,, \\
S &= -\frac{\partial F}{\partial T}
=- \beta^2\; \partial_\beta \left(\frac{1}{\beta}\,  \log \Tr{}{e^{-\beta H}} \right)
\end{split}
\label{}
\end{equation}
Thus modulo a simple rescaling of the result with respect to the inverse temperature, the modular entropy is appositely named as opposed to the R\'enyi entropy. It will turn out that the modular entropy has a clean geometric interpretation in the gravitational dual when we describe how these objects are realized in the holographic dual.

An important fact to keep in mind is that all of the aforementioned entropies are defined in terms of traces and thus entirely  determined by the eigenvalues of $\rhoA$. They  are therefore insensitive to separate unitary transformations on $\rhoA$ or on $\rhoAc$. The only way to change the entanglement is to simultaneously act with a unitary on $\regA \cup \regAc$.

As before, let us assume that the total density matrix is given by a pure state $\ket{\Psi} \bra{\Psi}$. The Schmidt decomposition $\ket{\Psi} =\sum_{i}\lambda_i |\ap_i\lb_{\regA}|\beta_i\lb_{\regAc}$ tells us that non-trivial eigenvalues of $\rhoA$ are the same as those of $\rhoAc$. Therefore we find
$\Tr{\regA}{\rhoA^q}=\Tr{\regAc}{\rhoAc^q}$ and thus the equalities of entanglement entropies
\begin{equation}
S^{(q)}_{\regA}=S^{(q)}_{\regAc},
\end{equation}
for any $q$. This clearly shows that the entanglement entropy does not show an extensive property as opposed to thermodynamical entropy (see also \S\ref{sec:heeprops}), despite the close similarities in the definition.

%~~~~~~~~~~~~~~~~~~~~~~~~~~~~~~~~~~~~~~~~~~~~~~~
\section{Continuum QFTs}
\label{sec:eeqft}
%~~~~~~~~~~~~~~~~~~~~~~~~~~~~~~~~~~~~~~~~~~~~~~

Having understood how to spatially bipartition a discrete lattice system, we now can proceed to take the continuum limit by sending $\ep \to 0$.  Furthermore, while the discussion above assumed that the Hilbert space at each site was discrete, it is clear that there is no obstruction to generalizing the analysis to non-compact Hilbert spaces at each site. We henceforth will assume that this has been done.

Thence passing into the realm of local quantum field theories, we find that given a wavefunctional $\Psi[\Phi(x)]$ for the instantaneous state of the system, we can mimic the previous construction to define $\rhoA$ and its associated entanglement measures.
Here, $\Phi(x)$ is a collective label for the collection of fields that characterize the system and
$x$ is a set of  spatial coordinates that describe the spatial location on a time-slice.

Intuitively it is clear that the construction involves ignoring the part of the wavefunctional that corresponds to the the spatial region $\regAc$. The process of tracing over the complementary region $\regAc$ amounts to integrating over all field configurations in that domain, i.e., for $x\in \regAc$ to obtain $\rhoA$. Once we have the reduced density matrix, we have to write down
the operator $\log \rhoA$ and attempt to compute $S_\regA$. This is clearly the trickiest proposition, since taking the logarithm of a continuum operator involves a host of technical complications. The primary strategy we will adopt is to learn how to  obtain the entanglement entropy through suitable analytic continuation of the R\'enyi entropies.

% Figure
\begin{figure}[htbp]
\begin{center}
\includegraphics[width=3.5in]{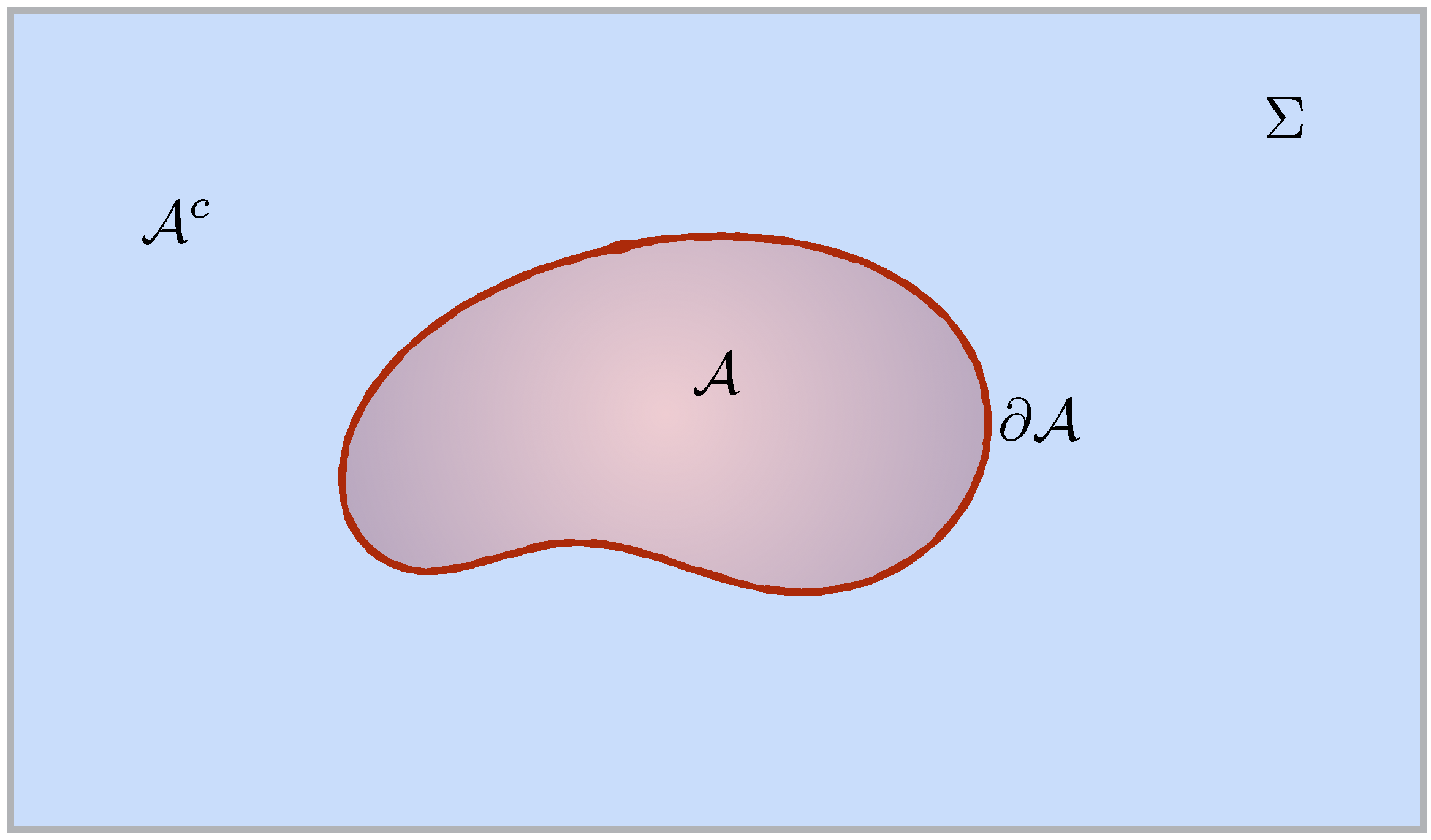}
 \caption{A continuum QFT which has been spatially bipartitioned into two components on a Cauchy slice $\Sigma$. We have indicated the region $\regA$ and its complement $\regAc = \Sigma\backslash \regA$. The separatrix is a spacetime codimension-2 surface, called the entangling surface.}
\label{f:lattice}
\end{center}
\end{figure}

Let us first try to set up the basic ingredients for constructing the matrix elements of the reduced density matrix. We  assume  that we have been handed a $d$-dimensional relativistic QFT on some Lorentzian spacetime $\bdy$, which  we take to be  globally hyperbolic.\footnote{ As we will be primarily interested in relativistic systems, we will use $d$ to indicate the total spacetime dimension of the field theory. When necessary to  explicitly distinguish the spatial dimension, we will resort to the notation $d = d_s+1$,  with $d_s$ denoting the number of spatial dimensions.} For the most part, we can just consider the case where the background is flat Minkowski spacetime, $ \bdy_d = {\mathbb R}^{1,d-1}$, but the abstraction allows for a general discussion. Since $\bdy_d$ is globally hyperbolic, we pick a Cauchy slice $\Sigma_{d-1}$ which is a (achronal) spacelike slice, defining in the QFT, a moment of simultaneity. On $\Sigma_{d-1}$ we then have a state of the system. This could be a pure state given by a wavefunctional $\Psi[\Phi(x)]$, or more generally a density matrix $\rho_\Sigma$ ($x$ now defines a  coordinate chart on $\Sigma_{d-1}$). We will give an explicit path integral construction of  such a state on $\Sigma_{d-1}$ in \S\ref{sec:pi}.

The rest of the construction is now straightforward. We pick some  spacetime codimension-1 region $\regA$ on the Cauchy slice, which allows for a spatial bipartitioning of the form $\regA \cup \regAc = \Sigma_{d-1}$. The boundary of the region is, as before, the entangling surface $\entsurf$, which we note is a codimension-2 hypersurface in $\bdy$. Since we are working in the continuum, we
anticipate UV singularities. These can be dealt with by introducing an explicit UV  regulator $\epsilon$. Geometrically we can view this in terms of working in a tubular neighbourhood of $\entsurf$ of width $\epsilon$, which will serve the purpose of regulating the short distance entanglement between the degrees of freedom inside and outside of the entangling surface.

Our previous discussion would suggest that we now go ahead and decompose the Hilbert space ${\cal H}$ of the QFT into ${\cal H}_{\regA} \otimes {\cal H}_{\regAc}$. For theories with no gauge symmetries, this is indeed sufficient. However, when we have gauge fields we have to face up to the problem of defining a separation of the Hilbert space into the tensor factors in a gauge-invariant way. Unfortunately no such decomposition exists -- this can be seen by considering the lattice description of gauge fields. The basic operators are
the link variables, as opposed to the site variables we have considered so far.
When we cut the links as in Fig.~\ref{f:lattice}, we have to decide where the broken link belongs, to ${\cal H}_\regA$ or to ${\cal H}_{\regAc}$, leading to an ambiguity. This has been much discussed in recent literature, cf., \cite{Buividovich:2008gq,Donnelly:2011hn,Casini:2013rba,Donnelly:2014gva}. Heuristically, one can imagine cutting the link variables along the entangling surface, but making a particular choice whilst doing so as to whether the said link degree of freedom belongs to ${\cal H}_\regA$ or to ${\cal H}_{\regAc}$. We will often assume that a particular such choice has been made on the lattice, leading to a specific prescription for the continuum path integral we are about to describe.

The reduced density matrix $\rho_{\regA} := \Tr{{\cal H}_{\regAc}}{\rho_\Sigma}$ captures the entanglement between $\regA$ and $\regAc$  as explained earlier. Once we have this operator, we can then give a quantitative measure of the entanglement by computing the von Neumann entropy as defined in \eqref{eq:Sadef}.

In local relativistic QFTs, there is an important physical statement about causality which will be  useful for our discussions. Since $\Sigma$ is a Cauchy slice, the future (past) evolution of initial data on it allows us to reconstruct the state of the QFT on the entirety of $\bdy$. In other words, the past and future domains of dependence of $\Sigma$ , $D^\pm[\Sigma]$, together make up the background spacetime on which the QFT lives, i.e.,
$D^+[\Sigma] \cup D^-[\Sigma] = \bdy$.  Likewise,  the domain of dependence of $\regA$,  $\domdA = D^+[\regA] \cup D^-[\regA]$,  is the region where the reduced density matrix $\rhoA$ can be uniquely evolved once we know the Hamiltonian acting on the reduced system in $\regA$.\footnote{ The domains of dependence are causal sets which are determined simply where a given set of points can communicate to or be communicated from, etc. For instance, $D[\regA]$ is defined as the set of points in $\bdy$ through which every inextensible causal curve intersects $\regA$.
A technical complication to keep in mind is that we take $\regA$ to be an open subset of $\Sigma$; consequently, $D[\regA]$ is an open subset of $\bdy$.
We refer the reader to \cite{Wald:1984ai} for a discussion of these concepts.
} So given a state or a density matrix in some spatial domain, be it $\Sigma_{d-1}$ or
$\regA$, there is a unitary operator which allows us to evolve this state within the corresponding domain of dependence.

\begin{figure}
\begin{center}
\includegraphics[width=5in]{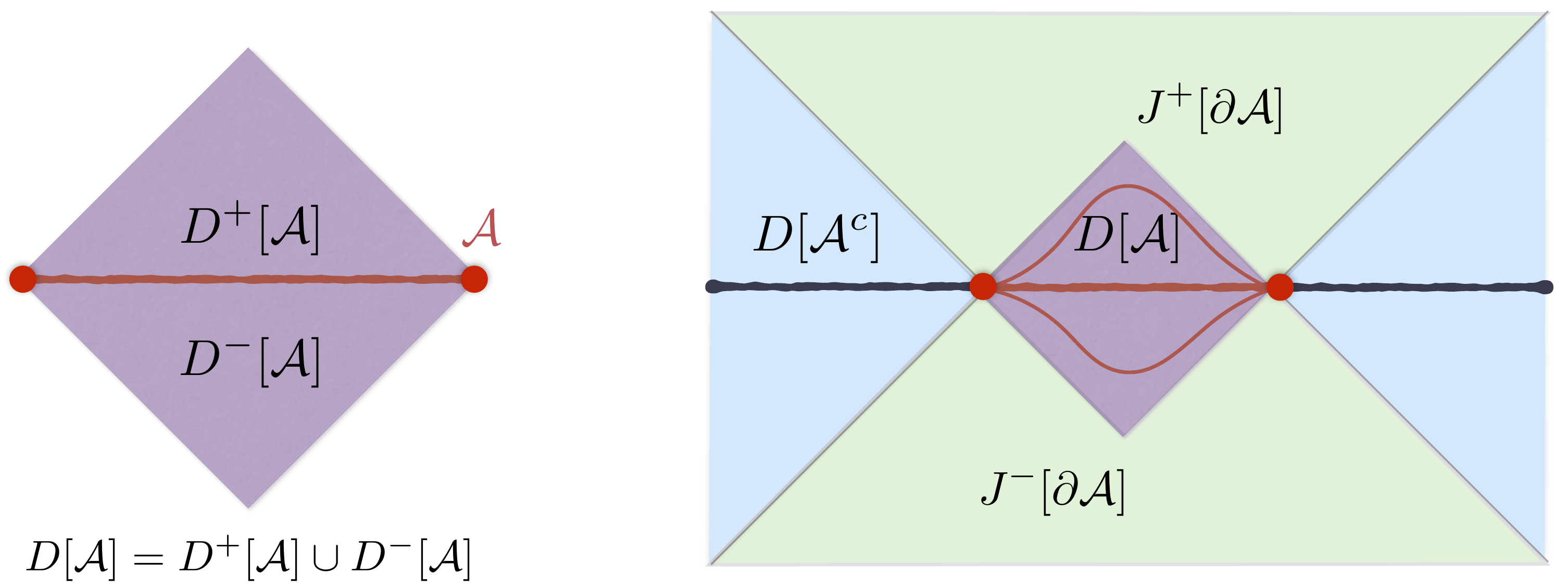}
\setlength{\unitlength}{0.1\columnwidth}
\caption{
An illustration of the causal domains associated with a region $\regA$, making manifest the decomposition of the spacetime into the four distinct domains indicated in \eqref{eq:bdy4d}.
Two deformations $\regA'$ are also included for illustration in the right panel.
}
\label{f:bdy4d}
\end{center}
\end{figure}

Now the domains of dependence of $\regA$ and $\regAc$ by themselves do not make up the full spacetime, $\domdA \cup \domdAc\neq \bdy$. We have to account for the regions which can be influenced by the entangling surface $\entsurf$. Denoting the causal  future (past) of a point $p\in\bdy$ by $J^\pm(p)$,  we find that we have to keep track of the regions $J^\pm[\entsurf]$, which are not contained in either $\domdA$ or $\domdAc$. As a result, the  full spacetime $\bdy$ decomposes into four causally-defined regions: the domains of dependence of the region and its complement, and the causal future and past of the entangling surface:
\begin{equation}
\bdy = \domdA \cup \domdAc \cup J^+[\entsurf] \cup J^-[\entsurf]\,.
\label{eq:bdy4d}
\end{equation}
This is illustrated in  Fig.~\ref{f:bdy4d}; the point to remember always is that codimension-2 spacelike surfaces like $\entsurf$ have a two-dimensional normal bundle with Lorentzian metric signature. We can therefore always visualize them as a point in a two-dimensional space, and then the concepts one is familiar with in spacetime diagrams drawn in two-dimensional Minkowski spacetime. Further details and some more formal statements can be found in \cite{Headrick:2014cta}.

% Figure

The decomposition \eqref{eq:bdy4d} is particularly convenient for formulating constraints on entanglement entropy that follow from relativistic causality. If we unitarily evolve the  reduced density matrix $\rhoA$, by transformations which are supported solely on ${\cal H}_{\regA}$ or on ${\cal H}_{\regAc}$, the eigenvalues of $\rhoA$ remain unaffected. Thus the R\'enyi and von Neumann entropies are invariant under such unitary transformations.  These could include perturbations of the Hamiltonian and local unitary transformations supported in the domains $D[\regA]$ or $D[\regAc]$.
Now consider a deformation of the spatial region $\regA$, onto another region $\regA'$ lying on a different Cauchy slice $\Sigma'$, such that $D[\regA]=D[\regAc]$ (as indicated in Fig.~\ref{f:bdy4d}). The state  $\rho_{\Sigma'}$ on the new slice is related by a unitary transformation to the state $\rho_\Sigma$. It is clear that such a transformation can be constructed from operators localized in $\regA$, and so does not change the entanglement spectrum of $\rhoA$. We can of course make similar arguments for the complementary region $\regAc$.

Furthermore, if we fix the state at $t\to-\infty$, and consider perturbation to the Hamiltonian supported in some region  $\mathfrak{r}_{\delta H}$, then by virtue of causality, we can only affect the state in the causal future of this region. In other words, in any region of the spacetime which does not intersect $J^+[\mathfrak{r}_{\delta H}]$, our change has no effect whatsoever.  More pertinently,  perturbations of this form can only affect the entanglement spectrum when $J^+[\mathfrak{r}_{\delta H}]$ intersects $J^-[\entsurf]$. These perturbations will influence both $\regA$ and $\regAc$ and thus can be used to modify the entanglement. In any other scenario, we can deform the region
to pass to the past of perturbation in $\mathfrak{r}_{\delta H}$, thus leaving $S_\regA$ unaffected. By reversing the time ordering, if we fix the state at $t\to+\infty$, the spectrum can be affected only by perturbations in $J^+[\entsurf]$. In summary, we have the following properties of $\rhoA$:
\begin{itemize}
\item
The entanglement spectrum of $\rhoA$ depends only on the domain $D[\regA]$ and not on the particular choice of Cauchy slice $\Sigma$. The spectrum is thus a so-called ``wedge observable'' despite the fact that it is not, of course, an observable in the usual sense.\footnote{ To belabor the obvious, $\rhoA \log \rhoA$ is not a linear operator on the Hilbert space.}
\item
Fixing the state in either the far past or the far future, the entanglement spectrum of $\rhoA$ is insensitive to any local deformations of the Hamiltonian in $\domdA$ or $\domdAc$.
\end{itemize}
These are the crucial causality requirements that entanglement (and R\'enyi) entropies are required to satisfy in any relativistic QFT.

%~~~~~~~~~~~~~~~~~~~~~~~~~~~~~~~~~~~~~~~~~~~~~~~
\section{Path integrals \& replica}
\label{sec:pi}
%~~~~~~~~~~~~~~~~~~~~~~~~~~~~~~~~~~~~~~~~~~~~~~

We have now given a formal definition of the reduced density matrix in continuum QFTs. For computational purposes, however, it is most useful to eschew the operator description in terms of wavefunctionals and pass directly to a functional integral perspective. We will do so in a couple of steps: We will first construct a path integral that computes  the matrix elements of $\rhoA$, taking care to ensure that we respect the causality requirements  described above. We will then see how to compute the R\'enyi entropies by considering a functional integral on a ``branched cover'' geometry, and therefrom pass to the entanglement entropy itself by invoking an analytic continuation.

\paragraph{Functional integral for reduced density matrix elements:} Usually path integrals in QFTs involve integrating over field configurations with various operator insertions for computing observables. One is most familiar with such constructs for Euclidean QFTs in which the observables we compute are Wightman functions. In the Lorentzian context, one has to make a choice of temporal ordering which leads to a multiplicity of correlation functions, and the path integrals should be engineered to reflect this freedom. The Euclidean framework turns out to be most appropriate when we consider static states for which the time evolution is trivial. More generally, one may also invoke the Euclidean construct when the observable is computed at a moment of time reflection symmetry, i.e., when we have instantaneous staticity.

% Figure
\begin{figure}[htbp]
\begin{center}
\includegraphics[width=3in]{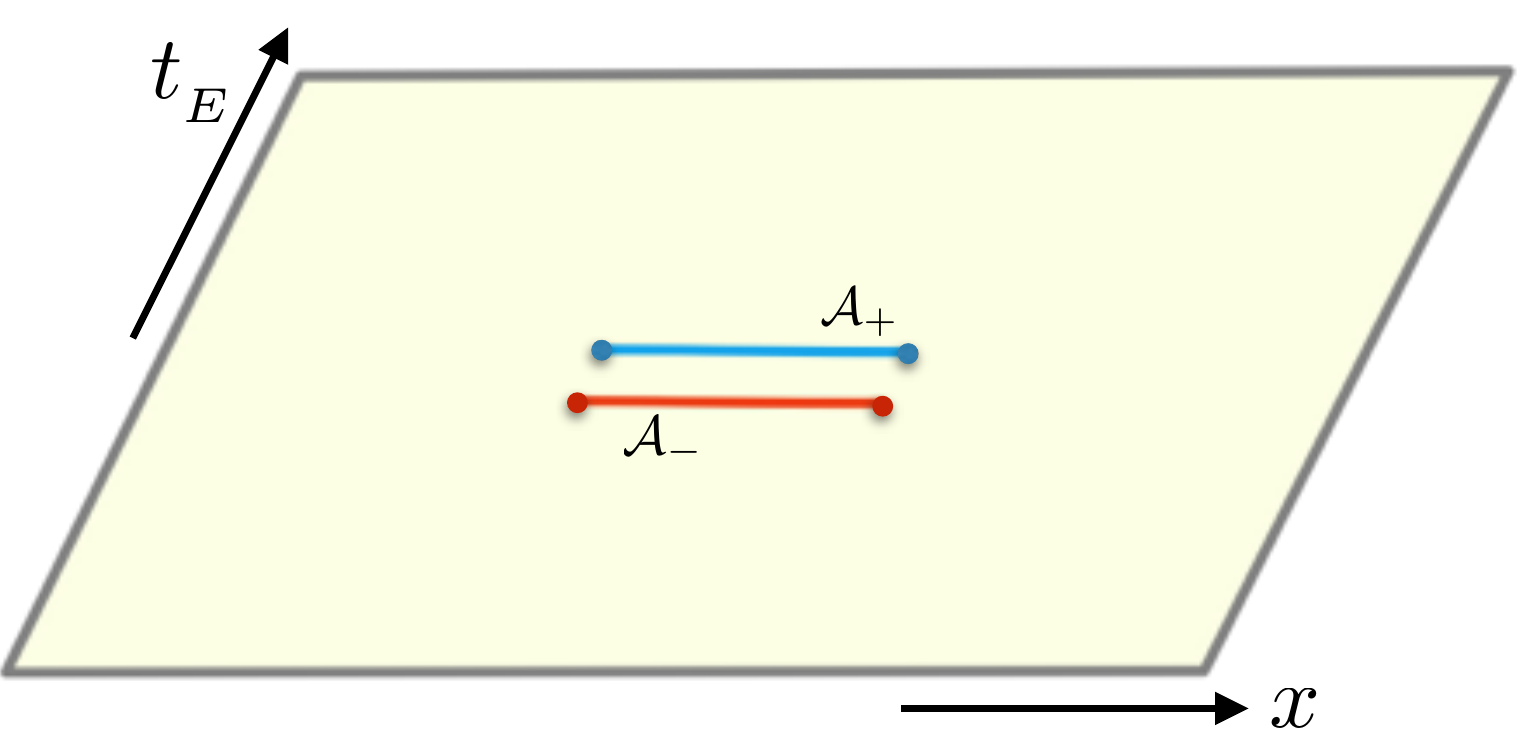}
\caption{ The Euclidean geometry for computing the matrix elements of the reduced density matrix $\rhoA$. We have sketched the situation in two-dimensional Euclidean space as indicated. The two cuts at $\regA$ have been separated in an exaggerated manner to indicate the boundary conditions we need to impose, cf., \eqref{eq:rhoAE}.}
\label{f:rhopmE}
\end{center}
\end{figure}

Say we wish to define $\rhoA$ on a Cauchy slice $\Sigma_{t=0}$ when there is no non-trivial time evolution. Since we are singling out a region $\regA$, we should demarcate fields into two sets $\Phi(x) =\{ \Phi_\regA(x), \Phi_{\regAc}(x)\}$ by restricting their domains of support. The reduced density matrix  acts as an operator on ${\cal H}_\regA$. Its matrix elements may be defined by their action on fields supported in
$\regA$. To see this, let us imagine regulating the path integral by imposing boundary conditions for
fields in $\regA$ as follows:
\begin{equation}
\Phi_{\regA} \big|_{t=0^-}= \Phi_- \,, \qquad  \Phi_{\regA} \big|_{t= 0^+} = \Phi_+ \,.
\label{eq:bcA}
\end{equation}
This is equivalent to cutting open the path integral in a restricted domain of space at $t=0^\pm$ and projecting the result onto definite field values. This is easily achieved by introducing a delta functional into the path integral. Thus,
\begin{align}
(\rhoA)_{-+}
&=
  \int\, [{\cal D} \Phi] \, e^{-S_\text{QFT}[\Phi]} \, \delta_\text{E}(\Phi_{\mp \regA})
\nonumber \\
\delta_\text{E}(\Phi_{\mp \regA})
& \equiv
   \delta\left(\Phi_{\regA} (t=0^-) -  \Phi_- \right)\, \delta\left(\Phi_{\regA} (t=0^+)- \Phi_+ \right) \,,
\label{eq:rhoAE}
\end{align}
where we  introduce, for convenience, a shorthand for the delta-function which is  inserted into the path integral to extract the elements of the density matrix -- it may equivalently be thought of as a functional representation of a projector. See Fig.~\ref{f:rhopmE} for an illustration of the geometry in two-dimensional theories for a time-independent state.

% Figure
\begin{figure}[htbp]
\begin{center}
\includegraphics[width=6in]{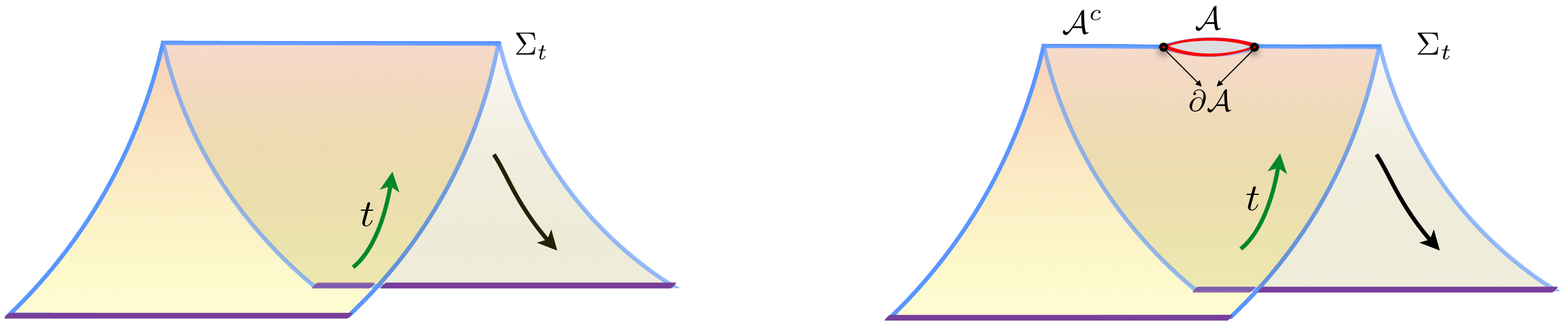}
 \begin{picture}(0,0)(0,0)
 \end{picture}
\caption{ The Schwinger-Keldysh geometry for computing the matrix elements of the reduced density matrix $\rhoA$ in time-dependent settings. On the left panel, we show the general contour which involves a time-fold at the Cauchy surface of interest. On the right panel, we illustrate the opening out of the path integral at $\regA$ to allow for the appropriate past/future boundary conditions  \eqref{eq:rhoAL}.}
\label{f:rhopmL}
\end{center}
\end{figure}

For states with non-trivial time dependence, the above needs modification. One cannot use the information about the entire spacetime without violating the causality requirement. If we want to know the elements of $\rhoA(t)$, we should not be making use of the wavefunctionals at later times $t' >t$. The canonical way to deal with this situation is to use the Schwinger-Keldysh framework \cite{Schwinger:1960qe,Keldysh:1964ud}.\footnote{ This is also known as the closed time-path formalism or the in-in formalism. A closely related discussion for open quantum systems appears in \cite{Feynman:1963fq}.} The essential idea is to consider evolving initial conditions from the time the state of the quantum system is prepared up until the instant we wish to compute the reduced density matrix. So one considers the causal past of the Cauchy slice $J^-[\Sigma_{t=0}]$, but instead of evolving forward from there on, one retraces the evolution back to the initial state. Intuitively, this forward-backward evolution serves to cancel out unknown information of the final state of the evolution from the computation.  See Fig.~\ref{f:rhopmL} for an illustration.

The Schwinger-Keldysh contour provides a path integral prescription for computing any real time process, and as such the path integral with two copies of $J^-[\Sigma_{t=0}]$ glued together on the Cauchy surface of interest constructs for us the instantaneous state of the system (by projecting as usual onto definite field configurations). One may either view this as a single copy of the system living on a complex contour, or more simply by viewing the forward and backward evolution as two different copies of the same system. This doubling of degrees of freedom is a central feature of Schwinger-Keldysh path integrals. The reader may find the classic references \cite{Chou:1984es,Landsman:1986uw} useful (see also \cite{Haehl:2015foa,Haehl:2016pec} for a novel perspective).

 To obtain the reduced density matrix elements, we cut open the functional integral around $\regA$ and impose boundary conditions just above and below as in \eqref{eq:bcA}. One can then write the time-dependent reduced density matrix elements as
\begin{align}
(\rhoA)_{-+} &= \int_{J^-[\Sigma_t]} \, [{\cal D} \Phi_\skR] [{\cal D} \Phi_\skL] \, \  e^{i\, S_\text{QFT}[\Phi_\skR] - i\,S_\text{QFT}[\Phi_\skL]}  \ \delta_\text{L}\left(\Phi_{\skR\skL;\; \regA}^{-+} \right)
\nonumber \\
\delta_\text{L}\left(\Phi_{\skR\skL;\;  \regA}^{-+} \right) & \equiv
 \delta\left(\Phi_{\skR,\regA} (t=0^-) -  \Phi_- \right)\, \delta\left(\Phi_{\skL,\regA} (t=0^+)- \Phi_+ \right) \,.
\label{eq:rhoAL}
\end{align}
We have adopted the view that we have two copies of  our system with fields labeled $L$ and $R$, respectively. The right fields are evolved forward in time from the initial state to the Cauchy slice, while the left fields are evolved backwards from the Cauchy slice back to the initial state. This is the origin of the relative sign in the action (in which we have also restored the factors of $i$ relevant for a real-time computation).

\paragraph{Functional integral for powers of the reduced density matrix:} Once we have constructed a functional integral to compute the matrix elements of $\rhoA$, it is straightforward to multiply these so  as to obtain the matrix elements of its powers
$(\rhoA)^q$. The method used to effect this computation is called the {\em replica method}, since we replicate the computation described above a number of times.

% Figure
\begin{figure}[t]
\begin{center}
\includegraphics[width=3in]{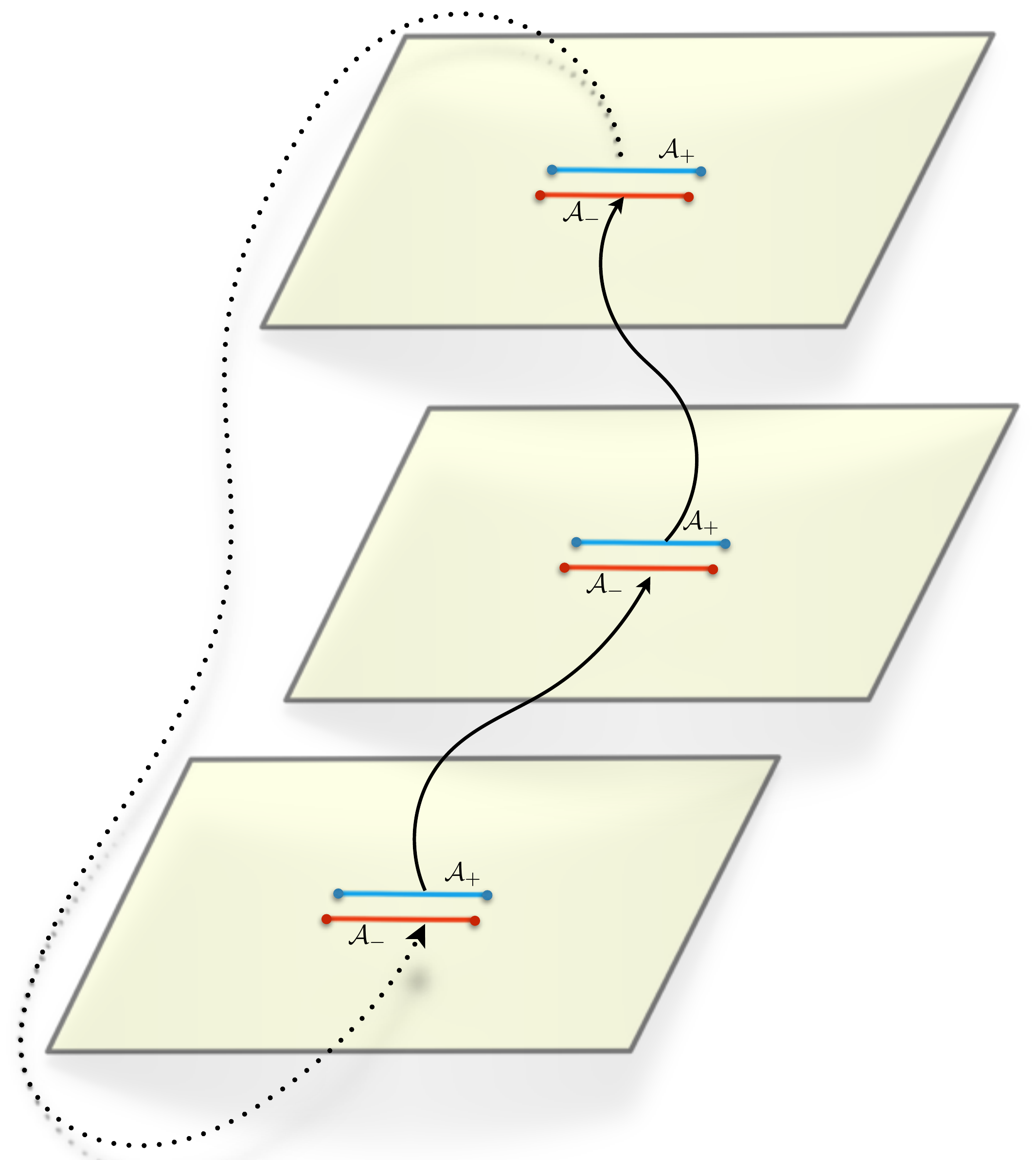}
\caption{ The Euclidean geometry for computing the matrix elements of powers of the  reduced density matrix $\rhoA$ and trace thereof, pictorially depicting \eqref{eq:replicaE}. We have illustrated the situation in which we glue three copies of the replicated path integrals to construct $\rhoA^3$ matrix elements with the identifications between boundary conditions on the replica copies indicated by the arrows. The final trace to compute the third R\'enyi entropy is indicated by the dotted line.}
\label{f:rho3E}
\end{center}
\end{figure}

The computation of matrix elements of $(\rhoA)^q$ is achieved by taking $q$-copies of the functional integral computing $\rhoA$ and making some identifications. Matrix multiplication requires that we integrate over the boundary conditions for the $+$ component in the $k^{\rm th}$ density matrix with the $-$ component of the $(k+1)^{\rm st}$ density matrix, i.e., $\Phi_+^{(k)} = \Phi_-^{(k+1)}$.
For the static situation, we may thus write
\begin{align}
(\rhoA)^q_{-+} &= \int\, \prod_{j=1}^{q-1} \; d\Phi_+^{(j)}\   \delta(\Phi^{(j)}_+ - \Phi^{(j+1)}_-)
\nonumber \\
& \hspace{2cm}
  \times \bigg[ \int \prod_{k=1}^q \; [{\cal D} \Phi^{(k)}] \;  \bigg\{ e^{-\sum_{k=1}^q \;
  S_\text{QFT}[\Phi^{(k)}]} \,  \delta_\text{E}(\Phi^{(k)}_{\mp \regA})  \bigg\}
\bigg]
\label{eq:replicaE}
\end{align}
The outer integral over the boundary conditions with the delta functions serves to perform the desired identifications of the reduced density matrix elements. The inner functional integral simply replicates the path integral computing the individual matrix elements. We illustrate this path integral contour in Fig.~\ref{f:rho3E}.

% Figure
\begin{figure}[htbp]
\begin{center}
\includegraphics[width=5in]{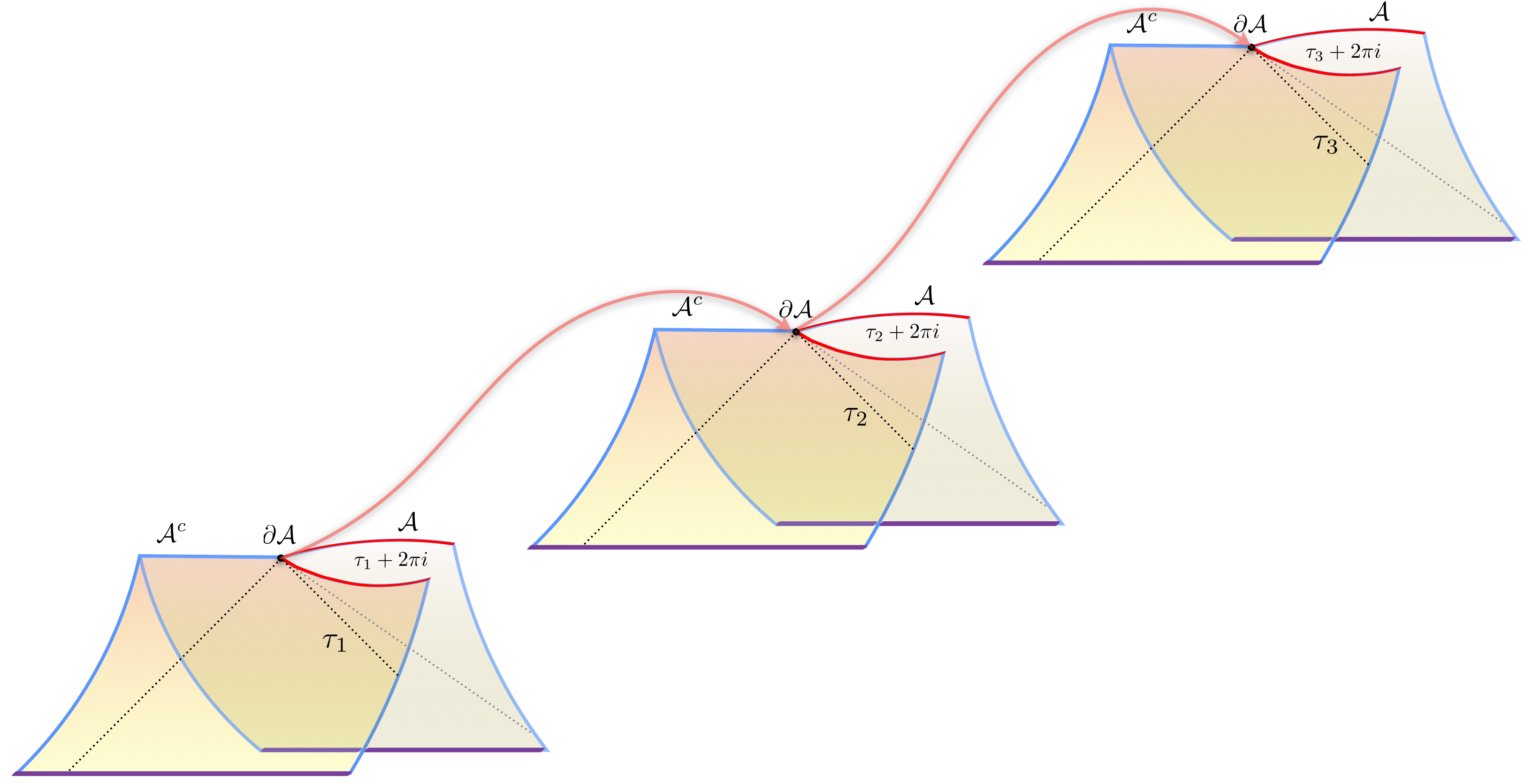}
\caption{ The generalized Schwinger-Keldysh geometry for computing the matrix elements of powers of the  reduced density matrix $\rhoA$ and trace thereof, pictorially depicting \eqref{eq:replicaL}. We have illustrated the situation in which we glue three copies of the replicated path integrals to construct $(\rhoA)^3$ matrix elements with the identifications between boundary conditions on the replica copies indicated by the arrows.}
\label{f:rho3L}
\end{center}
\end{figure}
Similarly for the time-dependent states, we would identify the boundary conditions at $t=0^+$ of the $k^{\rm th}$ density matrix with those of the $t=0-$ pertaining to the $(k+1)^{\rm st}$ density matrix. We should now allow for the fact that the $\pm$ components correspond to the left and right Schwinger-Keldysh fields, respectively. So instead of
\eqref{eq:replicaE}, we end up with a more complicated expression, cf., Fig.~\ref{f:rho3L}.
\begin{align}
(\rhoA)^q_{-+} &= \int\, \prod_{j=1}^{q-1} d\Phi_{\skL+}^{(j)} \   \delta(\Phi^{(j)}_{\skL+} - \Phi^{(j+1)}_{\skR-})
\nonumber \\
& \hspace{2cm}
  \times \bigg[ \int \prod_{k=1}^q \; [{\cal D} \Phi^{(k)}] \;  \bigg\{ e^{-\sum_{k=1}^q \;
  S_\text{QFT}[\Phi^{(k)}]} \,  \delta_\text{L}(\Phi^{-+ \ (k)}_{\skR\skL;\;  \regA})  \bigg\}
\bigg]
\label{eq:replicaL}
\end{align}

While the expressions looks quite complicated written out this way, it  is much simpler to visualize the path integral construction pictorially. We should view each copy of $\rhoA$ as being computed on a copy of the background spacetime. This is $\bdy $ for the Euclidean computation or two copies of $J^-[\Sigma_t] \subset \bdy$ joined together along the Cauchy slice for the real-time Lorentzian computation.\footnote{ When it is necessary to distinguish this construction, we will refer to the Lorentzian geometry with an explicit subscript, viz., $\bdy_\text{Lor}$.}
The geometric picture is then comprised of taking $q$ copies of these manifolds (or parts thereof) and making identifications across them as prescribed by \eqref{eq:replicaE}  and \eqref{eq:replicaL} respectively. This is illustrated for the two cases of interest in Figs.~\ref{f:rho3E} and \ref{f:rho3L}, respectively. One can equivalently think of the $q$-copies of $\bdy$ with the prescribed identifications as constructing a new manifold $\bdy_q$. Following a canonical construction in topology, we will refer to $\bdy_q$ as the $q$-fold branched cover over $\bdy$ (or $\bdy_\text{Lor}$).

In either case, we now can compute the path integral of the theory by integrating over all the fields living on the background $\bdy_q$. We define this as being
\begin{equation}
\mathcal{Z}_q[\regA] = {\rm Tr}\left( \rhoA^q\right) \equiv \mathcal{Z}[\bdy_q] \,.
\label{eq:zqdef}
\end{equation}
From here we can extract the R\'enyi entropies via
\begin{equation}
S_\regA^{(q)} = \frac{1}{1-q}\, \log\left({\rm Tr}\left( \rhoA^q\right) \right)=
\frac{1}{1-q} \,\log  \left(\frac{\mathcal{Z}_q[\regA] }{\mathcal{Z}_1[\regA]^q }\right)
\equiv \frac{1}{1-q} \, \log\left(\frac{\mathcal{Z}[\bdy_q]}{\mathcal{Z}[\bdy]^q}\right) .
\label{eq:sqZ}
\end{equation}

\paragraph{From replica to entanglement entropy:} We now have functional integrals that compute matrix elements of arbitrary integer powers of the density matrix. Taking the trace, which now simply involves identifying $\Phi^{(1)}_-$ with $\Phi^{(q)}_+$ for the Euclidean computation, we get the R\'enyi entropies defined in \eqref{eq:Rendef}.

One important element of the replica computation is that the cyclicity of the trace translates into a cyclic permutation symmetry amongst the various copies of the functional integral. The Euclidean path integral then has a cyclic ${\mathbb Z}_q$ symmetry acting on its components. The Lorentzian computation on the other hand has $2q$-copies of the background geometry, but these  are sewn together respecting the cyclic ${\mathbb Z}_q$ symmetry. We will refer to this symmetry as the \emph{replica symmetry}.

We will note here that the Euclidean computation also has a time translation symmetry, part of which is a time-reflection ${\mathbb Z}_2$ symmetry. In general, to work with the simpler Euclidean functional integral, it suffices that we have such a discrete symmetry; one does not require full time independence. More physically, we can employ \eqref{eq:replicaE} whenever the state in question is at a moment of time reflection symmetry. The reason to bring this up is that some authors choose to combine this with the cyclic symmetry and refer to the resulting dihedral group $\mathbb{D}_{2d} = \mathbb{Z}_q\ltimes \mathbb{Z}_2$ as the replica symmetry. We choose not to do so, since it is only the ${\mathbb Z}_q$ that remains relevant for a general real-time computation.

To get the entanglement entropy itself, we are going to invoke \eqref{eq:ree}, which requires us to take the $q\to1$ limit of the R\'enyi entropies thus computed.  Now, having a function only at integer values does not in general allow one to analytically continue the argument to real values.  A case in point is the function $\sin(\pi z)$ that  vanishes at integer $z$, which not only exemplifies the  situation, but also provides a resolution. Functions defined on integers which,  in a addition, are well-behaved as $z \to \pm i\, \infty$ allow for a unique analytic continuation away from the integers. This is the content of Carlson's theorem, which may be derived in turn from the Phragm\'en-Lindel\"of principle. The theorem requires that the functions do not grow rapidly at imaginary infinity; as long as the growth is sub-exponential, one is guaranteed a unique analytic continuation. By bounding the behaviour of the function in certain directions, these results assert that the function itself is bounded in the complex plane, which then allows for a unique analytic continuation.

So as long as we are able to physically argue that the R\'enyi  entropies are well-behaved, one could be assured of a well-defined entanglement entropy using the replica trick.\footnote{ There are some other pitfalls, which we will get to later. For instance, the replica symmetry may itself be broken dynamically.}
In any event, from a physical viewpoint one should attempt to take the analytic continuation seriously and see whether the results are sensible. Failure of the replica construction usually implies that there are some interesting physical phenomena to be understood. We will take the perspective that the replica trick suffices, and for the most part our discussion will focus on circumstances in which  the results thus derived make physical sense.

%~~~~~~~~~~~~~~~~~~~~~~~~~~~~~~~~~~~~~~~~~~~~~~~
\section{General properties of Entanglement entropy}
\label{sec:genee}
%~~~~~~~~~~~~~~~~~~~~~~~~~~~~~~~~~~~~~~~~~~~~~~

Having understood the basic definition of entanglement entropy in continuum quantum field theories, we can now ascertain some general features that one expects this quantity to have. We will give a brief discussion of the sensitivity of this quantity to the short distance physics first and then turn to a general set of inequalities that we expect it to uphold. We will later have occasion to contrast these with the corresponding behaviour in holographic systems.

%~~~~~~~~~~~~~~~~~~~~~~~~~~~~~~~~~~~~~~~~~~~~~~~
\subsubsection{1. UV and IR properties}
\label{sec:uvirprop}
%~~~~~~~~~~~~~~~~~~~~~~~~~~~~~~~~~~~~~~~~~~~~~~

Firstly, we should note that the quantity as defined is UV divergent and thus needs to be regulated.  This follows from the fact that any state in a local QFT has short-range correlations in the ultra-violet (UV). Furthermore, the definition of $\rhoA$ requires us to consider partitioning the system across an entangling surface, and we should thus anticipate the correlations of modes right across this to contribute to the entanglement entropy in a significant manner. One essentially needs to look at the modes within a cut-off distance from the entangling surface and ascertain their contributions which will be divergent.

Intuitively, we expect the divergence to be proportional to the number of EPR pairs that straddle the entangling surface. This would predict a sub-extensive behaviour of entanglement, for the leading contribution would be proportional to the area of the entangling surface. This intuition was one of the primary reasons for the initial focus on this quantity \cite{Bombelli:1986rw} to draw analogy with the behaviour of the Bekenstein-Hawking entropy for black holes.

In $d$-dimensional free field theories, we can indeed show that the leading divergent terms in the UV limit
$\ep\to 0$ obey the area law \cite{Srednicki:1993im,Bombelli:1986rw}:
\begin{equation}
S_\regA =\gamma\; \frac{\text{Area}(\entsurf)}{\epsilon^{d-2}}+\cdots ,
\label{areala}
\end{equation}
where we omitted less divergent terms denoted by the ellipses. As anticipated on physical grounds,  this leading divergent term is proportional to the area of the boundary of the region $\regA$ and not extensive in the size of $\regA$. The coefficient $\gamma$ depends on the specifics of field theory; one can argue that it is proportional to the number of fields, i.e., scales with the number of degrees of freedom.

One can give general arguments to show that the structure of the subleading terms depends on the intrinsic and extrinsic geometry of the entangling surface $\entsurf$. In general, one can argue that for states in the Hilbert space of a relativistic QFT, the UV behaviour takes the form:
\begin{align}
S_\regA =
\begin{cases}
a_{d-2} \left(\frac{L}{\epsilon}\right)^{d-2} + a_{d-4} \left(\frac{L}{\epsilon}\right)^{d-4} + \cdots +
a_1\, \frac{L}{\epsilon} + (-1)^{\frac{d-1}{2}} \,  {\sf S}_\regA + {\cal O}(\epsilon)
\,, \qquad d  \ \text{odd} \\
a_{d-2} \left(\frac{L}{\epsilon}\right)^{d-2} + a_{d-4} \left(\frac{L}{\epsilon}\right)^{d-4} + \cdots
+ (-1)^{\frac{d-2}{2}} \, {\sf S}_\regA \, \log\left(\frac{L}{\epsilon}\right) + {\cal O}(\epsilon^0) \,, \qquad d \ \text{even}
\end{cases}
\label{eq:Suvgend}
\end{align}
where $L$ is a proxy for the size of the region $\regA$.
The difference in the behaviour of odd and even-dimensional CFTs can be traced back to the structure of the UV divergences in the theory. While most of the coefficients $a_i$ in the above expansion are scheme-dependent and hence not individually meaningful, we should emphasize that non-trivial information is contained in the universal piece denoted by $\sf{S}_\regA$. This term captures useful information about the conformal anomalies in the theory and plays an important role in the entanglement-based results on renormalization group flow \cite{Casini:2004bw,Myers:2010tj,Jafferis:2011zi,Liu:2012eea,Casini:2012ei}. 

We should note that the above structure is inferred from holographic- and anomaly-based considerations. While it is rather intractable to do explicit computations of entanglement entropy of interacting field theories (with the exception of holographic theories), there is a rigorous derivation of area law for gapped interacting systems for $d=2$ \cite{Hastings:2007aa}.   However, holographic results strongly suggest that the area law \eqref{areala} should hold for any field theory with a UV fixed point for $d>2$.

A special case of note are two-dimensional conformal theories ($d=2$),  for which\eqref{eq:Suvgend} predicts only a  logarithmic divergence and therefore fails to follow an area law. In this exceptional case, it can be understood heuristically that the logarithm arises as a limiting case of a power law divergence and is consistent with the entangling surface that is comprised of a set of disconnected points.

To conclude this discussion, we also wish to highlight a semantic point regarding terminology. It is common in many-body systems to talk about states having area law entanglement versus those that have volume law entanglement. This statement refers to the infra-red properties, i.e., the scaling with $L$ in \eqref{eq:Suvgend} above with a fixed UV cut-off $\epsilon$. We will return to this issue in the course of our discussion, but for now make note of the fact that we would say that a state has
\begin{itemize}
\item area-law entanglement if $S_\regA \sim L^{d-2}$ at fixed $\epsilon$,
\item volume-law entanglement if $S_\regA \sim L^{d-1}$ at fixed $\epsilon$.
\end{itemize}
Typically, vacuum or ground states of a system exemplify the former behaviour, while highly excited or thermal states exemplify the latter. Note that in the latter case we would encounter an IR scale, e.g., in thermal states the $S_\regA = (LT)^{d-1}$ with UV cut-off held fixed. One can also have states with intermediate  behaviour (e.g., logarithmic scaling),. For instance, fermionic systems with Fermi surfaces $S_\regA \sim L^{d-2} \log (k_F L) $  are usually referred to as a logarithmic violation of the area law owing to the presence of a new IR scale, the Fermi momentum $k_F$. We refer the reader to \cite{Swingle:2011np} for a heuristic discussion of potential IR behavior in diverse physical systems.

%~~~~~~~~~~~~~~~~~~~~~~~~~~~~~~~~~~~~~~~~~~~~~~~
\subsubsection{2. Entropy inequalities}
\label{sec:eeineq}
%~~~~~~~~~~~~~~~~~~~~~~~~~~~~~~~~~~~~~~~~~~~~~

By virtue of its definition as the von Neumann entropy of a reduced density matrix, the entanglement entropy satisfies a set of very general inequalities. We will refer to these as the quantum inequalities. We give a brief discussion of these in the following, and refer the reader to the original references \cite{Araki:1970ba,Lieb:1973lr,Lieb:1973cp}  and the recent mathematical review \cite{Carlen:2010aa} for  details.

To describe these inequalities, we will need to consider a state which allows partitions into multiple sets. In quantum mechanics or lattice systems, we can imagine these sets to be components of a larger Hilbert space obtained by taking tensor products of simpler systems. In continuum QFT, we will imagine that each of the components refers to a particular subregion of the Cauchy slice.\footnote{ Once again modulo the fact that we need to supply suitable caveats to discuss theories with gauge invariance in which  spatial regions do not necessarily allow for such a factorization.} We will take these regions (or, in general, subsystems) to be labeled as
$\regA_i$ and $\regA = \cup_i \, \regA_i$.  With this understanding, we will talk about partitions of the state in the Hilbert space ${\cal H} = \otimes_i {\cal H}_{\regA_i}$ etc.
We will also adapt a shorthand whereby $\regA_1 \regA_2 = \regA_1 \cup \regA_2$. For example $\rho_{\regA_1 \regA_2}$  will be an operator on ${\cal H}_{\regA_1} \otimes {\cal H}_{\regA_2}$ obtained perhaps by tracing out the part of the system in $\left(\regA_1 \regA_2\right)^c$ and its entropy will be $S_{\regA_1 \regA_2}$ etc. We now list  the salient entropy inequalities that will be relevant for our discussion.

\begin{itemize}
\item The simplest of the inequalities is {\em subadditivity}. For a bipartite system ${\cal H}_{\regA_1} \otimes {\cal H}_{\regA_2}$  we have
\begin{equation}
S_{\regA_1} + S_{\regA_2} \geq  S_{\regA_1 \regA_2}  \,.
\label{eq:sa}
\end{equation}
In QFTs, this is often trivially satisfied for overlapping regions, i.e., $\regA_1 \cap \regA_2 \neq 0$, owing to the fact that the UV divergent area term on the l.h.s overwhelms that on the r.h.s. It  does however remain more generally true and  prompts the definition of {\em mutual information} for such a bipartite system
\begin{equation}
I(\regA_1: \regA_2 ) =S_{\regA_1} + S_{\regA_2} -  S_{\regA_1 \regA_2}  \geq 0 \,.
\label{eq:midef}
\end{equation}
\item A useful inequality which can be derived by appending a third system $\regA_3$ and purifying $\rho_{\regA_1 \regA_2}$ is the {\em Araki-Lieb} inequality. This is usually stated directly for bipartite systems in the form
\begin{equation}
\mid S_{\regA_1} - S_{\regA_2} \mid \leq S_{\regA_1 \regA_2} \,.
\label{eq:alineq}
\end{equation}
Note that if $\regA_1 \regA_2$ makes up the entire system which is known to be pure, i.e., $\regA_2 =  \regAc_1 $ then we can immediately conclude that $S_\regA = S_{\regAc}$.

Rather curiously, this inequality has never been directly proved; known proofs derive it as a consequence of the subadditivity inequality using purification. One can combine it with the latter to bound the entropy of the joint region $\regA_1 \regA_2$, i.e.,
\begin{equation}
\mid S_{\regA_1} - S_{\regA_2} \mid \leq S_{\regA_1 \regA_2} \leq S_{\regA_1} + S_{\regA_2}
\label{eq:classical}
\end{equation}
In this form, the two inequalities serve to bound $S_{\regA_1 \regA_2} \geq \text{max}\{S_{\regA_1}, S_{\regA_2} \}$ which should be familiar from classical monotonicity.

\item The most interesting quantum inequality is {\em strong subadditivity}  \cite{Lieb:1973lr,Lieb:1973cp}, which places interesting constraints on potential entropy functions.  It encodes in a certain heuristic sense the idea that $S_\regA$ is a concave function. There are many useful ways to state the inequality; we will choose to do it for a tripartite system
${\cal H}= {\cal H}_{\regA _1}\otimes {\cal H}_{\regA_2} \otimes {\cal H}_{\regA_3}$. One has
\begin{align}
 S_{\regA_1 \regA_2}+S_{\regA_2 \regA_3}  & \geq S_{\regA_1 \regA_2 \regA_3}+ S_{\regA_2},
 \label{eq:ssa1}\\
 S_{\regA_1 \regA_2}+S_{\regA_2 \regA_3}  & \geq
S_{\regA_1}+ S_{\regA_3} \,.
 \label{eq:ssa2}
\end{align}
The second of these follows from the first by purification.  These inequalities are valid as long as the inner product between the states of the Hilbert space is positive and thus relies on the underlying quantum system being unitary.
\end{itemize}

%~~~~~~~~~~~~~~~~~~~~~~~~~~~~~~~~~~~~~~~~~~~~~~~
\section{Relative entropy}
\label{sec:relent}
%~~~~~~~~~~~~~~~~~~~~~~~~~~~~~~~~~~~~~~~~~~~~~~

One other quantity of interest in our discussion is the notion of \emph{relative entropy}. Given two density matrices $\rho$ and $\sigma$, we can define an object  $S(\rho|\sigma)$, which provides a measure of distinguishability between them. It is  defined as
\begin{equation}
S(\rho||\sigma)=\Tr{}{\rho\log \rho}-\Tr{}{\rho\log \sigma}\,.
\label{eq:relent}
\end{equation}
 A detailed discussion of  relative entropy from a quantum information perspective can be found in \cite{Wehrl:1978zz,Vedral:2002zz}.

It satisfies two important properties: positivity and monotonicity. The former simply asserts that the relative entropy is non-negative for any two density matrices and vanishes only when the two are equal, i.e.,
\begin{equation}
S(\rho||\sigma) \geq 0 \,, \qquad S(\rho||\sigma) = 0 \;\; \Longrightarrow \;\; \rho=\sigma\,.
\label{eq:repos}
\end{equation}
 This basic property can be understood from the fact that
if we assume $f(x)$ is a concave function, i.e., $\frac{d^2f}{dx^2} \leq 0$, then we have
\begin{equation}
\Tr{}{ f(\rho)-f(\sigma)-(\rho-\sigma)f'(\sigma) } \leq 0\,.
\end{equation}
Indeed, if we  now set $f(x)=-x\log x$, we immediately find $S(\rho||\sigma)\geq 0$.

Monotonicity of relative entropy is the statement that relative entropy decreases under inclusion.
Say we start with a pair of reduced density matrices $\rho, \sigma$ and trace out the same  degrees of freedom  to obtain reduced density matrices $\rhoA,\sigma_\regA$. Under this process the relative entropy is reduced:
\begin{equation}
S(\rho_\regA || \sigma_\regA ) \leq S(\rho || \sigma) \,, \qquad \rhoA = \Tr{{\regAc}}{\rho}, \;
\sigma_\regA = \Tr{{\regAc}}{\sigma} \,.
\label{eq:relmon}
\end{equation}
 A proof of this statement can, for instance, be found in \cite{Nielsen:2010aa}.

There is a useful way to rewrite the relative entropy result in a manner reminiscent of thermodynamic formula.
Let us treat $\sigma$ as the reference state and introduce its modular Hamiltonians ${\cal K}_\sigma$ as defined in \eqref{eq:modHam}. We can use this to define the \emph{modular free energy}
\begin{equation}
F(\rho) = \Tr{}{\rho \, {\cal K}_\sigma} -S(\rho)\,,
\label{eq:modfree}
\end{equation}
where $S(\rho)$ is the von Neumann entropy of the density matrix. Using this, it is easy to see that
\begin{equation}
\begin{split}
S(\rho|| \sigma) &= \Tr{}{\rho \log \rho} - \Tr{}{\sigma \log\sigma}+\Tr{}{\sigma \log\sigma} -
 \Tr{}{\rho \log\sigma}\\
&= -S(\rho) + S(\sigma) - \vev{-\log\sigma}_\sigma + \vev{-\log\sigma}_\rho\\
& = F(\rho) - F(\sigma)
\end{split}
\label{eq:relfree0}
\end{equation}
We can further express this as the difference modular Hamiltonian expectation value and that of the entropies, viz.,
\begin{equation}
S(\rho|| \sigma)  =\Delta\vev{{\cal K}_\sigma} - \Delta\vev{S} \geq 0
\label{eq:relfree}
\end{equation}
where the positivity of relative entropy guarantees the last inequality.

While the relative entropy is not symmetric in its arguments, it can be used as a distance measure for states that are in the neighbourhood of each other. Consider a reference state $\sigma = \rho_0$ and let
$\rho=\rho_0+\epsilon\,\rho_1  + \epsilon^2\, \rho_2 + \cdots$ be a one-parameter family of states in its  neighbourhood. We can evaluate the relative entropy as a power series in $\epsilon$.

The first observation to make is that the relative entropy is at least quadratic in the deviation parameter $\epsilon$: $S(\rho||\sigma) = {\cal O}(\epsilon^2)$. The contribution to relative entropy at ${\cal O}(\epsilon)$  vanishes for any choice of $\rho_0$. This means:
\begin{equation}
\delta S = \delta \vev{{\cal K}_{\rho_0}}\,.
\label{eq:e1stlaw}
\end{equation}
Thus, while in general the change in entanglement entropy is only bounded by the change in the modular Hamiltonian, to linear order in the deformation, the inequality is saturated.  This statement is known as the \emph{first law of entanglement} \cite{Blanco:2013joa}, owing to its similarity to the thermodynamic expression $dE = T\, dS$. It has played an important role in holographic context, as we shall describe in \S\ref{sec:egeometry}.

At the quadratic order in $\epsilon$ we find the relative entropy can be used to define a positive definite inner product on the perturbations to the reference density matrix, via
\begin{align}
S(\rho_0 + \epsilon \, \rho_1||\rho_0 )
&\equiv \epsilon^2\; \vev{\rho_1\,, \rho_1}_{\rho_0}
\nonumber \\
&= \frac{1}{2}\epsilon^2 \,  \Tr{}{\rho_1 \,\frac{d}{d\epsilon} \log(\rho_0+\epsilon\, \rho_1)}
\label{eq:qfisher}
\end{align}
This quadratic function is non-negativity definite owing to the positivity of relative entropy and is known as the \emph{quantum Fisher information}. Recently various authors have used the relative entropy to derive interesting holographic constraints, cf., \S\ref{sec:egeometry}.

%~~~~~~~~~~~~~~~~~~~~~~~~~~~~~~~~~~~~~~~~~~~~~~~
\chapter{Entanglement entropy in CFT$_2$}
\label{sec:eecft2}
%~~~~~~~~~~~~~~~~~~~~~~~~~~~~~~~~~~~~~~~~~~~~~~

The description of the general methodology for computing entanglement entropy in \S\ref{sec:qft} gives a clean, albeit abstract prescription. As with any functional integral, it helps to develop some intuition as to where the computation can be carried out explicitly. For a general QFT in $d >2$ the computation appears intractable in all but the simplest of cases of free field theories \cite{Srednicki:1993im}. However, it turns out to be possible to leverage the power of conformal symmetry in $d=2$ ,to explicitly compute entanglement entropy in some situations \cite{Calabrese:2009qy}. In fact, the revival of interest in entanglement entropy can be traced to the work of Cardy and Calabrese \cite{Calabrese:2004eu} who re-derived the results of \cite{Holzhey:1994we} and went on to then explore its utility as a diagnostic of interesting physical phenomena in interacting systems. We will give a brief overview of this discussion, adapting it both to the general ideas outlined above and simultaneously preparing  for our holographic considerations in the sequel.\footnote{ For a review of analysis of entanglement entropy in free field theories, refer to \cite{Casini:2009sr}.}

Consider a two-dimensional theory with conformal invariance, i.e., a CFT$_2$. A good account of these theories can be found in the books \cite{Ginsparg:1988ui,DiFrancesco:1997nk}. Such theories can be described by giving\footnote{ Higher-dimensional CFTs are similarly described by their operator spectrum and OPE coefficients.}
\begin{itemize}
\item The central charge $c$,
\item A list of the quasi-primary operators ${\cal O}_{h,\bar{h}}$ which have definite weight, i.e., scaling dimensions, $\{h, \bar{h}\}$, under local Weyl rescaling, and
\item The OPE coefficients $C_{\alpha \beta}^\gamma$ which appear in the OPE: ${\cal O}_\alpha {\cal O}_\beta \sim C_{\alpha\beta}^\gamma\, {\cal O}_\gamma$, where we drop the dependence on the insertion point.
\end{itemize}
The Hilbert space of states can be obtained from this data: owing to the state operator correspondence
(cf., \cite{DiFrancesco:1997nk}), we can map a given local operator ${\cal O}_{h, \bar{h}} $ onto a state in the Hilbert space $\ket{h,\bar{h}} =
{\cal O}_{h,\bar{h}} \ket{0}$.
We will make use of some of this structure later on in our discussion, but for now we wish to show how to compute entanglement entropy in such theories.

The special feature of two-dimensional CFTs is that they possess an infinite dimensional global symmetry algebra, called the Virasoro algebra. While in any dimension we have the group of conformal transformations extending the $S(d,1)$ Poincar\'e symmetry in relativistic systems to $SO(d,2)$, when $d=2$, the $SO(2,2)$ algebra gets enhanced. The reason is that, in two dimensions, a local conformal transformation can be viewed as independent holomorphic and anti-holomorphic transformations. This can be seen by adapting complex coordinates $z = x+ i\,\tE$ and $\bar{z} = x-i\,\tE$ and noting that a local conformal map factorizes into $f(z) \, g(\bar{z})$.

%~~~~~~~~~~~~~~~~~~~~~~~~~~~~~~~~~~~~~~~~~~~~~~~
\section{A single-interval in CFT$_2$}
%~~~~~~~~~~~~~~~~~~~~~~~~~~~~~~~~~~~~~~~~~~~~~~

As discussed in \S\ref{sec:qft}, we have to pick a state and a region $\regA$ to talk about the entanglement entropy $S_\regA$. To start with we will consider simple static states for a CFT$_2$ on ${\mathbb R}^{1,1}$ and ${\mathbb R} \times {\bf S}^1$, respectively. We will exploit the time independence to work in Euclidean signature,  mapping the background geometry to the complex plane ${\mathbb C} = {\mathbb R}^2$ and the cylinder, respectively. The discussion below applies to both cases equivalently, so we will indicate the picture for the plane and generalize therefrom to the cylinder.

Consider  then the vacuum state $\ket{0}$ of the CFT$_2$ on ${\mathbb C}$. We pick an instant of time, say $t=0$ w.l.o.g., and define $\regA$ to be an interval $-a < x < a$. The entangling surface in this scenario is the two endpoints of the interval. The replica construction requires us to take $q$-copies of the complex plane with slits cut out along $\regA$ and to glue them cyclically to construct the manifold $\bdy_q$. This construction in two-dimensional geometry constructs a $q$-sheeted surface with prescribed branching at $\{x= \pm a, t =0\}$, as illustrated in Fig.~\ref{f:rho3E}.  What is clear for the complex plane is that the cyclic gluing of $q$ copies of that plane does not change the topology:  $\bdy_q$ is a genus-0 surface; we just have to deal with a function that is multi-branched.

We are required to compute the partition function $Z[\bdy_q]$  as a first step. Since $\bdy_q$ is a genus-0 surface, we should be able to conformally map it back to the complex plane. Equivalently, we can start with fields $\phi(x,t)$, which live on a single copy of the complex plane, and upgrade them to $\phi_k (x,t)$ with $k=1,2,\cdots,q$ which live on the $q$-copies. The gluing conditions for constructing $\bdy_q$ can be mapped to boundary conditions for the fields
\begin{equation}
\phi_k(x,0^+)  = \phi_{k+1}(x,0^-) \,, \qquad x \in \regA = \{ x| x \in (-a,a)\}
\label{eq:bcscft}
\end{equation}
These boundary conditions can be equivalently implemented by passing from the basis of $q$-independent fields to a composite field $\varphi(x,t)$ living on $\bdy$ obeying twisted boundary conditions. The map one seeks should thus implement the twists  by the cyclic ${\mathbb Z}_q$ replica symmetry. An  easy way to think about this construction is that we are  no longer working with the original CFT but rather with the cyclic product orbifold theory \cite{Headrick:2010zt}.\footnote{ See \cite{Haehl:2014yla} for an abstract discussion of how one can use orbifold technology to understand the computation of R\'enyi entropies. The original references on orbifolds \cite{Dixon:1986jc,Dixon:1986qv} are a great resource for learning about the technology we employ below.}

One introduces then,  as in any orbifold theory, a set of twist fields which implement the twisted boundary conditions. For the case at hand the twists are by $q^{\rm th}$ roots of unity, and the main property we need for the twist operator ${\cal T}_q$ is that it induces a branch-cut of order $q$ for the fields at its insertion point. Standard orbifold technology reveals that the scaling dimension of the twist operator is
\begin{equation}
h_q = {\bar h}_q = \frac{c}{24} \left(q - \frac{1}{q}\right) .
\label{eq:twistdim}
\end{equation}
The main advantage of introducing these fields is that we can write down the partition function of our theory on $\bdy_q$ in terms of correlation functions of the twist fields:
\begin{equation}
\mathcal{Z}[\bdy_q]  = \prod_{k=0}^{q-1} \, \vev{{\cal T}_q(-a,0)\, {\cal T}_q(a,0)}_\bdy
\label{eq:Ztwist}
\end{equation}
where we used the subscript $\bdy$ to indicate that the correlation function is meant to be computed on the original manifold. For our choice of $\regA$ being a single connected interval, the above computation is very simple. Treating the twist fields as conformal primaries with scaling dimension given by \eqref{eq:twistdim}, we learn that
\begin{equation}
\mathcal{Z}[\bdy_q] =\left(\frac{2a}{\epsilon}\right)^{-\frac{c}{6} \left(q- \frac{1}{q}\right)}
\end{equation}
where we introduced a UV regulator $\epsilon$ to write down the correlation function. We now find accounting for the normalization of the density matrix induced onto the region $\regA$ that
\begin{equation}
S_\regA^{(q)} =   \frac{c}{6}\, \left(1 + \frac{1}{q}\right)\, \log\frac{2a}{\epsilon} \,.
\label{}
\end{equation}
In this simple case it is in fact trivial to analytically continue from $q \in {\mathbb Z}_+$ to $q\sim 1$. One clearly obtains:
\begin{equation}
S_\regA = \frac{c}{3} \, \log\frac{2a}{\epsilon} \,.
\label{eq:hlwcc}
\end{equation}

The remarkable aspect of this answer is that the result is agnostic to the details of the CFT$_2$. It only cares about the overall central charge and thus one does not gain any deep insight into the nature of the degrees of freedom which are entangled. We should remark here that while unitary CFTs with $c<1$ have the spectrum determined  by the central charge, this does not happen for $c> 1$. In any event, the result for a single-interval entanglement entropy provides some overall information about the CFT in question, through its dependence on the central charge, even though it is unable to resolve finer details. Consequentially, one can use $S_\regA$ to provide an alternate measure for the central charge, and this line of thought is useful for providing an entanglement-based proof for the $c$-theorem \cite{Casini:2004bw}.

There are two other scenarios in which we can immediately write down the answer for the R\'enyi and entanglement entropy directly using the known two point functions of the twist operator. These correspond to situations where the geometry $\bdy = {\mathbb R}\times {\bf S}^1$. This can be interpreted in two ways: we can either talk about a CFT on a compact spatial geometry of size $\ell_{{\bf S}^1}$, or we can turn things around and view it as a thermal field theory on the real line with the circle parameterizing the Euclidean time direction with period set by the inverse temperature $\beta = \ell_{{\bf S}^1}$. We obtain in these two cases the results by a trivial generalization of the above computation, taking into account the natural distance measure on the cylinder.

For the finite spatial domain, we find
\begin{align}
S_\regA  = \frac{c}{3} \, \log \left(\frac{\ell_{{\bf S}^1}}{\pi \,\epsilon}\, \sin \left(\frac{2a}{\ell_{{\bf S}^1}}\right)\right)
\label{eq:svaccft2}
\end{align}
while for a thermal system in non-compact space, we end up with
\begin{equation}
S_\regA = \frac{c}{3} \, \log \left(\frac{\beta}{\pi \,\epsilon}\, \sinh \left(\frac{2\pi\,a}{\beta}\right)\right)
\label{eq:sthermcft2}
\end{equation}
In both of these cases, we are taking $\regA$ to have width $2 a$. We have expressed the result in terms of the physical  length scales, though one can equivalently express \eqref{eq:svaccft2} in terms of angular arc length.

%~~~~~~~~~~~~~~~~~~~~~~~~~~~~~~~~~~~~~~~~~~~~~~~
\section{Disconnected regions, multiple intervals}
%~~~~~~~~~~~~~~~~~~~~~~~~~~~~~~~~~~~~~~~~~~~~~~

The computation for multiple intervals is quite complex even in two-dimensional CFTs.  Consider for example a disjoint union of $m$ spatial intervals $\regA = \cup_{i=1}^m \regA_i$. The replica method says that we are required to construct the $q$-fold ${\mathbb Z}_q$  symmetric branched cover of the basic geometry with $2m$ branching points. Now each pair of regions will conspire to produce a single handle for every two copies of the replicated geometries. This implies that the branched cover spacetime is a higher genus Riemann surface with $g = (q-1)(m-1)$.

% Figure
\begin{figure}[htbp]
\begin{center}
\includegraphics[width=3in]{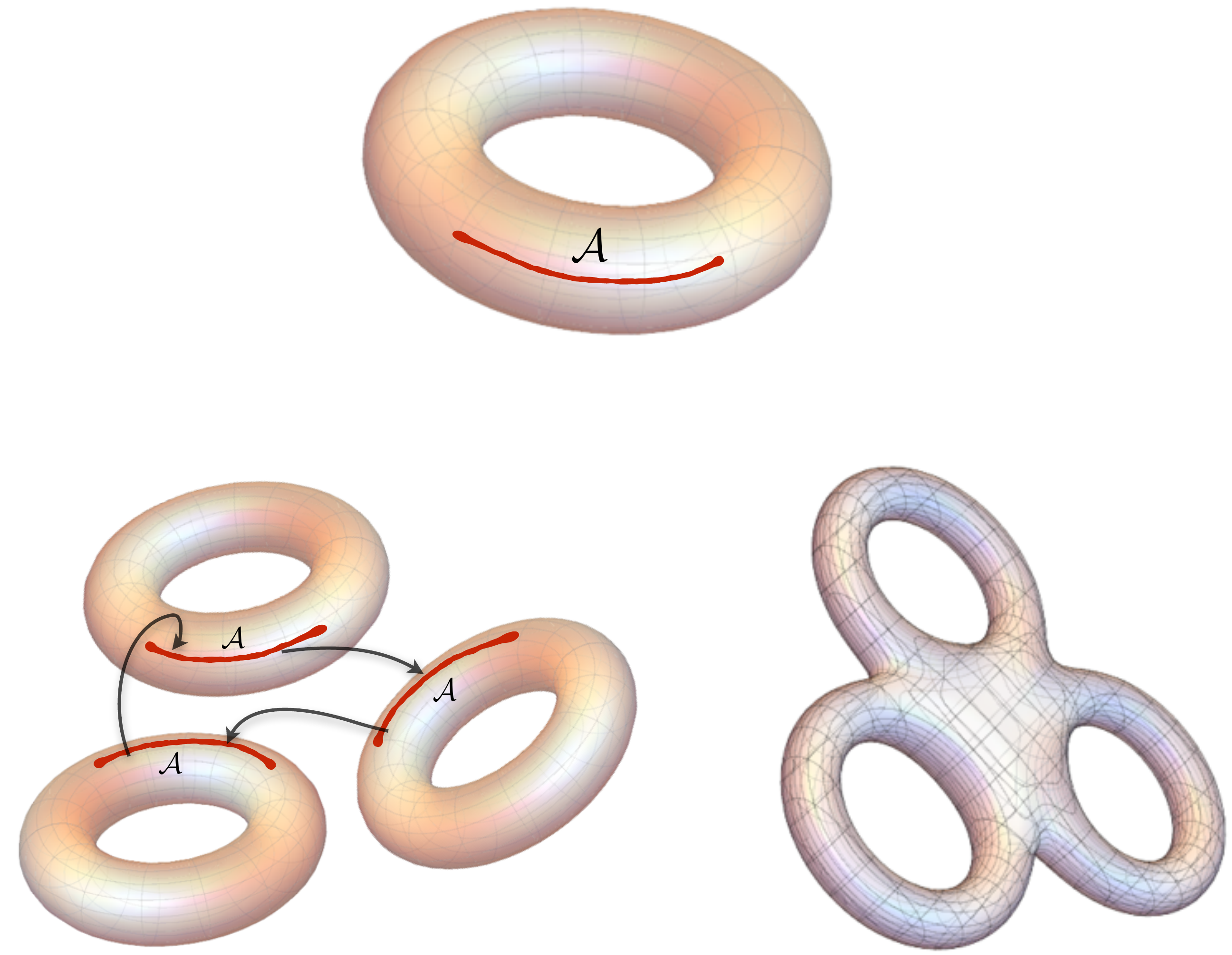}
 \caption{Construction of the branched cover geometry $\bdy_q$ involved in the computation of the $q^{\rm th}$ Renyi entropy in a  CFT$_2$ at finite temperature and in finite spatial volume. So $\bdy$ is a torus and we illustrate how 3 copies of this torus are cyclically sewn across the region $\regA$ to obtain a genus-3 surface required for $S^{(3)}$. }
\label{f:branch3}
\end{center}
\end{figure}

As a result, the computation of the R\'enyi entropy  involves evaluating the partition function of the CFT on this Riemann surface (or equivalently an appropriate correlator of $2mq$ twist operators). Neither of these is easy to compute explicitly in generic interacting CFTs. On general grounds, one can argue that the answer is a modular form of a certain degree, which is a function of the moduli of the Riemann surface.\footnote{ While a general genus-$g$ Riemann surface has $3g-3$ moduli, the branched cover R\'enyi geometries are a special subclass with the moduli being determined by the $2m-3$ cross-ratios.} The modular forms in question are functions of the moduli, and can be written naturally in terms of the $g \times g$ period matrix of the Riemann surface which encodes this data. In some special cases, we even happen to know what this object is, e.g., in the free compact boson CFT we can obtain the answer in terms of the Siegel theta functions; see for example \cite{Calabrese:2009ez} for computations in the free boson theory.

One however has an essential complication in analytically continuing the answer from
$q\in {\mathbb Z} _+ $ down to $q\to 1$. The reason is as follows: the R\'enyi index $q$ enters into the genus of the Riemann surface, and thence into the size of the period matrix. Analytic continuation would require us to suitably continue the theta functions to have non-integral-dimensional period matrices in their argument.  As a result, there is so far no explicit evaluation of entanglement entropy in these theories for disconnected regions.

There are a couple of exceptions to the above discussion. For $q=2, m =2$, the computation involves the torus partition function. This only involves the spectral data of the CFT and can thus be obtained directly. We can also obtain the result for the entanglement entropy for a free Dirac fermion as long as we don't sum over the spin structures in closed form \cite{Casini:2008wt}. We refer the reader to \cite{Headrick:2012fk} for a detailed discussion of these issues including a critical examination of the interplay with Bose-Fermi dualities.

Another interesting situation involves the thermal state of a CFT$_2$ in finite volume. In this case, even for a single-interval, one encounters a non-trivial higher genus computations, as explained in Fig.~\ref{f:branch3}.
Gluing torii across a single cut cyclically produces a surface of genus $q$ for the $q^{\rm th}$ R\'enyi entropy. Multiple regions may be analyzed similarly, with newer handles being generated from the gluing at each stage.

\newpage
%~~~~~~~~~~~~~~~~~~~~~~~~~~~~~~~~~~~~~~~~~~~~~~~
\part{Holography and entanglement}
\label{part:hee}
%~~~~~~~~~~~~~~~~~~~~~~~~~~~~~~~~~~~~~~~~~~~~~~

%~~~~~~~~~~~~~~~~~~~~~~~~~~~~~~~~~~~~~~~~~~~~~~~
\chapter{Holographic entanglement entropy}
\label{sec:hee}
%~~~~~~~~~~~~~~~~~~~~~~~~~~~~~~~~~~~~~~~~~~~~~~

The discussions in \S\ref{sec:qft} and \S\ref{sec:eecft2} make it rather clear that while we have to evaluate a sequence of functional integrals to compute the R\'enyi entropies, these are rather complex quantities which required us to work with QFTs on singular branched cover manifolds.  Apart from the case of CFT$_2$ discussed in \S\ref{sec:eecft2},  where the power of conformal invariance can be used to simplify the problem, this is a rather formidable task  for interacting QFTs, in general.

%~~~~~~~~~~~~~~~~~~~~~~~~~~~~~~~~~~~~~~~~~~~~~~~
\section{A lightning introduction to holography}
\label{sec:adscft}
%~~~~~~~~~~~~~~~~~~~~~~~~~~~~~~~~~~~~~~~~~~~~~~

We now turn to holographic field theories which are dual to gravitational field theories in a different spacetime using the gauge/gravity or AdS/CFT correspondence.  This remarkable correspondence was first discovered by Maldacena \cite{Maldacena:1997re}. It  asserts that a class of non-gravitational QFTs in $d$-dimensions are equivalently described in terms of a string  theory involving gravitational interactions. The correspondence was developed further in \cite{Gubser:1998bc,Witten:1998qj} and the reader can find detailed accounts of the basic statements in the classic reviews \cite{Aharony:1999ti,DHoker:2002aw}.

 The QFTs in question have two basic parameters which are relevant: a coupling constant $\lambda$, which measures the interaction strength between the constituents, and a measure of the effective number of degrees of freedom $\ceff$. For general values of  the parameters $\{\lambda, \ceff\}$ the dual picture is that of an interacting string theory. We are clearly oversimplifying here, by demanding that there be a single parameter controlling all interactions.

Things drastically simplify in the limit  $\ceff\to \infty$, which we will refer to as the \emph{planar limit}.
When the number of degrees of freedom are scaled to be large, the string interactions become weak;  one can then truncate to the tree level result,  thus obtaining the classical string limit. One way to view this statement is to note that in the planar limit the QFT path integral can be argued to localize around a new non-trivial saddle point which can be described by a master field configuration. The master field defines a new semiclassical limit with an effective Planck constant $\hbar \propto \frac{1}{\ceff}$, which dominates the configurations contributing to the functional integral.  The intuition that such a classical saddle could exist was developed by studying the large $N$ limit of gauge theories \cite{tHooft:1973jz,Witten:1979kh}, where $N$ captures the rank of the gauge group, say $SU(N)$. While master fields for pure gauge theories remain elusive, one now has, thanks to the gauge/gravity correspondence, a statement for many interacting QFTs.

 While we have achieved a great deal of simplification, mapping in the planar limit, an interacting quantum system to a classical dynamical system involving strings, in general, it is still  a formidable task to solve the classical string dynamics to extract interesting physical information. One can achieve more if we consider the strong coupling limit of the field theory $\lambda \to \infty$ whence the classical string dynamics further truncates to classical gravitational dynamics of the general relativistic form. Basically in this limit, the massive string states in the dual description become heavy and decouple, leaving only the dynamics of semiclassical  gravity. Thus in the combined limit $\lambda, \ceff \to \infty$ one can phrase complicated questions about the dynamics of quantum fields by studying semiclassical gravitational physics in a dual spacetime.
 We will see that this holographic map simplifies dramatically the computation of entanglement entropy.

 The prototype example of holographic field theories are supersymmetric gauge theories which are realized in the  low energy  limit of open string theories living on D-brane worldvolumes in string theory. The oft-mentioned case is that of
 ${\cal N}=4_{4d}$ $SU(N)$ Super  Yang-Mills, which is the maximally supersymmetric four-dimensional QFT comprising of $SU(N)$ gauge fields, six adjoint scalars, and adjoint Weyl fermions. This theory is a supersymmetric extension of pure Yang-Mills theory and enjoys exact conformal invariance for any value of the coupling $g_{YM}$. As a result, there are no dimensionful parameters in the theory (all fields are massless) and one may characterize the family of such theories by two parameters: $\lambda = g_{YM}^2\, N$, which is the dimensionless  't Hooft coupling, and $\ceff \propto N^2 -1$. In the large $N$ limit, which does correspond to planar gauge theory limit, we obtain the classical master field of this theory in terms of string theory on
 \AdS{5} $\!\times \,{\bf S}^5$. The strong coupling limit $\lambda \to \infty$ further truncates the dynamics to the two derivative Type IIB supergravity theory on the same background. The latter contains as a subsector the dynamics of Einstein-Hilbert gravity in \AdS{5} which will be of most interest to us.

We list a few other examples of well-known pairs of field theories and their gravity duals below:
\begin{itemize}
\item Two-dimensional CFTs with large central charge $c \gg 1$ are expected to be dual to classical theories on \AdS{3}. To ascertain whether they limit to classical Einstein-Hilbert gravity or something more complicated requires a more detailed analysis (see \cite{Hartman:2014oaa,Haehl:2014yla,Belin:2014fna,Belin:2015hwa} for some recent attempts to do so). The well-known example in this case is the $(4,0)_{2d}$ superconformal field theory (SCFT) that arises on the worldvolume of a bound state of D1 and D5 branes in string theory.
Given $Q_1$ D1-branes and $Q_5$ D5-branes wrapping ${\cal X}_4 \!\times \,{\bf S}^1$ with ${\cal X}_4$ being either $K3$ or ${\bf T}^4$, the worldvolume SCFT$_2$ is a symmetric orbifold theory with target space ${\cal X}^{Q_1Q_5}/S_{Q_1Q_5}$, see e.g., \cite{Seiberg:1999xz}. This theory is holographically dual to classical gravity on \AdS{3} $\!\times \,{\bf S}^3 \!\times \,{\cal X}_4$.
\item In three dimensions, Chern-Simons matter theories lead to conformal and superconformal field theories. The maximally supersymmetric theory in this case arises from the dynamics of M2-branes. It is a ${\cal N} = 8_{3d}$ Chern-Simons theory with gauge group $SU(N)_k \!\times \,SU(N)_{-k}$ with bifundamental matter,  called the ABJM theory \cite{Aharony:2008ug}. This theory is dual to classical gravitational dynamics on \AdS{4} $\!\times \,{\bf S}^7$.
\item The prototype example of a six-dimensional $(2,0)_{6d} $ SCFT (which is the maximum allowable dimension for superconformal invariance) is the worldvolume theory of M5-branes \cite{Seiberg:1997ax}. While we have a rather poor understanding of the microscopic description of this theory we know that in the large $N$ (which is the number of M5-branes) limit, it is dual to gravity on \AdS{7} $\!\times \,{\bf S}^4$.
\end{itemize}

%~~~~~~~~~~~~~~~~~~~~~~~~~~~~~~~~~~~~~~~~~~~~~~~
\section{The gravitational setup}
\label{sec:gravcon}
%~~~~~~~~~~~~~~~~~~~~~~~~~~~~~~~~~~~~~~~~~~~~~~

More generally, one can have many more examples of $d$-dimensional conformally-invariant field theories (CFT$_d$) with varying amounts of supersymmetry which may be argued to be dual to classical gravity on \AdS{d+1} $\!\times \,{\cal Y}$ with
${\cal Y}$ being some compact space. The latter is required for a consistent embedding into string theory and will play a minor role in what follows.

We will henceforth be reasonably agnostic about a particular field theory and look for statements that are valid across the gamut of the AdS/CFT correspondence. We assume that we have a CFT$_d$ which  satisfies the criterion for the existence of a holographic map $\lambda, \ceff \gg 1 $, and consider the class of these which may be studied using classical gravitational dynamics in an asymptotically \AdS{d+1} spacetime, which we henceforth call $\bulk_{d+1}$.

We view the classical gravity theory as a low energy effective field theory with a consistent derivative expansion. In addition to dynamics of gravitons in \AdS{d+1}, we will also allow for matter, whose presence will depend on the particulars of the field theory we study. One may be somewhat abstract and write the classical gravitational dynamics as being derived from an action:
\begin{equation}
S_{bulk } = \frac{1}{16\pi\, \GN} \; \int d^dx \, \sqrt{-g}\, \left(R +\frac{d(d-1)}{\lads^2} +  \sum_i  \alpha_k\,  \mathcal{D}^{2k}R + {\cal L}_\text{matter} \right)
\label{eq:gravact}
\end{equation}
The leading terms here are the Einstein-Hilbert action with a negative cosmological constant $\Lambda = -\frac{d(d-1)}{2\,\lads^2}$.  We have allowed for the possibility of higher derivative corrections which are schematically denoted as covariant derivatives of the curvature.\footnote{ ${\cal D}$ is the metric compatible covariant derivative operator on $\bulk_{d+1}$.} These terms have dimensionful coefficients $\alpha_i$ which in principle could be non-trivial functions of the matter fields. In general, one may argue that these terms are suppressed at large $\lambda$. One has in addition two explicit dimensionful parameters
$\lads$ and $\GN \propto \ell_P^{d-1}$ with $\ell_P$ being the $d+1$-dimensional Planck scale. In string theories, we have in addition a string scale $\ell_s$ which enters into the determination of the higher derivative corrections through $\alpha_k$.

Given the three dimensionful length scales, the field theory parameters  $\lambda$ and $\ceff$ can be expressed as the dimensionless ratio of pairs. One typically finds relations of the form\footnote{ We could have simply written $\ceff \propto \left(\frac{\lads}{\ell_P}\right)^{d-1}$, but have chosen to fix the normalization to be consistent with standard conventions in explicit holographic dual pairs.}
\begin{equation}
\ceff = \frac{\lads^{d-1}}{16\pi \GN} \,,  \qquad \lambda = \left(\frac{\lads}{\ell_s}\right)^\gamma
\label{eq:dumap}
\end{equation}
where $\gamma >0$. In the familiar example of ${\cal N}=4_{4d}$ SYM, $\gamma = \frac{1}{4}$. Given this dictionary, we should also note that $\alpha_k \propto (\ell_s)^{2k} \sim \lambda^{-2k/\gamma}$, so the higher derivative corrections to the gravitational interactions may be viewed as terms arising in a strong coupling perturbation theory.

To proceed, we will need a dictionary between the field theory and gravitational observables. Consider a CFT$_d$ on some background geometry $\bdy_d$, which we take to be timelike and globally hyperbolic.\footnote{ These latter condition allows us to define a well-posed initial value problem for the quantum fields which may be evolved using the Hamiltonian.}
The field theory can be in any of the states in the Hilbert space and the AdS/CFT correspondence at a general level asserts that each such state maps to an analogous state in the closed string Hilbert space. The isomorphism between Hilbert spaces is the central feature of the correspondence. Of interest to us will be a limited class of states, said to belong to the so-called {\em code subspace} which have geometric duals.\footnote{ We will revisit these ideas in \S\ref{sec:egeometry}. The code subspace derives from ideas in quantum error correction.}  This special class of states of the field theory on $\bdy_d$ are described geometrically in terms of a bulk spacetime $\bulk_{d+1}$ with $\partial\bulk_{d+1} = \bdy_d$. The spacetime
$\bulk_{d+1}$ satisfies Einstein's equations derived from \eqref{eq:gravact}. For the bulk of our discussion, we will work in the corner of parameter space where $\ceff \gg 1 $ and $\lambda \gg 1 $ so that we can effectively restrict attention to the two derivative theory of Einstein-Hilbert  gravity setting $\alpha_k = 0$.

Given $\bdy_d$ and information about the state on this background, say by prescribing expectation values of various gauge-invariant operators, we solve the equations of motion resulting from \eqref{eq:gravact}
\begin{equation}
R_{AB}  + \frac{d}{\lads^2}  \, g_{AB} = T_{AB}^\text{\tiny{matter}}
\label{eq:bulksaddle}
\end{equation}
subject to the boundary condition $\partial \bulk_{d+1} = \bdy_d$. All the matter fields will obey boundary conditions that can be explicitly specified. It is however useful to first record some examples in which we have solutions with no matter.

The CFT$_d$ vacuum which preserves the $S(d,2)$ conformal symmetry is dual to the vacuum \AdS{d+1} spacetime. There are two special choices for $\bdy_d$:
\begin{itemize}
\item Einstein static universe $\bdy_d = {\mathbb R}\times {\bf S}^{d-1}$, whence the bulk spacetime is the global \AdS{d+1} geometry with metric
\begin{equation}
ds^2 = -f_0(\rho)\, dt^2 + \frac{d\rho^2}{f_0(\rho)} + \rho^2 \, d\Omega_{d-1}^2 \,, \qquad f_0(\rho) =1+\frac{\rho^2}{\lads^2} \,.
\label{eq:gads}
\end{equation}
\item  Minkowski spacetime $\bdy_d = {\mathbb R}^{d-1,1}$, whence the bulk geometry is the Poincar\'e patch of \AdS{d+1} which can be coordinatized as (nb: ${\bf x}_{d-1} = \{x_1, x_2, \cdots , x_{d-1}\} \in {\mathbb R}^{d-1}$)
\begin{equation}
ds^2 = \frac{\lads^2}{z^2} \left(-dt^2 + d{\bf x}_{d-1}^2 + dz^2 \right) .
\label{eq:pads}
\end{equation}
\end{itemize}

Both of these geometries preserve the entire conformal symmetry, though different parts are manifestly visible in the chosen coordinates.
The boundary is attained in the limit $\rho \to \infty$ and $z\to 0$ for the two cases discussed above. One can see that the induced metric on the boundary is conformal to the natural metric on the two spacetimes considered above.

\begin{figure}[htbp]
\begin{center}
\includegraphics[width=4cm]{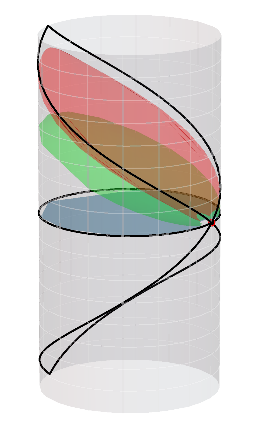}
\hspace{2cm}
\includegraphics[width=4cm]{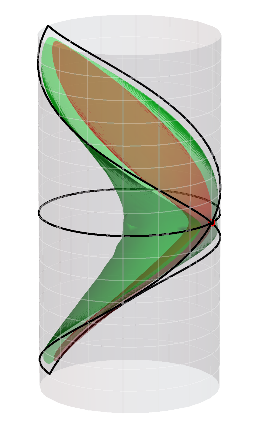}
\hspace{2cm}
\includegraphics[width=4cm]{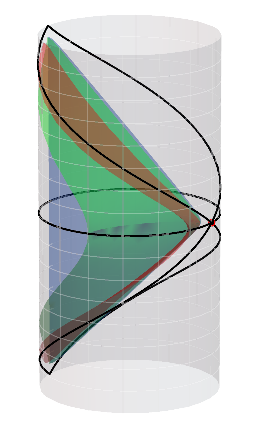}
\caption{The Poincar\'e coordinate chart illustrated within the global AdS spacetime. The three panels give slices of constant Poincar\'e time,  the radial coordinate, and the translationally invariant spatial coordinate respectively. }
\label{f:poincarecoords}
\end{center}
\end{figure}

The explicit coordinate transformation which maps between the two sets of coordinates can be obtained from the embedding the \AdS{d+1} spacetime as a hyperboloid in ${\mathbb R}^{d,2}$, i.e., the hypersurface:
\begin{equation}
-X_{-1}^2 -X_0^2  +\sum_{i=1}^d X_i ^2 = -\lads^2 \,, \qquad  ds^2_{{\mathbb R}^{d,2}} =  -dX_{-1}^2 -dX_0^2  +\sum_{i=1}^d \, dX_i^2 \,.
\label{eq:hyperd2}
\end{equation}
One finds the explicit set of transformations
\begin{equation}
\begin{split}
X_{-1 } &= \sqrt{\lads^2+\rho^2}\;\cos t_g=\frac{\lads^2+  z^2+{\bf x}_{d-1}^2-t^2}{2\,z} \,, \\
X_0 &= \sqrt{\lads^2+\rho^2}\; \sin t_g=\frac{\lads\, t}{z} \,,   \\
X_i & = \rho\, \Omega_{i}=\frac{\lads\, x_i}{z} \,, \qquad i=1,2,\cdots, d-1 \,, \\
X_d &= \rho\, \Omega_{d} =  \frac{z^2+{\bf x}_{d-1}^2 -\lads^2-t^2}{2\, z} \,.
\end{split}
\label{eq:gtopc}
\end{equation}
with $\Omega_i$ being direction cosines, $\sum_{i=1}^d \, \Omega_i^2 =1$. We illustrate the domain covered by the Poincar\'e coordinates in Fig.~\ref{f:poincarecoords}. The locus $z\to \infty$ which marks the boundary of the Poincar\'e coordinate chart is  referred to as the Poincar\'e horizon. It is a degenerate Killing horizon.

% Figure

Excited states of the field theory map to non-trivial asymptotically \AdS{d+1} geometries obtained as described above by solving Einstein's equations. The simplest example is provided by the thermal state of the CFT$_d$ which is dual to a Schwarzschild-\AdS{d+1} black hole spacetime. For the two choices of the boundary geometry as above, we get either
\begin{itemize}
\item The global Schwarzschild-\AdS{d+1} black hole spacetime:
\begin{equation}
ds^2 = -f(\rho)\, dt^2 + \frac{d\rho^2}{f(\rho)} + \rho^2 \, d\Omega_{d-1}^2 \,, \qquad f_0(\rho) =1+\frac{\rho^2}{\lads^2}  - \frac{\rho_+^{d-2}}{\rho^{d-2}} \left(1+ \frac{\rho_+^2}{\lads^2}\right).
\label{eq:gsads}
\end{equation}
\item The planar Schwarzschild-\AdS{d+1} black hole spacetime:
\begin{equation}
ds^2 = \frac{\lads^2}{z^2} \left(- f(z) \, dt^2 + d{\bf x}_{d-1}^2 + \frac{dz^2}{f(z)} \right) \,, \qquad f(z) = 1- \frac{z^d}{z_+^d}
\label{eq:psads}
\end{equation}
\end{itemize}
where $\rho_+$ and $z_+$ are the locations of the horizons in the two cases, respectively.

 The \SAdS{d+1} black hole temperature relates to the horizon radius via
\begin{equation}
\begin{split}
\text{global}: & \qquad T= \frac{1}{4\pi\,\lads} \, \left( d\,\frac{\rho_+}{\lads} + (d-2)\, \frac{\lads}{\rho_+}\right) \,, \\
\text{Poincar\'e}: & \qquad T = \frac{d }{4\pi\, z_+}
\end{split}
\label{eq:Tsads}
\end{equation}
The global black holes exist only above a minimum temperature, $T > \frac{\sqrt{d(d-2)}}{2\pi\, \lads} $.  Solutions with $\rho_+ < \sqrt{\frac{d-2}{d}}\, \lads$ are referred to as  small black holes; they are similar to their asymptotically flat cousins in their thermodynamic properties (they have negative specific heat). Those with $\rho_+ >\sqrt{\frac{d-2}{d}}\,\lads$ are called  large black holes. The planar solutions can be obtained from them in the scaling regime $\rho_+ \gg \lads$, whence the curvature of the sphere at the horizon becomes negligible.

A special case which is of interest owing to its analytic tractability is \AdS{3}, where, unlike the asymptotic flat case, black holes exist. The solutions are referred to as BTZ solution \cite{Banados:1992wn}. The metric takes the remarkably simple form:
\begin{equation}
ds^2 = -\frac{r^2 - r_+^2}{\lads^2}\, dt^2 + \frac{\lads^2\, dr^2}{r^2 -r_+^2} + r^2 d\varphi^2 \,,
\label{eq:btz}
\end{equation}
with $r_+ \in {\mathbb R}$ parameterizing the location of the horizon and $\varphi \in [0,2\pi]$. As written, this is the global BTZ solution, whose boundary is ${\bf S}^1 \times {\mathbb R}$. We can decompactify the circle and write the planar BTZ solution by replacing $\varphi \to \frac{x}{\lads}$. Notice that the solution for $r_+ = i\,\lads$ reproduces the global \AdS{3} solution. This is not  surprising, since BTZ solutions are obtained by quotienting the global \AdS{3} spacetime by an isometry. In fact, note  that solutions with
$r_+ = i\, \lads \sqrt{1-\mu}$ with  $\mu \in [0,1)$ describe horizon-free solutions called conical defects. They can be thought of as corresponding to geometries obtained by backreacting a point particle of mass $\propto \mu$. A point mass in three spacetime dimensions has a logarithmic Newtonian potential which impacts the fall-off in the radial direction, but this conspires with the exponential growth of the spatial volume in AdS to produce the above behaviour.

Let us also record that the Euclidean BTZ solution has the metric (setting $r = \lads \,\sinh\rho$):
\begin{equation}
ds^2 = \cosh^2\rho\;  d\tE^2 + d\rho^2 + \lads^2\, \sinh^2 \rho \; d\varphi^2\,.
\label{eq:eBTZ}
\end{equation}
This spacetime has a boundary ${\bf S^1}_{\tE} \times {\bf S^1}_\varphi$, which is  a two-torus ${\bf T}^2$. We would obtain the same solution for  the Euclidean \AdS{3} geometry with the roles of $\tE $ and $\varphi$ interchanged. This is the bulk analog of a modular transformation operation on a  CFT$_2$ on ${\bf T}^2$.

More general spacetimes involving matter can be obtained once we identify the states of the field theory we want to consider. It is particularly useful to know that a gauge-invariant local operator of the field theory maps via the AdS/CFT dictionary to a local bulk matter field. The asymptotics of the field in the geometry $\bulk_{d+1}$ can be mapped to the operators themselves and to the classical sources which couple to them. For instance, a minimally coupled (self-interacting) scalar field of mass $m^2$ in  \AdS{d+1} behaves asymptotically as
\begin{equation}
\phi(z,t,{{\bf x}}) \sim {\cal J}(t,{\bf x})\, z^{d-\Delta} + \vev{{\cal O}(t,{\bf x})} \, z^\Delta \,, \qquad \Delta = \frac{d}{2} + \sqrt{\frac{d^2}{4} + m^2\, \lads^2}
\label{eq:asyphi}
\end{equation}
The two fall-offs are referred to as the non-normalizable $z^{d-\Delta}$ which couples to the classical (non-fluctuating) source and  the normalizable mode $z^\Delta$ picks out the expectation value (in the presence of the source).
This formula is valid for $m^2\, \lads^2 \geq -\frac{d^2}{4}$ which is the Breitenlohner-Freedman bound for stability in \AdS{d+1}. Operators are relevant, irrelevant, or marginal, depending on whether the dual bulk field has negative, positive, or vanishing mass, respectively.\footnote{ There is one subtlety which is worth keeping in mind: for $-\frac{d^2}{4}\leq m^2\,\lads^2 \leq -\frac{d^2}{4}+1$ we both modes turn out to fall-off fast enough to be normalizable. One then can choose to swap the identification of sources and operators leading to what is sometimes referred to as alternate quantization (or Neumann boundary conditions instead of the conventional Dirichlet boundary conditions). This choice is relevant when we want to talk about operators that come close to saturating the unitarity bound in CFT$_d$.}

There is a useful heuristic way to motivate the AdS spacetimes as the geometric duals to CFTs in their vacuum state. Focus on the field theory in Minkowski spacetime with coordinates $(t,{\bf x}_{d-1})$. Scale invariance demands that under spatial and temporal scalings
$t\to \lambda \, t, x_i \to \lambda\, x_i$, the state be invariant. The AdS spacetime geometrizes this; for example,
\begin{equation}
 t\to \lambda \, t, x_i \to \lambda\, x_i \, \qquad z \to \lambda \, z \,,
\label{eq:scales}
\end{equation}
leaves the Poincar\'e metric invariant.  Indeed, the full isometry group of \AdS{d+1} is the group $SO(d,2)$ which is the group of conformal transformations of a CFT$_d$. Eq.~\eqref{eq:scales} furthermore accords the radial coordinate $z$ an interesting interpretation in the field theory \cite{Susskind:1998dq}: it can be viewed as a geometrization of the energy scales in the field theory  leading to the idea of the scale/radius duality. Microscopic scales in the field theory correspond by this dictionary to macroscopic scales in the bulk geometry; UV physics maps to IR physics and vice versa. In particular, a field theory excitation that is well localized on some scales of order $\epsilon$ translates to gravitational excitations that are supported near the boundary of the AdS spacetime, viz., $\rho \to \infty$ in global coordinates \eqref{eq:gads} or $z\to 0$ in Poincar\'e coordinates \eqref{eq:pads}.
 Macroscopic excitations in the field theory will correspond to gravitational effects deep in the interior of the spacetime, which would be confined near the center of global AdS $\rho \to 0$ or the Poincar\'e horizon $z\to \infty$. A localized excitation created in a scale-invariant field theory will expand out as time progresses, distributing its energy on larger and larger spatial scales. This may  simply be interpreted as the gravitational free-fall of a bulk particle under the influence of the attractive AdS potential.

 We will see how this scale/radius relation plays out in various ways in the course of our discussion. In practical applications, any field theory calculation that requires a UV regulator will translate into imposing a geometric IR cut-off in the AdS spacetime. We will for the most part choose to translate a field theory UV cut-off $\epsilon$ into a rigid cut-off $z = \epsilon$ in the bulk geometry.

We will henceforth focus on gravitational theories in \AdS{d+1} as a simple proxy for the holographic correspondence.

%~~~~~~~~~~~~~~~~~~~~~~~~~~~~~~~~~~~~~~~~~~~~~~~
\section{The holographic entanglement entropy}
\label{sec:hrrt}
%~~~~~~~~~~~~~~~~~~~~~~~~~~~~~~~~~~~~~~~~~~~~~~

Now that we have a dictionary between the states of the field theory and asymptotically \AdS{} geometries, we can turn to asking how entanglement entropy is captured holographically. This question was first addressed by Ryu and Takayanagi (RT) in \cite{Ryu:2006bv,Ryu:2006ef} in which they gave a prescription for static time-independent situations. This prescription was subsequently generalized by Hubeny, Rangamani, and Takayanagi (HRT) in \cite{Hubeny:2007xt} to general states, including arbitrary time dependence.

Given a holographic CFT$_d$ on a boundary geometry $\bdy_d$, we want to figure out how to compute the entanglement entropy of a given spatial region $\regA$. We will take this region to lie on some Cauchy slice $\Sigma \subset \bdy_d$, so that we are computing the entanglement entropy at some particular instant in time. Note that, as always, $\Sigma = \regA \cup \regAc$ and the entangling surface is $\entsurf$.

The holographic entanglement entropy prescriptions are very simple to state in the general time-dependent case. Firstly, one is instructed to find a surface $\extrA$ which is a codimension-2 \emph{extremal} surface in the bulk spacetime $\bulk_{d+1}$ anchored on $\entsurf$. By virtue of being extremal, the surface $\extrA$ is a local extremum of the area functional and is subject to the boundary conditions that $\extrA \big|_\bdy = \entsurf$.  Among all such surfaces, for there can be more than one such, we are required to only consider those that satisfy a \emph{homology constraint}. This demands that $\extrA$ is smoothly retractable to the boundary region $\regA$. More precisely, there should exist a spacelike, bulk codimension-1, 
smooth  interpolating surface  $\homsurfA \subset \bulk_{d+1}$
which is bounded by the extremal surface $\extrA$ and the region $\regA$ on the boundary. Finally, among the entire family of extremal surfaces satisfying the homology requirement, we should pick the one that has the smallest area.  The holographic entanglement entropy is then given by the area of this surface in Planck units in a manner similar to the black hole entropy formula of Bekenstein and Hawking. To wit,
\begin{align}
S_\regA = \min_X \frac{\text{Area}(\extrA)}{4\,\GN} \,, \qquad  X =  \extrA:
  \begin{cases}
    & \; \partial \extrA\equiv \extrA\big|_{\partial \bulk} = \entsurf
    \\
    &
    \exists \;\homsurfA \subset \bulk : \partial \homsurfA= \extrA \cup \regA \\
    \end{cases}
\label{eq:HRTprop}
\end{align}

Note that these statements extend in an obvious manner to general situations arising in the string theoretic context. Suppose the dual of QFT on $\bdy_d$ is string theory on $\bulk_d\times\, {\cal Y}$; then we should take $\extrA$ to be a codimension-2 surface in the full spacetime subject to the restrictions above and measure the area in the appropriate higher-dimensional Planck units (e.g., replace $\GN \to G_N^{(10)}$). For direct product geometries this simply extends the statement above to say that our surfaces wrap the internal space ${\cal Y}$ whilst being extremal in $\bulk_{d+1}$. This should be clear in simple examples such as \AdS{5} $\! \times \, {\bf S}^5$. More generally, the compact space ${\cal Y}$ may be non-trivially fibered over the base $\bulk_{d+1}$
and one needs to find a surface that is genuinely extremal in such geometries. For instance, the $\frac{1}{2}$-BPS states of ${\cal N}=4$ SYM which are dual to the LLM geometries \cite{Lin:2004nb} or the microstate geometries of the $D1-D5$ system \cite{Lunin:2002bj} fall into this general class.

While the specification of an extremal surface as the local extremum of the area function is sufficient, it is sometimes useful to give a more geometric characterization of it. We can do this by noting that the surface $\extrA$ is a codimension-2 spacelike surface. Hence, the space transverse to it in $\bulk$ has a timelike and a spacelike normal, i.e., its normal bundle has a local Lorentzian structure of ${\mathbb R}^{1,1}$. In such situations, it is helpful to pass to a basis of null normals by taking appropriate linear combinations. Let the two null normals to $\extrA$ be $N_{(1)}^A$ and $N_{(2)}^A$, respectively. We choose to normalize them by making the choice
\begin{equation}
N_{(1)}^A N_{(2)}^B \, g_{AB} = -1 \,, \qquad N_{(1)}^A N_{(1)}^B\, g_{AB} = N_{(2)}^A N_{(2)}^B\, g_{AB}  = 0\,.
\label{eq:ndef}
\end{equation}
The condition of extremality can be phrased in terms of the extrinsic curvature of these null normals. Define
the projector onto the surface $\extrA$
\begin{equation}
\gamma_{AB} = g_{AB}  + N^{(1)}_A \, N^{(2)}_B  + N^{(2)}_A \, N^{(1)}_B
\label{eq:gamdef}
\end{equation}
through the use of which we can obtain the extrinsic curvature tensors
\begin{equation}
K^{(i)}_{AB} = \gamma^C_A \, \gamma^D_B\;  \nabla_A N^{(i)}_B \,.
\label{eq:ecurvdef}
\end{equation}
The statement of extremality then simply asserts that
\begin{equation}
\gamma^{AB}\, K^{(i)}_{AB} = 0  \qquad \Longrightarrow \qquad K^{(1)} = K^{(2)} =0\,.
\label{eq:extrK0}
\end{equation}
Thus the statement of extremality can equivalently be phrased in terms of the vanishing of the null extrinsic curvatures in the two normal directions to the extremal surface.

In the case where the field theory state is static, or more generally if we consider states at a moment of time reflection symmetry,  then we can more simply focus on \emph{minimal surfaces} $\extrA$ which lie in the bulk on a constant time slice. This was the original RT proposal put forth in \cite{Ryu:2006ef}, while the general prescription given above is the HRT version.

% Figure
\begin{figure}[htbp]
\begin{center}
\includegraphics[width=4.5in]{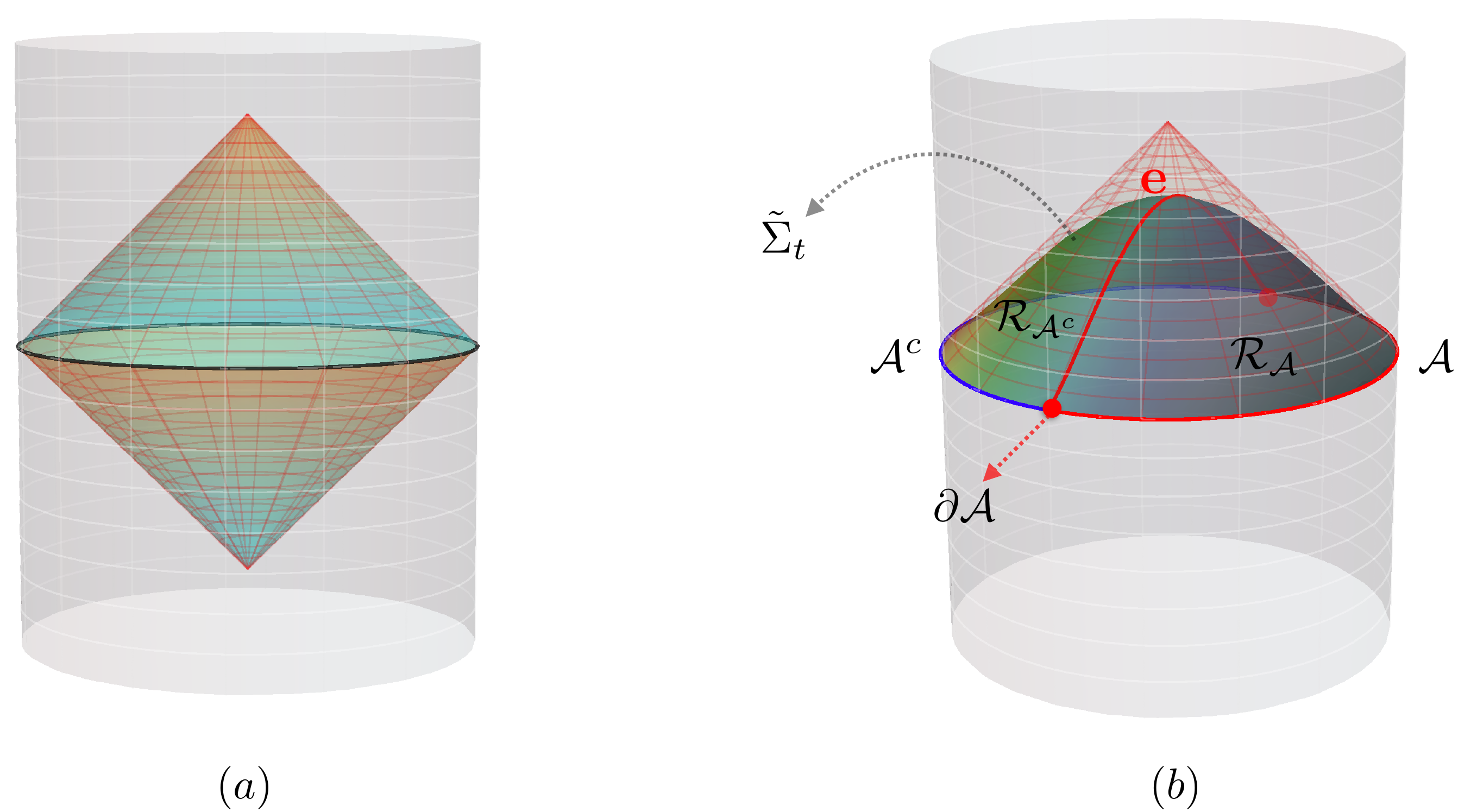}
\caption{(a) Depiction of the FRW wedge in the bulk and (b) a Cauchy slice $\bulkCS$ within that wedge that is divided by the boundary bipartition into two homology surfaces $\homsurfA$ and $\homsurfAc$.
}
\label{f:bulkfrw}
\end{center}
\end{figure}

To appreciate the distinction, note that in general the Cauchy slices  $\bulkCS $ in the bulk are not uniquely determined by a Cauchy slice $\Sigma$ in the boundary geometry $\bdy$. Given $\Sigma\subset \bdy_d$, one can pick any Cauchy surface of the bulk as long as each point on it remains spacelike separated from it. This defines a region which we refer to as the FRW wedge of the bulk spacetime, see Fig.~\ref{f:bulkfrw}. We could in principle expect the extremal surface to lie somewhere in this FRW wedge. We can clearly see that  it cannot lie outside -- for if it did, then it would be timelike related to the boundary Cauchy surface, which would lead to an inconsistency.\footnote{ As described in \cite{Headrick:2014cta}, one can use known causal features of entanglement entropy to argue that it lies within a rather restricted region of the FRW wedge called the causal shadow, see later.}

However, when we have a timelike Killing field of the boundary geometry which is respected by the CFT state then one can canonically extend the boundary Cauchy slice into the bulk as the corresponding equal time surface.  Heuristically, once we have  a unique mapping from a constant time slice of the boundary to a corresponding one in the bulk, we can argue that the entanglement entropy should be computed by a surface on that slice itself. Indeed, separating the null extrinsic curvatures into a basis of timelike and spacelike normals, we learn that \eqref{eq:extrK0} can be decomposed differently into $K^{(t)} = K^{(s)} = 0$.  If we have a bulk Cauchy slice that respects time reflection symmetry, then $K^{(t)}_{AB}  =0$ altogether, so we only need to find a surface on $\bulkCS$ which satisfies $K^{(s)} =0$. The reader can then convince themselves that this condition is equivalent to the Riemannian problem of finding minimal surfaces.

Now we have emphasized that the $c_\text{eff} \to \infty$, planar limit reduces the boundary QFT to an effective classical field theory. The astute reader may be left pondering why then are we describing an intrinsic quantum feature such as entanglement in this limit. A quantum state in the field theory Hilbert space has some spatially ordered entanglement, which can be explored in an  asymptotic expansion in $c_\text{eff}^{-1}$. When there is a macroscopic contribution at ${\cal O}(c_\text{eff})$, this tends to dominate the limit, and is  captured by a saddle point analysis. The RT/HRT proposals only capture this leading order contribution to the entanglement in the form of  geometric data. Contributions of ${\cal O}(1)$ are not described by bulk geometry, but rather require one to study the bulk entanglement entropy, as we shall explicitly see.

While the RT proposal was suitably covariantized by the HRT prescription of upgrading minimal surfaces to extremal surfaces, for certain considerations the description in terms of extremal surfaces proves sub-optimal. Aron Wall \cite{Wall:2012uf} came up with a reformulation of the HRT proposal in terms of a maximin construction, which provides a nice complementary perspective. We will see its great utility in later discussion when we prove the holographic entropy inequalities.

The construction proceeds as follows:  given a boundary region $\regA$, we pick a bulk Cauchy slice $(\bulkCS)_{guess}$ such that $\partial( \bulkCS)_{guess}= \regA \cup\regA^c = \Sigma$. On this slice we find a minimal surface; call it $\mathbf{m}_{guess}$. Note that this is a well-defined boundary value problem in Riemannian geometry, since $(\bulkCS)_{guess}$ is spacelike. One then varies the choice of the bulk Cauchy slice, finding minimal surfaces on the entire family. We can imagine this construction by filling out the FRW wedge of $\Sigma$ with Cauchy slices containing (spacetime codimension-2) minimal surfaces  drawn on them. One is instructed to compute the area of the minimal surface on each slice and take the one from this (infinite) family with the maximum area. This is the {\em maximin surface}. To wit,
\begin{equation}
{\cal E}_\text{maximin}^\regA = \mathbf{m}_{*} \in \{ \textbf{m}_{guess} \subset (\bulkCS)_{guess} \; \text{is minimal}\} \;\; \& \;\; \text{Area}{(\mathbf{m}_*)}  \;\; \text{is maximal}
\label{}
\end{equation}
While it is not obvious from this definition it transpires that a maximin surface is in fact an extremal surface ${\cal E}_\text{maximin}^\regA  = \extrA$. Note that there isn't a unique Cauchy surface that contains the extremal surface. As is hopefully clear from Fig.~\ref{f:bulkfrw}b, any spacelike slice of the FRW wedge that is pinned at $\regA \cup\regAc$ on the boundary and passes through the extremal surface $\extrA$, provides a Cauchy slice on which $\extrA$ is a minimal area surface. Since the slices are all pinned at $\extrA$ we don't see any temporal variation and the maximin condition enforces that that the surface is truly extremal in the full spacetime. For details of the proof, we refer the reader to \cite{Wall:2012uf}.

We now have at our disposal the holographic entanglement entropy prescriptions. We will first take note of their derivation, and thence proceed in subsequent sections to analyze the physical consequences of these ideas.

%~~~~~~~~~~~~~~~~~~~~~~~~~~~~~~~~~~~~~~~~~~~~~~~
\chapter{Deriving holographic entanglement proposals}
\label{sec:derivation}
%~~~~~~~~~~~~~~~~~~~~~~~~~~~~~~~~~~~~~~~~~~~~~~

The holographic entanglement entropy proposals described in \S\ref{sec:hrrt} were first inspired by drawing an analogy with black hole entropy. While one can argue that the various known  properties of entanglement entropy are satisfied by the holographic construction, this per se does not pin down a precise proposal. Furthermore, it does not explain how the
prescription for the computation of entanglement entropy relates to the dynamics of the gravitational theory in the bulk. For instance, we gave the prescription in \S\ref{sec:hrrt} for Einstein-Hilbert gravitational dynamics -- one would like to know how to take into account the finite $\alpha_k$ corrections as in \eqref{eq:gravact}.

These issues have been addressed in the literature over the past few years and we now have a reasonable understanding of the origins of the RT/HRT proposals and generalizations thereof. We will below give a  description of the elements of the proof in the context of Einstein-Hilbert gravity, and indicate the various generalizations.

Before we get into the details of the argument, let us record here some of the significant attempts in the literature to address a derivation of the proposals. The first attempt at a proof was provided in \cite{Fursaev:2006ih}. Here it was realized that the branched cover construction of $\bdy_q$ from $\bdy$ can be viewed in terms of a conical singularity on the boundary. If one makes the naive assumption that this boundary conical singularity extends into the bulk trivially, then an evaluation of the action on such a singular solution leads to the RT formula. However, this construction fails to respect the rules of the AdS/CFT dictionary. Given suitable boundary conditions, we are required to find bulk solutions which are consistent solutions to the gravitational equations of motion. The naive ansatz with the singularity extended into the bulk  does not satisfy Einstein's equations. This  was first explained in the analysis of \cite{Headrick:2010zt}, whose author went on to provide additional evidence for the RT formula.

A major step towards a proof was provided by the analysis of Casini, Huerta, Myers (CHM) \cite{Casini:2011kv}, who focused on spherically symmetric domains in the vacuum state of a CFT. Utilizing two facts, (a) the knowledge of the modular Hamiltonian for Rindler space, and (b) a conformal map relating the Rindler wedge to the domain of dependence of the circular region, they were able to argue that the reduced density matrices for such configurations are equivalent up to  a unitary transformation to a thermal density matrix. This enabled them to use the standard dictionary between thermal physics and black holes to determine the entanglement entropy in terms of the Bekenstein-Hawking entropy of the black hole. Recall that the latter is given for stationary black holes as the area of the  bifurcation surface of the horizon (the locus where the null Killing generator vanishes). The bifurcation surface turns out to satisfy the extremality condition, and thus one can see how the RT prescription could arise from this sequence of relations.

Subsequently, Lewkowycz and Maldacena (LM) \cite{Lewkowycz:2013nqa} gave a local version of this argument, described below, which allowed them to derive the RT proposal.\footnote{ The LM construction was originally generalized towards understanding higher derivative corrections in \cite{Dong:2013qoa,Camps:2013zua}.}  Recently \cite{Dong:2016hjy} have extended this argument to provide a derivation for the  HRT proposal. The distinction is that, in the covariant analysis, one has to employ  elements of the Schwinger-Keldysh formalism, as should be clear from the field theory discussion in \S\ref{sec:pi}.

%~~~~~~~~~~~~~~~~~~~~~~~~~~~~~~~~~~~~~~~~~~~~~~~
\section{Deriving the RT proposal}
\label{sec:rtderive}
%~~~~~~~~~~~~~~~~~~~~~~~~~~~~~~~~~~~~~~~~~~~~~~

The key step in deriving the RT proposal is the basic entry into the AdS/CFT dictionary which says that ,in the semiclassical limit provided by large $\ceff$, the bulk geometry is a saddle point configuration of the string theory path integral with prescribed boundary conditions; recall the discussion around Eq.~\eqref{eq:bulksaddle}.
The computation of R\'enyi entropies of quantum field theories has the advantage of being phrased in familiar geometric terms, described for instance in \S\ref{sec:pi}, which we now exploit to determine the bulk duals. It will be clear from the analysis that what we can access from the RT proposal is the macroscopic part of entanglement entropy, which is given in terms of a classical solution in the limit $c_\text{eff} \to \infty$. We will subsequently describe how to think about corrections to the result.

 Given a field theory on $\bdy$ with a chosen region $\regA$ lying on a Cauchy slice, the computation of the $q^{\rm th}$ R\'enyi entropy necessitates a replica boundary geometry,  $\bdy_q$, which is a $q-$fold `branched cover' over the original manifold $\bdy$. The branching is along the codimension-2 entangling surface $\entsurf$.  Given suitable boundary conditions for the fields on the branched cover geometry, we have a path integral prescription for the R\'enyi computation. Our strategy will be to evaluate this path integral by passing over to the bulk string theoretic path integral and use the aforementioned semiclassical approximation to evaluate the result as the on-shell action of a  gravitational saddle point geometry.

We first focus on time-independent static states in which the boundary geometry $\bdy$, and thence the replica space $\bdy_q$, both have a timelike Killing field $\partial_{t_{_E}}$. This is a useful starting case to consider as it allows us to analytically continue the bulk path integral from Lorentzian to Euclidean signature, which has some advantages. For one, we have a reasonable understanding of the Euclidean Quantum Gravity path integral, despite certain subtleties involved in its definition and evaluation (mostly due to the wrong sign for the conformal mode, which fortunately will not play a role in our analysis). For another, the bulk boundary conditions we need to impose are sufficiently straightforward to state and implement, both at the topological and geometric levels.

We are required to find a bulk manifold $\bulk_q$ which has as its boundary $\bdy_q$. This provides a well-posed boundary value problem for the Euclidean quantum gravity path integral. Once we have the solution, we should evaluate the on-shell action to obtain the saddle point value of the R\'enyi entropy.

Let us make these a bit more precise. We will work in the $\ceff \gg 1$ limit and simply take the bulk theory given by Einstein-Hilbert gravity. Thus we approximate, for the partition function of the string theory
\begin{equation}
\begin{split}
\mathscr{Z}_\text{string}
&=
  \int [D\Phi_\text{string}] e^{-S_\text{string}} \\
&
  \stackrel{c_\text{eff}\to \infty}{\approx}
  \int [Dg] \, \exp\left(-\;\frac{1}{16\pi\,\GN} \; \int_\bulk\, d^{d+1}x\, \sqrt{g}\, \left[R
  + \frac{d(d-1)}{\lads^2}\right] \right)
\end{split}
\label{eq:Zstr}
\end{equation}
We proceed to evaluate the r.h.s. in the saddle point approximation and write:
\begin{equation}
\mathscr{Z}_\text{string} [\bulk_q] \approx e^{-I[\bulk_q]} \,,
\label{}
\end{equation}
with $\bulk_q$ being a stationary point of the Einstein-Hilbert action, i.e., it solves \eqref{eq:bulksaddle} with the boundary condition $\partial\bulk_q = \bdy_q$. Given this on-shell action ,we can evaluate the R\'enyi entropy as
\begin{equation}
\begin{split}
S^{(q)} &=  \frac{1}{1-q}\, \log \left( \frac{\tr{(\rhoA^q)}}{(\tr{\rhoA})^q} \right)  \\
&
  = \frac{1}{1-q} \, \log\left(\frac{\mathcal{Z}[\bdy_q]}{\mathcal{Z}[\bdy]^q} \right) \\
&
  \approx \frac{1}{1-q} \,
  \log\left(\frac{ \mathscr{Z}_\text{string}[\bulk_q] }{\mathscr{Z}_\text{string}[\bulk_1]^q } \right) \\
&
  = \frac{1}{1-q}  \left( \log \mathscr{Z}_\text{string} [\bulk_q]  -q \log\mathscr{Z}_\text{string} [\bulk_1]  \right)
\end{split}
\label{}
\end{equation}
where $\bulk_1 = \bulk$ is the asymptotically AdS manifold with $\partial \bulk = \bdy$. The factor of $q$ in the difference arises from the normalization of the reduced density matrix.

We are thus far assuming $q\in {\mathbb Z}_+$. To get the entanglement entropy though, we will want to analytically continue to non-integral values of $q$. The major insight of Lewkowycz-Maldacena \cite{Lewkowycz:2013nqa} was to argue that the analytic continuation is simpler in the gravitational context. The argument can be distilled into two separate components:
\begin{itemize}
\item a kinematical part which provides the essence of how to implement the analytic continuation, and
\item a dynamical part wherein one actually ensures that the ansatz chosen satisfies Einstein's equation.
\end{itemize}
Once we solve these two, we should evaluate the contribution of the saddle point configuration to obtain $S^{(q)}$. In practice it will be easier to evaluate the modular entropy defined in \eqref{eq:modent} directly in gravity.

\paragraph{1. Kinematics:} The construction of the replica manifold $\bdy_q$ as a $q$-fold branched cover over $\bdy$ (branched at $\entsurf$) comes equipped owing to the cyclicity of the trace with a replica ${\mathbb Z}_q$ symmetry. This symmetry basically shuffles the individual copies of $\bdy$ in $\bdy_q$; we can furthermore take a quotient  $\bdy_q/{\mathbb Z}_q$ which is topologically equivalent to $\bdy$ itself (see Fig.~\ref{f:branch3}. We refer to the quotient space as the fundamental domain of the branched cover.

 The key assumption one makes in the argument of \cite{Lewkowycz:2013nqa} is to extend the replica ${\mathbb Z}_q$ symmetry  into the bulk. The bulk spacetime $\bulk_q$ has to be obtained by solving the field equations. We are thus restricting attention to those spacetimes that  admit a natural ${\mathbb Z}_q$ action inherited from the boundary conditions. Let us then consider the bulk quotient space ${\hat \bulk}_q = \bulk_q/{\mathbb Z}_q$.
  The replica symmetry will not act  smoothly in the bulk and thus ${\hat \bulk}_q $ will contain some singularities.
  While we are assuming that the bulk solution $\bulk_q$ admits a ${\mathbb Z}_q$ action, the symmetry does not have to act smoothly.  ${\hat \bulk}_q$ could and in general does contain ${\mathbb Z} _q$ fixed points. These are singularities which are typically of the orbifold type. A crucial assumption one makes in the construction is that the singular locus in
${\hat \bulk}_q $ is codimension-2 in the spacetime -- we will call this surface $\fixM_q$.

Intuitively, this boundary condition is natural. The boundary conditions involve a spacetime manifold branched over $\entsurf$, which is a codimension-2 surface in $\bdy$. The action of ${\mathbb Z}_q$ in the bulk, inherited from the boundary, should act so as to extend this. Locally near the boundary one expects therefore to see that
$\entsurf$ gets extended into a part of the bulk singular locus.  Hence one anticipates that $\fixM_q$ is a natural analog of $\entsurf$ in the bulk since the boundary conditions of the problem demand that it be anchored on $\entsurf$.

The argument above, whilst intuitive, is unfortunately too local and only valid in the near-boundary region of $\bulk_q$. It is indeed possible to conjure examples \cite{Haehl:2014zoa} in which the fixed point set in ${\hat \bulk}_q$ is not codimension-2. The question of when this happens can be addressed purely through topological considerations without detailed reference to the gravitational dynamics. The results of the aforementioned paper show that as long as a one has a family of replica symmetric geometries parameterized by some $q$, which furthermore are smooth for $q \in {\mathbb Z}_+$, then one can  extend the local statement of the singular locus to a global statement. We will assume this henceforth and take $\fixM_q$ to be a codimension-2 surface of ${\hat \bulk}_q$.

Having made these assumptions, we are now in a position to set up the gravitational problem. We will work in a single fundamental domain of the quotient space  ${\hat \bulk}_q$, which we have seen is topologically  isomorphic to the original bulk spacetime $\bulk$.  However, these two spacetimes have vastly different geometries (for example,  they have different metrics) owing to the boundary conditions. This difference can be accounted for by the singular locus $\fixM_q$. This codimension-2 surface can be treated as a source of energy-momentum which backreacts on the spacetime $\bulk$ to deform it to ${\hat \bulk}_q$. To ensure that  we have the correct geometry, we must require appropriate boundary conditions at the fixed point locus itself, for $\fixM_q$ is not a generic singularity but one that arises from a smooth spacetime $\bulk_q$ by an orbifold construction. The fact that we are taking an orbifold of a $(d+1)$-dimensional spacetime to get a codimension-2 singular locus suggests that the singular locus should be treated as a cosmic brane which carries a tension
\begin{equation}
T_q = \frac{1}{4\GN}\, \frac{q-1}{q}
\label{eq:ctesion}
\end{equation}
where we have reinstated all the factors of Newton's constant appropriate for a source of energy density localized in a codimension-2 surface of the spacetime.

The claim then is that we can compute the geometry of ${\hat \bulk}_q$, and thence $\bulk_q$, by starting with
$\bulk$ with the codimension-2 cosmic brane with the above value of tension. We solve Einstein's equations
\eqref{eq:bulksaddle} with $T^{\tiny{\text{matter}}}_{AB}$ arising from the cosmic brane tension. Having determined the solution for ${\hat \bulk}_q$ we compute the on-shell action of this part of the spacetime and exploit the locality of the gravitational action to infer that the action contribution of $\bulk_q$ should be $q$ times that of a single domain, viz.,
\begin{align}
I[\bulk_q] = q\, I[{\hat \bulk}_q]
\label{eq:lmI1}
\end{align}
While the quotient space has a conical singularity with defect angle $\frac{2\pi}{q}$, the covering space $\bulk_q$, we re-emphasize, is smooth; this observation will play a crucial  role in setting up the boundary conditions.

The advantage of the above manipulations becomes manifest when we have to consider analytic continuation in $q$ for purposes of computing entanglement entropy. In the gravitational computation involving the cosmic brane, the parameter $q$ simply appears as the tension of the brane. This suggests that we can compute R\'enyi entropies for non-integral values of the index by suitably tuning the cosmic brane tension.

This line of thought brings with it a very helpful bonus. We can separate the deformation of the geometry into two parts: tangential and normal to the cosmic brane. Let us adapt coordinates to the cosmic brane, whose worldvolume we parameterize by coordinates $y^i$ with $i=1,2,\cdots, d-1$. The normal directions will be coordinatized by $\{t_{_E} ,x \}$ since we are  still working in Euclidean space. In the local neighbourhood of the cosmic brane, we can adapt to Gaussian coordinates so that the metric can be written in the canonical form:
\begin{equation}
ds_{_ E}^2=dx^2+d\tE^2+\left(\gamma_{i j}+2\, K_{ i j}^x\, x+2\, K_{ i j}^t\, \tE \right)dy^i\, dy^j+ \cdots \,.
\label{eq:lm0}
\end{equation}
We have retained only the leading terms in the Taylor expansion about the surface located at $x=0, \tE=0$.  To this order, the Gaussian coordinate chart only sees the extrinsic curvature of the codimension-2 surface embedded in spacetime. Working to higher orders  would entail keeping track of the curvature contributions, as in the usual Riemann normal coordinates.

% Figure
\begin{figure}[htbp]
\begin{center}
\includegraphics[width=4.5in]{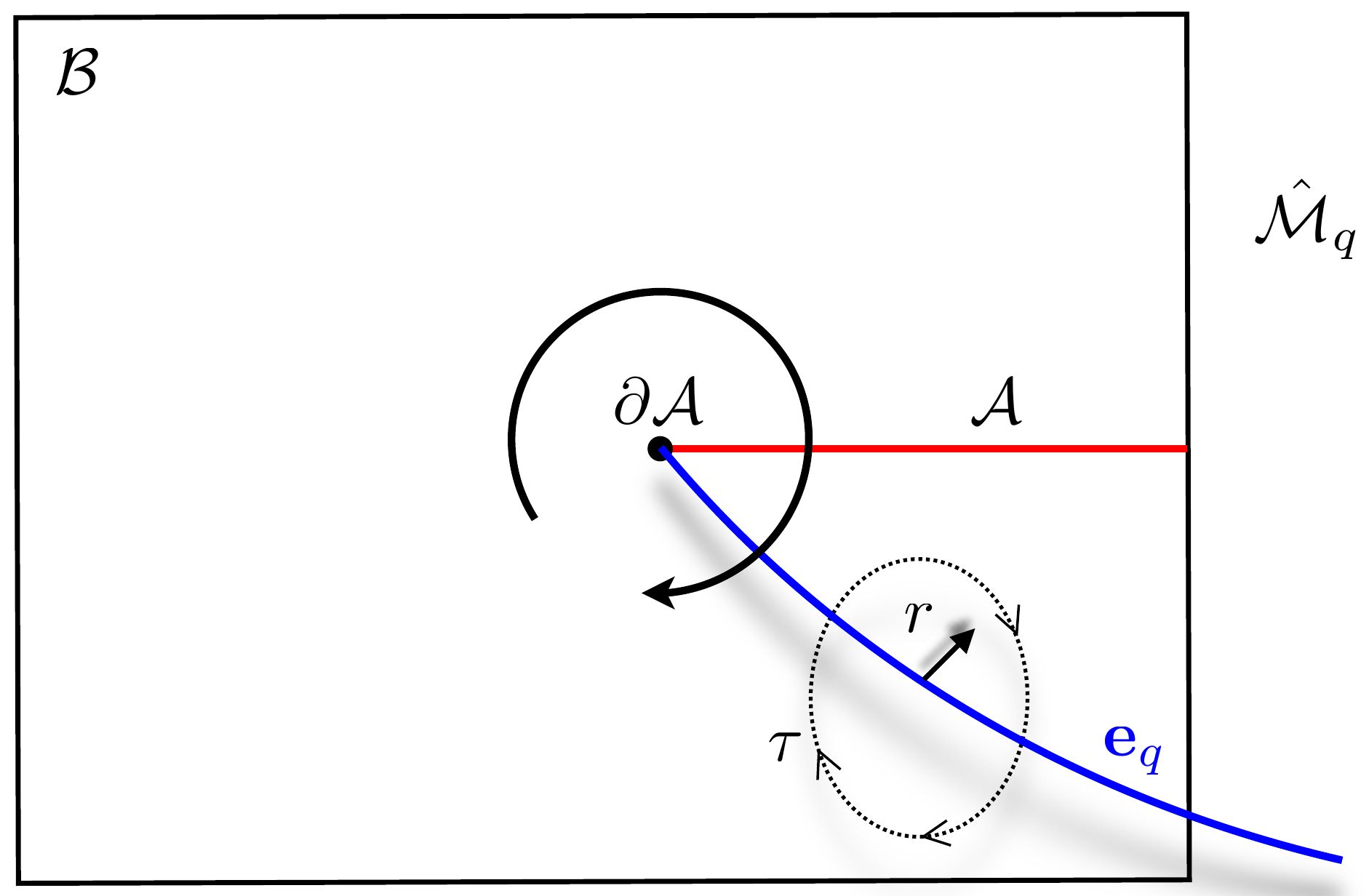}
\caption{The local geometry in the vicinity of the fixed point set $\fixM_q$ which extends out from the entangling surface. We have chosen to parameterize the normal plane to this set in polar coordinates \eqref{eq:lm2} with $r$ being the radial direction away from the fixed point and $\tau$ the angular coordinate that circulates around the codimension-2 locus.
}
\label{f:bdycondq}
\end{center}
\end{figure}

\paragraph{2. Dynamics:} Having set up the basic problem in the gravitational context, we now want to figure out what configurations dominate and thence compute their on-shell action.  To enforce the boundary conditions in the gravitational solution, let us examine the metric close to $\fixM_q$ in polar coordinate
 $x \pm i\,\tE= r \,e^{\pm i \,\tau}$.  The replica  ${\mathbb Z}_q$ symmetry implies that the action is invariant with respect to a global shift of the polar coordinate in the normal plane $\tau$, viz., $\tau \to \tau + 2\pi$. This feature is illustrated in  Fig.~\ref{f:bdycondq}. On the other hand, as we approach $\fixM_q$, the coordinate $\tau$ has to traverse through all the replica copies before reverting back to itself, i.e., it should be identified under $\tau \sim \tau + 2\pi\,q$. Using the global smoothness of the saddle point covering space geometry $\bulk_q$, we infer that the local spacetime near $\fixM_q$ in the quotient ${\hat \bulk}_q$ has be of the form\footnote{  This is heuristic, as the geometry is a nontrivial fibration of the normal bundle parameterized by the $(r,\tau)$ over the codimension-2 base.}
\begin{equation}
ds^2 =
         \left( q^2 \, dr ^2+ r^2 \, d\tau^2\right)+
        \left(\gamma_{i j}+2\, K_{ i j}^x\, r^q\cos \tau+2\, K_{ i j}^t\, r^q \sin \tau \right) dy^i \, dy^j  + \cdots
\,,
\label{eq:lm2}
\end{equation}
eliding over higher order terms. We wish to draw attention to the explicit $q$ dependence. Its presence implies that in order for  the metric to be smooth near $r=0$, we must encounter some non-trivial backreaction;  one cannot simply identify  $\tau \sim \tau+2\pi q$ in \eqref{eq:lm0}. The dependence on the normal coordinates, and in particular the factors of $r \, e^{\pm i\tau}$, are easily determined  by looking at which of the local mode solutions are smooth. The admissible solutions behave as   $(r^q \, e^{i\,\tau})^{\pm \omega }$ and
$(r^q \, e^{i\,\tau})^{\pm i\,\omega }$ in the vicinity of $r=0$.

Once we have an ansatz, we should simply compute the field equations to discern when they would be satisfied.
Evaluating the curvatures for the geometry \eqref{eq:lm2}, we find divergent contributions proportional to  $(q-1)\, \frac{K^a}{r}$ where $K^a \equiv K^a_{ij}\,\gamma^{ij}$ is the trace of the extrinsic curvature. Examining potential higher order terms, one learns that none of these can help compensate this contribution. The only way for the equation of motion to be satisfied by the ansatz \eqref{eq:lm2} is for the extrinsic geometry of $\fixM_q$ to be determined by the leading order analysis in the distance away from the locus.
It then follows that the set of admissible codimension-2 surfaces are required to have a vanishing trace of the extrinsic curvature in the normal directions!
Since we have a $t \rightarrow -t$ symmetry, we have trivally $K^t=0$ and one thus derives the minimal surface condition of \cite{Ryu:2006bv}:
\begin{equation}
\lim_{q\to 1} \,\fixM_q  \to \extrA \,, \qquad \extrA \in \bulk \; \text{with}\;\;  t=0, \;K^x =0\,.
\label{eq:lmrt}
\end{equation}

\paragraph{3. The on-shell action:}  Remarkably, a local analysis around the fixed point serves to determine the RT proposal involving minimal surfaces. Having obtained the right surface, we need to determine the on-shell action and see that the result for the von Neumann entropy is indeed given by the area formula \eqref{eq:HRTprop}. A-priori  this is a global computation, which depends on the solution everywhere. However,  diffeomorphism invariance ends up localizing the
result for the modular entropy to a codimension-2 surface.

There are many ways to do the computation, but one that is particularly useful is to employ an argument based on the covariant phase space approach in gravitational theories \cite{Iyer:1994ys}. In fact, as originally explained in \cite{Lewkowycz:2013nqa} and recently elaborated upon by \cite{Dong:2016fnf}, one can compute more readily the derived quantity $\partial_q I[\hat{\bulk}_q]$ for any value of $q$.  This directly leads to the modular entropy $\tilde{S}^{(q)}$; this turns out always to  localize  onto a codimension-2 surface, while $I[\hat{\bulk}_q]$ does indeed necessitate an integration over the entire manifold.  The end result will be that we have a geometric result for the modular entropy which will limit to the correct  von Neumann entropy in the limit  $q \to1$.

The main idea involves viewing the derivative with respect to $q$,  $\partial_q$, as a change in the bulk solution (and its boundary conditions). Standard variational calculus says that any variation of a classical action can be written as a combination of the equations of motion and boundary terms (using integration by parts where necessary). In gravity this takes the form:
 \begin{equation}
 \delta I[\hat{\bulk}_q]=\int_{\bulk_q} \left[\text{E}^{AB} \delta (g_q)_{AB}+d \Theta((g_q)_{AB},\partial_q (g_q)_{AB}) \right] \,.
 \label{eq:var0}
 \end{equation}
 where the boundary terms have be collected into a symplectic potential $\Theta$. We will review this formalism in some detail in \S\ref{sec:waldi}.

For a typical variation that appears in a standard AdS/CFT calculation, \eqref{eq:var0} would evaluate to a term at the asymptotic boundary  $\partial \bulk_q =\bdy_q$, as long as it does not have an internal boundary to the spacetime.
 However,  we wish to consider the variation of $q$, which instead changes the boundary condition near the fixed point set $\fixM_q$. For the choice $\delta g_{AB}=\partial_q g_{AB}$, the variation satisfies $\partial_q (g_q)_{AB} \big|_{\bdy_q}=0,\;\; \partial_q (g_q)_{AB}\big|_{\fixM_q} \not = 0$. Thus the change engendered by the replica index variation is localized at the fixed point locus and has no contribution from the asymptotic boundary of the spacetime.  Accordingly we should encounter a contribution that is localized on the fixed point locus.

Rather than evaluate the contribution from the fixed point $\fixM_q$, let us consider regulating the singularity. We excise a tubular neighbourhood of size $\epsilon$ around the locus and denote by $\fixM_q(\epsilon)$ the codimension-1 surface bounding this neighbourhood.
 One may therefore write
 \begin{equation}
 \partial_q I[\hat{\bulk}_q]=\int_{\fixM_q(\epsilon)} \Theta((g_q)_{AB},\partial_q (g_q)_{AB}) \,.
 \label{eq:var1}
\end{equation}
We now have to evaluate the symplectic potential on the solution and then take away the regulator by sending $\epsilon \to 0$. In this fashion it is clear that the result will indeed be a local functional of geometric data on the fixed point $\fixM_q$.

In the present case, we won't actually evaluate this integral; it can actually be done given the symmetries. There is a faster and equivalent way to the answer, wherein we simply pretend that $\fixM_q(\epsilon)$ is a physical codimension-1 boundary. In that case, we would have to prescribe boundary conditions for the gravitational fields, to make sure that the Einstein-Hilbert action in the second line of \eqref{eq:Zstr} gives the correct equations of motion under variations. The standard boundary terms which ensures this, is the Gibbons-Hawking functional, given in terms of the extrinsic curvature of the boundary. For the present case, this artificial boundary condition would involve a contribution of the form
\begin{equation}
I_{bdy}[\hat{\bulk}_q]=\frac{1}{8 \pi \GN}\int_{\fixM_q(\epsilon)} \, {\cal K}_\epsilon \,,
\label{}
\end{equation}
at the blown-up singular locus. Here $\mathcal{K}_\epsilon$ is the trace of the extrinsic curvature of the codimension-1 surface $\fixM_q(\epsilon)$ (for Einstein-Hilbert dynamics). It is much simpler to evaluate this quantity and subsequently remove the cut-off. The result we seek is then
\begin{equation}
\partial_q I[\hat{\bulk}_q]=-\partial_q I_{bdy}[\hat{\bulk}_q]
\,, \qquad
\label{eq:var2}
\end{equation}

Working in the local coordinates \eqref{eq:lm2} in an open neighbourhood of $\fixM_q$, one finds  ${\cal K}_\epsilon=\frac{1}{q \,\epsilon}$, and thus we get the result for the modular entropy
\begin{equation}
\tilde{S}^{(q)}  = \partial_q I[\hat{\bulk}_q]= \frac{\text{Area}(\fixM_q)}{4\, q^2 \GN}
\label{eq:renderq}
\end{equation}
which, as $q \rightarrow 1$, gives us the RT formula.  In obtaining the final answer, we used the variation of the metric
\eqref{eq:lm2} at $\fixM_q$; it is given to be $g^{r r} \partial_q g_{r r} \big|_{\fixM_q}=\frac{2}{q}$ and vanishes for the other components.

The orbifold picture allows us to analytically continue the on-shell action $I[\bulk_q]$ to non-integer $q$. The physical interpretation of the (parent space) solution for non-integer $q$ is unclear, but these geometries are just an intermediate step to compute the action.

%~~~~~~~~~~~~~~~~~~~~~~~~~~~~~~~~~~~~~~~~~~~~~~~
\section{Deriving the HRT prescription}
\label{sec:covgen}
%~~~~~~~~~~~~~~~~~~~~~~~~~~~~~~~~~~~~~~~~~~~~~~

Recently, a bulk derivation of the covariant HRT prescription was given in \cite{Dong:2016hjy}. We will now give a brief discussion of the salient features of their argument. It is also worth noting that attempts to understand this proposal led various authors to test elements of its consistency  with the general expectations in QFT over the years. We will review some of these when we discuss properties of the holographic entanglement entropy in \S\ref{sec:heeprops}.

The key issue we have face up to is that in genuine time-dependent circumstances, we cannot invoke the trick of passing to a path integral over an Euclidean manifold.\footnote{ In the absence of time-reflection symmetry, the analytic continuation of $t \to i\,\tE$ will lead to a complex manifold. Moreover we cannot in general assume that we can analytically continue, for we could involve physical non-analytic time-dependent sources.}
In the boundary field theory, we have already indicated in \S\ref{sec:pi} the necessary changes one needs to incorporate into the replica construction using the Schwinger-Keldysh  path integral construction. We evolve  from the initial state up until the moment of interest, say $t$, and then retrace our footsteps back to the far past. This forward-backward evolution induces a kink at the Cauchy slice $\CSA\equiv \regA \cup \regAc$  on the boundary $\bdy$, as we only retain the part of the geometry to its past $J^-[\CSA]$.

% Figure
\begin{figure}[tp!]
\begin{center}
\includegraphics[width=4.2in]{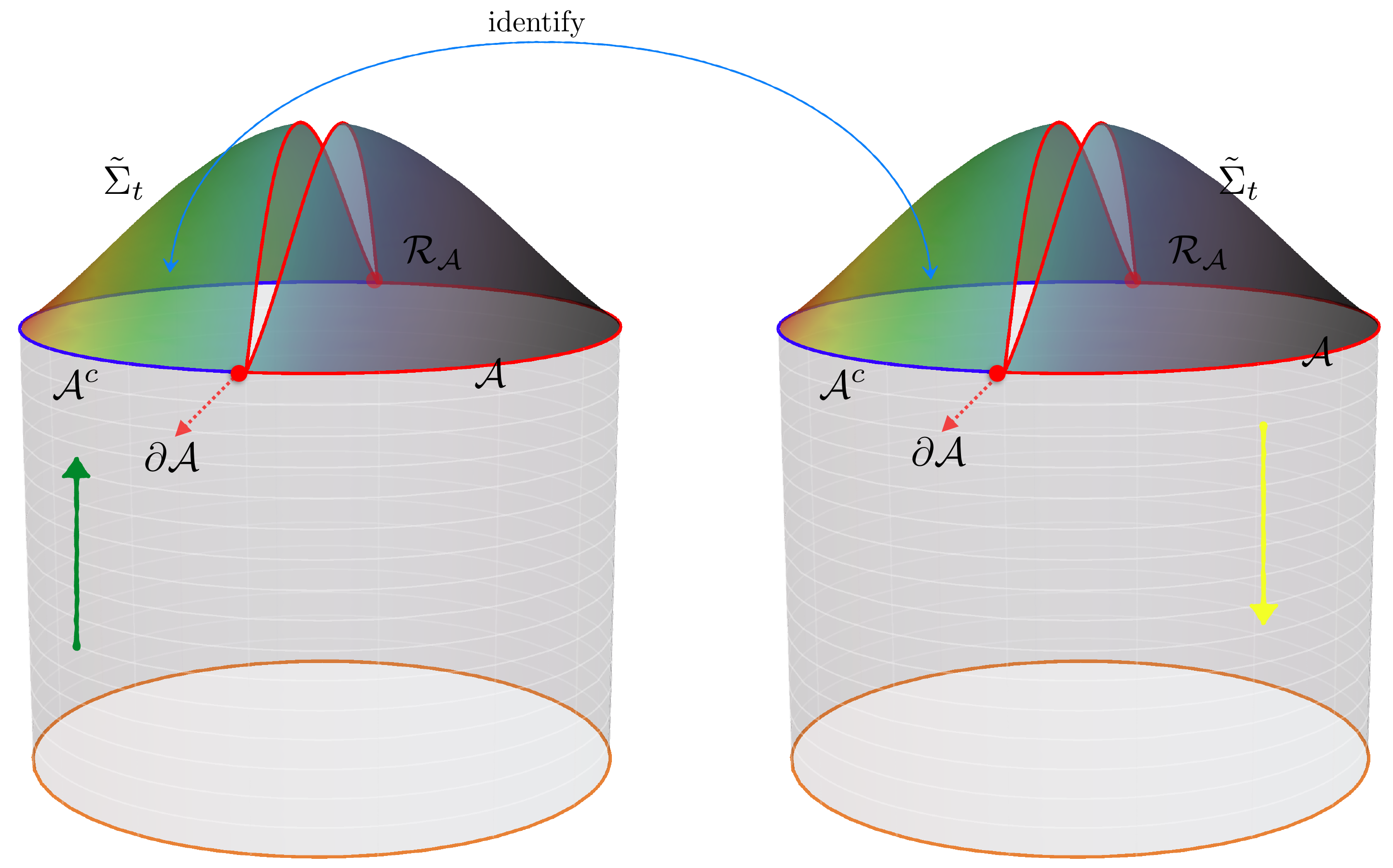}
\caption{The bulk construction of the reduced density matrix elements involves two copies of the spacetime in question, which are glued across the part of the bulk Cauchy slice associated with $\regAc$. With these parts identifies one is free to prescribe different boundary conditions for the fields in $\homsurfA$ for the forward and backward parts of the evolution to obtain the matrix elements of $\rhoA$ in the gravitational construction. Further gluing the two copies of spacetime along $\homsurfA$ will lead to the evaluation of $\Tr{}{\rhoA}$.
}
\label{f:bulksk}
\end{center}
\end{figure}

The question then is how to extend this field theory construction in the holographic context. A prescription for extending field theory Schwinger-Keldysh contours into the bulk gravitational theory was developed in
\cite{Skenderis:2008dh,Skenderis:2008dg}. The idea is to consider in the bulk an analogous fold along some Cauchy slice $\bulkCS$, with the proviso that the bulk evolution will proceed only in the part of the spacetime to the past of $\bulkCS$, i.e., in $\bulkJ^-[\bulkCS]$.\footnote{ We will use a tilde to distinguish bulk Cauchy surfaces and causal sets  from analogous quantities on the boundary.} In other words, the initial conditions are evolved forward from $t=-\infty$ up to $\bulkCS$ and then we evolve back to construct the bulk Schwinger-Keldysh contour. This forward-backward evolution through $\bulkCS$, across which two copies of the bulk manifold are glued together, is illustrated in Fig.~\ref{f:bulksk}. On the Cauchy slice as we reverse the evolution,  we have to provide appropriate boundary conditions.

Once we understand the interpretation of the Schwinger-Keldysh in the bulk gravitation theory, the rest of the argument splits up naturally into two steps. We first use the Schwinger-Keldysh replica trick to build the geometry $\bulk_q$ for computing $(\rhoA)^q$, which would involve gluing $2q$ copies of the bulk spacetime in a replica symmetric fashion. However, on this geometry we still have a natural action of the ${\mathbb Z}_q$ replica symmetry. By taking a quotient the replica spacetime $\bulk_q$ with this symmetry, we can  construct the orbifold spacetime $\hat{\bulk}_q = \bulk_q/{\mathbb Z}_q$ as before with a conical defect $\fixM_q$. The analysis then boils down to figuring out how the conical defect affects the bulk equations of motion.

% Figure
\begin{figure}[htbp]
\begin{center}
\includegraphics[width=3in]{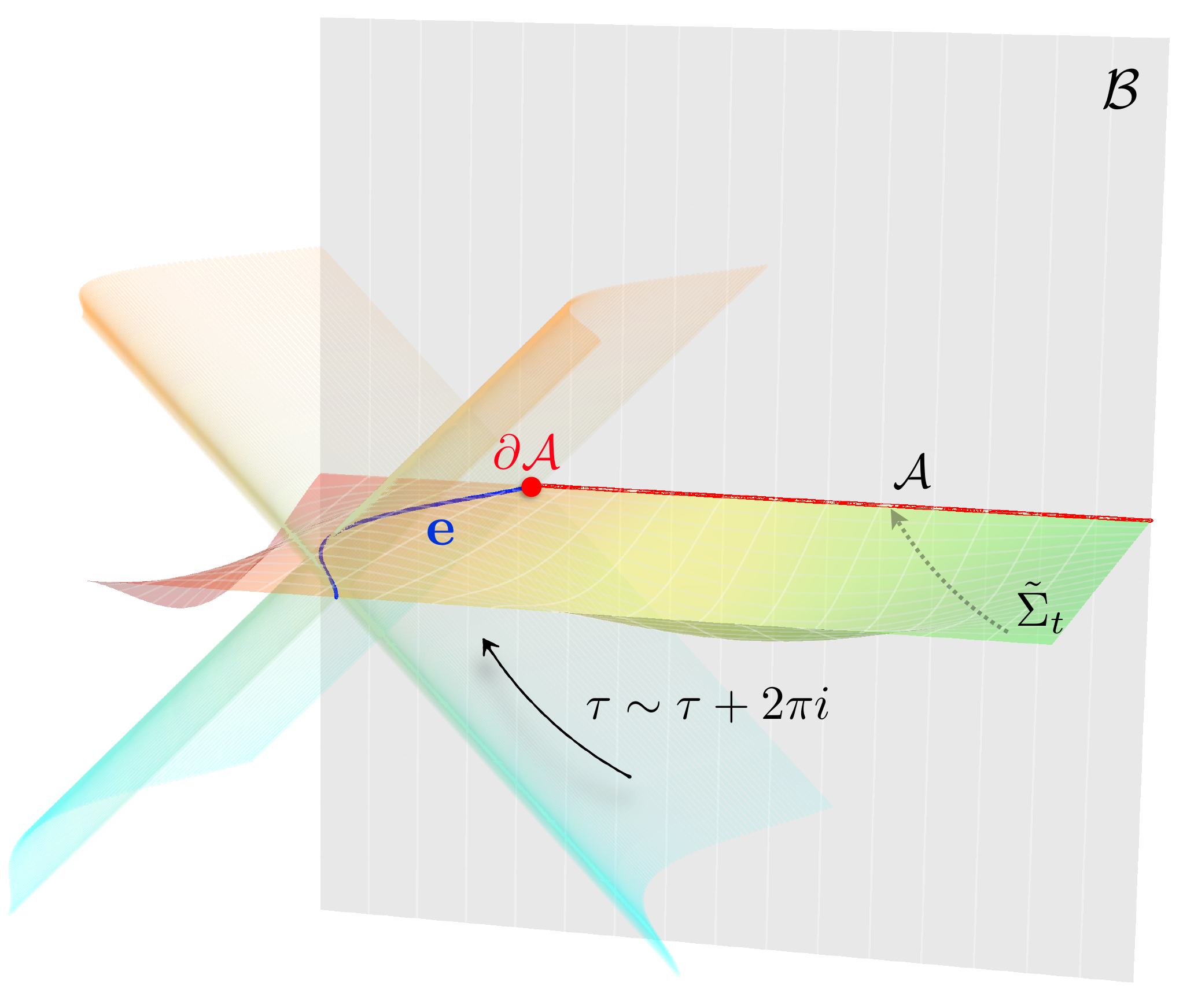}
\caption{The local geometry in the vicinity of the fixed point set $\fixM_q$  which extends out from the entangling surface
in the Lorentzian setting (analog of Fig.~\ref{f:bdycondq}). The normal plane is parameterized using Rindler-like coordinates
 \eqref{eq:lmL} with $r$ being the radial direction away from the fixed point. We have indicated the bulk Cauchy surface where we cut-off the spacetime and also the horizons emanating from the codimension-2 locus.
}
\label{f:bdycondqM}
\end{center}
\end{figure}

For purposes of understanding entanglement entropy, it suffices, as in the Euclidean case, to examine the behaviour in the weak defect $q\to 1$ limit. One can then argue that the local structure of the spacetime near this defect locally looks like the Lorentzian analog of \eqref{eq:lm2}, viz.,
\begin{align}
& ds^2 = \left(q^2 dr ^2-r ^2 \, d\tau^2\right)+
        \left(\gamma_{i j}+2\, K_{ i j}^x \,r^q
        \cosh \tau+2\, K_{ i j}^t \,r^q
        \sinh \tau\right) dy^i \, dy^j \nonumber \\& +\left[r^{f_q\,(q-1)}-1\right]
         \delta g_{\mu\nu} \, dx^\mu\, dx^\nu +\cdots
\label{eq:lmL}
\end{align}
where $f_q$ is a normalization factor, with $f_q (q-1) \in 2 {\mathbb Z}_+$ for $q\in {\mathbb Z}_+$. The logic, as before, is to solve the equations of motion with this ansatz. The geometry in the neighbourhood of the fixed point locus is illustrated in Fig.~\ref{f:bdycondqM}.

We need to determine the correction to the geometry $\delta g_{\mu\nu}$ owing to the backreaction from the defect and simultaneously constrain the locus $\fixM_q$. This is readily done and for the most part not too different from the Euclidean computation. Assuming Einstein-Hilbert dynamics in the bulk, we find the equations of motion reduce to
\begin{align}
\text{EOM} \propto (q-1) \, \frac{1}{r} \, K^a  + \text{regular}\left(\delta g\right).
\label{eq:eomL}
\end{align}
The extrinsic curvature terms in \eqref{eq:lmL} lead to singular behaviour in the neighbourhood of $r =0$. Correction terms to the metric in  $\delta g_{\mu\nu}$ are unable to cancel this divergent piece. Therefore one deduces  that the trace of the extrinsic curvature in the  two normal directions must vanish, viz., $K^t = K^x =0$.
Taking the linear combination of the  spacelike and timelike normals to work with the two null normals, we equivalently conclude that the  null expansions  vanish. Defining $x^\pm = \frac{1}{\sqrt{2}}\, \left(x^0 \pm x^1\right)$, we thus have the extremal surface condition postulated in \cite{Hubeny:2007xt}, viz.,\footnote{ Note here that $K^0_{ij}$ is the component of the extrinsic curvature in the timelike normal direction to a codimension-2 surface (likewise $K^1_{ij}$ is the corresponding spacelike component) and should not be confused with the extrinsic curvature for $\bulkCS$ (which has a timelike normal), denoted by ${\cal K}$.  For codimension-2 spacelike surfaces in  Lorentzian manifolds, these null normals are a natural basis for the normal bundle.}
\begin{equation}
\begin{split}
K^a =0 & \;\; \Longrightarrow \;\; \theta^\pm = \frac{1}{\sqrt{2}} \left(K^0 \pm K^1\right) = 0 \,, \\
& \;\; \Longrightarrow \;\;  \lim_{q\to 1} \fixM_q \to \extrA \,, \qquad \extrA \in \bulk \;\text{is extremal}.
\end{split}
\label{eq:extremalEom}
\end{equation}

Having determined that the surface which is the fixed point locus of the replica symmetry, using the Schwinger-Keldysh construction, is extremal, we now can also constrain the bulk Cauchy slice in the FRW wedge $\bulkCS $. These Cauchy slices are not only  anchored at the boundary at the appropriate slice, i.e., $\partial \bulkCS = \CSA$, but they also have to contain the extremal surface $\extrA \in \bulkCS$.   We have already indicated this in our depiction of the bulk domains in Figs.~\ref{f:bulkfrw} and  \ref{f:bulksk}. In a sense the construction of \cite{Dong:2016hjy} exploits some element of the maximin construction \cite{Wall:2012uf}.  A simple corollary is that the  homology condition is natural. The bulk Cauchy slice naturally admits a bipartite decomposition: $\bulkCS  = \homsurfA \cup \homsurfAc$ and give two spatial codimension-1 bulk regions with the appropriate boundary conditions.

The last step in the discussion involves demonstrating that the on-shell action of the gravitational theory obtained using this Schwinger-Keldysh construction reduces to the area of the extremal surface. The subtleties of this analysis are due to the fact that we have to actually evaluate an oscillatory path integral in Lorentzian signature directly, using a saddle point approximation. The calculation is however simplified by working with the Lorentzian analog of \eqref{eq:var2}, and evaluating the desired boundary terms. At the end of the day, one finds the expected answer
\begin{equation}
\begin{split}
\partial_q I[\hat{\bulk}_q]&= \frac{1}{8 \pi \GN} \; \partial_q \; \int_{\fixM_q(\epsilon)} \; {\cal K}_\epsilon = i \frac{\text{Area}(\fixM_q)}{4\, q^2 \GN} \,,  \\
&\;\Longrightarrow\;\;
S_\regA = \frac{\text{Area}(\extrA)}{4\,\GN}
\end{split}
\end{equation}
The extra factor of $i$ in the on-shell action offsets the $i$ in the definition of the Lorentzian path integral, to give a real answer for the entanglement entropy.

%~~~~~~~~~~~~~~~~~~~~~~~~~~~~~~~~~~~~~~~~~~~~~~~
\section{Higher derivative gravity}
\label{sec:hdgr}
%~~~~~~~~~~~~~~~~~~~~~~~~~~~~~~~~~~~~~~~~~~~~~~

Should we consider higher derivative gravitational theories, the general analysis can be carried through in a similar fashion as discussed in \cite{Fursaev:2013fta,Dong:2013qoa,Camps:2013zua}. What becomes clear is that the local analysis suffices to pin down the singular locus in the $q\to 1$ limit, but this does not in all cases determine the functional which we minimize to obtain the surface.

To describe the result, we need some notation. Let $\mathcal{L}(g,\mathcal{D})$ be the Lagrangian of a diffeomorphism-invariant theory of gravity, as in \eqref{eq:gravact}, with the overall normalization as indicated there. Let $\extrA$ be a codimension-2 surface in a bulk spacetime $\bulk$ which solves the equations of motion resulting from this action. The tangent space of $\bulk$ can be decomposed into the tangent space of $\extrA$ and its normal bundle. As before, $N^{(i)}_A$ are the normals to the codimension-2 surface and $K^{(i)}_{AB}$ the corresponding extrinsic curvatures. We let $M^a$ for $a,b=1,\cdots, d-1$ be the tangent vectors to the surface at a point.  The unit binormal to the surface is defined as
$\varepsilon_{AB} = N^{(i)}_A \, N^{(j)}_B\, \varepsilon_{ij}$, while the projector $P_{AB} = N^{(i)}_A N^{(j)}_B g_{ij} $ localizes us onto the normal directions.  Finally, we assemble the extrinsic curvatures into
$K_{ABC} = N_{A(i)} K_{BC}^{(i)} =  N_{A(i)}\, M_B^{(a)} M_C^{(b)} K_{ab}^{(i)}$.
These can roughly be seen as the antisymmetric and symmetric combinations of the induced measure on the normal bundle.

The analysis now proceeds along the same lines as described in \S\ref{sec:rtderive}, with the new ingredient being the changed equations of motion. At the end of the day, the expression derived for the entanglement entropy can be given as an integral over $\extrA$ of the following functional:
\begin{equation}
\begin{split}
{\sf D}_\mathcal{L} &= -\frac{\delta \mathcal{L}}{\delta R_{ACBD} }\, \varepsilon_{AC}\, \varepsilon_{BD} +
\sum_{\alpha} \;  \frac{2}{Q_\alpha+1} \left(\frac{\delta^2 \mathcal{L}}{\delta R_{A_1C_1 B_1D_1} \delta R_{A_2 C_2 B_2D_2 } }\right)_\alpha\; K_{H_1 C_1 D_1}\; K_{H_2 C_2 D_2}
\\
& \; \times
\bigg[(n_{A_1 A_2} n_{B_1 B_2} - \varepsilon_{A_1 A_2} \varepsilon_{B_1 B_2}) n^{H_1 H_2} \bigg]
+
\bigg[(n_{A_1 A_2} \varepsilon_{B_1 B_2} + \varepsilon_{A_1 A_2} n_{B_1 B_2}) \varepsilon^{H_1 H_2} \bigg]
\end{split}
\label{eq:dong}
\end{equation}
The first term in the above expression is the famous Wald functional, which computes the entropy of a black hole in the theory with Lagrangian density $\mathcal{L}$. We will have occasion to visit its derivation in \S\ref{sec:egeometry}. The second term involves the sum over the auxiliary index $\alpha$ which captures contributions from decomposing the spacetime Riemann tensor into components tangential and normal to the surface.  What matters given the index structure is the set of terms in which the Lagrangian density is varied with respect to  $R_{zazb} $ and $R_{{\bar z} c {\bar z} d }$ simultaneously, where $z$ and ${\bar z}$ are complex coordinates for the Euclideanized normal bundle $z = r\, e^{i\, \tau}$ from earlier. For each such contribution, one has to ascertain the strength of the singularity $\fixM_q$ which is captured by
$Q_\alpha$. One can estimate this by examining the powers of the  metric function $g_{rr}$ in the Gaussian normal coordinates. All in all, the answer for the entanglement entropy is then given by
\begin{equation}
S_\regA = \frac{1}{8\, \GN } \int d^{d-1} x \, \sqrt{h}\, {\sf D}_\mathcal{L}
\label{}
\end{equation}

The construction generalizes the Iyer-Wald construction \cite{Iyer:1994ys} of black hole entropy for higher derivative theories very nicely. For stationary black  holes, one has a bifurcation surface, which is a fixed point locus of the time translational symmetry. As a consequence the bifurcation surface has vanishing extrinsic curvature. In this situation, the second term vanishes. In a sense, the entanglement entropy function derived  in \cite{Dong:2013qoa} provides  a generalization of the black hole entropy formula. It also appears that this construction has a useful role to play in providing a definition of higher derivative black hole entropy in a dynamical setting. Evidence for this was provided to linear order in fluctuations away from stationarity in \cite{Bhattacharjee:2015yaa,Wall:2015raa}.

Thus for  higher derivative theories the functionals derived in \cite{Dong:2013qoa,Camps:2013zua} give us the geometric generalization of the area functional which computes the holographic entanglement entropy. However, as remarked earlier,
these functionals themselves are not to be extremized to compute the location of the surface $\extrA$ in all cases (we also refer the reader to \cite{Miao:2015iba,Dong:2015zba,Camps:2016gfs} for discussion of some subtleties with \eqref{eq:dong}). This remains an open question to date.

%~~~~~~~~~~~~~~~~~~~~~~~~~~~~~~~~~~~~~~~~~~~~~~~
\section{Implications of the bulk replica construction}
\label{sec:bulkee}
%~~~~~~~~~~~~~~~~~~~~~~~~~~~~~~~~~~~~~~~~~~~~~~

Let us pause to take stock of the replica construction in the gravitational theory as described in the previous two subsections. In both the Euclidean and the Lorentzian geometries, the idea has been to realize that the replica spacetime can be interpreted as the covering space of an ${\mathbb Z}_q$ orbifold. The latter allows us to view the geometry as being deformed owing to the presence of a conical singularity which we denoted $\fixM_q$ in our discussion. The  advantage of the gravitational story is that the tension of the cosmic brane which is responsible is a simple function of the R\'enyi index $q$, cf., Eq.~\eqref{eq:ctesion}. One can then easily dial this tension, effectively implementing the analytic continuation we desire quite simply in the gravitational description. Furthermore, the cosmic brane $\fixM_q$ is a codimension-2 object, which in the $q \to 1$ limit limits to the HRT surface $\extrA$ (in the original geometry $\bulk$).

The extremal surface $\extrA$ plays the same role in the bulk as the entangling surface $\entsurf$ does on the boundary. Recall that both surfaces are codimension-2 in their respective spacetimes $\extrA \subset \bulk$ and $\entsurf \subset \bdy$. Moreover, just as for the entangling surface, the extremal surface divides any bulk Cauchy surface it lies on into an inside and an outside. We can in fact take the bulk Cauchy surface to be the union of the homology surfaces $\homsurfA$ and $\homsurfAc$, respectively, which we recall connect $\extrA$ to $\regA$ on one side and to $\regAc$ on the other. We can use this fact to ascertain some interesting facts about the quantum corrections to the field theory entanglement entropy.

Firstly, note that the RT and HRT prescriptions only give us the leading large $\ceff$ answer to the field theory entanglement entropy, owing to the fact that we are only retaining the term proportional to $\frac{1}{\GN}$. In field theory, we expect
\begin{equation}
S_\regA = \ceff\, S_\regA^{saddle} + S_\regA^{1-loop} + {\cal O}\left(\ceff^{-1} \right)
\label{eq:slargec}
\end{equation}
In writing this expression, we have already implicitly used the fact that the leading answer arises from a saddle point analysis in the bulk gravitational description. The natural question is then, where does one get the 1-loop correction term from?

This question was answered in \cite{Faulkner:2013ana} in which it was argued that the 1-loop correction should be viewed as the regulated contribution arising from the entanglement entropy of bulk modes subject to the bipartitioning $\homsurfA \cup \homsurfAc$ across $\entsurf$. From the Euclidean quantum gravity path integral perspective, this is quite natural since the leading correction to the saddle point answer should arise from the 1-loop determinant around the saddle (as for any functional integral). Modes in the bulk that are on one side of the extremal surface are naturally correlated with those on the other side. For obtaining the leading order corrections, it suffices to treat the bulk theory perturbatively in $\GN$, which means that we can focus on quantum fields in a rigid background $\bulk$. One then anticipates that the bulk entanglement entropy will have a divergent contribution which goes like the area of the extremal surface along with subdominant terms which will lead to finite corrections. The divergent term can be viewed as a 1-loop renormalization of the bulk gravitational coupling $\GN$, since it can be combined with the leading RT/HRT answer. Thus suitably regulating the result, one obtains the contribution from the finite pieces correcting the boundary entanglement entropy. This result was independently verified in some situations by \cite{Barrella:2013wja} who exploited the power of two-dimensional CFTs to obtain the asymptotic expansion as in \eqref{eq:slargec}.

One therefore can write schematically:
\begin{equation}
\begin{split}
S_\regA &= \frac{\text{Area}(\extrA)}{4\, \GN} + S_{\homsurfA}^{bulk}  + {\cal O}\left(\GN\right)
 \\
&\equiv  \ceff\, \vev{\widehat{\text{Area}}(\extrA)} + S_{\homsurfA}^{bulk} + {\cal O}(\ceff^{-1})
\end{split}
\label{eq:1loop}
\end{equation}
The first line is meant to be read as a statement in semiclassical gravity, with geometric quantities and perturbative quantum gravitational fields thrown into the mix. The second line is an attempt to formalize this statement in the bulk quantum gravitational theory. Since AdS/CFT is a duality between a quantum theory on the boundary and a gravitational one in the bulk, we can attempt to relate operators in the two theories. The quantum gravitational theory at tree level admits an \emph{area operator} which is defined in the obvious manner -- expectation value of the area operator gives the classical area of the surface under consideration. The leading term in the answer is then interpreted as the expectation value of the area operator on the RT/HRT surface and the subleading term is the bulk entanglement entropy for the bipartitioning engendered by this surface. There are speculations about higher order corrections available in the literature \cite{Engelhardt:2014gca}, but unlike the 1-loop term above, they involve making assumptions about the behaviour of quantum gravitational dynamics.  It remains an open question whether one can use string theoretic considerations to pin down explicitly a perturbative expansion of $S_\regA$ in a large $\ceff$ perturbation series.

It has recently been argued that \eqref{eq:1loop} admits an interesting interpretation in terms of relative entropy \eqref{eq:relent}. Recall that the relative entropy can be expressed in terms of the modular Hamiltonian $\modA$. Under most circumstances,  the modular Hamiltonian is a complicated non-local operator since we are essentially defining it as the logarithm of a linear operator on the Hilbert space. However, the geometrization of the field theory entanglement entropy suggests that one can write a simple relation between the field theory modular Hamiltonian and that of the gravitational theory in the semiclassical limit \cite{Jafferis:2015del}
\begin{equation}
\modA =\ceff\, \vev{\widehat{\text{Area}}(\extrA)} +\;  {\cal K}_{\homsurfA}^\text{bulk} + {\cal O}(\ceff^{-1}) \,.
\label{eq:jlms1}
\end{equation}
This expression implies that the bulk and boundary relative entropies agree. Consider two field theory states $\sigma$  and $\rho$
and their corresponding bulk duals. We will use the former as our reference state and constrain the latter to be a small excitation about it. The restriction is to ensure that in the gravitational dual we can consider  the geometry dual to $\sigma$ as the background and the excitations in $\rho$ will be viewed as a few particle states atop this semiclassical background. We can then carry out the bulk semiclassical analysis as above and learn from the behaviour of the modular Hamiltonians that
\begin{equation}
S_\bdy(\rho_\regA||\sigma_\regA) = S_\bulk(\rho_{\homsurfA}||\sigma_{\homsurfA}) \,,
\label{}
\end{equation}
with the subscripts $\bdy$ and $\bulk $ referring obviously to the field theory and semiclassical gravity.
In other words, to the leading order semiclassical approximation, the bulk and boundary modular Hamiltonians agree. These observations will be very useful in understanding the reconstruction of the bulk geometry from the field theory, cf.,
\S\ref{sec:egeometry}.

%~~~~~~~~~~~~~~~~~~~~~~~~~~~~~~~~~~~~~~~~~~~~~~~
\chapter{Properties of holographic entanglement entropy}
\label{sec:heeprops}
%~~~~~~~~~~~~~~~~~~~~~~~~~~~~~~~~~~~~~~~~~~~~~~

The holographic RT and HRT prescriptions allow us to explore general properties of entanglement entropy in a class of QFTs. We will first examine the consistency of holographic entanglement entropy with expectations that follow from the basic definition as detailed in \S\ref{sec:genee}. We will also see that there are certain features that are peculiar to holographic systems, in part owing to the fact that we are working in the large $\ceff$ limit. We reiterate that the holographic entanglement entropy prescriptions are geared to capturing the leading semiclassical part of entanglement in terms of geometric data. Subleading corrections require ascertaining the bulk entanglement, as discussed in the previous section. All in all,  this leads to some unexpected features, which at first sight seem unconventional, but are easily understood once one fully appreciates the implications of the limit $\ceff\gg 1$ being effectively a semiclassical regime of the QFT.

%~~~~~~~~~~~~~~~~~~~~~~~~~~~~~~~~~~~~~~~~~~~~~~~
\section{An extremal surface primer}
\label{sec:extdeter}
%~~~~~~~~~~~~~~~~~~~~~~~~~~~~~~~~~~~~~~~~~~~~~~

Let us first discuss general strategies for determining extremal surfaces in an asymptotically AdS spacetime. Imagine that we have a QFT living on a background $\bdy_d$ and we are given the metric $h_{\mu\nu}$ on this boundary geometry. In addition, we will assume that
$\bdy$ admits a foliation by a timelike coordinate $t$ and we can describe the induced geometry on constant $t$ slices. On such Cauchy slices we demarcate a region $\regA$ with boundary $\entsurf$.

We pick coordinates  $\sigma^i$ on the entangling surface and view the embedding $\entsurf \subset \bdy$ being given by a set of mappings $x^\mu(\sigma^i)$. From this information, we can naturally deduce the geometry of $\entsurf$. The intrinsic geometry of the surface is determined by the induced metric, which is pulled back from the parent spacetime, viz.,
\begin{equation}
ds^2_{\entsurf} = h_{\mu\nu}\, \frac{\partial x^\mu}{\partial \sigma^a}\, \frac{\partial x^\nu}{\partial \sigma^b}\, d\sigma^a\, d\sigma^b
\equiv \tilde{\gamma}_{ab}  \,d\sigma^a\, d\sigma^b \,.
 \label{}
\end{equation}
The extrinsic geometry is obtained by examining the gradients of the normal directions to this entangling surface using the boundary analog of \eqref{eq:ecurvdef}.

In the bulk, we have a geometry $\bulk$, dual to the state of the field theory on $\bdy$. The precise details of the geometry will depend on the state in question. We should note however that the precise criterion states of the field theory which are well described by semiclassical bulk geometries is far from clear, though progress has been made in recent years. We will review some of these issues later in our discussion in \S\ref{sec:egeometry}.

 Assuming for the moment that the state of the QFT does have a classical gravity dual, we choose to parameterize the bulk geometry for definiteness in the Fefferman-Graham gauge \cite{Fefferman:1985aa} (see also \cite{deHaro:2000vlm} for its role in AdS/CFT), where one fixes a radial gauge $G_{zz} =\frac{1}{z^2}$ and $G_{z\mu}=0$. The bulk metric can thus be taken to be of the form
\begin{align}
ds^2_\bulk &= g_{AB}\, dX^A\, dX^B = \frac{dz^2}{z^2} + \frac{1}{z^2} \, \mathfrak{g}_{\mu\nu}(x,z)\, dx^\mu \ dx^\nu
\label{eq:bulkmet}
\end{align}
%

%~~~~~~~~~~~~~~~~~~~~~~~~~~~~~~~~~~~~~~~~~~~~~~
\subsubsection{Near-boundary geometry and energy-momentum tensor}
%~~~~~~~~~~~~~~~~~~~~~~~~~~~~~~~~~~~~~~~~~~~~~~

The metric function $\mathfrak{g}_{\mu\nu}(x,z)$ admits a Taylor series expansion (with zero radius of convergence) in the neighbourhood of the boundary. The result depends on the boundary spacetime dimension being  even or odd, since the latter case allows for the possibility of conformal anomalies.

One finds the schematic form (for explicit expressions, we refer the reader to \cite{deHaro:2000vlm})
\begin{align}
\mathfrak{g}_{\mu\nu}  &=
h_{\mu\nu} (x)+ z^2 \, h^{(2)}_{\mu\nu} (x)
+ z^4\, \,h^{(4)}_{\mu\nu} (x)+ \cdots + z^{2k}\, h^{(2k)}_{\mu\nu}(x) + \cdots
\nonumber \\
 & \qquad \qquad +\;
  z^d\, \log z \, h^{(d)}(x) + z^d\, T_{\mu\nu} (x)+ {\cal O}(z^{d+1})
  \label{eq:FGseries}
\end{align}

The expansion proceeds in even powers of $z$ due to the structure of Einstein's equations. The logarithmic term is only present in even $d$-dimensions and is related to the fact that the CFTs in even spacetime dimensions suffer from a conformal anomaly. The leading term $h_{\mu\nu}(x)$ and the `constant of integration' $T_{\mu\nu}(x)$ are sufficient data to determine the series solution completely. The terms $h^{(2k)}(x)$ with $0\leq k \leq  \frac{d}{2}$ are completely determined by the boundary metric $h_{\mu\nu}(x)$ and its derivatives (intrinsic curvatures). For instance, the first few terms in this expansion are:
\begin{equation}
\begin{split}
h^{(2)}_{\mu\nu} &=
  \frac{1}{d-2}\left( {}^{(h)} R_{\mu\nu} - \frac{1}{2(d-1)}\, {}^{(h)} R\, h_{\mu\nu}\right)
\\
h^{(4)}_{\mu\nu} & =
   \frac{1}{d-4} \, \Bigg(
   - \frac{1}{8\,(d-1)}\; {}^{(h)}\nabla_\mu\;  {}^{(h)}\nabla_\nu {}^{(h)} R
  + \frac{1}{4}\; {}^{(h)}\nabla^2  \; h^{(2)}_{\mu\nu}
\\
&
  -\frac{1}{2\,(d-2)}\; {}^{(h)} R^{\alpha\beta} \left[
    {}^{(h)} R_{\mu\alpha \nu \beta}
  - \frac{d-4}{d-2} \, \delta_{\beta\mu}\;  {}^{(h)} R_{\alpha \nu}
  - \frac{2}{(d-1)(d-2)}\, h_{\alpha\beta} \; {}^{(h)} R_{\mu\nu} \right]
\\
&
  + \frac{1}{4\, (d-2)^2} \left[ {}^{(h)} R^{\alpha \beta} \, {}^{(h)} R_{\alpha \beta}
  - \frac{3\,d}{4\,(d-1)^2} {}^{(h)} R^2\right]   h_{\mu\nu}
    \Bigg)
\end{split}
\label{}
\end{equation}

 The new piece of data in this expansion at ${\cal O}(z^d)$, $T_{\mu\nu}(x)$ corresponds to the expectation value of the energy-momentum tensor on the boundary. This cannot be determined by the local analysis and one needs detailed information of the state, in particular, to construct a geometry that is regular (everywhere outside putative horizons) to ascertain $\vev{T_{\mu\nu}}$. How this can be done is well explained in the literature, so we will assume henceforth that we have been handed the geometry of interest.

The boundary energy-momentum tensor is given as  \cite{Henningson:1998gx,Balasubramanian:1999re}
\begin{equation}
\vev{T_{\mu\nu}} = d\,\ceff\big( t_{\mu\nu} + C_{\mu\nu}[h]\big)
\end{equation}
where $C_{\mu\nu}[h]$ is a local functional of the boundary metric and its derivatives, capturing the contribution of the Weyl anomaly. It vanishes in $d=2n+1$ but depends non-trivially on the dimension for $d=2n$. For instance:
\begin{equation}
\begin{split}
d=2: & \quad C_{\mu\nu}[h]
  = h_{\mu\nu} \, \Tr{}{h^{(2)}}
\\
d=4: & \quad C_{\mu\nu}[h]
  = -\frac{1}{8} \left(
  \Tr{}{h_{(2)}}^2 -  \left(\Tr{}{h_{(2)}}\right)^2\right)  h_{\mu\nu}
  +\frac{1}{2} \left(h_{(2)}^2\right)_{\mu\nu} - \frac{1}{4}\, h^{(2)}_{\mu\nu} \;\Tr{}{h^{(2)}}
\end{split}
\label{eq:Canom}
\end{equation}

It is  more useful to record a covariant expression for the boundary energy-momentum tensor that is not tied to a specific gauge. Let the unit outward normal to the boundary be given by $n^\mu$. We define the extrinsic curvature of the boundary by
\begin{equation}
\mathcal{K}_{\mu\nu} = h_{\mu\rho} \, {\cal D}^\rho n_\nu
\label{eq:extrinsic}
\end{equation}
We then choose to regulate the spacetime with a rigid cut-off at $z = \epsilon_c$,  or equivalently, $r = \Lambda_c$ in global coordinates and find \cite{Balasubramanian:1999re}:
\begin{equation}
T^{\mu\nu} = \lim_{\Lambda_c \to \infty} \, 4\pi\, c_\text{eff}\, \Lambda_c^{d-2}\, \left[ \mathcal{K}^{\mu\nu} -\mathcal{K} \, h^{\mu\nu}
- (d-1)\, h^{\mu\nu} - \frac{1}{d-2} \left( {}^h R^{\mu\nu} - \frac{1}{2}\, {}^h R\, h^{\mu\nu} \right)\right]
\label{}
\end{equation}
%

%~~~~~~~~~~~~~~~~~~~~~~~~~~~~~~~~~~~~~~~~~~~~~~
\subsubsection{Extremal surface determination}
%~~~~~~~~~~~~~~~~~~~~~~~~~~~~~~~~~~~~~~~~~~~~~~

The HRT prescription requires that we find a bulk extremal surface in \eqref{eq:bulkmet}. We can parameterize the surface by intrinsic parameters $\xi^i$, and assume that $X^A(\xi^i)$ is the surface of interest. In analogy with the boundary discussion, we determine the induced metric whose area form we wish to extremize. One has
\begin{equation}
ds^2_{\extrA} = \frac{1}{z^2} \left(\frac{\partial z}{\partial \xi^i}\, \frac{\partial z}{\partial \xi^j} +
\mathfrak{g}_{\mu\nu}(x,z) \, \frac{\partial x^\mu}{\partial \xi^i} \; \frac{\partial x^\nu}{\partial \xi^j} \right) d\xi^i\, d\xi^j \,.
\label{eq:inducedE}
\end{equation}
We now can set up a variational problem defined by an action, which is nothing but the area functional of $\extrA$ in units of the AdS Planck length, viz.,
\begin{equation}
{\cal S}_{extremal}=\frac{\lads^{d-1}}{4\, \GN } \int \, d^{d-1}\xi\, \frac{1}{z^{d-1}} \, \sqrt{ \text{det}\left(\frac{\partial z}{\partial \xi^i}\, \frac{\partial z}{\partial \xi^j} +
\mathfrak{g}_{\mu\nu}(x,z) \, \frac{\partial x^\mu}{\partial \xi^i} \; \frac{\partial x^\nu}{\partial \xi^j} \right)}
\label{eq:extrarea}
\end{equation}
The Euler-Lagrange equations for this system with boundary conditions
\begin{equation}
\extrA \big|_{z \to 0} = \entsurf
\label{}
\end{equation}
sets up the problem of finding extremal surfaces. Once we  have found the surface of interest, we simply evaluate its area as the on-shell value of the action \eqref{eq:extrarea}. Note that we can use \eqref{eq:dumap} to write the result purely in terms of the boundary quantities. Various authors have studied a wide range of examples over the years and in most cases the analysis can be readily done using standard numerical techniques.

The standard way to proceed in static spacetimes with adequate symmetries is to exploit the symmetries, using the associated conserved charges, reducing the equations of motion following from \eqref{eq:extrarea} to a sufficiently amenable form, and then integrating them. Typically, one ends up considering situations in which the symmetries allow for reducing the equations of motion to a set of ordinary differential equations which can  usually be solved through a shooting method. This usually relies on ascertaining (again through symmetry) the deepest point in the bulk attained by the minimal surface and integrating out from there towards the boundary. One can then generally determine the boundary endpoints numerically as a function of the coordinates of the deepest  point and invert if necessary. This strategy works well as long as care is taken to ensure that we work with appropriately regulated boundary conditions. One then has to plug the solution into the action $S_{extremal}$ and evaluate its on-shell value. This can potentially be a source of errors, since one would like,  at the end of the day, to obtain a UV regulated area.

In some cases, it turns out to be efficient to adopt a gauge choice for $\xi^a$ that simplifies the action functional itself. In addition to picking a convenient gauge, one can also set up in the case of static spacetimes, a mean curvature flow, a relaxation algorithm that locates the minimal area surface directly; see Appendix A of \cite{Hubeny:2013gta} for details of this construction. There are also some sophisticated software packages, such as\emph{ Surface Solver} \cite{Brakke:1992aa}, developed for solving the Plateau problem in flat space, that  have been exploited to construct exotic minimal surfaces in \AdS{4} for a wide range of domains \cite{Fonda:2014cca,Fonda:2015nma}.

In the general time-dependent setting, one has to resort to either the shooting method described above or directly solve the resulting PDEs. It is interesting to contemplate exploiting the maximin construction to develop a Lorentzian analog of the mean curvature flow. A  naive attempt is guaranteed to fail due to the fact the Lorentzian problem does not involve elliptic  of PDEs courtesy of the temporal direction. We are not aware of an explicit implementation of such an algorithm to date, but this would  allow one to explore general properties of entanglement dynamics in inhomogeneous time-dependent backgrounds.

Let us illustrate this discussion with various examples, which will prove useful in our discussions to follow. We will restrict the domain of the bulk AdS spacetime to the region $z > \epsilon$ to regulate the computation of the area integrals. This will serve as a UV cut-off in the field theory.

% Figure
\begin{figure}[htbp]
\begin{center}
\begin{tikzpicture}
\draw[ultra thick,black] (-3,0) -- (3,0);
\draw[thick, red] (-2,0) -- (0,0) node [above] {$\regA$} -- (2,0) ;
\draw[thick,fill=red] (-2,0) circle (0.3ex);
\draw[thick, fill=red] (2,0) circle (0.3ex);
\draw[thick,fill=orange, opacity =0.1] (2,0) arc (360:180:2);
\draw[thick,orange] (2,0) arc (360:180:2);
\draw[thick,orange]  (0,-2) node [below] {$\extrA$};
\draw[thick,->] (3,0) -- (3.5,0) node [above] {$x$};
\draw[thick,->] (-3,0) -- (-3,-1) node [left] {$z$};
\end{tikzpicture}
\end{center}
\caption{Sketch of an extremal surface in pure \AdS{3} in Poincar\'e coordinates. }
\label{f:ads3semi}
\end{figure}
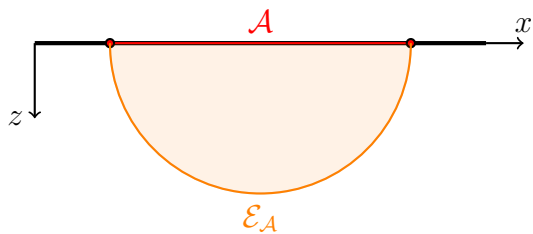

\paragraph{1. Vacuum state of CFT$_2$ on ${\mathbb R}^{1,1}$:} We take the region to be an interval of size $2a$ centered around the origin, viz.,  $\regA = \{x\in {\mathbb R} | x \in (-a,a)\}$. The dual geometry we need is the Poincar\'e-\AdS{3} spacetime \eqref{eq:pads} with $d=2$. Restricting attention to $t=0$ by virtue of staticity, we find that we need to find a spacelike geodesic in the $xz$ plane. This can be done by writing down the induced metric on a curve and the geodesic action:
\begin{equation}
{\cal S} = 4\pi\, \ceff \int \, \frac{\sqrt{x'(\xi)^2 + z'(\xi)^2}}{z}\, d\xi
\label{}
\end{equation}
Varying this action, one can check that the  resulting equations of  motion are solved by
a semi-circle in the $xz$ plane, cf., the illustration in Fig.~\ref{f:ads3semi}
\begin{equation}
x(\xi)  = a\, \cos \xi \,, \qquad z(\xi) = a\, \sin \xi
\end{equation}
and the length of this curve evaluates for us the entanglement entropy.
\begin{equation}
S_\regA  = 4\pi\, \ceff\; 2\, \int_{\frac{\epsilon}{a}}^\frac{\pi}{2} \, \frac{d\xi}{\sin \xi} = 8\pi \, \ceff\log \frac{2a}{\epsilon} = \frac{c}{3} \,\log\frac{2a}{\epsilon}
\label{eq:hol1cft2}
\end{equation}
In evaluating the integral, we converted the UV cut-off $z=\epsilon$ into a restriction on the domain of the affine parameter along the curve. In the final step we used the Brown-Henneaux result \cite{Brown:1986nw} that the asymptotic symmetry group of \AdS{3} is a Virasoro algebra with central charge $c= \frac{3\,\lads}{2\,\ell_P}$ to write the answer in terms of the true central charge (as opposed to $\ceff$). This simple computation agrees explicitly with the CFT$_2$ result  \eqref{eq:hlwcc}. This is no coincidence, for in both cases, the result is dictated purely by the conformal symmetry and we have indicated that the result is universally determined simply by the central charge.

% Figure
\begin{figure}[htbp]
\begin{center}
\begin{tikzpicture}
\draw[ultra thick,black] (-3,0) -- (7,0);
\draw[thick, red] (-2,0) -- (0,0) node [above] {$\regA_1$} -- (2,0) ;
\draw[thick, red] (4,0) -- (5,0) node [above] {$\regA_2$} -- (6,0) ;
\draw[thick,fill=red] (-2,0) circle (0.3ex);
\draw[thick, fill=red] (2,0) circle (0.3ex);
\draw[thick, fill=red] (4,0) circle (0.3ex);
\draw[thick, fill=red] (6,0) circle (0.3ex);
\draw[thick,fill=blue, opacity =0.1] (2,0) arc (360:180:2);
\draw[thick,blue] (2,0) arc (360:180:2);
\draw[thick,fill=purple, opacity =0.1] (6,0) arc (360:180:1);
\draw[thick,purple] (6,0) arc (360:180:1);
\draw[thick,blue]  (0,-2) node [below] {${\cal E}_{\regA_1}$};
\draw[thick,purple]  (5,-1) node [below] {${\cal E}_{\regA_2}$};
\end{tikzpicture}
\hspace{1cm}
\begin{tikzpicture}
\draw[ultra thick,black] (-3,0) -- (7,0);
\draw[thick, red] (-2,0) -- (0,0) node [above] {$\regA_1$} -- (2,0) ;
\draw[thick, red] (4,0) -- (5,0) node [above] {$\regA_2$} -- (6,0) ;
\draw[thick,fill=red] (-2,0) circle (0.3ex);
\draw[thick, fill=red] (2,0) circle (0.3ex);
\draw[thick, fill=red] (4,0) circle (0.3ex);
\draw[thick, fill=red] (6,0) circle (0.3ex);
\draw[thick,fill=orange, opacity =0.1] (6,0) arc (360:180:4);
\draw[thick,orange] (6,0) arc (360:180:4);
\draw[thick,fill=white, opacity =1] (4,0) arc (360:180:1);
\draw[thick,orange] (4,0) arc (360:180:1);
\draw[thick,black]  (2,-3) node [below] {${\cal E}_{\regA_1 \regA_2}$};
\end{tikzpicture}
\end{center}
\caption{Sketch of  the two potential  extremal surfaces for a disjoint union of two regions $\regA_1$ and $\regA_2$. We either have the union of the two individual extremal surfaces  ${\cal E}_{\regA_1} \cup {\cal E}_{\regA_2}$ or the surface  ${\cal E}_{\regA_1 \regA_2}$ which connects the two regions. Of these,  the one with minimal area gives the entanglement entropy for $\regA_1 \cup \regA_2$.  }
\label{f:sasurfs}
\end{figure}
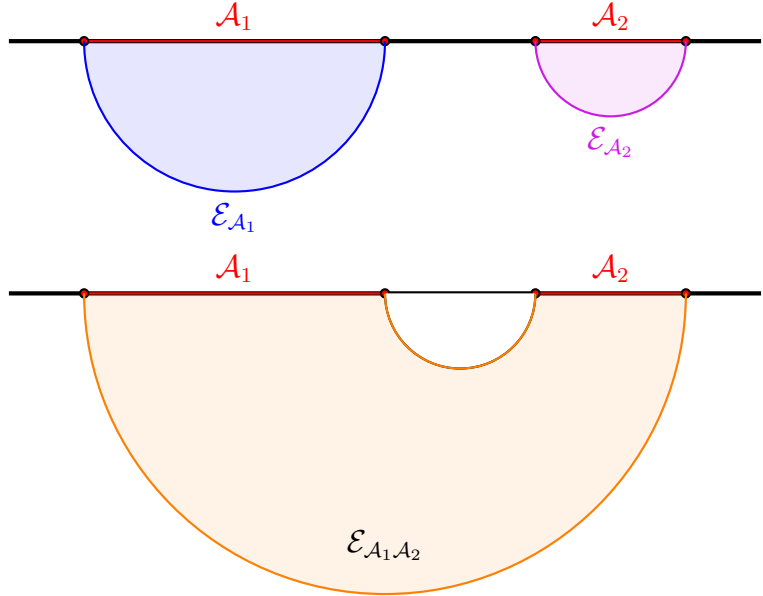

We have remarked in \S\ref{sec:eecft2} that the computation of entanglement entropy for multiple disjoint intervals in a CFT is a formidable task. The holographic answer however turns out to be very simple.  Let us consider $\regA = \cup_i \;  \regA_i$ with
$\regA_i = \{x\in {\mathbb R} | x \in (u_i,v_i)\}$. Then we can consider geodesics that connect the left endpoint of one-interval, say $\regA_i$, with the right endpoint of any other $\regA_j$ (including itself). The lengths of such geodesics are simply proportional
$2 \log \frac{|u_i - v_j|}{\epsilon} $. The holographic answer is then simply
\begin{equation}
S_\regA = \min \left(\frac{c}{3} \, \sum_{(i,j)} \log \frac{|u_i - v_j|}{\epsilon} \right) \,,
\label{eq:holkcft2}
\end{equation}
with the sum running over all pairs of choices from which we pick the globally minimum result. For instance, for two intervals, we have
\begin{equation}
S_\regA = \min \left(\frac{c}{3} \log \frac{|u_1 - v_1|}{\epsilon}  + \log \frac{|u_2 - v_2|}{\epsilon} ,
\log \frac{|u_1 - v_2|}{\epsilon} +\log \frac{|u_2 - v_1|}{\epsilon} \right)
\label{eq:hol2cft2}
\end{equation}
This is illustrated in Fig.~\ref{f:sasurfs};  basically one is instructed to draw  all the extremal surfaces subject to the homology constraint. We will explore the  implications of this result in the following.

\paragraph{2. Vacuum state of CFT$_d$ on ${\mathbb R}^{d-1,1}$:} In higher-dimensional field theories, the vacuum state on Minkowski space is  dual to the Poincar\'e-\AdS{d+1} geometry. However, we have a multitude of regions to choose from as we can pick any codimension-1 region sitting inside ${\mathbb R}^{d-1}$. There are two obvious regions of interest, which are worth analyzing in detail:
\begin{itemize}
\item Strips: These are regions preserving $(d-2)$-dimensional translational invariance. We can choose coordinates to describe the region as
\begin{equation}
\regA_{\parallel} = \{ {\bf x}_{d-1} \in {\mathbb R}^{d-1} | x_1 \in (-a,a),   x_i \in {\mathbb R} \;\; \text{for}\; i = 2, 3, \cdots d-1  \}
\label{eq:stripdef}
\end{equation}
Given the symmetries, we adapt the coordinates $\xi^a =x^a$ and find
\begin{equation}
\begin{split}
{\cal S} &= 4\pi\, \ceff \, \int d^{d-2}x \, dx_1\, \frac{\sqrt{1+ z'(x_1)^2}}{z^{d-1}}
 \\
\delta {\cal S} &=0 \Longrightarrow z'(x_1) = \frac{\sqrt{z_*^{2(d-1)} - z^{2(d-1)}}}{z^{d-1}} \,, \quad
z_* = a\, \frac{\Gamma\left(\frac{1}{2(d-1)}\right)}{\sqrt{\pi}\, \Gamma\left(\frac{d}{2(d-1)}\right)}
\end{split}
\label{}
\end{equation}
In deriving the equation of motion, we made use of the $x_1$ independence of the action to write down a conserved quantity, which we expressed in terms of  $z_*$, the turnaround point of the surface in \AdS. One can solve the for the surface explicitly in terms of hypergeometric functions; we give the expression for the two lobes of the  surface $x_1>0$ and $x_1<0$ which smoothly meet at $x_1 =0, z=z_*$:
\begin{align}
\pm x_1(z) =  \frac{z^d}{d\, z_*^{d-1}} \ _2F_1 \left(\frac{1}{2}, \frac{d}{2(d-1)}, \frac{3d-2}{2d-2} , \left(\frac{z}{z_*}\right)^{2(d-1)} \right) - \frac{\sqrt{\pi}}{d} \, \frac{\Gamma\left(\frac{3d-2}{2d-2}\right)}{\Gamma\left(\frac{2d-1}{2(d-1)}\right)}
\label{}
\end{align}
The area of the surface can be readily computed. Introducing an IR regulator $L$ for the translationally invariant directions, we have
\begin{equation}
S_{\regA_{\parallel}} = \frac{4\pi\, \ceff}{d-2} \, L^{d-2}  \left[\frac{2}{\epsilon^{d-2}} -
\left( \frac{2}{z_*}\right)^{d-1} \frac{1}{a^{d-2}} \right]
\label{}
\end{equation}
The leading  divergent term scales like the area of $\entsurf$; we will see that this is generic in holographic theories in due course. The absence of any subleading divergences is due to the fact that the entangling surface is both intrinsically flat and has no extrinsic curvature. This, in particular, guarantees the vanishing of the logarithmic term in even spacetime dimensions, which would have arisen due to the conformal anomaly.

\item Spherically symmetric ball-shaped domains: These are regions which preserve a $SO(d-2)$-dimensional spherical symmetry, viz.,
\begin{equation}
\regAB = \{x_i \in {\mathbb R}^{d-1}|\; \sum_{i=1}^{d-1} \; x_i^2 \leq R^2 \}
\label{eq:Aball}
\end{equation}
Now it is simpler to adapt coordinates $\xi $ to be the radial coordinate ($r$) of the ${\mathbb R}^{d-1}$ and take the remaining coordinates to be the angular directions of the ${\bf S}^{d-2} \subset {\mathbb R}^{d-1}$. The minimal surface is determined by the Euler-Lagrange equations of the action,
\begin{equation}
{\cal S}= 4\pi\, \ceff\, \omega_{d-2}\, \int \, d\xi \,
\frac{\xi^{d-2}}{z^{d-1}} \, \sqrt{1+z'(\xi)^2} \,.
\label{}
\end{equation}
Here $\omega_{d-2} = \frac{2 \, \pi^\frac{d-1}{2}}{\Gamma\left(\frac{d-1}{2}\right)}$
is the area of a unit ${\bf S}^{d-2}$.
The equations of motion are simpler than they appear at first sight; despite  their not being amenable to integration by quadratures, one can check that the minimal surface is a  hemisphere:
\begin{equation}
z^2  + \xi^2 = R^2 \,, \qquad \left\{z = R\, \cos\theta \,,\ \xi = R\, \sin\theta\right\} \,.
\label{eq:rtsphere}
\end{equation}
The entanglement entropy is evaluated by the integral:
\begin{equation}
S_{\regAB} = 4\pi\, \ceff\, \omega_{d-2}\, R^{d-2}\, \int_{-\frac{\pi}{2}}^{\frac{\pi}{2}} \, d\theta \,
\frac{(\sin\theta)^{d-2}}{(\cos\theta)^{d-1}}
\label{eq:Sdisc1}
\end{equation}
The final expression after performing the integral is given in Eq.~\eqref{eq:sball}.
We will give an alternate method to derive this answer later, one which exemplifies some important features of this geometry.
\end{itemize}

\paragraph{3. CFT$_2$ on ${\bf S}^1 \times {\mathbb R}$:} As discussed earlier,  this configuration can be used to describe the vacuum state of the CFT on a finite spatial domain (a circle) or a thermal state in non-compact space.

% Figure
\begin{figure}[htbp]
\begin{center}
\includegraphics[width=3in]{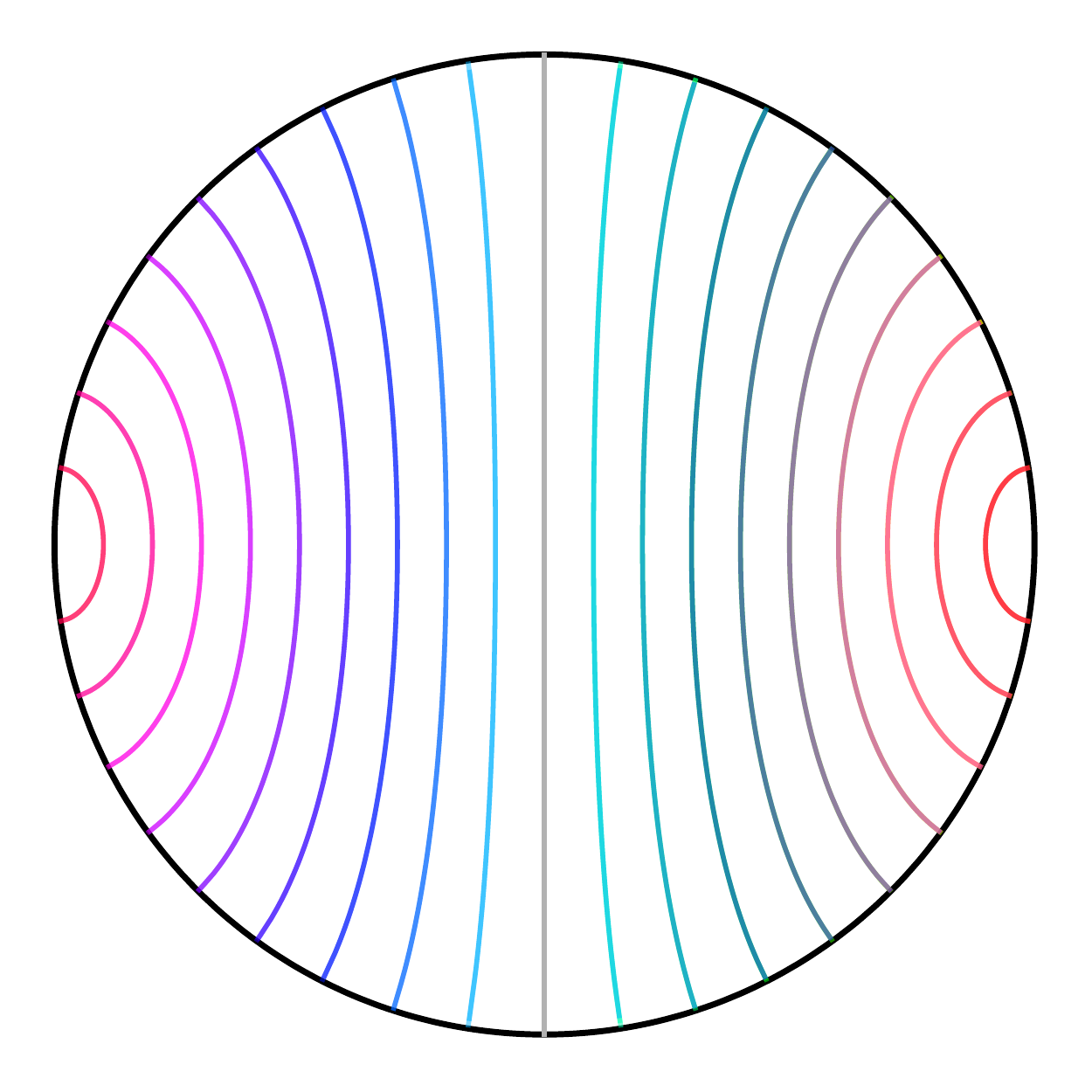}
\end{center}
\caption{Plot of minimal surfaces on the Poincar\'e disc \eqref{eq:Extads}. We have compactified the Poincar\'e disc by the map $r = \tan\varrho$ to bring the boundary to finite distance. The minimal surfaces are geodesics in this case and we have illustrated these for varying angular region size. Purity of state implies that $\regA$ and $\regAc$ coincide}
\label{f:EAads3}
\end{figure}

\begin{itemize}
\item  Let us first  discuss the vacuum state, whence the  bulk geometry has the metric \eqref{eq:gads} (with $d=2$), which we rewrite for convenience as
\begin{equation}
ds^2 = -(\frac{r^2}{\lads^2}+1)\, dt^2 + \frac{dr^2}{\frac{r^2}{\lads^2}+1} + r^2\, d\varphi^2
\end{equation}
We have to find a spacelike geodesic at $t=0$, which is easily done. We take the region
$\regA$ to be an arc of the $\varphi$ circle centered around the origin of angular width $2\varphi_\regA$.
The reader can verify that
\begin{equation}
r(\varphi) = \lads \left(\frac{\cos^2\varphi}{\cos^2 \varphi_\regA} - 1\right)^{-\frac{1}{2}}
\label{eq:Extads}
\end{equation}
is the locus of the geodesic.  These are plotted on the Poincar\'e disc in Fig.~\ref{f:EAads3}.
Upon evaluating the length  of the curve, we find
\begin{equation}
S_\regA
= \frac{c}{3}\, \log\left(\frac{\ell_{{\bf S^1}}}{\pi\, \epsilon} \sin\left(\frac{2a}{\ell_{{\bf S^1}}}\right) \right)
\label{}
\end{equation}
where we translated in terms of the arc-length $a$ of the region ($\ell_{{\bf S^1}}$ is the  proper radius of the circle) and used the Brown-Henneaux result again. This again agrees with \eqref{eq:svaccft2} for reasons outlined earlier.

\item The thermal state of the CFT$_2$ on non-compact space $x\in {\mathbb R}$ is described by the planar  BTZ geometry
\begin{equation}
ds^2 = -\frac{(r^2- r_+^2 )}{\lads^2}\, dt^2 + \frac{dr^2}{r^2-r_+^2} + \frac{r^2}{\lads^2}\, dx^2
\label{eq:pbtz}
\end{equation}
The extremal surface satisfies:
\begin{equation}
\frac{dr}{dx} = \frac{r}{\lads^2}\, \sqrt{(r^2 - r_+^2) \left( \frac{r^2}{r_*^2}-1\right)}\,,\qquad r_* = r_+\, \coth(a\, r_+)
\label{}
\end{equation}
where $r_*$ is  determined by restricting the range of $x \in (-a,a)$.  We can compute its length and obtain the answer for the entanglement entropy:
\begin{equation}
S_\regA = \frac{c}{3}\, \log\left(\frac{\beta}{\pi\,\epsilon}\; \sinh\left(\frac{2\pi\,a}{\beta}\right) \right)
\label{}
\end{equation}
which agrees with \eqref{eq:sthermcft2} as we anticipated. To write the answer in this form, we used the fact that BTZ black hole of radius $r_+$ corresponds to a thermal state of the field theory at $T = \frac{r_+}{2\pi\,\lads^2}$.
\end{itemize}

% Figure
\begin{figure}[htbp]
\begin{center}
\includegraphics[width=2.5in]{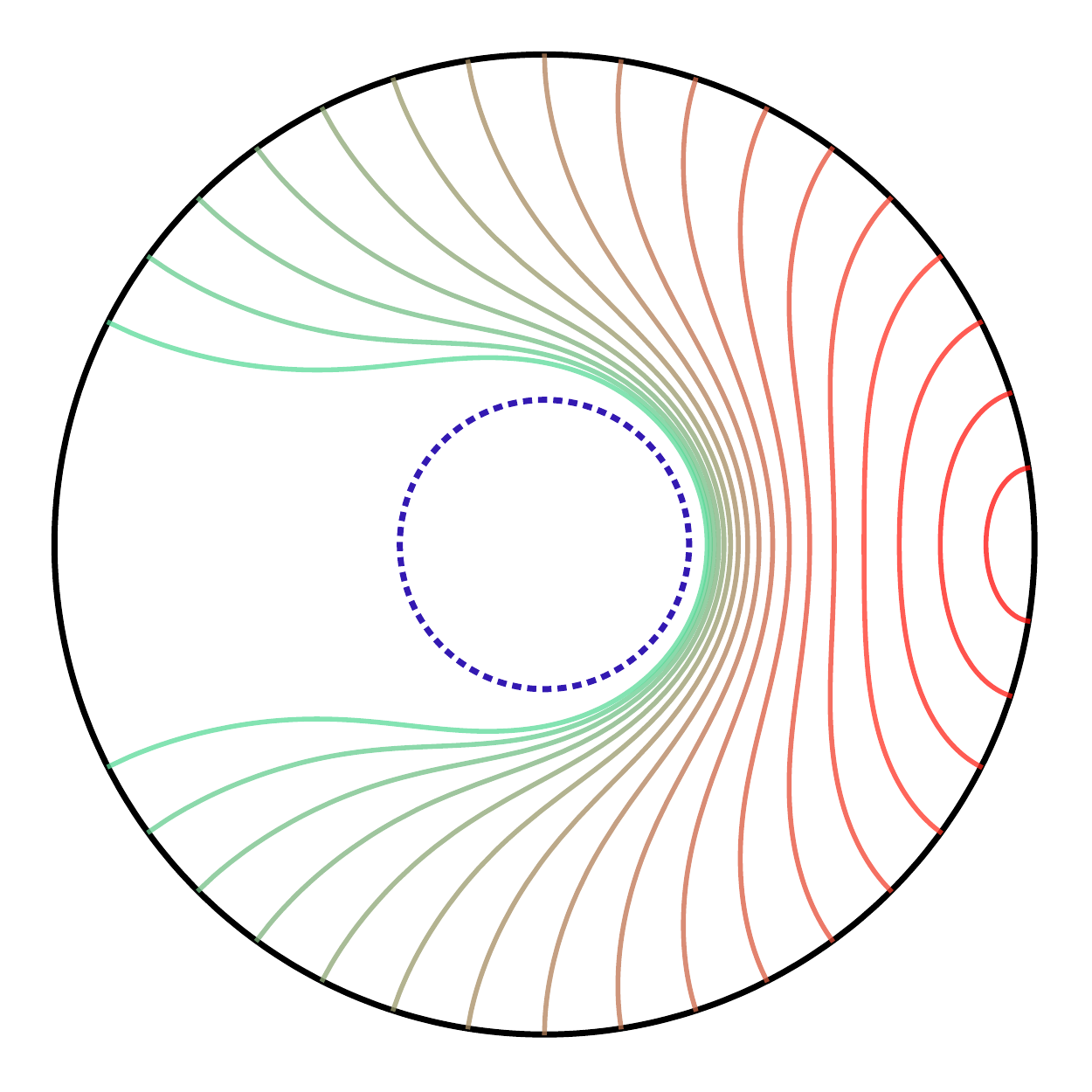}
\hspace{2cm}
\includegraphics[width=2.5in]{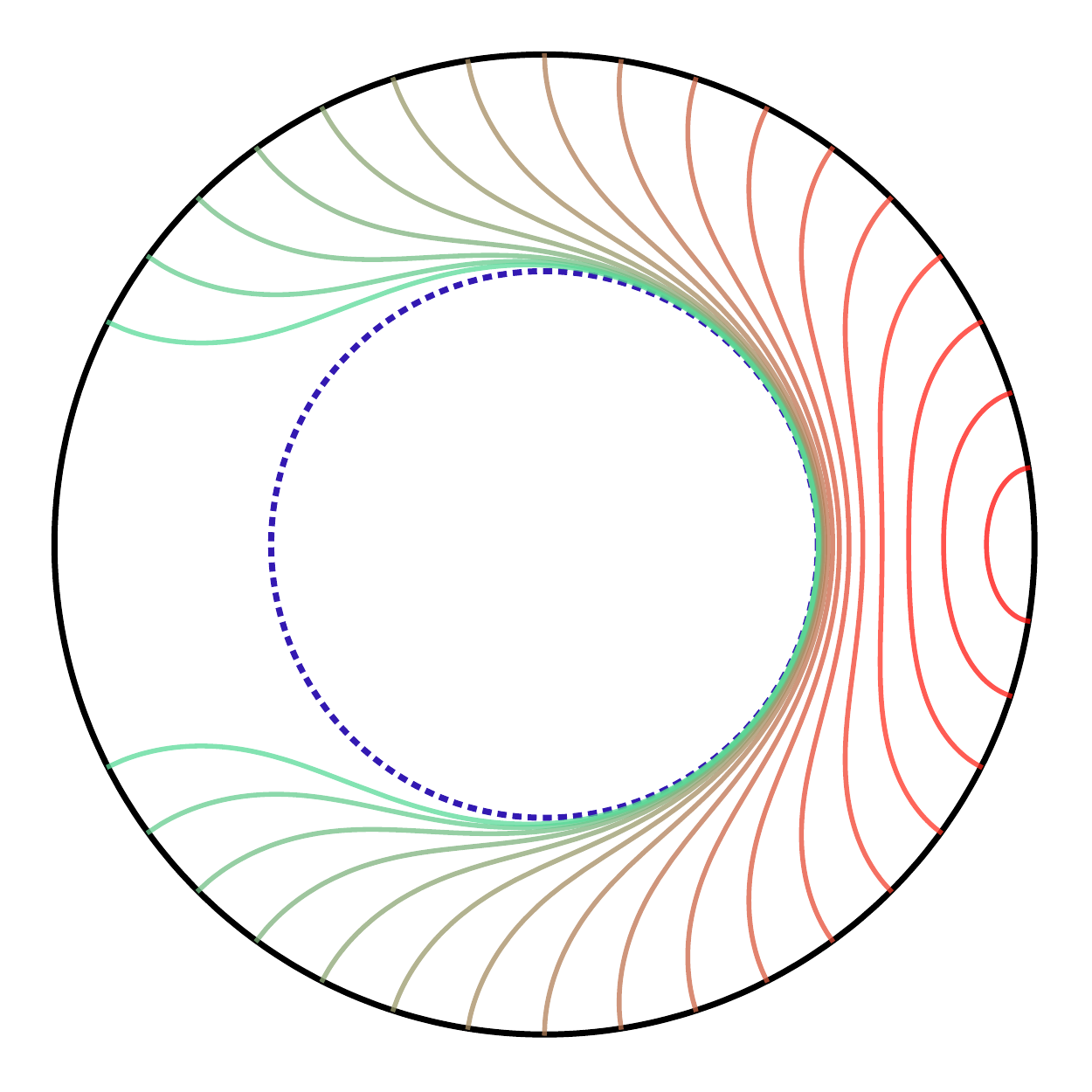}
\end{center}
\caption{Plot of minimal surfaces in the BTZ geometry the hyperbolic disc \eqref{eq:btzEA}. The conventions are as in
Fig.~\ref{f:EAads3}, with the horizon shown as the blue dashed curve. The plot on the left is for $\rh = 0.5\, \lads$
while that on the right is for $\rh =1.2 \,\lads$.}
\label{f:btzgeods}
\end{figure}

\paragraph{4. Thermal state of CFT$_2$ on ${\bf S}^1$:} The computation of entanglement entropy for a CFT$_2$ on a compact space at finite temperature is formidable. As indicated in \S\ref{sec:eecft2}, the R\'enyi entropies computing via replica require evaluating the partition function of the theory on arbitrary genus Riemann surfaces, cf., Fig.~\ref{f:branch3}. However, the holographic computation may be performed without much difficulty. The dual geometry is the global BTZ spacetime \eqref{eq:btz}. Setting $\lads=1$ for simplicity,  it is a simple matter to find the minimal surfaces for regions $\regA = \{\varphi: -\phA < \varphi| < \phA\}$. We simply need spacelike geodesics anchored at these boundary points which are given to be \cite{Hubeny:2012wa}
\begin{equation}
 \extrA^{(1)}: \quad \bigg\{t=0\,, \quad
 r = \gamma(\varphi,\phA,r_+) \equiv r_+  \, \left( 1 - \frac{\cosh^2 (r_+ \, \varphi)}{ \cosh^2 (r_+ \, \phA)} \right)^{\! \! -\frac{1}{2}} \ \bigg\}
\label{eq:btzEA}
\end{equation}
These curves are  plotted on the  Poincar\'e disc in Fig.~\ref{f:btzgeods}.

\begin{figure}[tp]
\begin{center}
\includegraphics[width=2.5in]{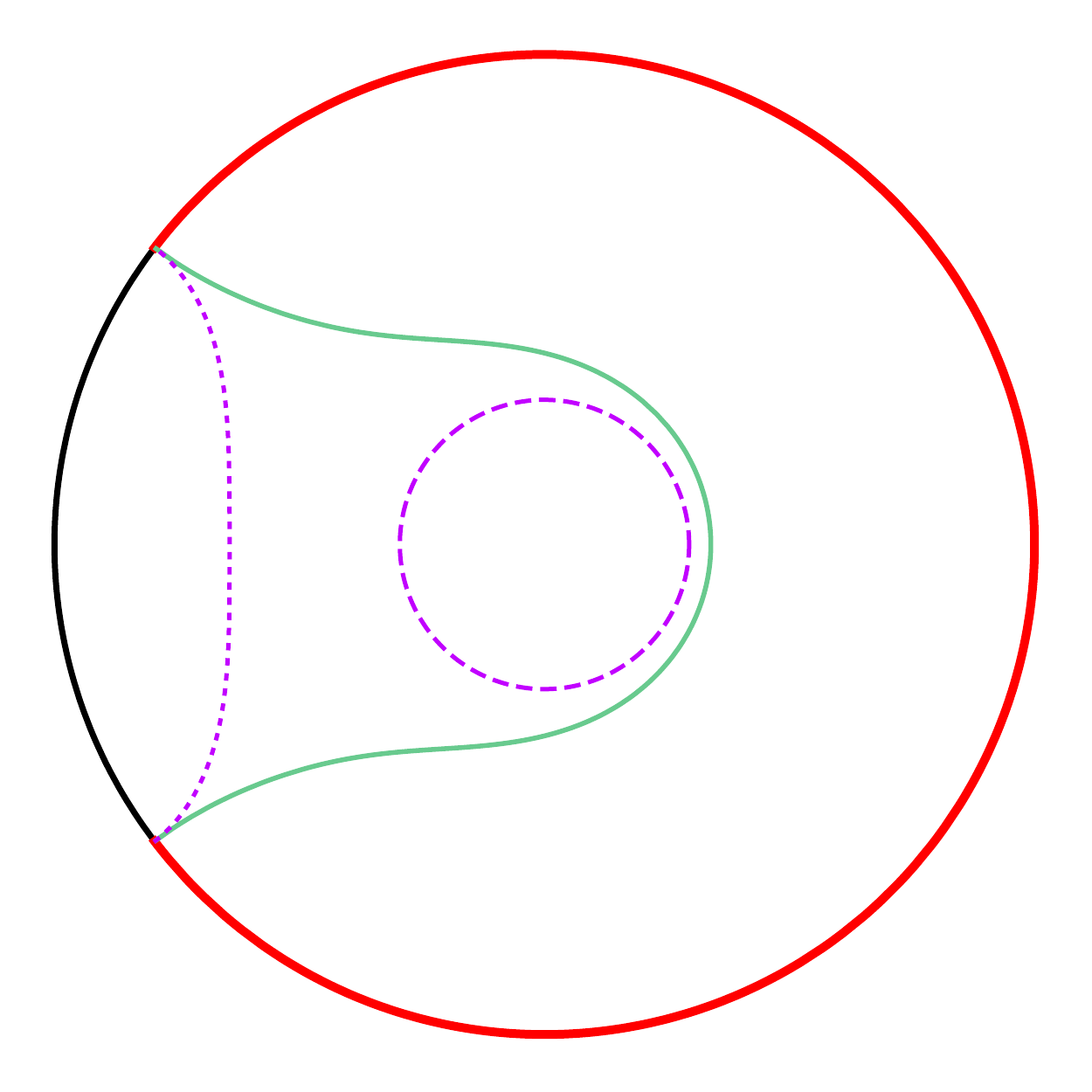}
\hspace{2cm}
\includegraphics[width=2.5in]{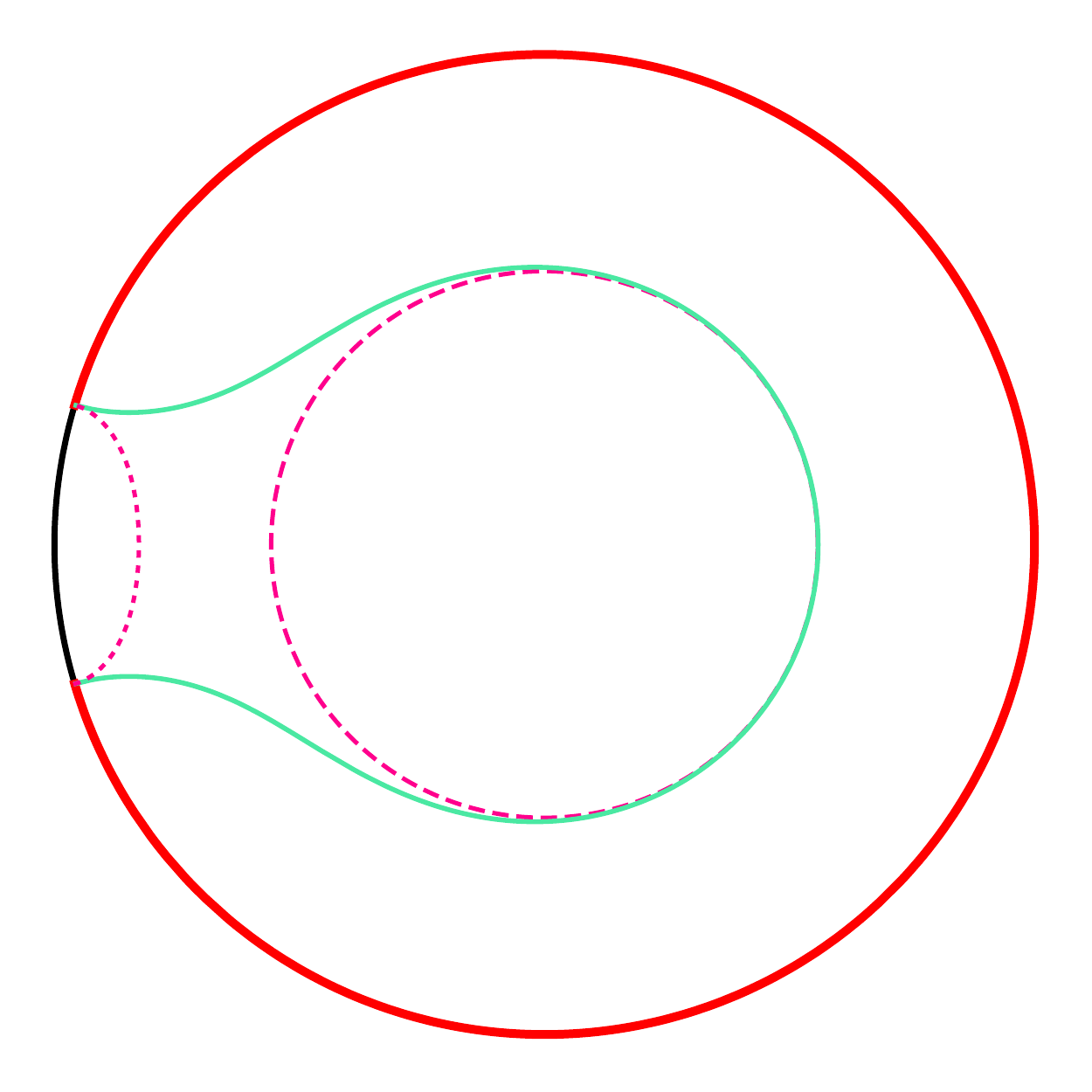}
\end{center}
\caption{ The transition between the connected $\extrA^{(1)}$ and disconnected  $\extrA^{(2)}$ minimal surfaces  in a BTZ black hole.
We illustrate the situation for $r_+ =0.5\, \lads$ and $r_+ = 1.2 \,\lads$. }
\label{f:ALtransition}
\end{figure}

For a given angular arc on the boundary, there are two potential minimal surfaces, one that stays homologous to the region ($\extrA^{(1)}$ above) and another that goes around the black hole, viz.,the curve $r =\gamma(\varphi, \pi-\phA,r_+)$ instead, as depicted in Fig.~\ref{f:ALtransition}. Accounting for the fact that we need to pick the globally minimal area surface in the homology class of the boundary region, we find that the holographic entanglement entropy is given by
\begin{align}
S_\regA(\phA) =\min \bigg\{ \frac{\text{Area}(\extrA^{(1)})}{4\, G_N^{(3)}}\;,\;\frac{\text{Area}(\extrA^{(2)})}{4\, G_N^{(3)}} \bigg\} \,,
\end{align}
where
\begin{align}
\extrA^{(2)} = \big\{t=0, \;\; r=r_+ \big\} \cup \big\{t=0,\;\; r =\gamma(\varphi, \pi-\phA,r_+) \big\} \,.
\label{}
\end{align}
The final answer upon evaluating the lengths reduces to
\begin{equation}
S_\regA =
  \begin{cases}
    &\frac{c}{3}\, \log \left(\frac{\beta}{\pi\, \epsilon} \, \sinh \left(\frac{R}{\beta}\,\phA\right)
    \right) \,,\hspace{3.5cm} \phA < \phA^\star
    \\
    & \frac{c}{3}\,\pi\,r_+ + \frac{c}{3}\, \log \left(\frac{\beta}{\pi\, \epsilon} \, \sinh \left(
    \frac{R}{\beta}\,(\pi - \phA) \right) \right) \,,\qquad \phA \geq \phA^\star
    \end{cases}
\label{eq:salargec}
\end{equation}
where we wrote the answer for a spatial circle of size $R$. We also introduced the critical angular scale $\phA^\star$  where the two saddles of the area functional exchange dominance; explicitly
\begin{align}
\phA^\star(r_+) = \frac{1}{r_+}\, \coth^{-1} \left(2\, \coth(\pi\, r_+) -1\right)  \,, \qquad \lim_{r_+ \to \infty}\; \phA^\star(r_+) = \pi \,.
\end{align}
%

% Figure

This phenomenon is indicative of a very general behaviour called the \emph{entanglement plateaux} in \cite{Hubeny:2013gta}. We will explain this more generally when we analyze holographic entropy inequalities in \S\ref{sec:heeineq}, for it corresponds to the saturation of the Araki-Lieb inequality. Note that the transition point approaches the size of the entire system as the temperature increases as noted above. Thus in the high temperature limit, the entanglement plateau transition scale approaches the size of the  system.

\begin{figure}[htbp]
\begin{center}
\includegraphics[width=5in]{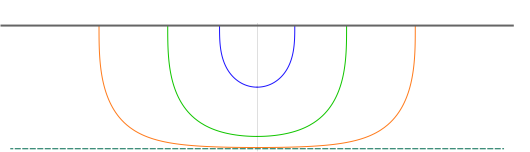}
\caption{Minimal surfaces in the planar \SAdS{5} spacetime for a black hole of size $z_+$. We pick the boundary region to be a strip of various widths, such that the turning point of the surface, $z_*$, occurs at
$z_* =0.5 z_+$, $z_* =0.9 z_+$ and $z_* = 0.99 \,z_+$, respectively.
}
\label{f:PSAdS5strip}
\end{center}
\end{figure}

\paragraph{5. Thermal state of CFT$_d$ on ${\mathbb R}^{d-1,1}$:} The geometry dual to the thermal density matrix of the CFT at temperature $T$ is given by the planar \SAdS{d+1} black hole \eqref{eq:psads} with the relation $ T = \frac{d}{4\pi\, z_+}$.\footnote{We now set $\lads =1$ to avoid cluttering up the notation. It can be reinstated through dimensional analysis.} The computation is easily done for either the strip-like regions \eqref{eq:stripdef} or the ball-shaped regions. The minimal surface  action can be easily seem to be,
\begin{equation}
\begin{split}
{\cal S}_{\regA_{\parallel}}  &= 4\pi\,\ceff L^{d-2}\, \int dx_1 \, \frac{1}{z^{d-1}}\, \sqrt{ 1+\frac{z'(x_1)^2}{f(z)}}
 \\
{\cal S}_{\regAB} &=  4\pi\,\ceff\, \omega_{d-2}\, \int d\xi \, \frac{\xi^{d-2}}{z^{d-1}}\, \sqrt{ 1+\frac{z'(\xi)^2}{f(z)}}
\label{}
\end{split}
\end{equation}
in the two cases of interest. We again use $L$  as the IR regulator of the translationally invariant directions for the strip. The equations of motion are easy enough to derive, but not trivial to solve. For  spherical domains, one can solve for the minimal surface in terms of Appel functions, but the resulting expression is unilluminating. It is more useful to examine the behaviour of the surfaces in the geometry, which are plotted in Fig.~\ref{f:PSAdS5strip}.

% Figure

For small regions $\regA$ on the boundary, the extremal surface lies close to the boundary. Using the scale/radius duality, we conclude that the holographic result only captures the UV sensitive part of the entanglement entropy. On the other hand, once we start to look at regions which are large compared to the thermal scale, then the extremal surfaces dip further down into the bulk. However, in static geometries they cannot penetrate the black hole horizon \cite{Hubeny:2012ry} (as long as they are anchored on the same boundary). This means that they get down nearly as far as the horizon, the turn-around point $z_* \simeq z_+$,  and straddle the horizon for almost the entire length of the region before returning to the boundary. This is clearly seen in the plots displayed in Fig.~\ref{f:PSAdS5strip}.

The behaviour of extremal surfaces is in accord with our expectation for entanglement entropy in a thermal state. For small regions $\regA$, the density matrix $\rhoA $ only carries the universal UV data which scales like the area of the region: $S_\regA \sim \frac{\text{Area}(\entsurf)}{\epsilon^{d-2}}$ as always. However, on macroscopic scales compared to the thermal scale, the area of the extremal surface exhibits extensive volume law behaviour from the IR; the contribution from the part that hugs the horizon scales like
$\text{Vol}(\regA) \, T^{d-1}$. This conforms to the general expectations elucidated in \S\ref{sec:uvirprop}.

\paragraph{6. CFT$_d$ on ${\bf S}^{d-1} \times {\mathbb R}$:} We can discuss both the vacuum state and the thermal state.  For the vacuum state, the dual geometry is the global \AdS{d+1} spacetime in Eq.~\eqref{eq:gads}.

The dual of the thermal state in finite spatial volume is more intricate. At high temperatures the dual geometry is the global \SAdS{d+1} black hole. For these solutions, the temperature of the black hole, which is the same as the field theory temperature, is given in terms of the horizon radius as noted in Eq.~\eqref{eq:Tsads}. Since they only exist above  a minimum temperature, $T > \frac{\sqrt{d(d-2)}}{2\pi\, \lads} $, it is clear that the low temperature phase has to be dominated by some other configuration. The only other solution satisfying the boundary conditions is the thermal \AdS{} spacetime. This has the same metric as the global \AdS{d+1} solution except that the Euclidean time circle is periodically identified.

It turns out that the global black holes dominate  only when their horizon size is larger than the AdS scale, i.e.,  $\rho_+ \geq \lads$, and not at the point where they come into existence. The system is characterized by a first order phase transition in the CFT   \cite{Witten:1998zw} at  $T_c = \frac{d-1}{2\pi\,\lads}$  called the Hawking-Page transition \cite{Hawking:1982dh}.
The low temperature  phase of the thermal CFT$_d$, with  $T < T_c $ is thermal \AdS{d+1} geometry, while the high temperature phase is always dominated by the  black hole.

\begin{figure}[htbp]
\begin{center}
\includegraphics[width=2.5in]{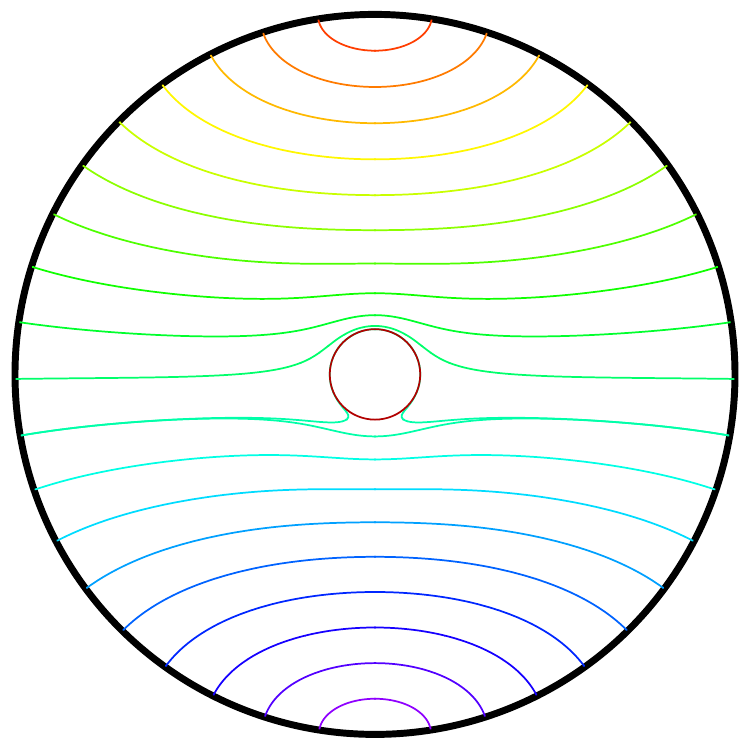}
\hspace{1cm}
\includegraphics[width=2.5in]{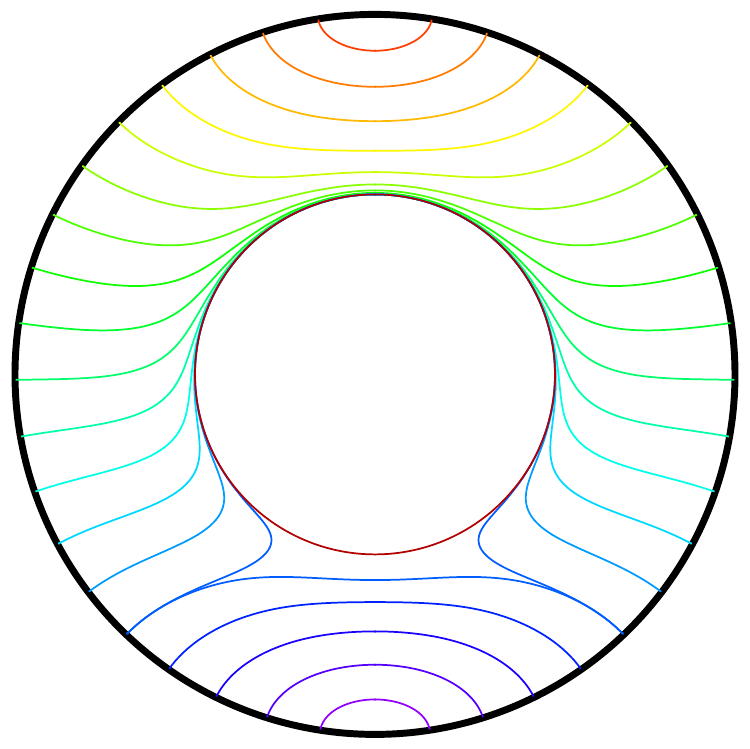}
\end{center}
\caption{Minimal surfaces in the global \SAdS{5} spacetime (figure from \cite{Hubeny:2013gta}). The left panel shows a small black hole $r_+ = 0.2\,\lads$ while the right panel is for $r_+ = \lads$. Apart from the fact that surfaces are repelled by the horizon, the absence of connected minimal surfaces for large regions is worth noting. We don't display the multiply wrapped surfaces around the horizon discovered in \cite{Hubeny:2013gta}.
}
\label{f:sads5}
\end{figure}

We take the region $\regA$ to be a polar-cap of the boundary ${\bf S}^{d-1}$. Picking coordinates  $\{\theta \in [0,\pi\, \Omega_{d-2}\} $ on the ${\bf S}^{d-1}$ such that the metric takes the form $d\Omega_{d-1}^2 = d\theta^2 + \sin^2\theta\, d\Omega_{d-2}^2$ we have the $SO(d-2)$ symmetric region
\begin{equation}
\regA_{polar-cap} = \{\theta, \Omega_{d-2} | 0\leq \; \theta \leq \theta_\regA\}
\label{}
\end{equation}
The minimal surface can be found from the action:
\begin{equation}
{\cal S} = 4\pi\, \ceff\, \omega_{d-2} \, \int d\xi\, \; (\rho\sin\theta)^{d-2} \,
\sqrt{\frac{1}{f(\rho)} \, \left(\frac{d\rho}{d\xi}\right)^2  + \rho^2 \, \left(\frac{d\theta}{d\xi}\right)^2 }
\label{}
\end{equation}
where we have left $\xi$ as the coordinate along the surface without gauge fixing it. The surfaces have to be found numerically in this case.
Computationally, it turns out to be simplest to work in a gauge where
$\sqrt{\frac{1}{f(\rho)} \, \left(\frac{d\rho}{d\xi}\right)^2  + \rho^2 \, \left(\frac{d\theta}{d\xi}\right)^2 }
= 1$, so that the evaluation of the on-shell action becomes less prone to numerical errors.  In the global \AdS{d+1} case, the extremal surfaces are analogous to those in global \AdS{3}  in Fig.~\ref{f:EAads3}. The black hole spacetime deforms the surfaces away from the horizon; a set of surfaces for the \SAdS{5} black hole are depicted in Fig.~\ref{f:sads5}.

% Figure

There are some salient features  worth noting in the high temperature phase described by the global \SAdS{d+1} spacetime. For small regions, the minimal surfaces are similar to that in the global \AdS{5} geometry, albeit with small deformations to account for the presence of the black hole. The deformations get larger  with the surfaces wanting to stay away from the horizon as in the planar case. Rather curiously, for sufficiently large regions, we find that there is no connected minimal surface! More precisely, for a given black hole size $r_+$, there exists a critical boundary region size, beyond which one finds no single connected minimal surface which satisfies the homology constraint. The only minimal surface is the surface corresponding to the smaller complementary region.

As in the BTZ discussion above, to satisfy the homology constraint one needs to take into account this surface for the smaller region and the bifurcation surface of the black hole event horizon. Once again this exemplifies  the entanglement plateaux phenomenon, which we discuss in detail in \S\ref{sec:heeineq}. The absence of connected minimal surfaces for large regions can be inferred from the causality constraints on the RT/HRT construction. As explained in \cite{Hubeny:2013gba}, the causal domains for  finite boundary regions in the global \SAdS{d+1} black hole can have non-trivial topology. These in turn lead to a restriction on where the extremal surfaces can lie, following from the fact that the causal wedge of the boundary domain of dependence has to be contained within the entanglement wedge. This forces the extremal surfaces to split. We will explain these concepts in  \S\ref{sec:egeometry}.

 There are also  some other peculiar properties of minimal surfaces in the global black hole spacetime. One finds subdominant saddle point solutions with the surfaces  wrapping the horizon multiple times  \cite{Hubeny:2013gta}. They play no role in the study of entanglement, but their presence points to the non-trivial interplay of the minimality/extremality condition with steep gravitational potential wells.

In the low temperature phase, we are in the thermal \AdS{d+1} geometry. Since we are looking at entanglement entropy at a fixed time slice, the identification of the Euclidean thermal circle is irrelevant. The extremal surfaces are the same as in the vacuum case, so we learn that for any boundary region, the result for the entanglement entropy coincides with its value in the vacuum. While this a-priori sounds bizarre, we are so far only talking about the leading part of the answer in the $c_\text{eff} \to \infty$ limit. There are corrections to the semi-classical result coming from the bulk entanglement entropy at $\mathcal{O}(1)$ explained in \S\ref{sec:bulkee}, which pick out the thermal contributions. This example will be helpful in building intuition about the connection between field theory entanglement and bulk geometry.

In the thermal state of the field theory, we have a density matrix for the entire system and our considerations involve looking at subregions $\regA$. We could however take the full system, in which case the entanglement entropy will actually compute the thermal entropy. This has a nice geometric interpretation using the thermofield double construction. The thermal entropy is the entanglement entropy of two copies of the system in the thermofield double state. In the black hole phase, the two copies are the two causally disconnected asymptotic regions, while in the low temperature phase, they are two copies of the \AdS{} geometry with no macroscopic entanglement.

\paragraph{7. Spherical domains of a CFT vacuum:} Finally, let us return to the example of spherical domains in the vacuum state of a CFT. We have already seen how to compute the minimal surfaces directly using the RT prescription. One can however make some general observations based on symmetry considerations as explained in Casini, Huerta, and Myers \cite{Casini:2011kv}. These domains preserve a $SO(d-2)$ rotational symmetry which we exploit fully below.

\begin{figure}[htbp]
\begin{center}
\includegraphics[width=5in]{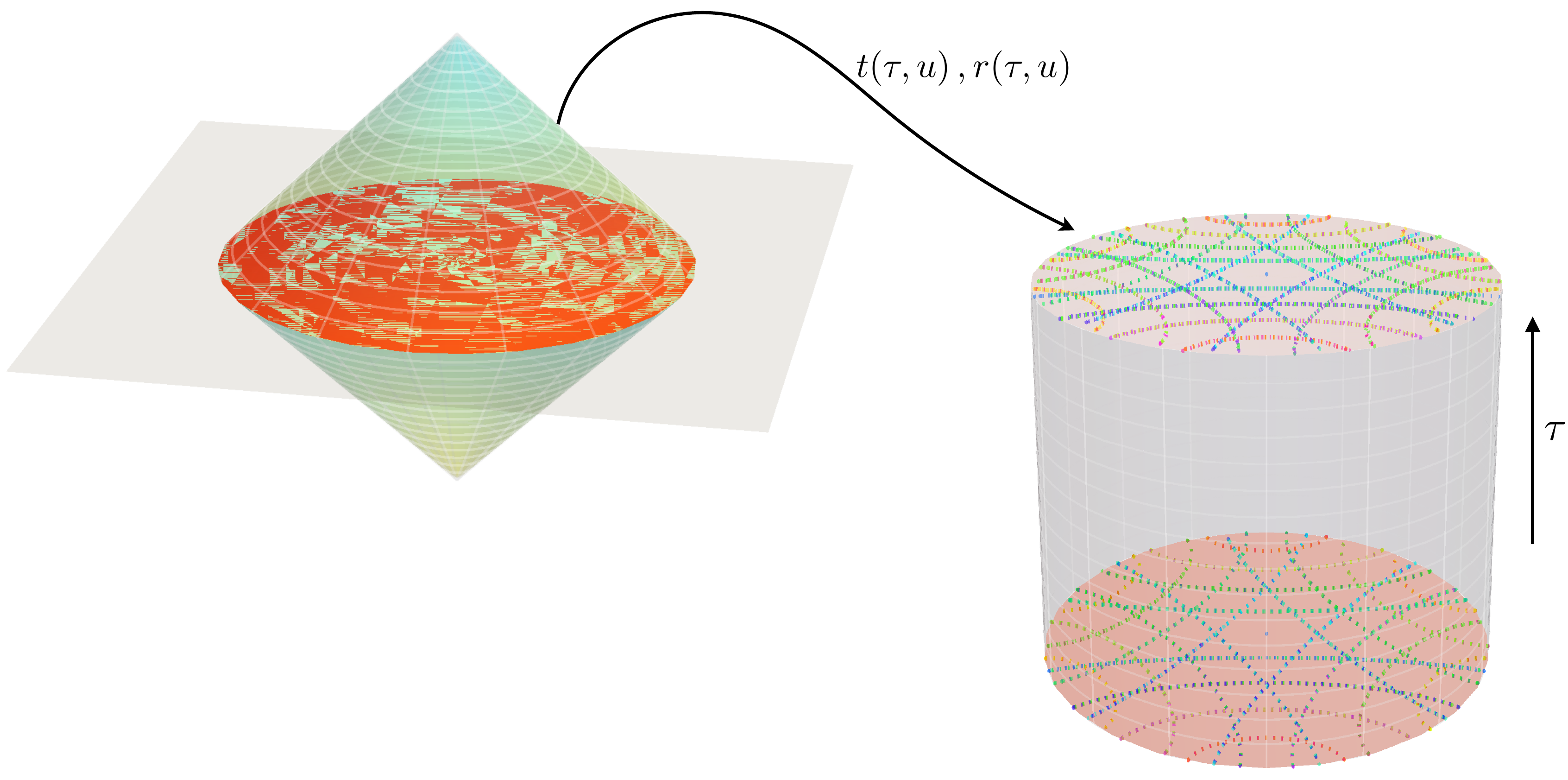}
\end{center}
\caption{The domain of dependence of a disc in in ${\mathbb R}^2 \subset {\mathbb R}^{1,2}$. By the conformal map
\eqref{eq:hycoord} the interior of this domain is mapped to the hyperbolic cylinder $\mathbb{H}_2 \times {\mathbb R}$.}
\label{f:chmmaps}
\end{figure}

\subparagraph{A sequence of conformal maps:}
Let us consider a ball-shaped region $\regA_{ball} \equiv \regAB \subset {\mathbb R}^{d-1,1}$.\footnote{ With minor changes, we can make similar observations for the polar-cap regions of ${\bf S}^{d-1} \times {\mathbb R}$.} The domain of dependence $D[\regAB]$ is a double-cone, with two apices  $p^\pm = \{t=\pm R, \; {\bf x}_{d-1} = 0\}$ at the future and past, respectively. We have already noted that the entanglement entropy is a wedge observable, and takes the same value of any Cauchy slice contained in the domain of dependence of the region in question. A geometric fact which is useful is the realization that
$D[\regAB]$ can be conformally mapped to a hyperbolic cylinder $\mathbb{H}_{d-1} \times \mathbb{R}$, as depicted in Fig.~\ref{f:chmmaps}.

To see this, start with flat space in polar coordinates adapted for the spherically symmetric ball $\regAB$:
\begin{equation}
ds^2  = -dt^2 + dr^2 + r^2\, d\Omega_{d-2}^2\,,
\label{eq:flatd}
\end{equation}
and consider the coordinate transformation:
\begin{align}
t = R\, \frac{\sinh\left(\frac{\tau}{R}\right)}{\cosh u + \cosh\left(\frac{\tau}{R}\right)} \,, \qquad r = R\, \frac{\sinh u }{\cosh u + \cosh\left(\frac{\tau}{R}\right)}\,.
\label{eq:hycoord}
\end{align}
Under this transformation, the metric \eqref{eq:flatd} becomes
\begin{align}
ds^2 = \frac{1}{\left[\cosh u + \cosh\left(\frac{\tau}{R}\right) \right]^2 }{}
\Bigg(-d\tau^2 +  R^2 \left(du^2  + \sinh^2 u \, d\Omega_{d- 2}^2\right)\Bigg)
\label{eq:hypc}
\end{align}
which one recognizes to be conformally related to the metric on the $(d-1)$-dimensional hyperbolic space ${\mathbb H}_{d-1}$ (i.e., Euclidean \AdS{d-1}) direct product with a timelike direction parameterized by $\tau$. We refer to this spacetime as the Lorentzian hyperbolic cylinder ${\mathbb R} \times {\mathbb H}_{d-1}$.

% Figure

Now the metric on the hyperbolic cylinder can be related by a second conformal mapping to a more familiar spacetime, the Rindler geometry, which is flat space written in boosted coordinates. This second set of transformations is simply achieved by writing the metric of $\mathbb{H}_{d-1}$ in Poincar\'e coordinates adapted to translational symmetry; we have
\begin{align}
ds^2 &= \frac{1}{\left[\cosh u + \cosh\left(\frac{\tau}{R}\right) \right]^2 }
\Bigg(-d\tau^2 +  \frac{dz^2  + \sum_{i=2}^{d-1} dX^i dX^i}{z^2}\Bigg) \nonumber \\
&\equiv
\frac{1}{z^2 \left[\cosh u + \cosh\left(\frac{\tau}{R}\right) \right]^2 }
\Bigg(-z^2 \, d\tau^2 +  dz^2  + \sum_{i=2}^{d-1} dX^i dX^i \Bigg)
\label{eq:rindlerhc}
\end{align}
While we could have directly attained this geometry starting from the flat metric, the intermediate step of having the hyperbolic cylinder will be useful momentarily.

In any event, we can follow the spherical domain through the coordinate transformations. We learn that the region $r \leq R$ is mapped under \eqref{eq:hycoord} to the entire hyperbolic geometry $\mathbb{H}_{d-1}$. One can think of this map as zooming out  the entangling surface  to infinity and zooming into the region $\regA$. Direct information about the complementary region $\regAc$ gets lost in the process. Its fate should become clear as we consider the nature of the density matrix and the holographic picture.

The hyperbolic space is the region $z \geq 0$ in the Poincar\'e coordinates. The final metric we have written is (conformally equivalent to) the Rindler coordinates   $\{z,t,X^i\}$ on flat space, with $\tau$ being the Rindler time coordinate. We can pass to Cartesian coordinates by the simple expedient of setting
\begin{equation}
X^1 \pm X^0 = X^\pm = z\, \exp\left(\pm \frac{\tau}{R}\right) \,.
\label{eq:Rcart}
\end{equation}
The entangling surface is now mapped to $z=0$, which is the bifurcation surface of the Rindler wedge in ${\mathbb R}^{d-1,1}$,
i.e., $X^pm=0$. The Rindler wedge can be thought of as the casual development of the half-space $X^1>0$ and what we have established is that this is conformally equivalent to $\domd{\regAB}$. This sequence of maps is illustrated in Fig.\ref{f:chmmaps}.

To take stock: the domain of dependence of a spherical ball in Minkowski space is conformal to both the Rindler geometry and the hyperbolic cylinder. In the first two descriptions, the domain of dependence of the complementary region remains explicitly visible, but it gets pushed out by the conformal mapping in the third.

\subparagraph{The Rindler modular Hamiltonian:} The rationale for this rigamarole can now be made transparent. Consider the Rindler geometry and  focus on the reduced density matrix  $\rho_{Rindler} $ obtained on the half-space
\begin{equation}
\regA_{Rindler} =\{ X^\mu \in {\mathbb R}^{d-1,1} | X^0 = 0\,, X^1 > 0\} \,.
\label{eq:rindler}
\end{equation}
  A salient result in algebraic QFT due to Bisognano-Wichmann \cite{Bisognano:1975ih,Bisognano:1976za} states that the modular Hamiltonian corresponding to this density matrix ${\cal K}_{Rindler}$ is just the Minkowski boost generator in the direction $X^1$. It implements a modular evolution as a  Rindler time translation
\begin{equation}
{\cal K}_{Rindler}: \tau \mapsto \tau + 2\pi\, R\, s \,,
\label{}
\end{equation}
which, by passing to standard Minkowski coordinates $X^\pm = z\, e^{\pm \frac{\tau}{R}} $, can be seen to be equivalent to a boost;
$X^\pm(s) = e^{\pm 2\pi\, s}$. This can equivalently be understood by noting that the Minkowski vacuum appears thermally populated to a uniformly accelerated observer as noted by Unruh \cite{Unruh:1976db}. All in all this implies that the vacuum modular Hamiltonian $\regA_{Rindler}$ in any QFT can simply be written as
\begin{equation}
{\cal K}_{Rindler} = 2\pi\, \int_{\regA_{Rindler}} \; d^{d-1}X\; X^1\, T_{00}(0,{\bf X})\,.
\label{eq:RindlermodH}
\end{equation}

\subparagraph{The modular Hamiltonian for the spherical ball:} We were interested in the density matrix $\rho_{\regAB}$ but what we have learnt so far is that $\rho_{\regA_{Rindler}}$ is amenable to general treatment in any relativistic QFT. However, should the theory in question be conformal, then by virtue of the fact that the vacuum state in Minkowski spacetime $\ket{0}$ is conformally invariant, we can infer properties of $\rho_{\regAB}$ through the sequence of mappings. As we will have many an occasion to refer to this density matrix, let us give a new notation
\begin{equation}
\rho_{\regAB}^\text{vac} = \rho_{_{\ball}}
\label{}
\end{equation}

Since the conformal group leaves the state in question invariant, one concludes that the reduced density matrices are related by a unitary transformation $U$: $\rho_{_{\ball}}= U\, \rho_{\regA_{Rindler}}\, U^\dagger$.  Tracing through the sequence of coordinate transformations, one can then show that
the modular Hamiltonian for the ball-shaped domains in the vacuum state of a CFT takes the form:
\begin{equation}
{\cal K}_{_{\ball}}= 2\pi\, \int_{\regAB} \, d^{d-1} x\; \frac{R^2-r^2}{2\,R} \,T_{00}(x)
\label{eq:BallmodH}
\end{equation}
Furthermore, since the various density matrices are related by unitary transformations, we end up with the same von Neumann entropy.

\subparagraph{Thermal state on the hyperbolic cylinder:}  While we now have the expressions for the modular Hamiltonians, which can be exponentiated to obtain the density matrices, one still has to learn to compute various entropies from them. This can be achieved most efficiently by invoking the intermediate element in our mapping sequence: the hyperbolic cylinder.

A Rindler observer sees the Minkowski vacuum as a thermal state in any field theory. So the reduced density matrix for one Rindler wedge is simply a thermal density matrix.  Using  the map to the hyperbolic cylinder,  the reduced density matrix induced $\mathbb{H}_{d-1}$ must again be simply the thermal density matrix.  In terms of the CFT Hamiltonian $ {\cal H}_{\mathbb{H}_{d-1}} $ we can therefore write:
\begin{equation}
{\cal K}_{\mathbb{H}_{d-1}} =2\pi\, R\; {\cal H}_{\mathbb{H}_{d-1}} \,,
\qquad \rho_{\mathbb{H}_{d-1}} = e^{-\beta\,  {\cal H}_{\mathbb{H}_{d-1}}}\,.
\label{eq:HCmodH}
\end{equation}
The  temperature is aligned with the curvature scale of the hyperbolic space
\begin{equation}
T = \frac{1}{\beta} =  \frac{1}{2\pi\,R}
\label{eq:hyT}
\end{equation}

As noted earlier, the interior of $\regAB$ got mapped onto the hyperbolic space $\mathbb{H}_{d-1}$, and we are learning from this exercise that $\rho_{_{\ball}}$ is unitarily equivalent to a simple thermal density matrix, viz.,
\begin{equation}
\rho_{_{\ball}} = \tilde{U} \, e^{-\beta\,  {\cal H}_{\mathbb{H}_{d-1}}}\, \tilde{U}^\dagger\,,
\end{equation}
for some unitary $\tilde{U}$ that implements the geometric conformal map on the density matrices.

Ergo, finding the von Neumann entropy in this particular case is tantamount to studying CFT thermodynamics on a uniformly negatively curved hyperbolic space. Moreover,  taking powers of the thermal density matrix is achieved by dialing the temperature. One can get $S^{(q)}_\regA $ by simply tuning $T\to \frac{1}{2\pi\,q\,R}$.
This fact has been exploited in various free field computations \cite{Klebanov:2011uf} and holography \cite{Hung:2011nu}.

\subparagraph{Holography and hyperbolic black holes:} Finally, let us turn to the holographic context. The RT minimal surfaces that compute the entanglement entropy for the spherical domains are simply hemispheres in \AdS{d+1} respecting the $SO(d-2)$ symmetry, cf., \eqref{eq:rtsphere}. We should now be able to arrive at the same result by studying the thermal density matrix on the hyperbolic cylinder.

The AdS/CFT correspondence relates thermal states of the boundary CFT to black hole geometries. Since the CFT has to be on a spatial $\mathbb{H}_{d-1}$, our boundary conditions require an asymptotically locally \AdS{d+1} spacetime whose boundary is $\mathbb{H}_{d-1} \times {\mathbb R}$ at the specified temperature.  Spacetimes  satisfying these boundary conditions are the \emph{hyperbolic black holes} \cite{Emparan:1999gf}. Using the coordinatization of the hyperbolic cylinder as in \eqref{eq:hypc},
we can write the metric of the one-parameter family of hyperbolic-\AdS{d+1} black holes as
\begin{equation}
\begin{split}
ds^2 &= -\frac{\lads^2}{R^2} \; f_{\mathbb{H}}(\varrho)\, d\tau^2 + \frac{d\varrho^2}{f_{\mathbb{H}}(\varrho)} + \varrho^2 \, \left(du^2 + \sinh^2u\; d\Omega_{d-2}^2\right) \,, \\
f_{\mathbb{H}}(\varrho) &\equiv \left( \frac{\varrho^2}{\lads^2} -1 - \frac{\varrho_+^{d-2}}{ \varrho^{d-2}}\left(\frac{\varrho_+^2}{\lads^2}-1\right)\right) .
\end{split}
\label{eq:hypbhsol}
\end{equation}
The parameter $\varrho_+$ corresponds to the black hole mass. As always,  it determines the location of the horizon and fixes the black hole temperature:
\begin{equation}
T = \frac{\lads}{4\pi\, R} \left[ \frac{d \, \varrho_+}{\lads^2} - \frac{(d-2)}{\varrho_+} \right] .
\label{eq:Thypbh}
\end{equation}

The reduced density matrix for the spherical region is related to the thermal density matrix at a particular temperature given by \eqref{eq:hyT}. This is achieved by choosing $\varrho_+  = \lads$.

For this choice, the black hole solution simplifies considerably. In fact, one can check that for this choice of the horizon radius the solution \eqref{eq:hypbhsol} is simply the \AdS{d+1} spacetime in hyperbolic coordinates! An easy way to do this is to compute the Riemann tensor and find it to be that of a maximally symmetric spacetime, viz., $R_{ABCD}  \propto g_{AC} g_{BD} - g_{AD} g_{BC}$.

One can physically interpret the above result as follows: the conformal transformation makes our reduced density matrix $\rho_{_{\ball}}$ unitarily equivalent to a thermal density matrix. Moreover, the same transformation also blows up $\regA$ at the expense of sending $\entsurf$ to asymptotic infinity.  In the holographic dual, we retain the interior of a causal domain associated with the extremal surface mapping the rest of the spacetime to infinity. This converts the spacetime into a black hole geometry, with the minimal surface becoming the bifurcation surface of the black hole horizon. The purifying complement $\regAc$ becomes the second asymptotic region in the spacetime, since $\rho_{\mathbb{H}}$ admits a nice thermofield double construction.

Once we have the solution, we can immediately extract the entanglement entropy as the black hole entropy for the hyperbolic black hole. The Bekenstein-Hawking formula requires the area of the bifurcation surface. This is the hypersurface $\varrho = \varrho_+$ at $\tau =0$ which is indeed extremal in the spacetime.
\begin{equation}
\begin{split}
S_{\mathbb{H}_{d-1}} = S_{BH} &= \frac{\omega_{d-2} \; \varrho_+^{d-1}}{4\, \GN}\; \int_{u=0}^{u_\text{umax}} \, \sinh^{d-2} u \, du   \\
& = 4\pi\,\ceff\, \omega_{d-2} \left[-i^{1-d}\; \cosh u_\text{max} \  _2F_1\left(\frac{1}{2},\frac{3-d}{2};\frac{3}{2};\cosh ^2u_\text{max}\right) \right]
\end{split}
\label{eq:}
\end{equation}
One can also relate this expression to our previous answer \eqref{eq:Sdisc1}. The transformation $\tan\theta = \sinh u$ will convert that expression to the first line of the above.

The last step in relating this to the CFT data involves us matching the UV cut-off $\epsilon$ employed to regulate the entanglement of the spherical ball to an IR cut-off $u_\text{max}$ in $\mathbb{H}$.  This can be worked out by following the coordinate transformations, obtaining
\begin{equation}
u_\text{max} = -\log\left(\frac{\epsilon}{2\,R}\right)
\label{eq:uvirH}
\end{equation}

At the end of the day, one then finds
\begin{equation}
S_{\regAB} = \frac{2}{\pi^{\frac{d}{2}-1}}\, \frac{\Gamma(\frac{d}{2})}{d-2} \, a_d\, \frac{\text{Area}(\entsurf)}{\epsilon^{d-2}} + \cdots +
\begin{cases}
& 4\, (-1)^{\frac{d}{2}-1} \, a_d\, \log\frac{2R}{\epsilon} \,, \qquad d = 2m\,,\\
& (-1)^{\frac{d-1}{2}}\, 2\pi\, a_d\,, \qquad\qquad  \;\,d = 2m+1\,.
\end{cases}
\label{eq:sball}
\end{equation}
We have written the final answer for the entanglement entropy in terms of the area of the entangling surface
 $\text{Area}(\entsurf) = \omega_{d-2}\, R^{d-2}$ and a parameter
 $a_d\equiv \frac{2\pi^\frac{d}{2}}{\Gamma(\frac{d}{2})} \, \ceff$, which in even-dimensional CFTs coincides with the $a$-type component of the trace anomaly. We will have more to say about the trace anomaly in \S\ref{sec:cthm}.

%~~~~~~~~~~~~~~~~~~~~~~~~~~~~~~~~~~~~~~~~~~~~~~~
\section{Holographic UV and IR properties}
\label{sec:holuvir}
%~~~~~~~~~~~~~~~~~~~~~~~~~~~~~~~~~~~~~~~~~~~~~~

In \S\ref{sec:uvirprop}, we gave general arguments for the behaviour of entanglement entropy in continuum QFTs. One of the issues we highlighted there was the area law behaviour of the UV divergent term. One invoked the result by
heuristically appealing to the local nature of the QFT vacuum, and  was well supported by evidence from free field computations. We also saw there the imprint of IR scales in the entanglement entropy, in particular, their contributions to the finite or universal terms of entanglement entropy.

Let us now ask what the corresponding statements are in the holographic descriptions. The discussion applies equally to the static and the time-dependent scenarios captured by the RT and HRT prescriptions, so we will discuss them simultaneously in what follows.

\begin{figure}[htbp]
\begin{center}
\begin{tikzpicture}
\draw[ultra thick,black] (-3,0) -- (3,0);
\draw[thick, red] (-2,0) -- (0,0) node [above] {$\regA$} -- (2,0) ;
\draw[thick,fill=red] (-2,0) circle (0.3ex);
\draw[thick, fill=red] (2,0) circle (0.3ex);
\draw[thick,fill=blue, opacity =0.1] (2,0) arc (360:180:2);
\draw[thick,blue] (2,0) arc (360:180:2);
\draw[thick,blue]  (0,-2) node [below] {$\extrA$};
\draw[ultra thick,->] (-2,0) -- (-2,-0.5);
\draw[ultra thick,->] (2,0) -- (2,-0.5);
\draw[thick,->] (3,0) -- (3.5,0) node [above] {$x$};
\draw[thick,->] (-3,0) -- (-3,-1) node [left] {$z$};
\end{tikzpicture}
\end{center}
\caption{Sketch of an extremal surface indicating that it ends normally on the boundary. }
\label{f:perpext}
\end{figure}
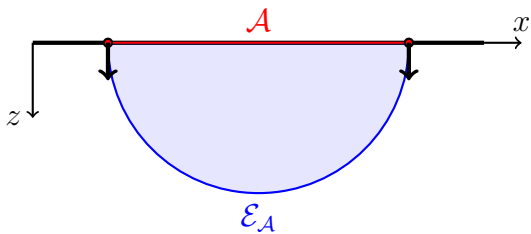

The first statement which is worth recording is that the extremal surfaces in AdS end normally on the boundary. This is very intuitive for minimal surfaces: they like to minimize their area, but the boundary of AdS extracts a steep gravitational penalty owing to the conformal factor. So the surfaces of interest try to exit this region as rapidly as possible. Thinking about the construction as a shooting-problem should suffice to convince oneself that the surface departs into the bulk from $\entsurf$ perpendicularly; this is  clearly visible in the various examples discussed in \S\ref{sec:extdeter} and is also illustrated  in Fig.~\ref{f:perpext}.

% Figure

More explicitly, this statement can be confirmed rather explicitly by working in the Fefferman-Graham coordinates
\eqref{eq:inducedE}.  To leading order in the small $z$ expansion, we gauge fix one of the $\xi$ to be the coordinate $z$, with the rest being tangential to $\entsurf$. Then  it is straightforward to check that the induced metric on  extremal surfaces in the vicinity of $z=0$ behaves as
\begin{equation}
ds^2_{\extrA} = \frac{1}{z^2} \left(dz^2 + \tilde{\gamma}_{ab}\, d\sigma^a \, d\sigma^b  + \cdots\right) \,.
\label{}
\end{equation}
in the  coordinate charts as described in \S\ref{sec:extdeter}. With this information, we can immediately see that:
\begin{equation}
\text{Area}(\extrA) = \int_{\epsilon}^{z_{IR}} \, \frac{dz}{z^{d-1}} \,
\int d^{d-2}\xi\, \sqrt{\tilde{\gamma}} + \cdots = \frac{\text{Area}(\entsurf)}{\epsilon^{d-2}} + \cdots
\label{}
\end{equation}
i.e., the area of the extremal surface diverges as expected, with the leading behaviour being determined by the area of the entangling surface as in \eqref{areala}. These statements follow essentially from the local near-boundary behaviour of the surface and capture the UV properties of the entanglement.

One can be a bit more systematic and decipher the subleading divergences explicitly; see \cite{Liu:2012eea} for some explicit expressions. In fact, the analysis of \cite{Henningson:1998gx,Graham:1999pm} can be used to immediately extract the universal terms in the entanglement entropy. In even-dimensional field theories, the coefficient of the logarithmic divergent term can be determined from the conformal anomaly of submanifolds immersed in the asymptotically AdS spacetime. The explicit results given for various examples in \S\ref{sec:extdeter} provide clear illustration of this fact. One can indeed go on and extract the subleading divergences which can be expressed in terms of the geometric data of $\entsurf$. In general, the subleading non-universal contributions are expressed  in terms of the intrinsic and extrinsic data of the entangling surface \cite{Solodukhin:2008dh}.

While that takes care of the UV properties, we can also understand geometrically the IR features. Recall our  dictionary between CFT states and dual geometries in \S\ref{sec:gravcon}. The UV features in generic states are the same as in the vacuum; geometrically this is clear, since the states in the Hilbert space of the QFT correspond to geometries with the same AdS asymptotics. Different states will however have very different bulk geometries and the differences will be most pronounced in the core  IR region (away from the boundary). As discussed earlier, a clear example of such a situation is the geometry dual to a thermal state of the QFT which is described by a black hole in AdS,  cf., Fig.~\ref{f:PSAdS5strip}.

In any state other than the vacuum, we expect there to be non-trivial expectation values of some field theory operator; at the very least we would have non-zero energy-momentum in the state. In the dual geometry, this amounts to the presence of gravitational or matter fields permeating the AdS spacetime, and deforming the geometry, in particular, by giving rise to gravitational potential wells.

For example, in the case of a planar-\SAdS{d+1} black hole, we have only the metric degrees of freedom excited as in \eqref{eq:psads}. The boundary thermal state acquires a non-zero stress tensor
\begin{equation}
\vev{T_\mu^{\;\nu}} \propto \ceff \, T^d \left(\delta_\mu^{\;\nu} + (d-1) \delta_\mu^{\ t}\, \delta_t^{\;\nu}\right) \,,
\label{}
\end{equation}
and the geometry has a steep gravitational potential owing to the presence of the black hole. The behaviour of minimal surfaces anchored on some region $\regA \subset {\mathbb R}^{d-1}$ has been described in Fig.~\ref{f:PSAdS5strip}; the presence of a black hole in the bulk deforms the  surfaces explicitly.

 In general, positive energy sources in the bulk push the surfaces closer towards the boundary,  usually resulting in the increase of the entanglement entropy. In case of the black hole we see the  effect is to make the RT surface for a large spatial regions  straddle the horizon. It is then clear that when we consider regions $\regA$ with $L_\regA \, T \gg 1$,  the dominant contribution to the entanglement entropy will arise from the part of the surface lying close to the horizon. As there is no variation in the radial direction, this will give a contribution which is proportional to the $\text{Vol}(\regA)$. In other words, the IR contribution to entanglement entropy will be the the macroscopic volume law term. On  the  other hand, the UV contribution will arise from the part of the surface connecting the horizon to the boundary, which leads to the usual area law divergent term. What we see here is again the UV/IR correspondence at work. The geometric picture makes clear that the origins of the IR contributions can be traced to the geometry deep in the interior, while the asymptotic AdS structure always ensures that we have the area law UV divergence.

%~~~~~~~~~~~~~~~~~~~~~~~~~~~~~~~~~~~~~~~~~~~~~~~
\section{Holographic entropy inequalities}
\label{sec:heeineq}
%~~~~~~~~~~~~~~~~~~~~~~~~~~~~~~~~~~~~~~~~~~~~~~

Let us now turn to the inequalities satisfied by the holographic entanglement entropy. As reviewed in \S\ref{sec:eeineq}, the very definition of $S_\regA$ as a von Neumann entropy of a normalized Hermitian density matrix implies non-trivial inequalities that must be satisfied. These must be upheld by the holographic prescriptions. For otherwise, we would talking about quantities which have no intrinsic meaning in the boundary quantum field theory, despite their geometric elegance.

Let us now take stock of the entropy inequalities in the holographic context. It is useful to separate the discussion into the ${\mathbb Z}_2$ time-reflection symmetric case in which the RT prescription suffices and the general story in time-dependent scenarios. An excellent discussion of the geometric properties in the former case can be found in \cite{Headrick:2013zda}.

\paragraph{Positivity of entanglement entropy:} This is obvious from both the RT and HRT prescriptions, which relate the von Neumann entropy to an area of a spacelike surface. The latter by definition has a positive definite area.

\paragraph{Subadditivity:} This too is straightforward considering that the leading divergent term is given by the area of the entangling surface. Thus  the leading terms already suffice to show that the mutual information defined in \eqref{eq:midef} is non-negative definite.

One can ask if the subadditivity inequality is saturated, which would correspond to vanishing mutual information. Generically, this cannot happen since the mutual information bounds the correlations between the two domains which for well-behaved quantum states cannot be strictly zero.

However, it turns out that, to leading order in $\ceff$, holographic theories can have vanishing mutual information. The simplest configuration realizing this is the case of two disjoint regions $\regA_1$ and $\regA_2$ which are spatially separated on a scale much larger than the individual regions themselves. There are then two potential extremal surfaces for $\regA_1 \regA_2$. One is simply the disjoint union ${\cal E}_{\regA_1} \cup {\cal E}_{\regA_2}$ but there is a second non-trivial surface ${\cal E}_{\regA_1 \regA_2}$ that bridges the two regions as depicted in Fig.~\ref{f:sasurfs}. The entanglement entropy for this configuration is given in  \eqref{eq:hol2cft2}.

From the explicit expression, one can check that the surface ${\cal E}_{\regA_1 \regA_2}$ has a smaller area when the regions are close together, but turns out to have a greater area in comparison to the disconnected surface ${\cal E}_{\regA_1} \cup {\cal E}_{\regA_2}$ for larger separations. For simply-connected domains on the boundary, these are the only two possibilities that respect the homology constraint. Thus, when the two regions are far enough apart, we naively predict that $I(\regA_1:\regA_2) =0$. This statement however should be qualified, since the area contribution only captures the leading large $\ceff$ part of the mutual information.  This is one of the peculiarities of the semi-classical limit.

This holographic result is also obtained from large $c$ CFTs \cite{Hartman:2013mia}; we will explain this computation in \S\ref{sec:largec}. As discussed in \S\ref{sec:bulkee}, the bulk entanglement entropy gives the leading order correction to this result and one indeed finds that mutual information is ${\cal O}(1)$ in the planar limit. This was explicitly verified in \cite{Barrella:2013wja} for two-dimensional CFTs at large central charge.

\paragraph{Strong subadditivity:} The strong subadditivity inequality is an important constraint on the von Neumann entropy. The standard proof of this inequality \cite{Lieb:1973cp} hinges on some fundamental matrix identities for finite-dimensional systems. The proof for continuum systems is considerably involved. One might hope that the geometrization of entanglement entropy in the holographic context helps elucidate some basic features. This indeed turns out to be the case.

The holographic proof of strong subadditivity was first given for the RT proposal by Headrick and Takayanagi \cite{Headrick:2007km} and only much later was extended to the HRT proposal by Wall \cite{Wall:2012uf}. We will now sketch the essential elements of the two proofs which are extremely simple, and illustrate the power of geometrization. Since the elements are slightly different for the two cases, we first start with the static RT case.

% Figure
\begin{figure}[htbp]
\begin{center}
\begin{tikzpicture}
\draw[ultra thick,black] (-4,0) -- (8,0);
\draw[thick, red] (-3,0)  -- (2,0) ;
\draw[thick, red] (-1,0) -- (6,0) ;
\draw[thick,fill=red] (-3,0) circle (0.3ex);
\draw[thick, fill=red] (2,0) circle (0.3ex);
\draw[thick, fill=red] (-1,0) circle (0.3ex);
\draw[thick, fill=red] (6,0) circle (0.3ex);
\draw[thick,fill=orange, opacity =0.1] (6,0) arc (360:180:4.5);
\draw[thick,orange] (6,0) arc (360:180:4.5);
\draw[thick,fill=teal, opacity =0.1] (6,0) arc (360:180:3.5);
\draw[thick,teal] (6,0) arc (360:180:3.5);
\draw[thick,fill=red, opacity =0.1] (2,0) arc (360:180:2.5);
\draw[thick,red] (2,0) arc (360:180:2.5);
\draw[thick,fill=purple, opacity =0.1] (2,0) arc (360:180:1.5);
\draw[thick,purple] (2,0) arc (360:180:1.5);
\draw[thick, fill=black] (0,-2.45) circle (0.3ex);
\draw[thick,red]  (-2,0) node [above] {$\regA_1$};
\draw[thick,red]  (0.5,0) node [above] {$\regA_2$};
\draw[thick,red]  (4,0) node [above] {$\regA_3$};
\draw[thick,orange]  (0,-5) node [above] {${\cal E}_{\regA_1 \regA_2 \regA_3}$};
\draw[thick,red]  (-1,-3) node [above] {${\cal E}_{\regA_1 \regA_2}$};
\draw[thick,teal]  (2.5,-4.25) node [above] {${\cal E}_{\regA_2 \regA_3}$};
\draw[thick,purple]  (0.65,-2.15) node [above] {${\cal E}_{\regA_2}$};
\end{tikzpicture}
\end{center}
\caption{Sketch of  the configuration for proving strong sub-additivity. We show the three boundary regions $\regA_1$, $\regA_2$ and $\regA_3$  and the extremal surfaces ${\cal E}_{\regA_1 \regA_2}$, ${\cal E}_{\regA_2 \regA_3}$,
${\cal E}_{\regA_1 \regA_2 \regA_3}$, and ${\cal E}_{\regA_2}$
corresponding to the regions $\regA_1\cup \regA_2$, $\regA_2\cup \regA_3$, $\regA_1\cup \regA_2 \cup\regA_3$ and $\regA_2$, respectively. To prove the desired inequality, we perform local surgery at the point indicated by the black dot (note that it is a codimension-2 surface). Rejoining the red and green surfaces at this point to be homologous to $\regA_2$ and $\regA_1 \regA_2 \regA_3$, we arrive at the inequality \eqref{eq:ssa1A}.
 }
 \label{f:ssaproof}
\end{figure}

Consider the strong subadditivity inequality in the form \eqref{eq:ssa1} which we reproduce here for convenience:
\begin{align}
 S_{\regA_1 \regA_2}+S_{\regA_2 \regA_3}  & \geq S_{\regA_1 \regA_2 \regA_3}+ S_{\regA_2},
 \label{eq:ssa1A}
 \end{align}
 For each of the regions $\regA_1 \regA_2$, $\regA_1\regA_3$, etc., appearing in the inequality, we have corresponding bulk minimal surfaces ${\cal E}_{\regA_1\regA_2}$,  ${\cal E}_{\regA_2\regA_3}$, respectively, whose areas compute the entanglement entropies of interest. All of the four minimal surfaces of interest lie on a single bulk time-slice. We sketch in Fig.~\ref{f:ssaproof} the minimal surfaces  for a particular configuration of the regions, which we take to be contiguous for the sake of simplicity.

At this point, it is important to keep in mind the homology constraint, whereby we require $\extrA$ in the bulk to be homologous to $\regA$ on the boundary. Given the configuration, we realize that we can recombine the minimal surfaces for the regions on the l.h.s. viz., ${\cal E}_{\regA_1\regA_2}$,  ${\cal E}_{\regA_2\regA_3}$ by performing a local surgery, i,e. piecewise cutting and gluing, to construct two new surfaces ${\cal F}_{\regA_1\regA_2\regA_3}$ and ${\cal F}_{\regA_2}$ that are homologous to the regions $\regA_1 \regA_2 \regA_3$ and $\regA_2$ appearing on the r.h.s. Now while ${\cal F} _{\regA_2}$ is homologous to $\regA_2$, it is clearly not the minimal area surface anchored on $\entsurf_2$. This would of course be true even if we smoothed out the kink originating from our piecewise construction, for by assumption
${\cal E}_{\regA_2}$ is the appropriate minimal area surface. Note that we are by construction assuming that ${\cal E}_{\regA_2}$ provides the global minimum of the area functional with the prescribed boundary conditions. It is then trivial to see that
\begin{equation}
\begin{split}
\text{Area}\left({\cal E}_{\regA_1\regA_2}\right) + \text{Area}\left({\cal E}_{\regA_2\regA_3}\right)
& \geq
\text{Area}\left({\cal F}_{\regA_1\regA_2\regA_3}\right) + \text{Area}\left({\cal F}_{\regA_2}\right)
 \\
&\geq
\text{Area}\left({\cal E}_{\regA_1\regA_2\regA_3}\right) + \text{Area}\left({\cal E}_{\regA_2}\right)
\end{split}
\label{}
\end{equation}
establishing at leading order in $\ceff$ the strong sub-additivity inequality. While we have illustrated the essence of the argument presented in \cite{Headrick:2007km}, there are some subtleties that have to be dealt with to complete the argument in a watertight manner. These are discussed in the original paper referenced above, and further commentary can be found in \cite{Headrick:2013zda}. We refer the reader to these sources for further discussion and also for the proof of the alternate form of the inequality \eqref{eq:ssa2}.

The simplicity of the proof of  strong subadditivity for the RT proposal stems from the fact that there is a single bulk Cauchy slice which contains all four surfaces of interest. This makes it simple to see that the cutting/gluing construction we use guarantees us a surface satisfying the homology requirement that has a larger area than the true minimal surface.
If we consider a similar configuration in the generic time-dependent situation, we run into an essential difficulty. There is no single Cauchy surface of the bulk that contains all four extremal surfaces. In general, the surfaces span out a non-trivial codimension-0 region of the bulk, making it difficult to implement a version of the above procedure.

The proof of strong subadditivity of the HRT proposal is greatly simplified by resorting to the maximin reformulation of \cite{Wall:2012uf}. As explained at the end of  \S\ref{sec:hrrt}, the maximin construction proceeds by picking a bulk Cauchy slice $\bulkCS$ corresponding to a given region $\regA \subset \CSA$ on the boundary. One then finds a minimal surface on this slice, and subsequently maximizes the area of minimal surfaces across a complete set of Cauchy slices inside the FRW wedge of $\CSA$.

To see how this is useful for our present purposes, we first need the following two results:
\begin{enumerate}
\item[(i).] If two regions $\regA_1$ and $\regA_2$ are related by an inclusion, say $\regA_2 \subset \regA_1$, then there exists a bulk Cauchy slice which contains both ${\cal E}_{\regA_1}$ and ${\cal E}_{\regA_2}$, with the latter surface being spacelike to the former. This nesting property of extremal surfaces was proved in \cite{Wall:2012uf}.
\item[(ii).] If we start from an extremal surface $\extrA$ corresponding to some region $\regA$ and follow a bulk null congruence of light rays, then the cross-sectional areas of the sections of the congruence are necessarily bounded from above by the area of the extremal surface in sensible theories of gravity. In Einstein-Hilbert theory, this follows from the Raychaudhuri equation, assuming that the matter satisfies a sensible energy condition, such as the null energy condition.
\end{enumerate}

These suffice to give a proof of the HRT proposal satisfying the strong subadditivity requirement. We  consider the extremal surfaces ${\cal E}_{\regA_1\regA_2\regA_3}$ and ${\cal E}_{\regA_2}$, which by (i) lie on some common bulk Cauchy slice, say $\bulkCS$ for definiteness. However, nothing tells us that ${\cal E}_{\regA_1\regA_2}$ and ${\cal E}_{\regA_2\regA_3}$ also lie on this slice, but that is immaterial. Irrespective of where these surfaces are, we are free to project them onto $\bulkCS$ by following the null congruence emanating from them. Viewing this as a projection map ${\cal P}$, we have two new surfaces
${\cal P} {\cal E}_{\regA_1 \regA_2}$ and ${\cal P}{\cal E}_{\regA_2 \regA_3}$ which are also now on $\bulkCS$. The second result (ii) guarantees  that the area of the thus projected extremal surfaces is smaller than the true result. We now have all the ingredients necessary to rerun the local surgery argument, since all four surfaces are confined to a single slice. Putting all the pieces together, we arrive at:
\begin{equation}
\begin{split}
\text{Area}\left({\cal E}_{\regA_1 \regA_2}\right) + \text{Area} \left({\cal E}_{\regA_2\regA_3}\right)
& \geq \text{Area}\left({\cal P}{\cal E}_{\regA_1 \regA_2}\right) + \text{Area} \left({\cal P}{\cal E}_{\regA_2\regA_3}\right)
\\
& \geq \text{Area}({\cal E}_{\regA_1 \regA_2 \regA_3}) + \text{Area}({\cal E}_{\regA_2}) \,,
\label{}
\end{split}
\end{equation}
where the first inequality hinges on gravity being attractive and the second follows from the local surgery argument. Altogether this establishes the strong-subadditivity result as desired for the HRT proposal.

\paragraph{Araki-Lieb inequality:} The Araki-Lieb inequality \eqref{eq:alineq} bounds the difference of the entanglement entropies of a system and its complement in terms of that of the total density matrix. However, as remarked in \S\ref{sec:eeineq}, its status as a fundamental inequality is unclear. Since it follows from subadditivity via purification, it continues to hold in holographic theories.  Perhaps more intriguingly, it can actually be saturated in these theories (at least to leading order in $\ceff$).

The simplest situation illustrating this feature  is the computation of entanglement entropy of a thermal state of a CFT$_d$ on ${\bf S}^{d-1} \times {\mathbb R}$.  We consider dividing the ${\bf S}^{d-1}$ which is a Cauchy surface at an instant of time, say $t=0$ w.l.o.g., into $\regA$ and $\regAc$.  Further, let $\alpha = \frac{\text{Vol}(\regA)}{\text{Vol}({\bf S}^{d-1})}$ denote the fractional size of the region $\regA$ relative to the entire system. The holographic dual of this state is a global \SAdS{d+1} black hole whose spatial section is topologically non-trivial, owing to the presence of the black hole horizon. On the $t=0$ slice, the bifurcation surface which is a ${\bf S}^{d-1} \subset$ \SAdS{d+1} is a non-contractible codimension-2 sphere.

% Figure
\begin{figure}[htbp]
\begin{center}
\includegraphics[width=3.5in]{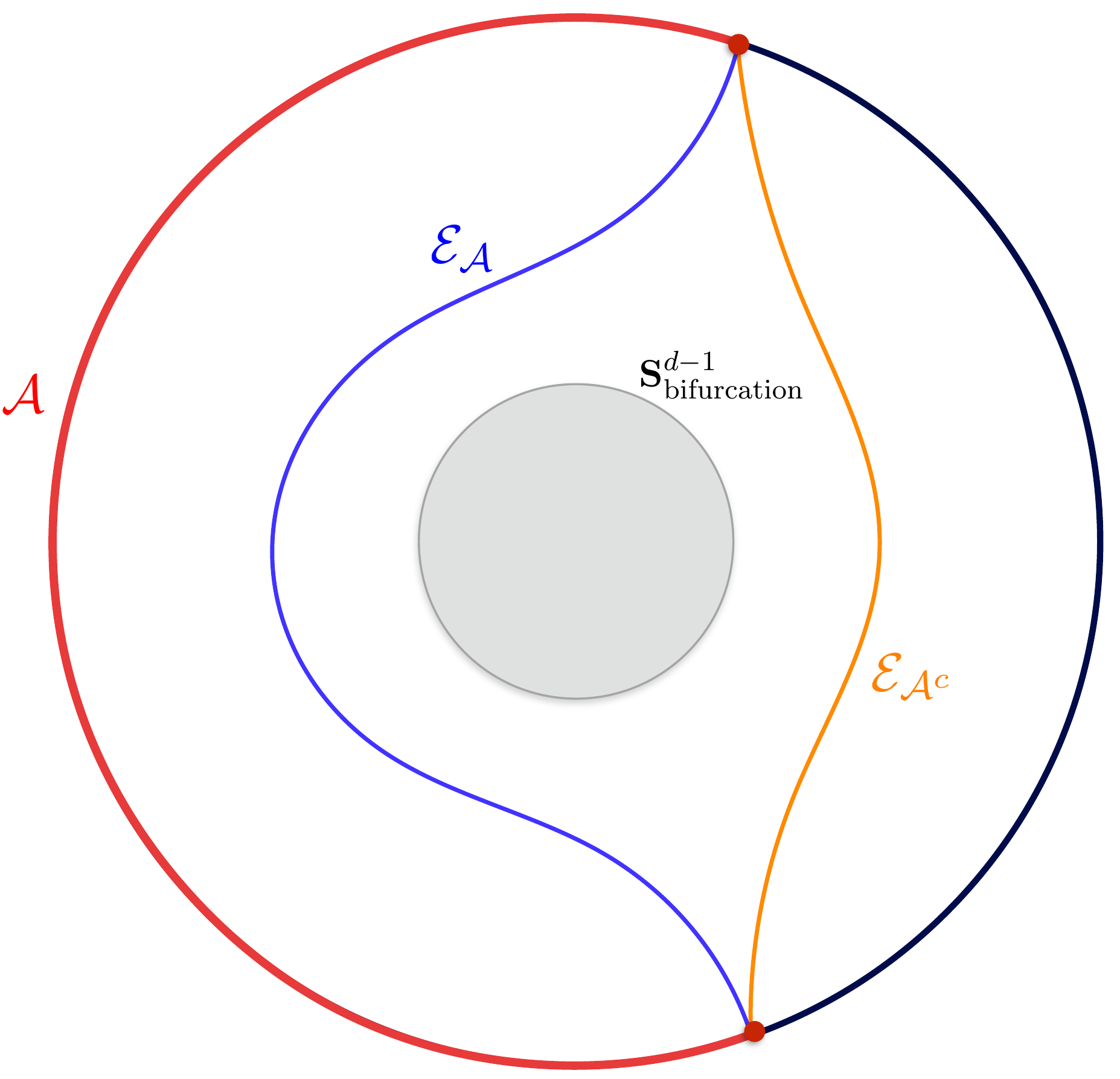}
\end{center}
\caption{The extremal surfaces around a global AdS black hole exhibiting the situation in which the Araki-Lieb inequality may be saturated. For the region $\regA$ which is the greater fraction of the boundary, there are two extremal surfaces satisfying the homology constraint: $\extrA$ and ${\cal E}_{\regAc} \cup {\bf S}^{d-1}_{_\text{bifurcation}}$ respectively. }
 \label{f:aldepict}
\end{figure}

For small regions $\regA$, viz., $\alpha \ll 1$, the entanglement entropy is given by the area of the minimal surface which stays on one side of the black hole.  However, for regions which are sufficiently large (for $\alpha$ exceeding $\frac{1}{2}$ at least), we have two potential contributions which are illustrated in Fig.~\ref{f:aldepict}:
\begin{enumerate}
\item[(a)] a single connected surface that is homologous to $\regA$, or
\item[(b)] a disconnected surface which comprises of the geodesic homologous to $\regAc$ and the bifurcation surface of the horizon.
\end{enumerate}
Moreover, as noted in \S\ref{sec:extdeter}, in $d=2$, we have an explicit \eqref{eq:salargec} exchange of saddle, while for $d>2$ and large enough regions,  the \SAdS{} geometry no longer admits connected minimal surfaces homologous to $\regA$.

Therefore, we find that once the boundary region $\regA$ exceeds a critical size $\alpha > \alpha_*$, we have
$\extrA = {\cal E}_{\regAc} \cup {\bf S}^{d-1}_{_\text{bifurcation}}$ leading to
\begin{equation}
S_\regA = S_\regAc + S_{thermal}
\label{}
\end{equation}
which we recognize as saturation of the Araki-Lieb inequality. This phenomenon was described in some detail in \cite{Hubeny:2013gta} in which the authors called it the {\em entanglement plateaux}, owing to the saturation of entanglement entropy for a large region.

While we have argued for this effect in a particular state of a holographic CFT, it turns out to be quite general and independent of the specific details of the regions and the symmetries preserved by the state in question  \cite{Hubeny:2013gta,Hubeny:2013gba}.  One way to understand the situation is as follows: We can associate two bulk domains corresponding to the domain of dependence $\domdA$ of the region $\regA$ on the boundary. The first of these is the causal wedge $\CW{\regA}$ which is simply the region of the bulk which can receive causal communications from or causally communicate to $\domdA$. On the other hand, the extremal surface construction motivates the idea of an entanglement wedge $\EW{\regA}$ which is the bulk domain of dependence of the homology surface $\homsurfA$.  We will describe these constructs in greater detail in \S\ref{sec:egeometry}.

The argument involves noting that the entanglement wedge has to contain the causal wedge. Furthermore, it can independently be shown that causal wedges in spacetimes can be topologically non-trivial. Formal arguments and explicit examples in terms of black holes were given in \cite{Hubeny:2013gba}, while examples involving causally trivial spacetimes were constructed in \cite{Gentle:2013fma}. Since the entanglement wedge has to contain the causal wedge, the presence of the holes in the latter  forces the extremal surfaces to become disjoint.  In other words,  phenomena such as the entanglement plateaux phenomenon will generically occur in holographic field theories.

\paragraph{Other entropy inequalities:} The class of holographic field theories being a subset of all quantum field theories leads one to ask if there are certain features of entanglement that are specific to them. This is indeed the case, for holographic entanglement entropy appears to satisfy a set of inequalities that are known to not hold in other quantum systems. Many of these appear to hold in the semiclassical limit. This behaviour is intimately tied to the fact that the holographic answer is given by an extremization procedure subject to some global conditions.

\subparagraph{a) Tripartite information inequality:} The prototype example of such holographic inequalities is the so-called {\em monogamy of mutual information} \cite{Hayden:2011ag}. Firstly, note that in  general a quantum information theoretic function $f$ is said to be \emph{monogamous} if
\begin{equation}
f(\regA_1:\regA_2)+f(\regA_1:\regA_3)\leq f(\regA_1 : \regA_2 \regA_3) \,.
\label{eq:mono}
\end{equation}
One can read \eqref{eq:mono} as saying that for an entanglement measure $f$, if  subsystem $\regA_1$ is almost maximally entangled, both with subsystem $\regA_2$ and a larger one $\regA_2\regA_3$, then there is almost no entanglement between $\regA_1$ and $\regA_3$, viz., $f( \regA_1: \regA_3)=0$. Monogamy of entanglement is then simply the statement of subaddivity, which asserts the positivity of mutual information defined in \eqref{eq:midef}. To ascertain the monogamy properties of mutual information, one defines the \emph{tripartite information} ${\sf I}_3$:\footnote{ This combination of entropies is also what appears in the computation of topological entanglement entropy for $2+1$-dimensional theories, as originally described  in
\cite{Kitaev:2005dm}. }
\begin{equation}
\begin{split}
\tmi(\regA_1 : \regA_2 : \regA_3)&= I(\regA_1: \regA_2) +I(\regA_1: \regA_3) - I(\regA_1: \regA_2 \regA_3)
\\
& = S_{\regA_1} + S_{\regA_2} + S_{\regA_3} - S_{\regA_1 \regA_2} - S_{\regA_1 \regA_3}  -S_{\regA_2 \regA_3}  + S_{\regA_1 \regA_2 \regA_3} \,.
\end{split}
\label{eq:mmi}
\end{equation}
Monogamy of mutual information would then require that the tripartite information be non-positive definite,
$\tmi \leq 0$ or equivalently.,
\begin{align}
 S_{\regA_1} + S_{\regA_2} + S_{\regA_3} + S_{\regA_1 \regA_2 \regA_3}  \leq  S_{\regA_1 \regA_2} + S_{\regA_1 \regA_3}  +S_{\regA_2 \regA_3}
\label{eq:mmi2}
\end{align}

Now it is known in simple quantum systems that the mutual information is not monogamous. It is easy to find states of qubit systems that have positive $\tmi$, the simplest example being provided by the GHZ state for 4 qubits. Note that we need at least 4 components to write down a pure state with $S_{\regA_1 \regA_2 \regA_3} \neq 0$. For the three qubit GHZ state, the $\tmi$ trivially vanishes; see \cite{Rangamani:2015qwa}  for a discussion of $\tmi$ and other measures of entanglement in simple qubit systems.

On the other hand, the holographic entanglement entropy has $\tmi \leq0$, as proved for the RT proposal in \cite{Hayden:2011ag} and for the HRT proposal in \cite{Wall:2012uf}. The basic idea of the proof in the two cases is similar to the discussion of the strong subaddivity, viz., one examines the surfaces contributing to the l.h.s. of \eqref{eq:mmi2} and shows by local surgery that they can be rearranged into contributions that can be associated with the regions on the r.h.s., cf., \cite{Hayden:2011ag,Wall:2012uf}.

\subparagraph{b) Holographic Entropy Cone:} More recently, using properties of minimal surfaces \cite{Bao:2015bfa} has derived an infinite set of entropy inequalities that are satisfied in holographic theories (in time reversal symmetric situations). These inequalities generalize the monogamy of mutual information, and carve out a convex polyhedron in the space of entropies called the \emph{holographic entropy cone}.

 To understand this concept, consider a partitioning of a Cauchy slice of a QFT into $(n+1)$-parts, in which $\{\regA_1, \regA_2, \cdots \, \regA_n\} $ are disjoint regions and the final region $\regA_{n+1} = \left(\cup_{i=1}^n\, \regA_i\right)^c$ is the purifier of the first $n$ regions. From the $n$-fundamental regions, we can form $2^n -1 $ disjoint unions. Letting
 $I \in \{1, 2, \cdots ,n \}$, we define $\regA_{_I} = \cup_{i\in I} \, \regA_i$ to denote elements of this collection.  Associated with each of these regions  is an entropy $S_{\regA_{_I}}$. One wishes to ask what constraints this collection is required to satisfy in order for the entropies to arise out of a holographic theory.

 The authors of \cite{Bao:2015bfa} address this issue when the entropies are all obtained from a RT minimal surface prescription for time-symmetric states of a holographic QFT.\footnote{ The corresponding question for classical entropies has been successfully addressed in \cite{Zhang:1998aa,Matus:2007aa}, while that for quantum entanglement is as yet undetermined, though partial progress has been made in \cite{Cadney:2012aa,Linden:2013aa}. } For fewer than five regions $n \leq 4$, the only relevant inequalities are the strong subadditivity inequality and the monogamy of mutual information. For larger numbers of regions, new inequalities arise, though the full set of inequalities for $n=5$ is also as yet not fully determined. One infinite class of inequalities is the {\em cyclic inequality}. Given $n \geq 2\,k+l$ regions $\regA_i$, and with $S(\regA_I|\regA_J)$ being the {\em conditional entropy} $S(\regA_I:\regA_J) = S_{\regA_I \regA_J} - S_{\regA_J} $, these inequalities can be expressed as
\begin{equation}
\sum_{i=1}^n\, S(\regA_i \,\cdots\, \regA_{i+l-1}: \regA_{i+l} \cdots \regA_{i+k+l-1}) \geq S_{\regA_1 \regA_2\, \cdots\, \regA_n}
\label{}
\end{equation}
This family contains the previous known inequalities: the choice $(n,k,l) = (2,0,1) $ gives strong subadditivity and the choice $(n,k,l) = (3,1,1)$ gives the monogamy of mutual information. It is clear from the structure of the conditional entropy that we will get alternating sums of regions and their partial unions of varying degrees. The strongest inequalities are argued to occur for the choice $(n,k,l) = (2m+1, m,1)$.

There are a few other inequalities obtained in \cite{Bao:2015bfa}, which along with the cyclic inequality share a basic property: a region $\regA_i$ appears in a balanced form, i.e., it occurs the same number of times on both sides of the inequality. This feature enables the proof to proceed by a suitable surgery argument as in the earlier discussion. The actual proofs and the determination of the inequalities is done by mapping to a graph theory problem. We refer the reader to the original paper for further discussion.

\subparagraph{c) Open Questions:}
There are several open questions in the context of entropy inequalities.
\begin{enumerate}
\item For one, it would be interesting to address whether the holographic entropy cone obtained from the RT prescription agrees with that obtained from the HRT prescription.
As of now, the full set of inequalities that are valid for arbitrary time dependence remains unclear and naive adaptations of the proofs presented in \cite{Wall:2012uf} do not address all the inequalities obtained in \cite{Bao:2015bfa}.

\item A-priori it is remarkable that the holographic entropy cone is a polyhedral cone, while the quantum entropy cone is in general not expected to be so. Is the polyhedrality a special feature of holography or that of time-independent states therein?

\item It would be interesting to understand the particularities of the entanglement structure in states that satisfy the holographic inequalities.

\end{enumerate}
These issues are important for ascertaining which states of a QFT could have semiclassical gravitational duals.

\newpage
%~~~~~~~~~~~~~~~~~~~~~~~~~~~~~~~~~~~~~~~~~~~~~~~
\part{Quantum Gravity}
\label{part:qg}
%~~~~~~~~~~~~~~~~~~~~~~~~~~~~~~~~~~~~~~~~~~~~~~

%~~~~~~~~~~~~~~~~~~~~~~~~~~~~~~~~~~~~~~~~~~~~~~
\chapter{Prelude: Entanglement builds Geometry}
\label{sec:pegeo}
%~~~~~~~~~~~~~~~~~~~~~~~~~~~~~~~~~~~~~~~~~~~~~~

As we have remarked earlier, it is rather remarkable that an intrinsically quantum concept such as entanglement has a very simple geometric dual. Part of the reason of course is that for planar field theories with $\ceff \gg 1 $, one essentially attains a classical limit. Nevertheless, it is intriguing that there is a close connection between geometric concepts in the bulk and quantum features of the boundary theory. One therefore naturally wonders whether this fact can be leveraged to learn how the holographic map between quantum field theories and gravitational dynamics actually works.

The first attempt to articulate this philosophy was in the work of Swingle \cite{Swingle:2009bg} and Van Raamsdonk \cite{VanRaamsdonk:2009ar,VanRaamsdonk:2010pw}. Swingle's main thesis was to draw analogy between tensor networks and the geometry of spatial sections of AdS/CFT, which will be discussed in the next section in more detail. To appreciate this, note that in the AdS/CFT context, the radial direction into the bulk geometry is naturally viewed as corresponding to the energy scale in the QFT \cite{Susskind:1998dq}; probing deeper into the bulk corresponds to probing the quantum state at lower and lower energy scales. One can easily see this by examining the behaviour of extremal surfaces for regions of increasing size, cf., \S\ref{sec:extdeter}. The idea is to relate this behaviour to that seen in tensor network constructions for ground states of interacting many-body systems. For lattice systems, one starts with a underlying UV state and proceeds to perform a series of coarse-graining transformations which aim to remove short-range entanglement and enable to one write down a suitable variational wavefunction for the state. The structure of the network encodes pictorially the entanglement pattern inherent in the state. This qualitative picture was supported by the behaviour of entanglement entropy. In tensor networks, the amount of entanglement for a segment of the lattice is captured by the minimum number of links of the network that one has to disconnect, which is highly suggestive of the RT construction using minimal surfaces.

Van Raamsdonk's main idea was to address a very basic question about the AdS/CFT correspondence: under what circumstances can a field theory state be dual to a smooth semiclassical geometry? What is a-priori clear is that while the correspondence gives a map from states in the  Hilbert space of the boundary QFT to that of the bulk string theory, not all states in the latter will have a nice realization in terms of semiclassical spacetimes. Most of the states of the QFT will correspond to highly stringy geometries wherein geometric concepts cease to be meaningful. The new ingredient was to exploit entanglement as a crucial diagnostic for  the emergence of geometry.  Let us take an extreme example: an unentangled product state of a QFT. This is rather atypical state and the vanishing entanglement suggests that there should be no connection between different parts of the state. One example of such a direct product state is the boundary state \cite{Miyaji:2014mca}.\footnote{ It is helpful to view this state in terms of a lattice discretization of the field theory.} Consequently, such a state should not have a geometric dual.

A more clear example is provided by the thermofield double state. Take two copies of the QFT, which we call the left (L) and right (R) theories. We construct the thermofield or Hartle-Hawking state by entangling energy eigenstates  $\ket{\mathfrak{r}_i} \in {\cal H}_R$ and $\ket{\mathfrak{l}_i} \in {\cal H}_L$, respectively, weighted by a Boltzmann factor, viz.,
$$ \ket{{\sf TFD}} = \frac{1}{\sqrt{Z(\beta)}} \, \sum_i \, e^{-\frac{1}{2}\,\beta\, E_i}\,
\ket{\mathfrak{r}_i\, \mathfrak{l}_i}\,.$$
 Tracing out one of the copies  leaves the other in a mixed thermal state, say
  $$\rho_R = \frac{1}{Z(\beta)}\, \sum_i \, e^{-\beta\, E_i} \, \ket{\mathfrak{r}_i} \, \bra{\mathfrak{r}_i} \,,$$
  at inverse temperature $\beta$. For small $\beta$ or equivalently large temperatures, the state is highly entangled; indeed as $\beta \to 0$, we obtain the maximally entangled state. At any non-vanishing $\beta$, we have the entanglement entropy being given as the thermal entropy of a single theory. At low temperatures, $\beta \gg 1$, however we expect the ground state to dominate.

As described in \S\ref{sec:extdeter}, the thermal state of a holographic field theory is dual to a large \SAdS{} black hole in the high temperature limit, but, owing to the Hawking-Page phase transition, becomes dual to the thermal AdS geometry at low temperatures. The two phases are characterized by the entropy or, equivalently, the free energy. In the high temperature limit, this scales with the number of degrees of freedom $\ceff$, while at low energies, it is
${\cal O}(1)$.

We can equivalently phrase this observation in terms of the entanglement entropy of the thermofield double state in ${\cal H}_R \otimes {\cal H}_L$. Measuring the entanglement per degree of freedom for the subsystem that is one of the copies of the two CFTs (say the right one), we note that $S_R = {\cal O}(1)$ for $\beta \ll 1$, but $S_R = {\cal O}(\ceff^{-1})$ for $\beta \gg 1 $. In other words, the low temperature theory is characterized by vanishing small entanglement in the semiclassical limit $\ceff \gg 1$. If we look to the dual geometries, the low temperature phase is described by the (Euclidean) thermal AdS solution, while the high temperature phase is dual to the \SAdS{} black hole. One can view the Lorentzian geometry for the thermal AdS as two copies of AdS spacetime, which are disconnected at leading order in $\ceff$ to capture the two Hilbert spaces ${\cal H}_L $ and ${\cal H}_R$.
The Lorentzian \SAdS{}  solution  also has two asymptotic regions corresponding to the two copies, but these are connected in the spacetime via a spatial Einstein-Rosen bridge.

In other words, a field theory state with macroscopic entanglement, as in the high temperature regime, is characterized  by a geometric dual where the entangled parties are spatially connected. It is imperative to note that this spatial connection does not imply temporal/causal connection. Indeed, in the \SAdS{} spacetime, the two copies of the CFT are not in causal contact, as they lie separated by the black hole horizon. The spatial Einstein-Rosen bridge which links ${\cal H} _R $ and ${\cal H}_L$ provides an information conduit which encodes the entanglement pattern, but no communication is possible across it. In other words, the ER bridge is a non-traversable wormhole.  On the contrary, in the absence of macroscopic entanglement,  there is no spatial connection between the two Hilbert spaces, as is clear from the two disjoint copies of AdS being the dual spacetime in the low temperature limit.

The above observation can be codified into an elegant slogan ``entanglement builds bridges''. The presence of sufficient entanglement is indicative of spatial connectivity in the dual holographic theory. This idea underlies the structures seen in the tensor network approach to constructing ground states as mentioned. Note that the tensor network constructions per se are not well suited for obtaining wavefunctions of highly excited states, while the geometry construction in terms of the gravitational description does not suffer from this handicap.

The essential idea of the connection between geometry and entanglement and the prototype example provided by the EPR/Bell-like entangled thermofield double state and its dual avatar in the form of the Einstein-Rosen bridge prompted Maldacena and Susskind to argue for a more general relation, dubbed ``ER = EPR''. The idea is that any quantum state with Bell-type bipartite entanglement is naturally viewed in terms of a spatial connection between the entangled parties. When the amount of entanglement is minuscule, one may only have a quantum wormhole connecting the pairs. As the entanglement builds up to macroscopic amounts, these quantum structures coalesce into correspondingly larger spatial entities which end up creating new geometric connections in the dual spacetime. This is a highly intriguing picture that suggests a deep connection between the nature of entanglement and the origin of geometry. A lot of effort has been devoted in the recent years to understanding this connection better and much remains to be understood. The rest of this chapter focuses on the salient results to date and outlines some important issues that deserve further scrutiny.

%~~~~~~~~~~~~~~~~~~~~~~~~~~~~~~~~~~~~~~~~~~~~~~~
\chapter{Entanglement at large central charge}
\label{sec:largec}
%~~~~~~~~~~~~~~~~~~~~~~~~~~~~~~~~~~~~~~~~~~~~~~

Much of our analysis thus far has been either purely in the realm of field theory or in holographic systems in which we exploit the gravitational description to compute the physical observables.  A general question one might ask is what are the necessary and sufficient conditions for holography to work? Could we recover universal results in a class of field theories that are well approximated by holographic computations?

For the present, we will adopt a set of criteria that are known to be sufficient for a QFT to have a holographic dual and compute entanglement entropy directly using field theoretic methods to contrast with the geometric computations. Our aim is to build up some intuition to address the general set of questions, as  raised above,  with regard to the holographic map. We will come back to address these general issues in the context of our discussion relating gravity and entanglement in \S\ref{sec:egeometry}.

Let us first spell out the set of criteria we are after, which we have already foreshadowed in \S\ref{sec:adscft}. In order for a field theory to be holographic, it must admit a planar or large central charge limit $\ceff\to \infty$. This by itself does not suffice; the theory also needs to have a sparse spectrum of light states with the low-lying spin $s \leq 2$ states being parametrically lighter than their $s >2 $ counterparts \cite{Heemskerk:2009pn}.  We will focus on the case of CFT$_2$ for much of the discussion below, where we can be a bit more precise. It was described in \cite{Hartman:2014oaa} that such a holographic CFT$_2$, in addition to admitting a sensible $c\to\infty$ limit,
must also have a large gap in the spectrum of its Virasoro primaries $\Delta_{gap} \sim {\cal O}(c)$ and the low-lying spectrum be constrained from growing too fast. We will adopt these criteria and see what we can learn about the entanglement entropy in such theories.

%~~~~~~~~~~~~~~~~~~~~~~~~~~~~~~~~~~~~~~~~~~~~~~~
\section{Universality features of CFT entanglement}
\label{sec:nhee}
%~~~~~~~~~~~~~~~~~~~~~~~~~~~~~~~~~~~~~~~~~~~~~~

Before getting into the specifics of the large central charge CFTs in $d=2$, let us take a moment to review a series of computations which, whilst naively appearing to be holographic, are not quite what we are after. This will also be helpful in clarifying some of the results later, setting the stage for our abstract discussion in \S\ref{sec:egeometry}.

First, recall that if we were to consider the entanglement entropy of any CFT$_2$ in its vacuum state on the plane/cylinder, or the thermal state in non-compact space, we would obtain the celebrated results described in \S\ref{sec:eecft2}  as quoted in \eqref{eq:hlwcc}, \eqref{eq:svaccft2} and \eqref{eq:sthermcft2}, respectively. In each of these three cases, we notice that the answer depends on the central charge in a rather simple fashion. The results are furthermore universal irrespective of the details of the theory. For instance, the spectral information about the field theory data is completely missing. Indeed, as explained earlier, these results simply capture a rather coarse feature of the CFT, and are not a good diagnostic of whether the theory is holographic or not. They follow pretty much immediately, as a consequence of the symmetry preserved by the states, and the subregions chosen. They therefore fail to stand as a  good test of holography. Indeed, one reproduces precisely the same expressions from holographic modeling by taking the dual to be Einstein gravity in \AdS{3}, for the very same reasons of symmetry.

One might imagine the situation in higher dimensions to be rather different, no matter what state we pick or which region we choose. However, there exist special situations in which once again one obtains universal results independent of the details of the field theory. This is well exemplified by the behaviour of entanglement entropy for spherical ball-shaped domains $\regAB$ in the vacuum state a CFT$_d$. We have described how one may relate the density matrix in this case to the thermal density matrix using a conformal map  in \S\ref{sec:extdeter}. From these computations, we can see, for example, from \eqref{eq:sball} that the entanglement entropy $S_{\regAB}$ is simply characterized by the $a$ central charge  of the CFT$_d$. We can write the answer a bit more suggestively  as \cite{Casini:2011kv}:
\begin{equation}
S_{\regAB}^\text{vac} =
\frac{\Gamma\left(\frac{d}{2} \right)}{\pi \, \Gamma\left(\frac{d-1}{2}\right)}\, \omega_{d-2} \, a_d \;
\frac{\text{Vol}(\mathbb{H})_{d-1}}{\lads^{d-1}} \,.
\label{eq:chm2}
\end{equation}
 Thus, despite appearances, the vacuum entanglement for ball-shaped regions also only depends on a single number $a_d$.

Not only is the vacuum entaglement $S_{\regAB}$ universal, but it turns out even perturbations of the vacuum state end up giving universal results. Consider the following question, which was recently addressed quite nicely in \cite{Faulkner:2014jva}. Let us say that we have a CFT$_d$ in its vacuum state.  We deform the theory by turning on an infinitesimal source of strength
$J_\delta$ for a relevant scalar operator $\mathcal{O}_\Delta$ of dimension
$\frac{d}{2} \leq \Delta \leq d$. Generically, such a deformation will induce an energy-momentum tensor of ${\cal O}  (J_\delta^2)$; it will also change the entanglement entropy. The change $\Delta S_{\regAB}$ was computed in \cite{Liu:2012eea} using holographic modeling. They found a rather simple result to leading order in the perturbation engendered by the source:
\begin{equation}
\Delta S_{\regAB} =  -J_\delta^2 \, R^{2(d-\Delta)} \, \frac{\pi^\frac{d+1}{2} (d-\Delta) \, \Gamma\left(1+ \frac{d}{2} -\Delta\right)}{2\, \Gamma\left(\frac{3}{2} + d-\Delta\right)} + \mathcal{O}( J_\delta^3)
\label{eq:liumrefined}
\end{equation}

Remarkably, this result also holds in any CFT$_d$; this was explicitly demonstrated in conformal perturbation theory by \cite{Faulkner:2014jva}! The analysis carried out therein was to compute this quantity directly in field theory. Using the replica method and exploiting the known modular Hamiltonian for the ball-shaped regions  in \eqref{eq:BallmodH}, the result for $\delta S_{\regAB}$ can be expressed as a combination of correlation functions $\vev{\mathcal{O}_\Delta \mathcal{O}_\Delta}$ and $\vev{T_{\mu \nu} \mathcal{O}_\Delta\mathcal{O}_\Delta}$.
The amazing fact was that the computation of these correlators was most efficiently organized in terms of an auxiliary gravitational problem in a Einstein-scalar theory. This is identical to the setup considered in  \cite{Liu:2012eea}, though now one is working in a regime in which the CFT is not necessarily holographic, for instance, $c_\text{eff}\sim \mathcal{O}(1)$.

Thus despite our naive expectations, if we focus on a class of regions and states in which the entanglement features are universal, then we fail to distinguish holographic theories from non-holographic ones. The gravitational analog of these results was explained in \cite{Haehl:2015rza}. In particular, it was shown there that the data we are considering is incapable of distinguishing between various gravitational interactions. It is always possible to again conjure up an auxiliary Einstein theory which reproduces all of this data. Notwithstanding these observations, a lot of information has been extracted from the spherical entangling regions in the context of holography, as we review in \S\ref{sec:egeometry}.

These examples illustrate that under certain circumstances, the entanglement entropy may carry little information about the QFT. The auxiliary gravitational problem we write down is purely kinematic, in that it is engineered to do a field theory computation. While it may appear that in some practical applications the AdS/CFT correspondence also  works the same way,  this similarity is illusory. The profound conceptual difference is that in the latter case, gravity is dynamical.

These examples serve to caution us  in the diagnosis of potential holographic implications. We need to choose appropriate field theory data for purposes of estimating whether the quantity simplifies in the holographic limit to reveal signatures of geometry.  Results that appear to hold for an arbitrary central charge generically ought to  be viewed as kinematic coincidences. While they may provide useful starting points for a discussion, owing to their simplicity, it is important to explore other observables that are sensitive to the dynamical aspects of holography.

Thus when we discuss CFT$_2$ entanglement entropy, we would have to conjure up alternate scenarios from single-interval entanglement. We can continue to work with the vacuum state, provided we generalize to pick $\regA$ to be a disjoint union of multiple intervals. Likewise in higher dimensions, we should be focusing on regions other than the spherical domains. In the rest of this chapter, we will see how to obtain answers for large $c$ CFT$_2$s and its implications for the holographic map.

%~~~~~~~~~~~~~~~~~~~~~~~~~~~~~~~~~~~~~~~~~~~~~~~
\section{CFT$_2$ at large $c$}
\label{sec:}
%~~~~~~~~~~~~~~~~~~~~~~~~~~~~~~~~~~~~~~~~~~~~~~

To set the stage for our analysis of large central charge CFTs in two dimensions and their putative AdS duals, we need to review some basic facts about conformal field theory correlation functions. The basic data for a CFT$_d$ as mentioned in \S\ref{sec:eecft2} are the scaling dimensions of the primaries and the OPE coefficients. The conformal symmetry fixes the functional form of two- and three-point functions, leaving these as the only parameters. Higher point functions may be obtained by using the operator product expansion (OPE) as we now review.

% Figure
\begin{figure}[htbp]
\begin{center}
\begin{tikzpicture}[scale=0.7]
\draw[thick, black] (-2,-0.5) node [right] {$\sum_p$};
\draw[thick,black] (-1,0.5)  node [left] {${\cal O}_1$} -- (0,-0.5) -- (1,-0.5) -- (2,0.5) node [right] {${\cal O}_4$};
\draw[thick,black] (-1,-1.5)  node [left] {${\cal O}_2$} -- (0,-0.5);
\draw[thick, black] (1,-0.5) -- (2,-1.5) node [right] {${\cal O}_3$};
\draw[thick, black] (0.5,-0.5) node [above] {${\cal O}_p$};
\draw[thick, black] (3.5,-0.5) node [right] {$=$};
\draw[thick, black] (6,-1) node [above] {$\sum_r$};
\draw[thick,black] (7,1)  node [left] {${\cal O}_1$} -- (8,0) -- (8,-1) -- (9,-2) node [right] {${\cal O}_3$};
\draw[thick,black] (7,-2)  node [left] {${\cal O}_2$} -- (8,-1);
\draw[thick, black] (8,0) -- (9,1) node [right] {${\cal O}_4$};
\draw[thick, black] (8.5,-1) node [above] {${\cal O}_r$};
\end{tikzpicture}
\end{center}
   \caption{The expression of a four-point function in terms of the summation over conformal blocks. There are two potential channels for the expansion; on the left is the $s$-channel for $z\to0$, while on the right is the $t$-channel with $z\to1$. Crossing symmetry relates the two expressions using associativity of the OPE.}
   \label{f:fptfn}
\end{figure}

Consider a four-point function of  scalar primary operators ${\cal O}_{i}(x_i)$ of dimensions $\Delta_i$ with $i=1,2,3,4$ in CFT$_d$. Using the conformal symmetry, we can simplify to a single function of the cross-ratios. To wit, in terms of
\begin{equation}
\vev{{\cal O}_1(x_1)\, {\cal O}_2(x_2)\,{\cal O}_3(x_3)\,{\cal O}_4(x_4)} = \left(\frac{x_{24}^2}{x_{14}^2}\right)^{\frac{1}{2} \Delta_{12}}
 \left(\frac{x_{14}^2}{x_{13}^2}\right)^{\frac{1}{2} \Delta_{34}} \,
\frac{g(u,v)}{(x_{12}^2)^{\frac{1}{2} (\Delta_1+\Delta_2)}   \, (x_{34}^2)^{\frac{1}{2} (\Delta_3+\Delta_4)}   }
\label{eq:genfpt}
\end{equation}
in which $\Delta_{ij} = \Delta_i-\Delta_j$ and likewise $x_{ij} = x_i - x_j $. The function $g(u,v)$ is a function of the two independent conformal cross-ratios of the four-points:
\begin{equation}
u \equiv z\, \bar{z} = \frac{x_{12}^2\, x_{34}^2}{x_{13}^2 \, x_{24}^2} \,, \qquad
v \equiv (1 -z)\, (1-\bar{z})= \frac{x_{14}^2\, x_{23}^2}{x_{13}^2\, x_{24}^2}
\label{eq:cratios}
\end{equation}
The cross-rations $(u,v)$ are independent in Lorentz signature, but in Euclidean signature, $z$ and $\bar{z}$ are complex conjugates of each other. The simple way to think about these is to use the conformal symmetry to fix three insertion points, say $x_1 =0$, $x_3=1$ and $x_4 = \infty$. Then $z$ and ${\bar z}$ are simply complex coordinates on the two-plane common to the four operators. We have written the decomposition in the $s$-channel assuming the operators ${\cal O}_1$ and ${\cal O}_2$ are proximate, which is valid as long as $u\ll1$ (equivalently $z\to 0$). Analogous expressions can be written down for other channels by exchanging the operators to which we apply  the OPE first, as depicted in Fig.~\ref{f:fptfn}.

The key point is that using the OPE, the function $g(u,v)$ may be expanded into conformal blocks:
\begin{equation}
g(u,v) = \sum_{{\cal O}_p} \, C_{12p} \, C_{p34} \, G_{\Delta_p,s_p} (u,v)
\label{}
\end{equation}
where ${\cal O}_p$ are primary operators of dimension $\Delta_p$ and spin $s_p$. This allows us to write the correlator in terms of the OPE coefficients as a sum over conformal partial waves by decomposing the block:
\begin{equation}
\begin{split}
&\vev{{\cal O}_1(x_1)\, {\cal O}_2(x_2)\,{\cal O}_3(x_3)\,{\cal O}_4(x_4)} = \sum_{{\cal O}_p} \; C_{12p} \, C_{p34} \, W_{\Delta_p,s_p}(x_i ) \\
& W_{\Delta,s}(x_i ) = \left(\frac{x_{24}^2}{x_{14}^2}\right)^{\frac{1}{2} \Delta_{12}}
 \left(\frac{x_{14}^2}{x_{13}^2}\right)^{\frac{1}{2} \Delta_{34}} \,
 \frac{G_{\Delta,s} (u,v) } {(x_{12}^2)^{\frac{1}{2} (\Delta_1+\Delta_2)}   \, (x_{34}^2)^{\frac{1}{2} (\Delta_3+\Delta_4)}   }
\end{split}
\label{}
\end{equation}
These conformal partial waves and the blocks can be viewed as an efficient basis of functions that are adapted to the conformal group $SO(d,2)$, analogous to the spherical harmonics for rotational symmetry.

We will primarily be interested in $d=2$, in which the conformal symmetry is enhanced to the full Virasoro symmetry. In this case, we can employ the fact that the Hilbert space of the theory decomposes into a set of primary states $\ket{{\cal O}_p}$  related to their operator counterparts ${\cal O}_p$ via the state operator correspondence and their Virasoro descendants, viz., operators obtained by acting with $L_{-n}$ and $\tilde{L}_{-n}$ with $n \geq1$. This allows us to write the four-point function in terms of Virasoro conformal blocks, such that
\begin{equation}
g(u,v) = \sum_{h,\bar{h}} \, P_{h, \bar{h}} \; {\cal V}_{h,\bar{h}}(u,v)
\label{}
\end{equation}
with $P_{h, \bar{h}} $ being the block coefficients which are theory-dependent, and ${\cal V}_{h,\bar{h}}(u,v)$ the Virasoro blocks built from the representation theory. As usual,  $\Delta_i = h_i+\bar{h}_i$ and $s_i = |h_i - \bar{h}_i|$.

 Alternately we may write an expression setting ${\cal V}_{h,\bar{h}}(u,v) = u^{\Delta-s} \, {\cal F}_{h,\bar{h}} (u,v)$
that makes explicit the OPE origins:
\begin{equation}
\vev{{\cal O}_1(x_1)\, {\cal O}_2(x_2)\,{\cal O}_3(x_3)\,{\cal O}_4(x_4)}_{d=2}
= \sum_p \, C_{12p} \, C_{p34} \, {\cal F}(h_i,h_p,c; z)
\,  \bar{{\cal F}}(\bar{h}_i, \bar{h}_p,c; {\bar z})
\label{eq:virblocks}
\end{equation}

Let us now specify to a particular case in which the operators are from widely different regimes in the spectrum. We pick operators
$\OL$  which lie in the low-lying spectrum, $\DL \sim {\cal O}(1)$, and $\OH$ which are heavy, $\DH \sim {\cal O}(c)$. We will take them to be spinless, so that following the above logic, we may express the four-point function as
\begin{equation}
\vev{\OL(x_1)\, \OL(x_2)\,\OH(x_3)\,\OH(x_4)} = \frac{(z\,{\bar z})^{2\,\DL}}{(x_{12}^2)^{\DL}\, (x_{34}^2)^{\DH}}  G(z,\bar{z})\,,
\label{eq:fpex}
\end{equation}
where $z = \frac{x_{12} x_{34}}{x_{13} x_{24}}$ and $\bar{z} = \frac{\bar{x}_{12} \bar{x}_{34}}{\bar{x}_{13} \bar{x}_{34}} $.
In writing the expression, we have performed the block expansion in the s-channel where we assumed that $z \leq \frac{1}{2}$, which implies that the light and heavy operators are closer to each other. In general, the Virasoro conformal blocks do not admit a closed form expression (global conformal blocks do).

Let us now motivate the concept of \emph{vacuum block dominance}, which we will use to simplify the computations. The block decomposition of the correlation function suggests that we fuse a pair of operators, say ${\cal O}_1$ and ${\cal O}_2$, with each other into a set ${\cal O}_p$, which then mediates the interaction with ${\cal O}_3$ and ${\cal O}_4$. Per se, this involves a scan over all intermediate states that contribute.  If we can truncate the contribution of the intermediate operators ${\cal O}_p$, then we may have some hope of making the computation tractable. This is not generically achieved, but there are some special features of the planar limit which allow for this possibility.

First of all, the very existence of the planar limit hinges on  the OPE coefficients simplifying to admit a large $c$ factorization; we require that $C_{ijp} \sim {\cal O}(\frac{1}{c})$. In other words, factorization implies that higher point functions can be obtained through Wick contractions.  This by itself is not sufficient to allow the full simplification we need, but suppose further that there are very few low-lying states. Then one might imagine that the number of channels for the OPE is rather restricted. In fact, in the extreme limit, we may go so far as to suggest that the only channel that contributes corresponds to the lightest operator exchange.
 This clearly has to be the  identity channel, which includes the identity operator $\mathbb{I}$ and its descendants (which in the Virasoro block includes the energy-momentum tensor). Since the identity is the lightest primary having the  smallest conformal dimension $h=\bar{h} =0$,  one expects it to give the leading contribution to the correlator. The vacuum block dominance assumes this to hold and proceeds to derive the consequences therefrom.

We now assume that the identity block is the only relevant one. Furthermore, we concentrate on computing correlation functions in the planar limit ($c \to \infty$) with the choice of external operators $\OL$ and $\OH$ satisfying
\begin{equation}
 1 \ll \DL \ll c \,, \qquad \text{and} \;\; \frac{\DH}{c} \;\; \text{fixed}\,.
\label{eq:lhscaling}
\end{equation}
In this limit, one can obtain an analytic expression for the Virasoro identity block in the small $u$ limit \cite{Fitzpatrick:2014vua,Fitzpatrick:2015zha}:
\begin{equation}
{\cal V}_{0,0}(u,v)  \overset{u\to 0}{\approx}  \, (\aH)^{\DL} \; v^{-\frac{1}{2}\, \DL (1-\aH)} \,
\left(\frac{1-v}{1-v^{\aH}}\right)^{\DL} \,, \qquad \aH \equiv \sqrt{1-\frac{12\, \DH}{c}}
\label{eq:vir0}
\end{equation}

This vacuum block dominance has been used in recent years to derive many properties of two-dimensional CFTs that are shared by holographic theories in \AdS{3}. We now proceed to review some of the applications of this with a primary view towards entanglement entropy.

%~~~~~~~~~~~~~~~~~~~~~~~~~~~~~~~~~~~~~~~~~~~~~~~
\subsubsection{Entanglement phase transitions}
\label{sec:ephases}
%~~~~~~~~~~~~~~~~~~~~~~~~~~~~~~~~~~~~~~~~~~~~~~

One important feature that is exhibited by the RT/HRT prescriptions is the existence of entanglement phase transitions. Recall that in $d=2$, if we consider multiple intervals, then there are potentially multiple extremal surfaces available, all satisfying the homology constraint, see, for example, Fig.~\ref{f:sasurfs}. Of these we are required to take the surface with globally minimal area in the correct homology class, which then leads to exchange of dominance among these available extremal surfaces. This was one of the main arguments put forth in \cite{Headrick:2010zt}. It was further demonstrated there that such behaviour is characteristic of theories with large central charge.

This idea was made explicit in a beautiful calculation by Tom Hartman that initiated the study of vacuum block dominance in large $c$ CFTs  \cite{Hartman:2013mia}. We will now review this result to illustrate how the large $c$ approximations work, making it clear that the answer is essentially holographic. In fact, there is more to say regarding the connection to gravity, which we shall explain at the end of  the discussion.

Consider the case of two intervals $\regA = \regA_1 \cup \regA_2$, as depicted in  Fig.~\ref{f:sasurfs},  for simplicity. We let
$\regA_1 = [x_1,x_2]$ and $\regA_2 = [x_3,x_4]$. To compute the
 $q^{\rm th}$ R\'enyi entanglement entropy in the vacuum state, we need to construct the branched cover geometry $\bdy_q$ which will end up being a genus $(q-1)$ Riemann surface, as each gluing will generate a handle. The computation of the partition function $\mathcal{Z}_q$ on this surface is however equivalent to computing the four-point function of twist operators inserted at the endpoints of $\regA$. So we need to compute  the four-point function $\vev{{\cal T}_q(x_1)\bar{{\cal T}_q}(x_2){\cal T}_q(x_3)\bar{{\cal T}_q}(x_4)}$. The idea is to evaluate this correlation in the large $c$ limit taking into account the conformal dimension of the twist operators scales like the central charge.

Now while all the operators are the same, we need to exercise care in employing the vacuum dominance to evaluate these correlation functions.  In drawing  Fig.~\ref{f:fptfn}, we assumed that we were expanding in the $s$-channel, but that is only valid for the cross-ration $z \to 0$.
On the contrary, when $z \to 1$, we should be expanding in the alternate channel. The choice of $z\to0$ and $z\to1$ refers to behaviour of the cross-ratio \eqref{eq:cratios}, but it is equally intuitive pictorially. In the $z \to 0$ limit, $x_{21} \to 0$, implying that
$\regA_1$ and $\regA_2$ are well separated from each other. On the other hand, when $z\to 1$, we see that the right endpoint of $\regA_1$ is proximate to the left endpoint of $\regA_2$, i.e., the intervals are closer to each other. This is precisely the situation illustrated in Fig.~\ref{f:sasurfs}. In the CFT, the choice we get to make is which two operators we choose to first perform the OPE.

The two choices amount to the identifications of the twist operators into our choice of two light and two heavy operators for purposes of employing \eqref{eq:vir0}. These correspond to:
\begin{equation}
\begin{split}
&(i).\quad  {\cal T}_q(x_1),\bar{{\cal T}}_q(x_2),{\cal T}_q(x_3),\bar{{\cal T}}_q(x_4))=(\OL,\OL,\OH,\OH) \,, \\
& (ii). \quad ({\cal T}_q(x_1),\bar{{\cal T}}_q(x_2),{\cal T}_q(x_3),\bar{{\cal T}}_q(x_4))=(\OL,\OH,\OH,\OL) \,.
\end{split}
\label{}
\end{equation}
Now we can apply the vacuum dominance and compute the correlator.   In both cases, we keep only the identity in the sum \eqref{eq:virblocks} and take the $q\to 1$ limit to obtain the von Neumann entropy. This effectively amounts to setting $(z\,{\bar z})^{2\,\DL} G(z,{\bar z} )=1$ in \eqref{eq:fpex}. One can  convince onself using the explicit expression \eqref{eq:vir0} for the identity block.  We are however left with the kinematical factor in the denominator of \eqref{eq:fpex} to contend with, and this contributes differently owing to our choice of OPE expansion. The entire answer for the entanglement entropy comes from this  piece :
\begin{equation}
\begin{split}
& (i). \quad \;S_\regA=\frac{c}{3}\log\frac{|x_2-x_1|}{\epsilon}+\frac{c}{3}\log\frac{|x_4-x_3|}{\epsilon} \,,  \qquad  z< \frac{1}{2}\,,\\
& (ii). \quad S_\regA =\frac{c}{3}\log\frac{|x_4-x_1|}{\epsilon}+\frac{c}{3}\log\frac{|x_3-x_2|}{\epsilon} \,,  \qquad z> \frac{1}{2}\,.
\end{split}
\label{eq:Sa12phase}
\end{equation}
The choice of which surface to pick is determined by the configuration that dominates the saddle point approximation using the vacuum block to determine the four-point function of the twist operators. This unsurprisingly is given by the minimum of $S_\regA$ in the two cases described in \eqref{eq:Sa12phase}. As indicated above, the configuration (i) dominates for $z<\frac{1}{2}$, while (ii) dominates for $z >\frac{1}{2}$. Essentially we can view the vacuum block approximation as an effective saddle point estimate with $c$ playing the role of the saddle point parameter.

One can readily appreciate the result: this corresponds to the two obvious ways to connect the endpoints of $\regA$ consistent with the homology requirement. The lengths of geodesics in \AdS{3} connecting two boundary points being simply $\log\frac{\ell}{\epsilon}$ for a spatial interval of proper size $\ell $. Moreover we see that the two configurations are precisely the ones admissible by the homology requirement. Thus,  assuming large $c$ vacuum dominance,
we derived the result depicted in Fig.~\ref{f:sasurfs}. We see a sharp phase transition at $z=\frac{1}{2}$ which is only possible because of the planar limit $c\to \infty$.

While it seems remarkable that the large $c$ result so simply reproduces the holographic answer, there is in fact a greater level of concordance than our brief discussion indicates. Paralleling the development of the computation in CFT by Hartman in \cite{Hartman:2013mia} was an analogous development in the gravitational side. Faulkner \cite{Faulkner:2013yia} attacked the same problem of computing entanglement R\'enyi entropies in CFT$_2$ though now using the holographic map to \AdS{3}. The idea was to essentially obtain the partition function on the branched cover geometry $\bdy_q$,  by a semiclassical approximation of the quantum gravity path integral in \AdS{3}. The computations can be done in Euclidean signature since we are in the vacuum state. Now, given a genus $g$ Riemann surface $\bdy_q$ one can construct a handlebody geometry, a Euclidean 3-manifold which admits a constant negatively curved metric, obtained simply by filling in $g$ mutually commuting cycles (in the homotopy sense) of the non-trivial $2g$ cycles. Furthermore, using known results in Liouville theory, one can  compute the action for these handlebodies; we refer the reader to the original paper and references therein for how this can be done (cf., also \cite{Krasnov:2000zq}).
Now the choice of cycles that one gets to fill is in one-to-one correspondence with the choice of OPE expansions at our disposal in the computation of the twist operator correlation functions. The assumption of vacuum block dominance is tantamount to singling out $g$ of the $2g$ cycles and requiring that only the identity and its descendants propagate along that cycle. This leads to a very satisfying picture of the connection between the large $c$ CFTs and gravitational theories in \AdS{3}, and makes extremely explicit the necessity of the approximations chosen above.

%~~~~~~~~~~~~~~~~~~~~~~~~~~~~~~~~~~~~~~~~~~~~~~~
\subsubsection{Excited state entanglement}
\label{sec:Hcft2}
%~~~~~~~~~~~~~~~~~~~~~~~~~~~~~~~~~~~~~~~~~~~~~~

We now turn to understanding how we can use the vacuum block dominance to compute the entanglement entropy for a highly excited state in a large $c$ CFT$_2$. For now, we will assume that the state is spatially homogeneous, and revert to the case with spatial inhomogeneity in the sequel. Say that our excited state, $\ket{\psi_h}$ was  created by a heavy (spinless) primary operator
${\cal O}_h$ of conformal  dimension $\Delta_h \sim {\cal O}(c)$,
\begin{equation}
\ket{\psi_h}=\lim_{z\to 0}\, z^{\Delta_h} {\cal O}_h(z)\ket{0} \,.
\end{equation}

We would like to compute the entanglement entropy for this excited state, taking $\regA$ to be a single connected interval on the cylinder. We can use the replica method to replicate the  state  $\ket{\psi_h}$  $q$-fold, and then obtain the $q^{\rm th}$ R\'enyi entropy by computing  the correlation function of the twist operators in this replicated state. The replica construction involves insertion of the ${\mathbb Z}_q$ twist operators  ${\cal T}_q$ at the entangling points $\entsurf$. These have conformal dimension $\Delta_q = \frac{c}{12} \, \left(q-\frac{1}{q}\right)$. Thus we need to compute the correlation function of twist operators in the replicated state $\ket{\psi}_q$. This seems formidable, since all operators  have conformal dimension of order $c$.

However, for computing the entanglement entropy in the limit $q \to 1$, we can make the following approximation. We note that the twist operator $\mathcal{T}_q$ starts to get light since $(q-1) \ll 1$ in this limit, so we can choose it to be the light operator $\OL$ and take
$\DL = \Delta_q$.
We further let the heavy operator be the one that creates the state ; this involves the $q$-th power of $O_h$:  $\OH=(O_h)^q$, so $\DH = q\, \Delta_h$. Working in this limit, we need only compute the four-point function of these operators. We insert the twist operator at $1$ and $z$, while we put the heavy operator at $0$ and $\infty$, corresponding to the four-point function
\begin{equation}
\vev{ \OH(0)\, \OL(z)\, \OL(1)\, \OH(\infty)}  \,, \qquad \OH = ({\cal O}_h)^q \,, \; \OL = {\cal T}_q \,,\,\;\; q \to 1^+
\end{equation}
in which we have chosen to regard the operator creating the state to act at $0$ and $\infty$ and suitably picked coordinates to bring the endpoints of $\regA$ to $z$ and $1$, respectively.

We can now use the vacuum block dominance since \eqref{eq:lhscaling} is satisfied. Using \eqref{eq:vir0}, we can obtain the following expression for the von Neumann entropy:
\begin{equation}
S_\regA=\frac{c}{6}\log\left(\frac{|1-z^{\ah}|^2 \; |z|^{1-\ah}}{\ah^2 \;\epsilon}\right)  , \qquad \ah = \sqrt{1-\frac{12\,\Delta_h}{c} }
\label{eq:eelc}
\end{equation}
where $\epsilon$ is the UV cut off as usual. Performing the conformal map from the plane to the cylinder using the exponential map  $z=e^{iw}$, we find from \eqref{eq:eelc} the result for the entanglement entropy for a region of size  $2a$ on a circle of size $\ell_{{\bf S}^1}$
\begin{equation}
S_\regA=\frac{c}{3}\log \left(\frac{2\, \ell_{{\bf S}^1}}{\pi\, \epsilon\, \ah} \, \sin\left(\pi\,\ah\, \frac{a} {\ell_{\bf S}^1}  \right) \right).
\label{eq:eefgh}
\end{equation}
Since the state in question $\ket{\psi_h}$ is a pure state, the above is valid only for regions that are small. For larger regions, we need to replace $2a \to \ell_{{\bf S}^1} -2a$.

Let us contrast this with the expression quoted earlier \eqref{eq:svaccft2} for a finite region of length. What the above expression
\eqref{eq:eefgh} suggests is that we are measuring a rescaled length for the spatial circle $\ell_{{\bf S}^1} $ that is effectively rescaled by
$\frac{\ah}{2}$.  This is indeed what we expect from gravity: an excited state in \AdS{3}, which is heavy enough to backreact and modify the spacetime, but light enough not to form a BTZ black hole, is a conical defect. These geometries are created by  a massive point particle in three dimensions. They are quotients of \AdS{3} by a discrete group, whose action is precisely to orbifold of the circle, thereby introducing a conical defect with deficit angle $2\pi(1-\ah)$.

One can see this explicitly from the metrics \eqref{eq:btz} where we set $\rh = \lads (1-\mu) $ with $\mu \in (0,1)$ to obtain 
\begin{equation}
ds^2 = -\left(\frac{r^2}{\lads^2} + 1- \mu\right) \, dt^2 + \frac{dr^2}{\left(\frac{r^2}{\lads^2} + 1- \mu\right) } + r^2\, d\varphi^2
\end{equation}
We can absorb the factor of $1-\mu$ by working in rescaled radial coordinate $\tilde{r} =\frac{r}{\sqrt{1-\mu}}$. This will bring the first two metric functions to be of the global \AdS{3} at the expense of modifying the angular component of the metric. Examining the latter, we learn that we need  to quotient the circle parameterized by $\varphi$ accordingly, which results in the orbifold. The holographic computation of the minimal surface length will reproduce \eqref{eq:eefgh} with the identification $\mu = 12\,\frac{\Delta_h}{c}$.

In the computation above, we took $\Delta_h < \frac{c}{12}$. However, we can analytically continue the answer \eqref{eq:eefgh} when
$\Delta_h >  \frac{c}{12}$ simply by replacing $\ah \to T_h$ where $T_h$ is an effective temperature and write
\begin{equation}
S_\regA=\frac{c}{3}\log \left(\frac{1}{\pi\,\epsilon\, T_h} \, \sinh\left(2\pi\,a\, T_h\right)\right) \,, \qquad
T_h = \frac{1}{2\pi\ \ell_{{\bf S}^1}} \, \sqrt{\frac{12\, \Delta_h}{c}-1} \,.
 \end{equation}
This is the form of the entanglement entropy in a thermal state at temperature $T_h$, cf., \eqref{eq:sthermcft2}.  It also corresponds to the answer we get by a holographic computation in the BTZ background \eqref{eq:salargec}  (the result is quoted for small regions relative to the circle size).  What this reveals is that due to large degeneracy of states at high energy $\Delta_h > \frac{c}{24}$, the primary state $\ket{\psi_h}$ is indistinguishable from a typical state at the same energy. The latter, by statistical reasoning, is well approximated by a thermal state, which in the dual picture is the   BTZ black hole.

These results, which were discussed in \cite{Fitzpatrick:2014vua,Asplund:2014coa,Caputa:2014eta}, illustrate that the vacuum block dominance is the operative mechanism that allows the CFT computation to reproduce the holographic answer. To some extent, these examples are still somewhat governed by the symmetry, but less trivially so than the examples discussed in \S\ref{sec:nhee}.

%~~~~~~~~~~~~~~~~~~~~~~~~~~~~~~~~~~~~~~~~~~~~~~~
\subsubsection{Local quenches}
\label{sec:lclq}
%~~~~~~~~~~~~~~~~~~~~~~~~~~~~~~~~~~~~~~~~~~~~~~

As another illustration of the large $c$ computations, we will exhibit results for a locally-excited state  $\ket{\!\Psi}$ defined by acting a primary operator $\mathcal{O}(x)$ on the CFT vacuum 
\cite{Nozaki:2014hna,Nozaki:2014uaa,He:2014mwa,Caputa:2014vaa}:
\begin{equation}
\ket{\!\Psi}={\cal N}e^{-\epsilon_* H}{\cal O}(x)\ket{\!0},
\end{equation}
where ${\cal N}$ is the normalization factor and $\epsilon_{*}$ is the UV regularization parameter. Below we review the results for local excited states in large c CFTs derived in \cite{Asplund:2014coa}.

To evaluate the entanglement entropy in the locally excited state, we take the twist operator to be the light operator
$\OL ={\cal T}_q$ and the heavy operator to be again the $q^{\rm th}$ power of the operator inducing the excitation $\OH = {\cal O}$. Thus $\DL = \frac{c}{12}\left(q-\frac{1}{q}\right)$ and $\DH = q\, \Delta_{\cal O}$. We need to evaluate the correlation function
$\vev{{\cal T}_q(w_2)\, \bar{{\cal T}}_q(w_3)\, {\cal O}^q(w_1)\, {\cal O}^q(w_4)}$, where we choose
\begin{equation}
w_1=-w_4=-i\, \epsilon_*\,, \qquad  w_2=x_1+i\, t\,, \qquad w_3=x_2+i\,t \,.
\end{equation}
The notation is as before, with $\epsilon_*$ being a UV regulator of the disturbance and $t = i\, \tE$ the Lorentzian time coordinate.

Adopting the conformal block decomposition \eqref{eq:fpex} and thence assuming vacuum dominance,  we end up with
\begin{equation}
\begin{split}
\Tr{}{\rhoA^q} = \frac{
 \vev{{\cal O}^q(w_1)\, {\cal O}O^q(w_4)\, {\cal T}_q(w_2)\, \bar{{\cal T}}_q(w_3)}}
 {\vev{{\cal O}(w_1)\, {\cal O}(w_4)}^q}  =  x_{21}^{-2\, q\, \Delta_{\cal O}} \, (z\,\bar{z})^{2\,\Delta_{\cal O}} \, G(z,{\bar z})
\end{split}
\end{equation}
Taking the $q\to 1$ limit, we obtain the entanglement entropy \cite{Asplund:2014coa}
\begin{equation}
S_\regA=\frac{c}{6}\log \left[(x_2-x_1)^2\, \frac{|1-(1-z)^{\aO}|^2|1-z|^{1-\aO}}{\aO^2|z|^2\epsilon_*^2}\right] \,, \quad
\aO=\sqrt{1-\frac{12\,\Delta_{\cal O}}{c}}
\label{eq:saal}
\end{equation}
When $\epsilon_*$ is infinitesimally small, the cross-ratio behaves as
\begin{equation}
\begin{split}
& z=-\frac{2i\,(x_2-x_1)}{(x_1+t)\, (x_2+t)}\, \epsilon_*+ {\cal O}(\epsilon_*^2) \,,\\
& \bar{z}=\frac{2i\,(x_2-x_1)}{(x_1-t)\, (x_2-t)}\,\epsilon_*+{\cal O}(\epsilon_*^2) \,.
\end{split}
\label{}
\end{equation}

To evaluate the final answer as suggested by \eqref{eq:saal}, we need to be careful about the phase information in the cross-ratio
$1-z$ and $1-\bar{z}$.
In the early time limit, $0<t<x_1$, as well as in the  late time region $t>x_2$, we find $(1-z,1-\bar{z})\to (1,1)$ in the $\epsilon_*=0$ limit. Therefore we find that the entanglement entropy for these times reduces to the ground state entropy and thus
\begin{equation}
\Delta S_\regA= 0 \,, \qquad \text{for}\; \ 0<t<x_1 \;\ \text{and}\;\  t>x_2 \,.
\end{equation}
This trivial behavior is indeed what is predicted by the causal propagations of entangled particles, reproducing earlier results.

On the other hand, in the intermediate region $x_1<t<x_2$, we have $(1-z,1-\bar{z})\to (1,e^{2\pi i})$. Using this expression, we find we obtain the time-dependent entanglement entropy:
\begin{equation}
S_\regA=\frac{c}{6}\log \left[\frac{(x_2-x_1) \, (t-x_1)\, (x_2-t)}{\epsilon^2\, \epsilon_*}  \;\frac{\sin(\pi\aO)}{\aO}\right].
\end{equation}
If we choose $x_2-x_1$ much larger than $t$ and focus on the late time region $t\gg \epsilon_*$, we find
\begin{equation}
\Delta S_\regA\simeq \frac{c}{6}\log\frac{t}{\epsilon_*}+\frac{c}{6}\log\frac{\sin(\pi\aO)}{\aO}.
\end{equation}
This perfectly reproduces the  holographic result in \cite{Nozaki:2013wia}, once we identify $\epsilon_*=w_\epsilon$ and $\aO=\sqrt{1-\mu}$.
In other words, the quench, as for the excited state entanglement entropy, involves choosing a point particle in \AdS{3} with mass parameter set by $\mu =  \frac{12\,\Delta_{\cal O}}{c}$, consistent with our expectations.

%~~~~~~~~~~~~~~~~~~~~~~~~~~~~~~~~~~~~~~~~~~~~~~~
\chapter{Geometry from entanglement}
\label{sec:egeometry}
%~~~~~~~~~~~~~~~~~~~~~~~~~~~~~~~~~~~~~~~~~~~~~~

A-priori one can make the following observation: Let us say we are given the entanglement entropies of a collection of regions in the boundary field theory. Assuming that this data arises from areas of surfaces in the gravitational dual, one can ask what is the corresponding geometry? In particular,  we can seek the metric of the bulk spacetime, which leads to the given entanglement data. To appreciate the question better, note that  spatial bipartitioning of a field theory Cauchy slice is described by two functions worth of data in $d$ dimensions;  the entangling surface is a codimension-2 surface. We are assuming that we have a collection of entanglement entropies for various choices of regions $\regA$, which is far more data than that necessary to describe a metric in $(d+1)$-dimensional asymptotically AdS spacetime. After all, the latter is completely specified by the $\frac{(d+1)(d+2)}{2}$ functions of $d$-variables, while we have data indexed by two functions of $d$-variables. This is a vastly overdetermined problem.

This suggests that not all field theory entanglement patterns are geometry. To be sure, for any state of a holographic field theory, we may obtain the entanglement entropy of an arbitrary region. However, requiring that these be attained as a geometric construct puts further restrictions on the entanglement realized in holographic field theories. The question is not only which patterns are amenable to being obtained from geometric functionals like the area of a codimension-2 bulk surface, but also whether the field theory data is consistent with the semiclassical picture for a geometric description. For instance:
\begin{itemize}
\item Phase transitions of entanglement entropy  which occur at large $\ceff$, e.g., for thermal states or disjoint regions should be captured by the entanglement structure.
\item General holographic entanglement inequalities discussed in \S\ref{sec:heeineq} should be respected. Amongst these, in particular, we can require that the mutual information is monogamous,  $\tmi <0 $, in the holographic setting.
\end{itemize}

These criteria are not satisfied by generic states in the Hilbert space of a QFT. They only hold just in special limits of parameter space (e.g., strong coupling, planar limit), and even then only for certain subset of states does one expect that there is a nice geometric prescription.  We will refer to such states in the field theory as states in the {\em code subspace}. The nomenclature will become clear when we describe recent attempts to understand the holographic map in the language of quantum error correction. However, we can now give a qualitative picture of the code subspace which should be useful for building intuition for the discussion that follows.

\paragraph{The code subspace states:}
The vacuum state $\ket{0}$ of a CFT$_d$ on ESU$_d$ is the only state invariant under the full  $SO(d,2)$ conformal group in $d >2 $ (or the Virasoro symmetry in $d=2$).  For this state, symmetries dictate the dual geometry to be the  global \AdS{d+1} spacetime.  We trust the semiclassical physics on this background to describe accurately the dynamics of the CFT as long as  $\lads \gg \ell_s $ and $\lads \gg \ell_P$. Translating to CFT data, as long as $\ceff\gg1$ and $\lambda\gg 1$, the entanglement structure inherent in the vacuum for arbitrary regions $\regA$ is captured accurately by geometry. We can make similar observations should we consider the field theory on Minkowski spacetime instead.

We can also consider neighbouring states which are obtained by acting on the vacuum with a bounded number of creation operators, ${\cal O}_i$, for $i = 1, 2, \cdots k$, as long as $k \ll \ceff$. These excitations will add some excess energy but the total amount of energy is bounded by the conformal dimensions $\Delta = \sum_{i=1}^k \, \Delta_k$. As long as $\Delta_i \ll \ceff$, the backreaction on the dual spacetime geometry is vanishingly small. This follows from our earlier observation that the bulk gravity theory has
$\GN \sim \ceff^{-1}$. Since the strength of gravity is weak, the backreaction is controlled by $\GN\, \Delta \ll 1$, by assumption. What this means is that for such states, we can approximate the bulk dual in terms of a few particle excitations on an undeformed \AdS{d+1} background. So by working perturbatively in $\GN$, we can capture the effects of the local excitations atop the vacuum. The code subspace around the vacuum can thus be defined as the set of these low-lying states of the Hilbert space.

We can similarly  start with an excited state, which happens to admit a geometric dual with a non-trivial dual geometry.  For the former to pertain, we need to ensure that the energy in the state is macroscopically larger than in the vacuum, and thus require that the state has ${\cal O}(\ceff) $ energy. Such geometric states are not generic in the Hilbert space, since the generic high energy state would be expected to behave analogously to a black hole geometry, thanks to microcanonical typicality. Given such a state, we can further consider a few particle excitations atop such states, to define the code subspace around this high energy state.

States in the code subspace about a particular geometry have exponentially small overlap $\mathcal{O}(e^{-\ceff})$ with the states about any other geometry. In other words, insofar as the space of geometric states is concerned, we can decompose them into effective superselection sectors in the semiclassical regime of interest. This implies that we can without loss of generality, extend the code subspace about each such geometric state, into the generalized code subspace, which formally is the union of all such states. While the notion of a subspace about a given geometry makes sense, we should hasten to add that the extended definition does not, strictly speaking, give rise to a subspace of the Hilbert space.

 In this manner, we can chart out the generalized code subspace as an archipelago of states in the Hilbert sea of states. As described above, we will take this to be  the union of the the set of states which admit a geometric dual. They are classified by a central state, like the vacuum or the highly excited states with energies of ${\cal O}(\ceff)$, each being described by a smooth semiclassical metric,  together with a set of low-lying excitations around them. We caution though that in much of the literature, the code subspace only refers to the island centered around the vacuum state. The number of excitations one is allowed about each island is bounded, to avoid ending up with a typical black hole-like state. Restricting to the domain around the vacuum, we note that  code subspace is the low energy subspace of states. It must however be borne in mind that the space of states is a not a proper subspace of the Hilbert space, but a convenient label for referring to the class of geometric states.

A precise characterization of the code subspace exists only in the vicinity of the vacuum state of the CFT, since by symmetries, this has a unique gravitational dual. There are a handful of other states, such as the eternal black hole which is dual to the thermofield double state (in a doubled CFT Hilbert space), but a complete characterization of the code subspace is lacking. In part, the problem is to ascertain the full set of criteria which are necessary and sufficient for one to trust the geometric description. We know of a few necessary conditions which we outline below, but it remains to be seen whether this list is sufficient.

%~~~~~~~~~~~~~~~~~~~~~~~~~~~~~~~~~~~~~~~~~~~~~~~
\section{Criteria for geometric duals}
\label{sec:critgeom}
%~~~~~~~~~~~~~~~~~~~~~~~~~~~~~~~~~~~~~~~~~~~~~~
The set of QFT states that have geometric duals is not exhaustively known as of this writing. What we do know is a list of incomplete criteria that appear to be sufficient empirically. We will try to give a flavour of what these conditions are, breaking them into three inter-related but thematically distinct categories.
We should first ask ourselves under what conditions do certain field theories allow themselves to be described holographically. Once we ascertain this, we can ask if specific states in the Hilbert space of such QFTs admit a semiclassical geometric dual. Distinct from this list of criteria is our need to assume that we have a geometric dual for a particular QFT state and ask what such a description would entail for the field theory state in question. Let us address these three issues in turn.

\paragraph{I. Sufficient criteria for QFTs to have a semiclassical gravitational dual:}  We have explained some of the  conditions which we know to be necessary in  \S\ref{sec:adscft}. The field theories are required to have a large number of degrees of freedom, and admit a sensible analog of a planar limit, which we characterized by the requirement of a large central charge $\ceff \gg 1$. In addition, we need a second hierarchy which ensures that the curvatures scales of the gravitational solution are larger than the string scale, so that we can relate to Einstein-Hilbert dynamics (as opposed to a classical string theory). One consequence of this is that operators with spin $s >2 $ are required to be parametrically heavier than those with $s \leq 2$ \cite{Heemskerk:2009pn}. In known examples, this is ensured by a hierarchy between the string  scale and the AdS curvature scales. At a heuristic level, one may say that this can be ensured by the QFT degrees of freedom being strongly coupled, but overall it is hard to quantify this statement.

In 2-dimensional CFTs, one can argue for a precise bound on the spectrum based on known geometric phenomena such as the Hawking-Page transition. This is the statement that the thermal density of states of the theory at low temperatures scales is ${\cal O}(1)$, while at high temperatures, it is of ${\cal O}(\ceff)$, with a phase transition at some temperature $T_c \sim {\cal O}(1)$. Per se, this requirement by itself does not guarantee that the dual field theory be described by semiclassical Einstein gravity. Nevertheless, this particular constraint is easy to  quantify  explicitly in terms of a bound on the spectral density.  In \cite{Hartman:2014oaa}, it is argued based on modular invariance that the number of states with energy $E = \Delta - \frac{c}{12} = h+\bar{h} - \frac{c}{12}$ is bounded as
\begin{equation}
\rho(E) \leq e^{2\pi(E + \frac{c}{12})} \,, \qquad E \sim {\cal O}(\epsilon)
\label{}
\end{equation}
While this bound is satisfied by known holographic field theories,  it also turns out to be upheld in a wider class of theories, such as the symmetric orbifold theory (or in more general permutation orbifolds \cite{Haehl:2014yla,Belin:2014fna,Belin:2015hwa}).

In general, since QFT density of states grow as $\rho(E) \sim e^{E^\alpha}$ with $\alpha <1$, one would want to rephrase this constraint in more precise terms. Specifically one would like to translate the spectral bound into a statement about a statement of the QFT parameter/moduli space. This remains as of this writing an interesting open question.

We have further seen in \S\ref{sec:largec} that the holographic results in \AdS{3} are reproduced by large $c$ computations once we assume  the sparseness of the low-lying spectrum to imply the vacuum block dominance. The latter condition is however stronger and it is likely that sparseness by itself is insufficient to guarantee the existence of a holographic dual. There are already indications that some features, such as the bound on Lyapunov exponents \cite{Maldacena:2015waa}, requires more than sparseness \cite{Perlmutter:2016pkf}. While explaining the chaos bound will take us far afield, let us briefly mention here that this bound focuses on the exponential intermediate time growth of certain out-of-time-ordered correlation functions. The initial inspiration for studying such objects originated from attempts to understand the link between entanglement and gravity  \cite{Shenker:2013pqa} in which the authors were interested in analyzing how entanglement can be disrupted by disturbing an initial entangled state and thence monitoring the state.   It is unclear at present whether there are constraints from an entanglement perspective that are stronger than the spectral sparseness  criterion.

\paragraph{II. Field theory constraints on geometry:} There are several requirements that we would infer from the observables in any state belonging to the QFT Hilbert space. For instance, one requires that the time-ordered correlation functions and even non-local objects such as entanglement entropy respect causality. This turns out to place restrictions on the gravitational theory and solutions therein.  We will focus here on the conditions that have been inferred from the various known features of observables and entanglement entropy.

\subparagraph{A. Causality:} The basic requirement that all physical observables respect causality has a profound implication for the gravitational solution. Naively, in the holographic setting, we have two distinct causal structures: the first is the boundary causal structure which we can choose at will, since this is just the arena where our QFT dynamics takes place. There is however  also the bulk causal structure which is determined post-facto by the duality. For the duality map to be sensible, we require that the bulk causal structure be compatible with the boundary causal structure. More strongly one would note that the boundary causal structure is the fundamental object and the bulk causal structure in asymptotically AdS spacetimes would have to reduce on physical observables to just the boundary one.

In Einstein-Hilbert gravity coupled to matter, a very useful theorem in this regard was proved by Gao and Wald \cite{Gao:2000ga}, who showed that for matter satisfying the null energy condition (NEC), the aforementioned requirement was satisfied in smooth dual geometries. Extensions of the statement to other forms of gravitational dynamics are unknown, but it is clear that the operative feature one needs is for ``gravity to be attractive''. The NEC is relevant for this in the simplest case, for it ends up ensuring that null geodesic congruences which start contracting due to gravitational attraction continue to do so. Once the dual geometry's causal structure is subsumed within the boundary causal structure, it follows that all local correlation functions computed holographically will respect the field theory requirements. This follows from the general observations in \cite{Marolf:2004fy}.

Causality  also places restrictions on entanglement entropy, as discussed in \S\ref{sec:eeqft}.  This has to be upheld in the dual gravitational construction. It turns out the NEC again suffices to show that the HRT/RT proposals are consistent with the field theory causality requirements \cite{Wall:2012uf,Headrick:2014cta}.  The argument which may be viewed as a generalization of  the Gao-Wald theorem to codimension-2 surfaces relies on showing that the extremal surface which computes the holographic entanglement entropy lies in the so-called {\em causal shadow} region of the bulk. This is the region of the bulk spacetime that is spacelike separated from the domains of dependences of both the boundary subregion and its complement, viz., it is the causal complement of the bulk future and past of the boundary $\domdA$ and $\domdAc$.
\begin{equation}
\text{Causal shadow:} \left(\tilde{J}^+[\domdA] \cup \tilde{J}^-[\domdA] \cup \tilde{J}^+[\domdAc] \cup \tilde{J}^-[\domdAc] \right)^c
\label{eq:cshadow}
\end{equation}
%

% Figure
\begin{figure}[h]
\begin{center}
\includegraphics[width=2.3in]{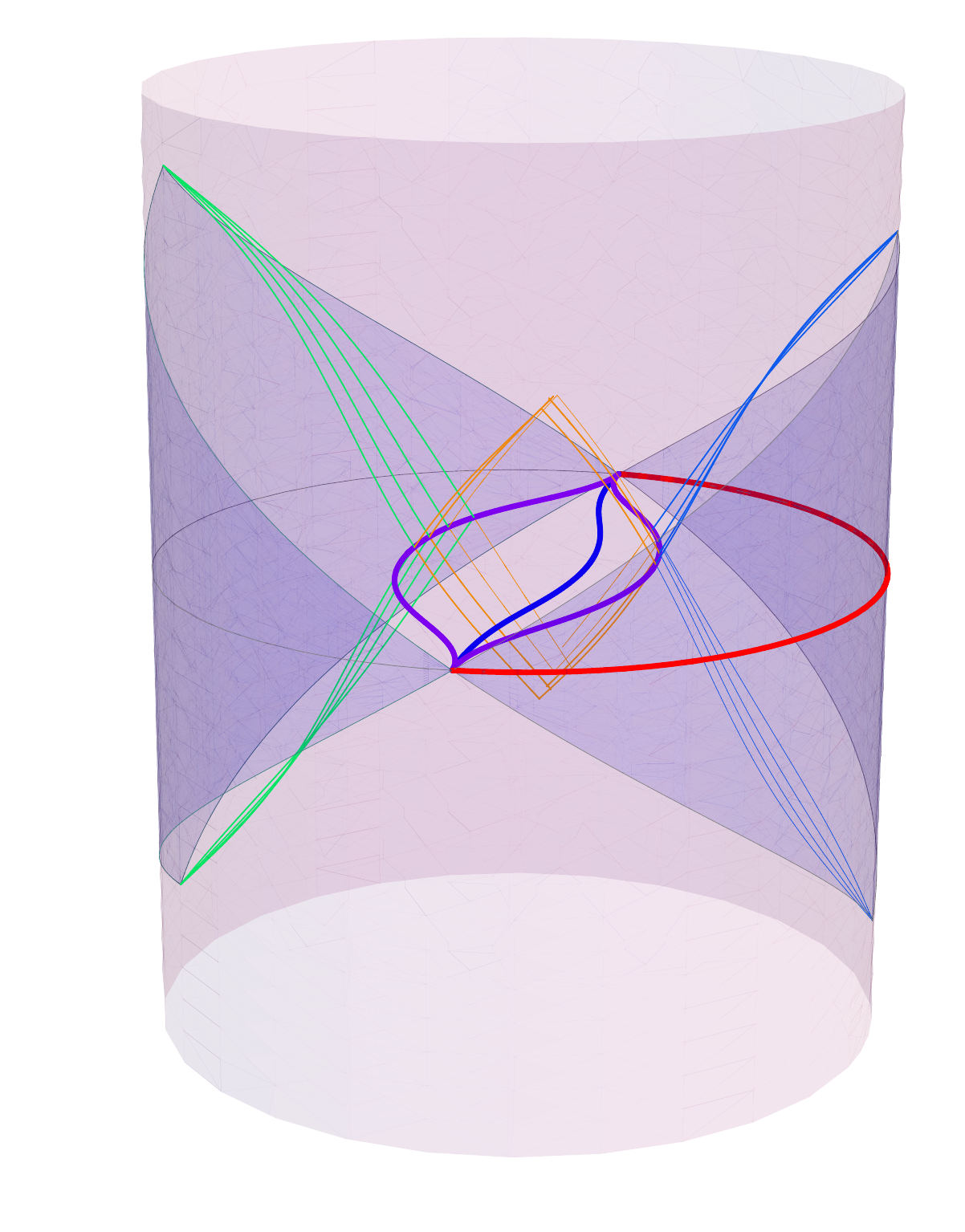}
 \caption{An asymptotically \AdS{d+1} spacetime with a non-trivial causal shadow. We have depicted regions $\regA$ and $\regAc$ along with their respective domains of dependence (shaded regions on the boundary), and causal wedges. The latter are only indicated by a few geodesics to avoid cluttering up the figure. The central region inside the spacetime bounded by the two purple curves is a spatial region on the initial slice that is spatial to both $\domdA$ and $\domdAc$. Its domain of dependence in the bulk is the causal shadow region, which alternately can be expressed as \eqref{eq:cshadow}. Figure taken from \cite{Headrick:2014cta}.}
\label{f:cshadow}
\end{center}
\end{figure}

In pure AdS spacetime for the ball-shaped regions, the causal shadow turns out to collapse to a codimension-2 hypersurface of the spacetime, thus uniquely singling out the RT/HRT surface. However, in more general situations, the causal shadow ends up being a bulk codimension-0 region as described in \cite{Headrick:2014cta}. This in particular implies that causality alone does not in general uniquely pin down the bulk surface responsible for computing entanglement entropy. Any codimension-2 surface located in the causal shadow would be an acceptable candidate for the construction; it is the bulk gravitational dynamics that truly pin down the particular surface of interest.

\subparagraph{B. Entanglement inequalities:} We now turn to the constraints on geometric states arising from entanglement structure of the QFT. A given collection of field theory entanglement entropies will of course satisfy the various quantum entropy inequalities such as those  listed in \S\ref{sec:eeineq}. These would of course have to be respected by the geometric duals, as we have explained in \S\ref{sec:heeineq}. One can make the following general observations:

\subparagraph{1. Strong sub-additivity:} The interesting quantum entropy inequality that does not follow straightforwardly from the holographic prescription is strong subadditivity. While the proof for the RT prescription goes through very easily without any input about the gravitational solution, the corresponding proof in general time-dependent situations relies on the gravitational dynamics respecting the NEC. It was first demonstrated that gravitational solutions supported by matter violating the NEC would lead to a failure of the strong subadditivity inequality in the dual state of the QFT \cite{Callan:2012ip}. The explicit proof  for the HRT prescription using the maximin construction \cite{Wall:2012uf} explicitly relies on this condition to show that areas of the extremal surfaces cannot increase under Lie drag along  orthogonal null congruences. These statements, as in the previous discussion, hold for bulk Einstein-Hilbert dynamics, and perturbatively away from it in higher order derivative corrections.

% Figure
\begin{figure}[htbp]
\begin{center}
\includegraphics[width=3in]{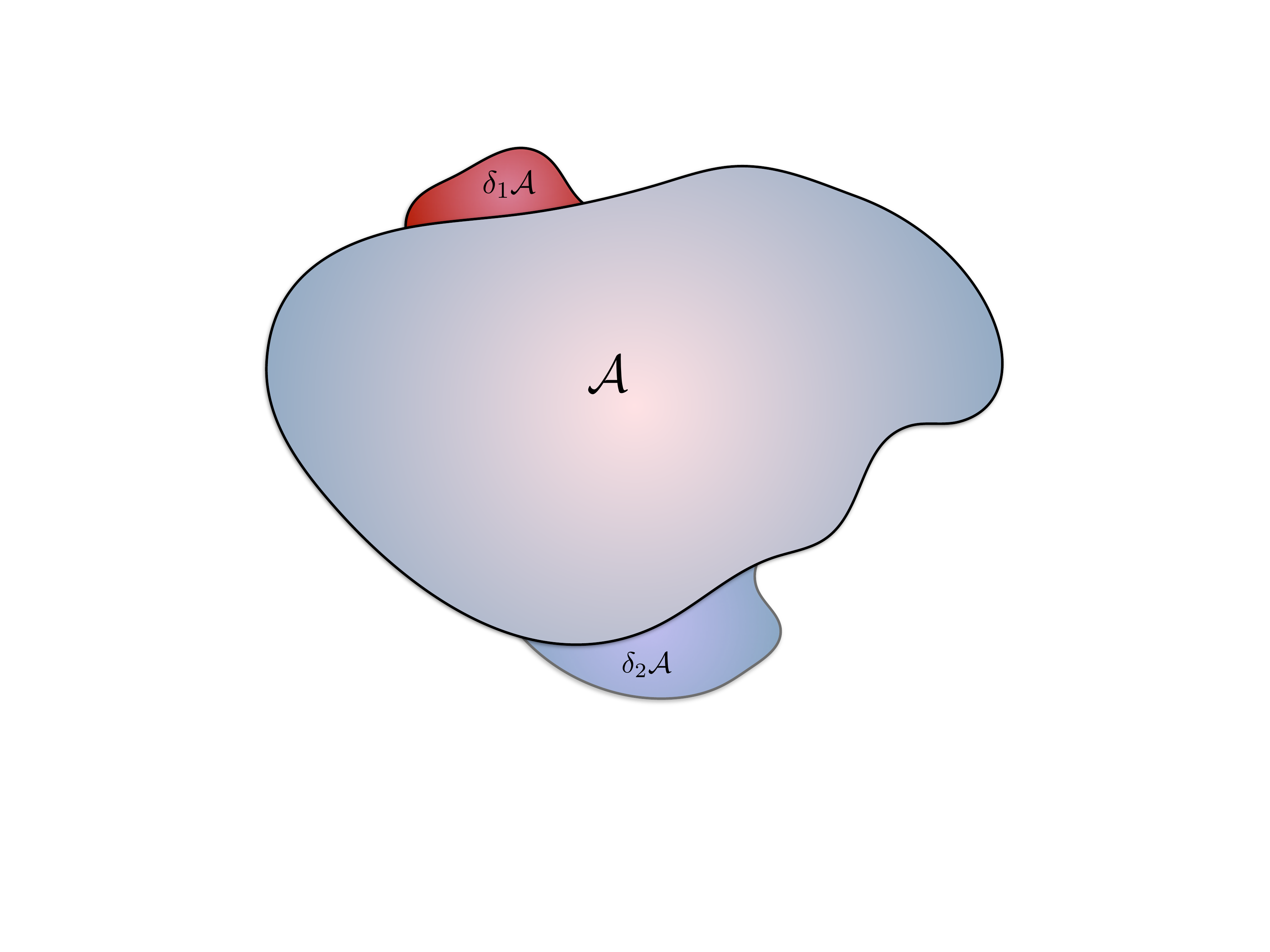}
 \caption{Illustration of a macroscopically large region $\regA$ which is perturbed in two independent directions infinitesimally to construct regions $\regA \cup \delta\regA_1$ and $\regA \cup \delta \regA_2$, respectively. These three regions, together with $\regA\cup \delta\regA_1 \cup\delta \regA_2$, form the spatial regions to which we apply the strong subadditivity inequality to derive the second variational formula \eqref{eq:eedensity}.}
\label{f:infssa}
\end{center}
\end{figure}

\subparagraph{2. Entanglement density and integrated NEC:} The statement of the strong additivity for spatially-organized entanglement can be converted into an infinitesimal (functional) variational statement.  Let us say that we have three regions $\regA$, $\delta\regA_1$ and $\delta \regA_2$ with the latter two adjoining the former and moreover being small deformations of the (macroscopically) larger region $\regA$, as illustrated in Fig.~\ref{f:infssa}. Applying the strong subadditivity inequality  \eqref{eq:ssa1} to the four regions $\regA$, $\regA \cup \delta\regA_1$, $\regA \cup \delta \regA_2$, and
$\regA \cup \delta\regA_1 \cup\delta \regA_2$ and expanding out to leading order in the infinitesimal variations $\delta\regA_1$ and $\delta\regA_2$, we are led to the variational statement
\begin{equation}
\hat{n}_\regA\left(\delta\regA_1, \delta\regA_2\right) \equiv-\delta_{\regA_1} \delta_{\regA_2} S_\regA \geq 0\,.
\label{eq:eedensity}
\end{equation}
What this says is that for a fixed quantum state, the entanglement density $\hat{n}_\regA(\delta \regA_1, \delta \regA_2)$ defined through this variational formula is non-negative definite. We have indicated explicitly the dependence of the regulating region in the subscript and view the small deformations thereof as the arguments of this density.  One can show that satisfying this infinitesimal version of the strong subadditivity is tantamount to the more general relations \eqref{eq:ssa1} and \eqref{eq:ssa2} being upheld for the state in question (for any choice of regions involved).

While the relations we have derived are valid quite generally, they are most interesting in two-dimensional QFTs.  It was argued that for certain states in such theories, the non-negative definiteness of the entanglement density is holographically dual to an integrated form of the NEC \cite{Lashkari:2014kda,Bhattacharya:2014vja}. One finds that
\begin{equation}
\hat{n}_\regA \geq 0 \;\; \Longrightarrow \;\; \int_{\extrA} \, \sqrt{\gamma_{\extrA}} \, N^A_{(i)}\, N^B_{(i)} \,E_{AB} \geq 0 \,.
\label{}
\end{equation}
In the above expression, $N^\mu_{(i)}$ with $i=1,2$ are null normals to the extremal surface \eqref{eq:ndef}, $\gamma_{AB}$  is the induced metric on the extremal surface $\extrA$, and $E_{AB}$ denotes the geometric tensor appearing in Einstein's equations. Using the bulk equations of motion $E_{AB} = 8\pi\, \GN \, T_{AB}^\text{matter}$, we would conclude that the matter stress tensor $T_{AB}^\text{matter}$ supporting the solution should have to satisfy the NEC in its integrated form over the extremal surface.

In \cite{Lashkari:2014kda}, this statement was demonstrated for Lorentz invariant states that geometrically encode renormalization group flows away from the AdS fixed point, while \cite{Bhattacharya:2014vja} showed this for perturbative states in the vicinity of the CFT vacuum. Should such a statement prove to be true in greater generality, one would have a very direct link between the strong subadditivity inequality and the NEC. Unfortunately this appears not to hold in the form stated in higher-dimensional field theories. It is an interesting open question to ascertain what is the connection between the infinitesimal form of strong subadditivity encoded in the entanglement density and constraints on geometry.

\subparagraph{3. Relative entropy constraints:} We have already explained how relative entropy \eqref{eq:relent} serves as a useful discriminator between quantum states, and furthermore obeys a positivity and monotonicity constraint. Working in the code subspace of the vacuum, can can therefore compare neighbouring states and infer properties that they should satisfy. These constraints have been analyzed in some detail in \cite{Lashkari:2014kda,Lashkari:2015hha,Lashkari:2016idm}, whose results we summarize below.

Much of the discussion below assumes the reference state to be the CFT vacuum $\ket{0}$. One also considers spherical regions $\regAB$ in order to be able to have explicit access to the modular Hamiltonian. So $\sigma = \rhoB$ is the vacuum reduced density matrix induced onto the region, and we will take $\rho = \rho_{\regAB}$ is the density matrix corresponding to some  excited state.  The caveats explained in \S\ref{sec:nhee} do pertain, but only insofar as perturbative excitations around the vacuum density matrix are concerned. Large deviations away from the vacuum are no longer constrained by the symmetry. Overall the use of relative entropy leads to an interesting set of gravitational statements, which can be used to rule certain configurations from being geometries dual to sensible quantum states.

\begin{itemize}
\item Firstly, for states that are infinitesimally apart, we know that the relative entropy vanishes to leading order \eqref{eq:e1stlaw}. This follows from the positivity requirement, cf., \S\ref{sec:genee}. It turns out that this statement implies the linearized Einstein's equations in the dual gravitational theory \cite{Lashkari:2013koa,Faulkner:2013ica}. We will describe this in some detail in \S\ref{sec:einstein}.
\item Going beyond the linear order, the relative entropy implies an inequality \eqref{eq:relfree}. As with strong subadditivity, this places constraints on the gravitational side. In \cite{Lashkari:2015hha}, it was argued that the quadratic correction to the relative entropy, i.e., the quantum Fisher information \eqref{eq:qfisher} requires that a natural notion of energy defined within the entanglement wedge be positive.
\item The positivity of relative entropy beyond the perturbative limit implies that the difference of the quasi-local energy between the two dual states must similarly be non-negative definite. From here one can derive a family of positive energy theorems for gravitational theories in AdS spacetimes.
\item Monotonicity of relative entropy likewise translates to the statement of positivity of the gravitational symplectic flux.
\end{itemize}

We will postpone the discussion of the canonical energy until we have built up some necessary machinery relating to the covariant phase space of gravitational theories in \S\ref{sec:einstein}.

\paragraph{III. Constraints on field theory states to admit geometric dual:} Thus far we have explained the general conditions under which a given geometry may be interpretable as being dual to a state in the Hilbert space of the dual QFT. Our discussion is by no means exhaustive, as there presumably are still other constraints waiting to be discovered. What we know so far is that
\begin{itemize}
\item In a wide-ranging set of states, owing to the asymptotic $\ceff \to \infty$ planar limit, the mutual information between two widely separated regions becomes sub-dominant, $I(\regA_1: \regA_2 ) \sim {\cal O}(1)$ when $\regA_1$ and $\regA_2$ are macroscopically apart. Such a phase transition ends up being guaranteed by the planar limit once the field theory in question has a sparse low-lying spectrum \cite{Headrick:2010zt}. This per se is therefore not a strong diagnostic of the existence of a geometric dual, but more of a useful first check.
\item Geometric states are required to satisfy the monogamy of mutual information $\tmi(\regA_1 : \regA_2 :\regA_3) \leq 0$. This, for instance, precludes states like the GHZ state from admitting geometric holographic duals. One however can check that randomly chosen pure states of just a few qubits appear  in general to have $\tmi < 0$ \cite{Rangamani:2015qwa}, as is the case with random tensor network states described in \cite{Hayden:2016cfa}. Neither of these examples are close to being holographic in other respects, so again the relative strength of this criterion remains unclear.
\item As with the tripartite information $\tmi$, one might wonder whether there are further constraints we should impose based on the general inequalities derived in the form of the holographic entropy cone \cite{Bao:2015bfa}. In order for these inequalities to provide a useful constraint, one should first generalize them to the covariant setting. At present we lack the knowledge of  the covariant entropy cone. Therefore further investigation is necessary prior to formulating a useful constraint.
\end{itemize}

%~~~~~~~~~~~~~~~~~~~~~~~~~~~~~~~~~~~~~~~~~~~~~~~
\section{The dual of a density matrix}
\label{sec:dendual}
%~~~~~~~~~~~~~~~~~~~~~~~~~~~~~~~~~~~~~~~~~~~~~~

Let us now consider a variant of the question: ``Which states in the QFT admit geometric duals''? Suppose we are given both a global state of the QFT and in addition a spatial subregion $\regA$. By tracing out the state in the complement, we obtain $\rho_\regA$, and then can ask: \emph{``Given the knowledge of $\rhoA$, is it possible to associate a specific region of the bulk geometry that is dual to it?''}

The first set of investigations to focus on this question as phrased was in the works  \cite{Bousso:2012sj,Czech:2012bh}, which was subsequently elaborated upon in \cite{Wall:2012uf,Headrick:2014cta}. The current understanding is that the region of the bulk spacetime that is dual to the density matrix is the so-called \emph{entanglement wedge}. To explain this concept, we need to understand first how we propose to relate bulk and boundary data in the holographic correspondence and thence explain the rationale for the entanglement wedge. Evidence in favour of the entanglement wedge comes from viewing the holographic map as a quantum
error-correcting code \cite{Dong:2016eik}.

%~~~~~~~~~~~~~~~~~~~~~~~~~~~~~~~~~~~~~~~~~~~~~~~
\subsection{Local bulk operators in holography}
\label{sec:hkll}
%~~~~~~~~~~~~~~~~~~~~~~~~~~~~~~~~~~~~~~~~~~~~~~

We have focused our attention in the previous section on asking, when is a given QFT state describable by geometry? Supposing we have one such state wherein the classical gravitational description is an excellent approximation, we should wonder: how does one see that the bulk theory admits approximate locality? This is a feature of classical gravitational theories that we ought to recover from the QFT data alone. A constructive way to proceed on this front would be to construct local bulk operators that would capture the essence of what we seek in the semiclassical gravitational picture. One would  hope that systematic perturbative corrections in $(\ceff)^{-1}$ would allow for a determination within the remit of quantum gravitational perturbation theory.

The first discussion of these local bulk operators dates back to the early days of the AdS/CFT correspondence; \cite{Banks:1998dd} describe the construction of the boundary to bulk map by invoking the relation between bulk fields and boundary QFT operators. This was subsequently developed in a series of works by Hamilton, Kabat, Lifschytz, and Lowe \cite{Hamilton:2005ju,Hamilton:2006az,Hamilton:2006fh}, who gave a nice characterization of the bulk operators in terms of boundary QFT data explicitly using a Green's function technique. We will follow here a modern treatment of  this discussion following \cite{Heemskerk:2012mn}.

The basic piece of data we need for the construction is the ``extrapolate map'' entry in the AdS/CFT dictionary \cite{Banks:1998dd,Balasubramanian:1998de,Harlow:2011ke}, which allows us to recover boundary correlation functions by extrapolating the insertion points of bulk correlation functions. Recall that the GKPW construction \cite{Gubser:1998bc,Witten:1998qj} for the bulk to boundary map relates bulk fields to corresponding CFT operators.

\paragraph{1. Global reconstruction:} We will first work in global AdS \eqref{eq:gads} and then indicate how we can pass onto various other domains. This particular construction will therefore be referred to as the \emph{global reconstruction}.  To simplify the discussion, it is helpful to bring the radial coordinate to a finite domain which can be done by the coordinate transformation $\rho = \cot\varrho$ so that the metric \eqref{eq:gads} simplifies to
\begin{equation}
ds^2 = \frac{1}{\sin^2 \varrho} \left(dt^2 + d\varrho^2 + \cos^2 \varrho \; d\Omega_{d-1}^2 \right)
\label{eq:gads2}
\end{equation}
The boundary now is at $\varrho = 0$ and the origin of \AdS{d+1} is attained at $\varrho  = \frac{\pi}{2}$.

Since we need to refer to bulk and boundary coordinates in the same breath, let us introduce some notation.
Henceforth $X$ will be a shorthand for bulk coordinates, $X \equiv \{x^\mu, \varrho\}$
where $x^\mu$ (typically abbreviated as $x$) denotes the boundary coordinates $x^\mu = \{ t, \Omega_{d-1}\}$.
Given a bulk field $\phi(X)$,  we infer from the bulk dynamics that the fields behave asymptotically as \eqref{eq:asyphi}
\begin{equation}
\lim_{\varrho \to 0} \, \varrho^{-\Delta} \, \phi(x^\mu, \varrho)  = {\cal O}(x^\mu)
\label{eq:extrapol}
\end{equation}
We have dropped the expectation value around the operator and choose to interpret this equation now as a statement between operators of the bulk semiclassical gravitational theory and the QFT on the boundary. This identification comes with an essential subtlety, for we are conflating the bulk and boundary Hilbert spaces; these are dual to each other, but in reading \eqref{eq:extrapol}, we are going to pretend that we can extend the bulk Hilbert space to include the  boundary.   One can think of  \eqref{eq:extrapol} as the solution to the solution to the bulk dynamical equations in the absence of boundary sources.

Given this solution, we can formally attempt to write down a Green's function that inverts the relation above and reconstructs $\phi(X) $ given ${\cal O}(x)$. This is what is referred to as the bulk reconstruction programme, since here we start from the boundary data which is well understood and obtain approximately local operators in the dual which can then be used as probes of the local geometry. The difficulty in implementing this at a naive level is simply that, as stated, we do not have a standard Cauchy evolution problem. The boundary of AdS is a timelike hypersurface and we would be required to relate data defined therein onto regions that are spacelike separated from it. Nevertheless it is possible to show that within the $(\ceff)^{-1}$ perturbation theory, this is possible to do. To leading order in $\ceff$, this simply follows from the fact that the bulk fields are essentially free (all interactions are suppressed in the large $\ceff$ expansion). In this case, we simply work in a basis of Fourier modes and invert the relation directly.

% Figure
\begin{figure}[tp]
\begin{center}
\includegraphics[width=1.75in]{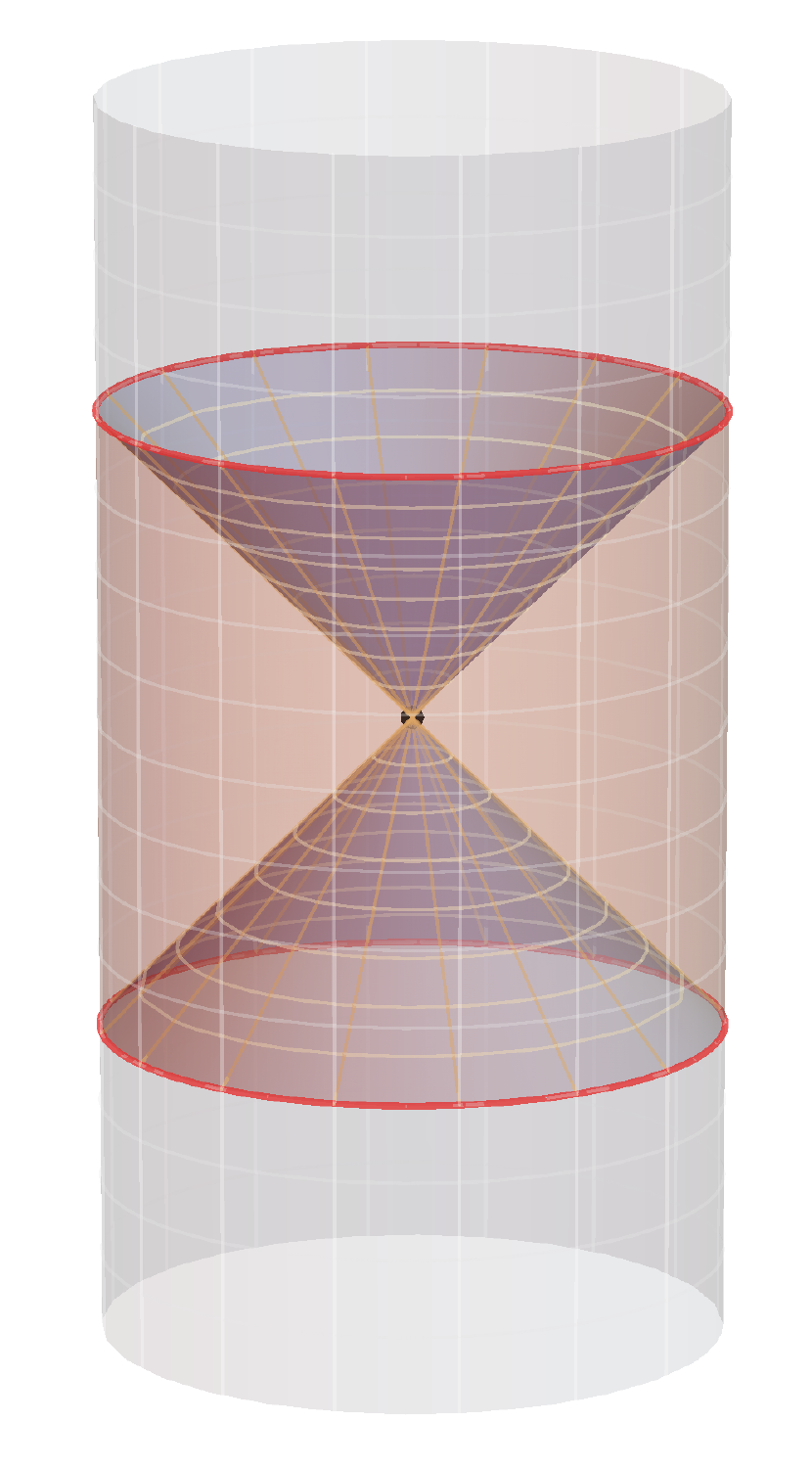}
\hspace{1cm}
\includegraphics[width=1.75in]{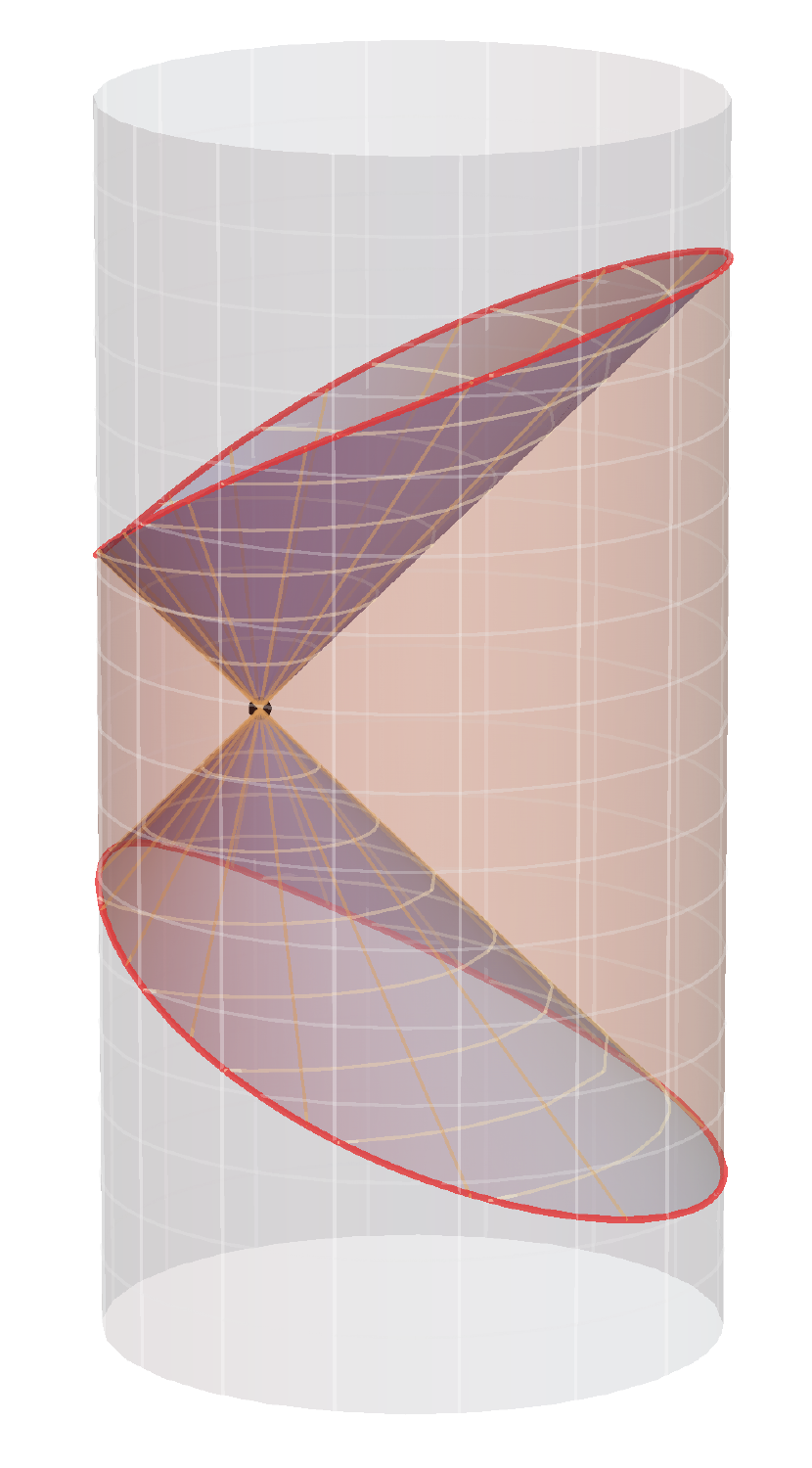}
\hspace{1cm}
\includegraphics[width=1.75in]{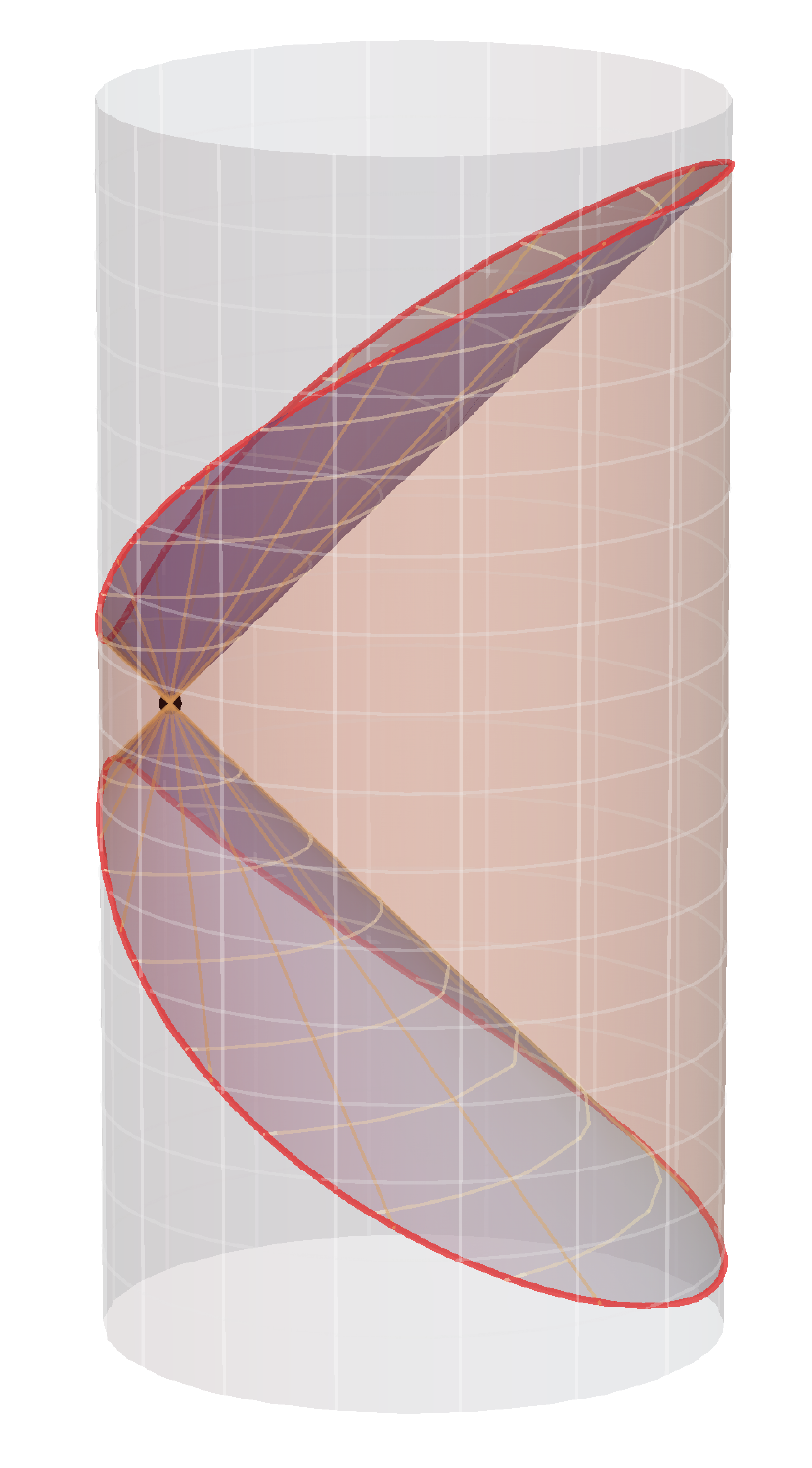}
\setlength{\unitlength}{0.1\columnwidth}
\begin{picture}(0.3,0.4)(0,0)
\put(-3.28,2.92){\makebox(0,0){$\bullet$}}
\put(-3.1,2.9){\makebox(0,0){$p$}}
\put(-4.47,3.3){\makebox(0,0){${\cal T}_p$}}
\put(-0.35,2.93){\makebox(0,0){$\bullet$}}
\put(-0.15,2.9){\makebox(0,0){$p$}}
\put(1.03,3.3){\makebox(0,0){${\cal T}_p$}}
\put(2.84,2.93){\makebox(0,0){$\bullet$}}
\put(3.04,2.9){\makebox(0,0){$p$}}
\put(4,3.3){\makebox(0,0){${\cal T}_p$}}
\end{picture}
\caption{
The time strip ${\cal T}_p$ associated with a single bulk point $p$ illustrated for various locations of the bulk point. A local bulk field inserted at $p$ is given as the integral of local boundary operators smeared over ${\cal T}_p$ weighted by a kernel \eqref{eq:hkll}.
}
\label{f:hkllstrip}
\end{center}
\end{figure}
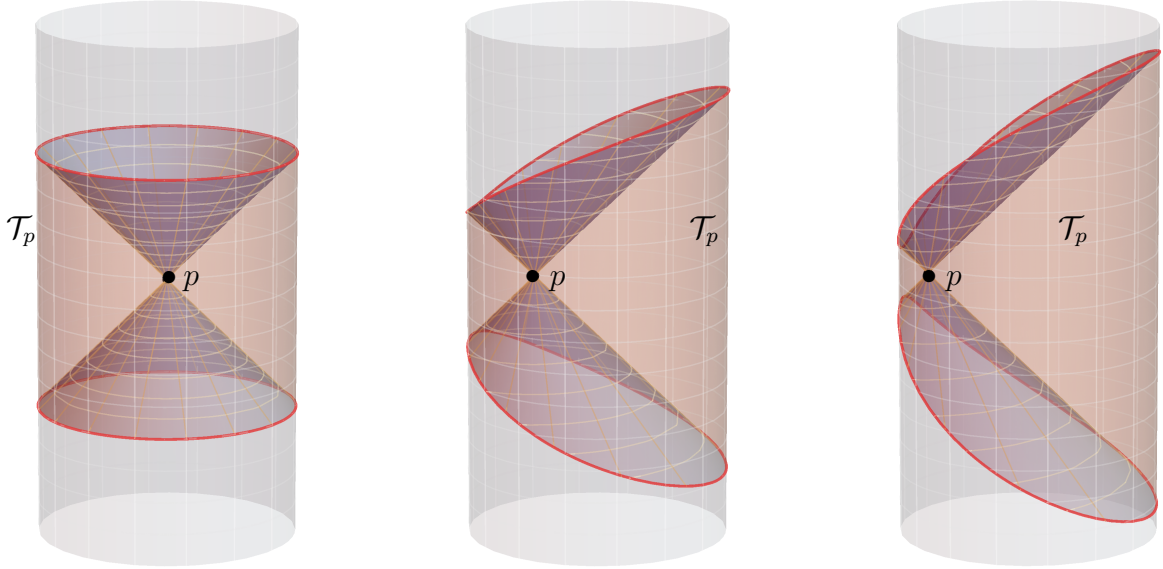

In order to write down a formula for the local bulk operator thus obtained, let us introduce some notation. Given a point $p$ with coordinates  $X $ in the bulk, we introduce a boundary \emph{time strip} ${\cal T}_p$ which is defined to be the set of all boundary points that are non-timelike (i.e., spacelike or null) related to $p$.
\begin{equation}
{\cal T}_p =\Big\{ y \in \bdy_d \; \Big| \; p= X \in \bulk \; ,\; p\notin \tilde{I}^\pm(y)  \Big\}
\label{}
\end{equation}
We illustrate this for the pure \AdS{d+1} geometry in Fig.~\ref{f:hkllstrip}.

The global reconstruction result can now be stated quite simply by noting that the explicit inversion of \eqref{eq:extrapol} can be achieved by a boundary-to-bulk kernel $K(y^\mu|x^\mu,\varrho)$ such that
\begin{equation}
\phi(p) \equiv \phi(X) = \int_{{\cal T}_p} d^dy \; K(y^\mu|x^\mu,\varrho)\; {\cal O}(y^\mu)
\label{eq:hkll}
\end{equation}
For the pure \AdS{d+1} geometry, one can give an explicit form for this kernel in terms of special functions. In even-dimensional spacetimes $d+1 =2n$, it is simple to express the result in terms of a spacelike Green's function in AdS spacetime. We can write for any asymptotic AdS spacetime a near boundary decomposition of the Green's function as:
\begin{equation}
G(X|X') \xrightarrow{\varrho' \to 0} \frac{1}{(2\Delta -d)\, \lads^{d-1}} \, \left(\varrho'^\Delta\, G_{source}(X|x') + \varrho'^{d-\Delta}\, G_{vev} (Y|x') \right)
\label{eq:GKrel}
\end{equation}
We want to localize the latter term which only involves the normalizable modes. This can be done by suitably choosing the boundary conditions for the Green's function in \AdS{d+1}.

In even-dimensional AdS spacetimes, the relevant spacelike Green's function can be expressed in terms of the AdS invariant distance $\sigma$ between two bulk points, say $X$ and $X'$.
\begin{equation}
\sigma = X\cdot X' = \lads^2 \; \frac{\cos^2(t-t') - \sin\varrho \,\sin\varrho'\; \Omega_{d-1} \cdot \Omega'_{d-1}}{\cos\varrho \, \cos\varrho'}
\label{eq:adsinvd}
\end{equation}
For scalar operator ${\cal O}$ of conformal dimension $\Delta$, we find \cite{Heemskerk:2012mq}
\begin{equation}
G_{d=2m-1}(\sigma) = \frac{\pi}{2}\, (\sigma^2-1)^\frac{d-1}{4}\; \mathbf{P}^\mu_{\nu}(\sigma) + \Im\left( (\sigma+i\epsilon)^2 - 1\right)^{-\frac{d-1}{4}}\;  \mathbf{Q}^\mu_{\nu}(\sigma + i\epsilon)
\label{eq:eadsG}
\end{equation}
where $\mathbf{P}^\mu_\nu$ and $\mathbf{Q}^\mu_\nu$ are Legendre polynomials of the third kind. One can check that this is consistent with \eqref{eq:GKrel} with a  vanishing source Green's function. The odd-dimensional case is more complicated, as there appears to be no spacelike Green's function that is expressible in terms of the AdS invariant distance alone (see \cite{Heemskerk:2012mq}).

The kernel $K(y|x,\varrho)$ in general is a solution to the bulk equations of motion subject to the normalizable boundary conditions of the extrapolate map. As a result,  its specific form depends on the geometry of the bulk spacetime $\bulk$. Since different QFT states correspond to different  backgrounds, we have to determine the kernel independently for a given state. States that have macroscopically distinct properties will have widely differing kernels, but those that only involve perturbative excitations can be understood within the planar  (inverse $\ceff$) perturbation theory.
One sees here the relevance of the code subspaces. In any one component of the code subspace, we can determine the kernel $K(y|x,\varrho)$  using its form in the the geometry of the parent state atop which we consider a few particle excitations.

\begin{figure}[tp]
\begin{center}
\includegraphics[width=1.9in]{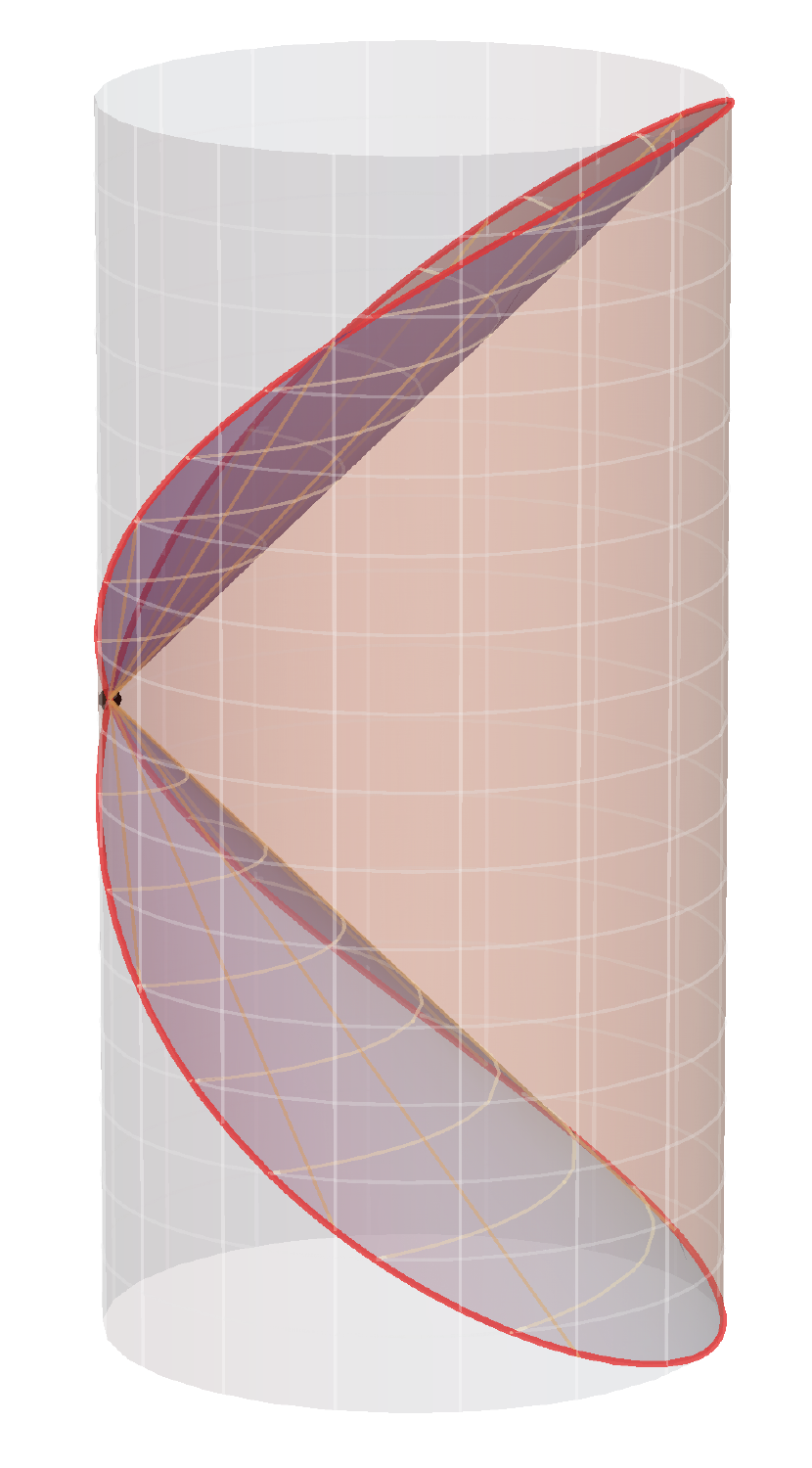}
\hspace{2cm}
\includegraphics[width=3in]{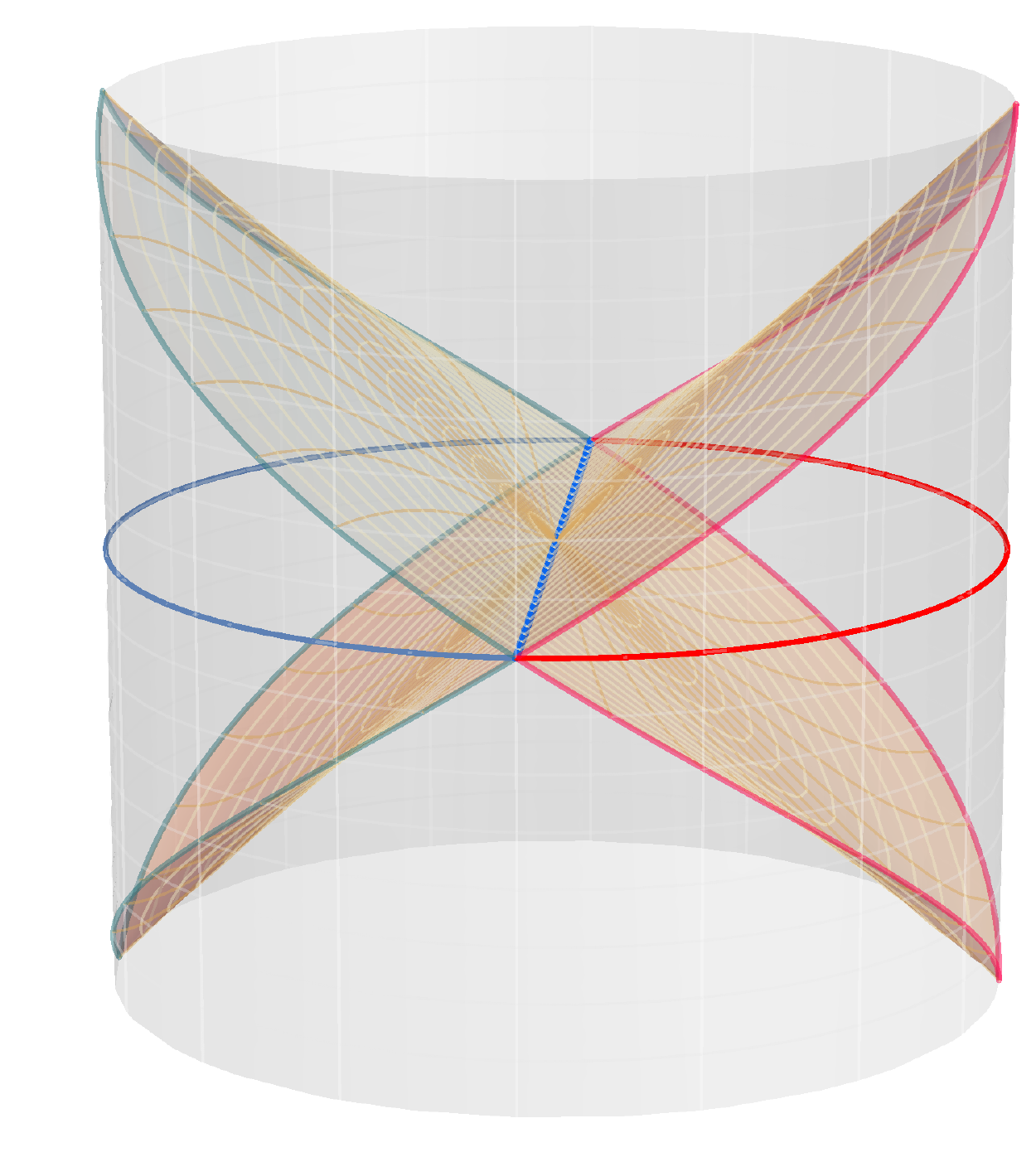}
\setlength{\unitlength}{0.1\columnwidth}
\begin{picture}(0.3,0.4)(0,0)
\put(-8.7,2.75){\makebox(0,0){$\bullet$}}
\put(-8.95,2.8){\makebox(0,0){${\cal I}^0$}}
\put(-7.2,2.5){\makebox(0,0){${\mathbb R}^{d-1,1}$}}
\put(-5.,2.7){\makebox(0,0){Rindler$_L$}}
\put(0.3,2.7){\makebox(0,0){Rindler$_R$}}
\end{picture}
\caption{
The boundary domains for the (a) Poincar\'e patch and (b) Rindler reconstructions. A local bulk field inserted at $p$ is given as the integral of local boundary operators smeared over ${\cal T}_p$ weighted by a kernel \eqref{eq:hkll}.
}
\label{f:hklllocal}
\end{center}
\end{figure}
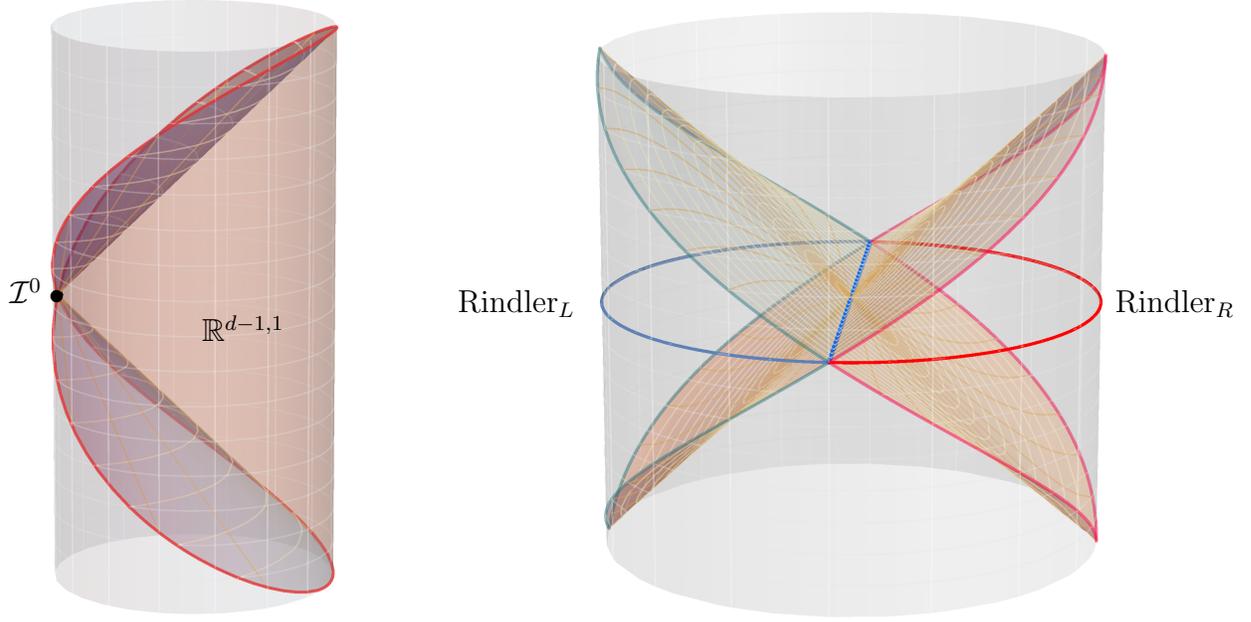

\paragraph{2. Local reconstruction:} Having understood how global reconstruction works, let us turn to the local version; by this we simply mean that we restrict attention to part of the bulk spacetime. We already know that consideration of the QFT on Minkowski spacetime results in the restriction of the bulk geometry to the Poincar\'e patch of AdS. We can be even more ambitious and restrict attention to further subregions. For example, a Rindler observer in ${\mathbb R}^{d-1,1}$ will only see part of the Minkowski spacetime. The corresponding bulk domain will similarly get truncated from the Poincar\'e wedge to the AdS-Rindler wedge, cf., Fig.~\ref{f:hklllocal}.

% Figure

In these situations, one should be able to obtain a representation of the bulk operators analogous to
\eqref{eq:hkll}. However, now the support of the integral cannot be restricted to the time-strip, since the coordinate patch of choice in AdS may not encompass this boundary domain. This is clear already from the Poincar\'e patch whose boundary is a Minkowski diamond illustrated above in Fig.~\ref{f:hklllocal} (cf., also Fig.~\ref{f:poincarecoords}). This issue was  addressed initially in \cite{Hamilton:2005ju} and certain subtleties in this discussion were clarified recently in \cite{Morrison:2014jha}.

For the field theory on Minkowski spacetime, by employing the explicit coordinate transformation between the global and Poincar\'e coordinates one can show that there is a representation of the form \eqref{eq:hkll} except now the integration is not carried out over the time-strip, but rather over the Minkowski diamond:
\begin{equation}
\phi(X) = \int_{{\mathbb R}^{d-1,1}} d^dy \; K(y|x,z)\; {\cal O}(y) \,, \qquad X = \{x^\mu,z\} \in \text{Poincar\'e patch}
\label{eq:hkllMink}
\end{equation}
 This statement pretty much follows from the global reconstruction if we allow ourselves the freedom to translate the bulk point $p$ out towards the boundary of \AdS{d+1}, so that it becomes the spatial infinity ${\cal I}^0$ of ${\mathbb R}^{d-1,1}$.

The Rindler patch of Minkowski spacetime provides a much more interesting example. In this case, we have to restrict the operator to lie within a sub-domain of the Poincar\'e patch of the bulk spacetime. To ascertain what the sub-domain is, let us try to re-express the results for the global \eqref{eq:hkll} and Poincar\'e patches  \eqref{eq:hkllMink} somewhat differently. In writing the expressions \eqref{eq:hkll} and \eqref{eq:hkllMink}, we focused on a particular bulk point and then worked out which corresponding region should we use to smear operators. Let us invert this picture and ask how to go from a given boundary region to a bulk point. While we defined the  point $p$ to be non-timelike related to all the points in the timestrip ${\cal T}_p$, we can equivalently view $p$ as belonging to the intersection of the bulk causal future and bulk casual past of the boundary time-strip. Moreover it is clear that any point which lies within this region can be represented by smearing boundary operators in the time-strip.

% Figure
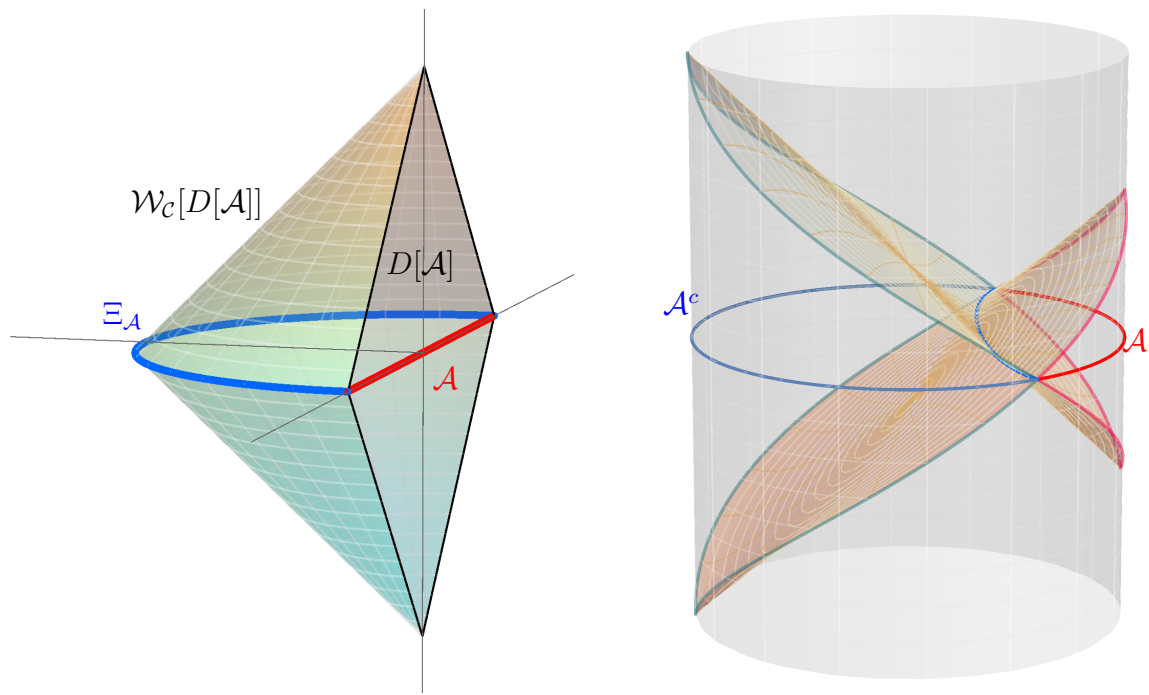
\begin{figure}[htbp]
\begin{center}
\includegraphics[width=3in]{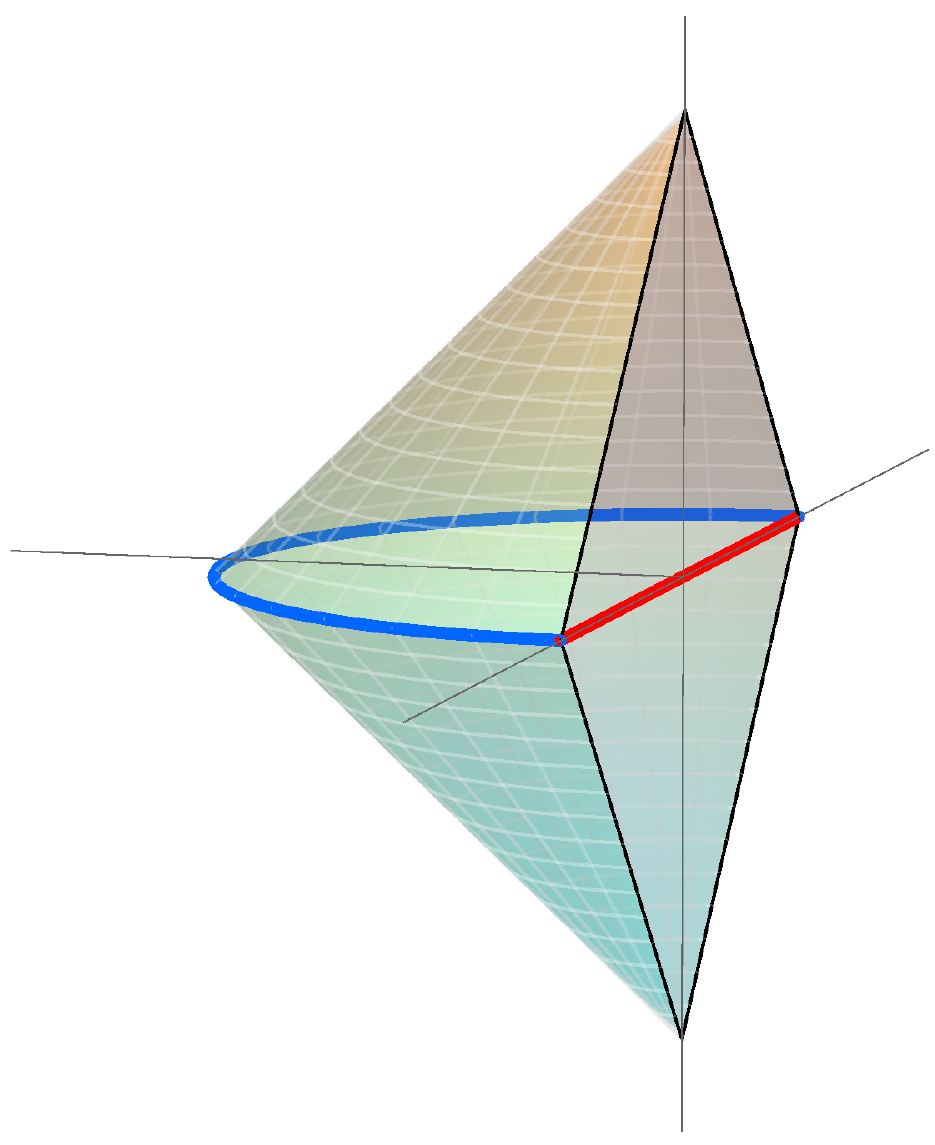}
\hspace{1cm}
\includegraphics[width=2.6in]{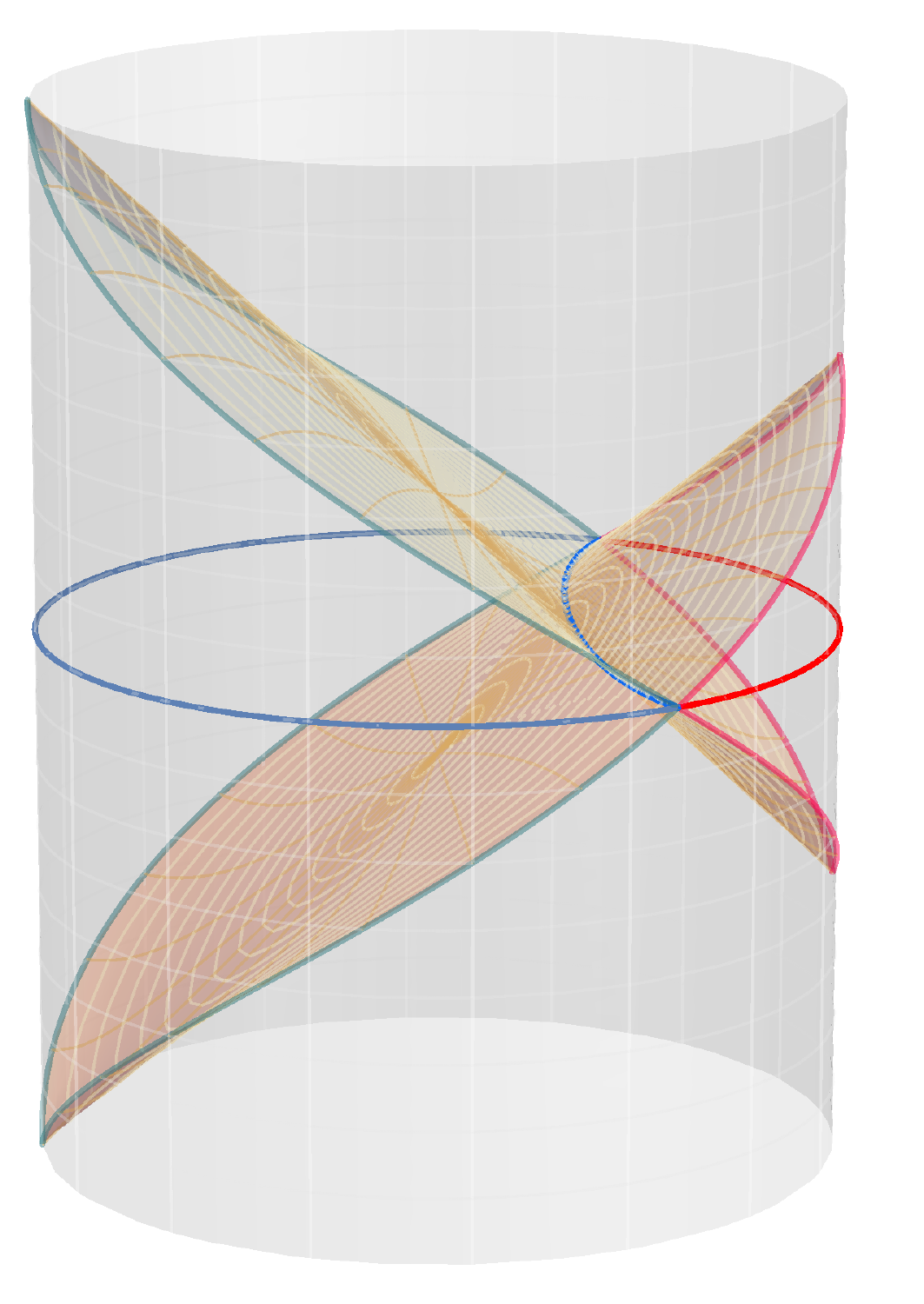}
\setlength{\unitlength}{0.1\columnwidth}
\begin{picture}(0.3,0.4)(0,0)
\put(-8.6,3.1){\makebox(0,0){$\color{blue}{\Xi_\regA}$}}
\put(-8.,4){\makebox(0,0){$\color{black}{\CW{\domdA}} $}}
\put(-6.,2.6){\makebox(0,0){$\color{red}{\regA}$}}
\put(-6.2,3.5){\makebox(0,0){$\color{black}{\domdA}$}}
\put(-4.1,3.2){\makebox(0,0){$\color{blue}{\regAc}$}}
\put(-0.45,2.9){\makebox(0,0){$\color{red}{\regA}$}}
\end{picture}
\caption{
Depiction of the causal wedge $\CW{\domdA}$ for a boundary domain of dependence. On the left, we plot the causal wedge in Poinca\'e coordinates, and on the right, we depict the domains $\CW{\domdA}$ and $\CW{\domdAc}$ in global AdS spacetime. In pure
\AdS{d+1}, the latter is coincides with a Rindler decomposition, but as we shall see later, this does not always have to hold.}
\label{f:cwedges}
\end{center}
\end{figure}

This brings us to the definition of a bulk domain that is determined purely on causal grounds. Consider a boundary causal domain ${\cal D}$.  This can be a time-strip ${\cal T}$ or even the  domain of dependence  $\domd{\regA}$ of some subregion $\regA$ which forms its Cauchy surface. Utilizing the causal structure of the bulk spacetime, we can define a bulk codimension-0 region, which we call the \emph{causal wedge} of this boundary domain, denoted $\CW{{\cal D}}$. In the case of a boundary domain of dependence, we will rely on the fact that the prescription of the Cauchy surface suffices to equivalently use the notation $\CWA$.

The causal wedge of a boundary domain ${\cal D}$ is simply the set of bulk points that are able to communicate with and receive communication from the said domain. To wit,
\begin{equation}
\CW{{\cal D}} = \tilde{J}^+[ {\cal D}] \cap  \tilde{J}^-[ {\cal D} ]
\label{eq:cwedge}
\end{equation}
We illustrate the causal wedges for boundary regions in both Poincar\'e and global AdS in Fig.~\ref{f:cwedges}.
A detailed discussion of causal wedges and their potential role in AdS/CFT can be found in \cite{Hubeny:2012wa,Hubeny:2013gba}; further generalizations and related constructs are critically examined in \cite{Hubeny:2014qwa}.
One can immediately check that, for the two simple cases of a point, $p$ in global \AdS{d+1} is the (degenerate) causal wedge of the time strip ${\cal T}_p$. Likewise the Poincar\'e patch may be viewed as the causal wedge of the boundary Minkowski diamond.
In these cases, all local operators within the causal wedge can be therefore reconstructed from the boundary operators suitably smeared out on the boundary time strip or Minkowski diamond.

Let us also record that the boundary of the causal wedge is the union of bulk null surfaces (it is a causal set after all) and ${\cal D}$.  The null surfaces are generated by ingoing bulk null geodesics emanating from the boundary of ${\cal D}$. Intersecting the past and future going null geodesics, we generically find a bulk codimension-2 surface which has been named the \emph{causal information surface } $\CIS$ \cite{Hubeny:2012wa}. It was further speculated that the area of this surface in Planck units ought to have some intrinsic and useful meaning in the dual QFT -- the mysterious dual has been named \emph{causal holographic information} $\chi_\regA$. One can show by the same arguments that establish the causality property of the HRT construction that $\chi_\regA$ bounds the holographic entanglement entropy $S_\regA$.  An interesting proposal is to interpret $\chi_\regA$ as a one-point entropy \cite{Kelly:2013aja}: we hold fixed the expectation values of all operators in $\domdA$ and maximize the von Neumann entropy over the space of density matrices.

The generalization to the Rindler space and more general situations is now straightforward. We define a local bulk operator in the causal wedge of some boundary domain ${\cal D}$ through a formula analogous to \eqref{eq:hkll} and \eqref{eq:hkllMink}. One writes
\begin{equation}
X \in \CW{{\cal D}} \;\; \Longrightarrow \;\; \phi(X) = \int_{{\cal D}} \, d^dy \; K(y|X)\; {\cal O}(y)
\label{eq:localhkll}
\end{equation}

In our discussion, we have focused primarily on scalar operators for which the smearing kernels $K(y|X)$ are easy to write down. This can be generalized to other kinds of tensor fields, though some care must be taken while dealing with gauge invariance. This issue arises both for bulk gauge fields which correspond to  conserved flavour currents on the boundary, as well as for the bulk gravitational degrees of freedom which map onto the boundary energy-momentum tensor.

Another issue that needs to be handled carefully is the backreaction of any local operator. We have so far pretended that we can work in a fixed background spacetime and describe operators inserted locally thereon.  This is true to leading order in $\ceff$ but any field insertion in the bulk will eventually cause backreaction. The strength of this backreaction is set by the energy carried by the field; in gravity, we would note that this scales like $\frac{m}{m_P}$ in Planck units. In field theory terms the relevant parameter is  $\frac{\Delta}{\ceff}$. As long as $\Delta \ll \ceff$, we can treat the backreaction perturbatively, which is the general intention in the code subspace. Once the field backreacts, we will induce gravitational dressing to the operator, which equivalently can be seen as the non-vanishing of the energy-momentum tensor's expectation value on the boundary. The gravitational field can be chosen to be suitably collimated in this perturbative expansion, so that the boundary energy-momentum tensor is non-vanishing in a well-localized region  inside the boundary domain bounding the causal wedge.

%~~~~~~~~~~~~~~~~~~~~~~~~~~~~~~~~~~~~~~~~~~~~~~~
\subsection{Subregion-subregion duality}
\label{sec:subregions}
%~~~~~~~~~~~~~~~~~~~~~~~~~~~~~~~~~~~~~~~~~~~~~~

 Now that we understand how to relate bulk and boundary operators, let us return to our question: for a given region $\regA$, what is the bulk ``dual'' of the reduced density matrix $\rhoA$?  We can interpret this question to mean that  we fix  $\rhoA$ and allow all compatible density matrices for the full state $\rho$ on $\regA \cup \regAc$.  With this understanding, we would like to know whether there is a natural  bulk spacetime region which is determined by $\rhoA$ independently of the choice of the global density matrix $\rho$.  A-priori we could also consider other questions such as those which examine whether the bulk region  is \emph{sensitive to} or is \emph{affected by} $\rhoA$. While these are interesting questions in their own right, we will focus attention on the following  specific question. Given $\rhoA$, in what region of the bulk can we uniquely reconstruct the geometric data (components of the metric and other fields)?

One proposal put forth in \cite{Bousso:2012sj} based on light-sheet arguments argued that the causal wedge was the correct dual. From one standpoint, this seems natural: bulk locality is manifest in the causal wedge thanks to the local reconstruction result described above. Any operator in the causal wedge of a region $\regA$ can be mapped back to the boundary domain $\domdA$ using \eqref{eq:localhkll}. One may indeed argue that this is the minimum bulk region that should be reconstructible using the boundary data in $\domdA$.

However, note that the local operator reconstruction simply picks a suitable local combination of QFT operators with support in $\domdA$. Per se, it does not have any information about the reduced density matrix itself. Consequently,  \cite{Czech:2012bh}, as well as \cite{Hubeny:2012wa, Wall:2012uf}, argued that the requisite bulk region should contain more than the causal wedge.  The first of these works discussed various criteria such a region ought to satisfy, noting that at the very least, the region in question ought to be cognizant of the  entanglement inherent in $\rhoA$.

Motivated by these discussions, one can  argue that the bulk region dual to the information contained in the reduced density matrix ought to be the \emph{entanglement wedge} $\EWA$. To define $\EWA$, we start with the observation that, given  the reduced density matrix, one can compute the entanglement entropy. Geometrically this implies that we should start with the extremal surface $\extrA$ and construct $\EWA$ therefrom. This can indeed be done quite simply by realizing the extremal surface, being codimension-2 in the bulk spacetime $\bulk$, naturally splits the bulk into four distinct regions: the future and past of the extremal surface, and two regions that are spacelike related to it in the direction of $\regA$ and $\regAc$,  respectively. We seek the latter set of regions which can be defined in terms of the homology surfaces $\homsurfA$ and $\homsurfAc$. Recall that the homology surface is a bulk codimension one surface which is bounded by the extremal surface and the boundary region cf., \eqref{eq:HRTprop}. Given this, we can simply define the entanglement wedge as the bulk domain of dependence of the homology surface, viz.,
\begin{equation}
\EWA \equiv \tilde{D}[\homsurfA] \,.
\label{eq:ewedge}
\end{equation}
Note that the decomposition of the bulk spacetime across the extremal surface can thus be simply expressed as
\begin{equation}
\bulk =  \EWA \cup \EWAc  \cup \tilde{J}^+[\extrA] \cup \tilde{J}^-[\extrA] \,,
\label{eq:bulk4d}
\end{equation}
 which is the bulk analog of \eqref{eq:bdy4d}. Indeed one can imagine illustrating these in the same vein as in Fig.~\ref{f:bdy4d} in one higher dimension.

 This picture is in fact naturally suggested by our earlier discussion involving the quantum corrections to entanglement entropy \cite{Faulkner:2013ana}. We have argued in \S\ref{sec:bulkee} that the leading $\frac{1}{\ceff} $ corrections to the boundary entanglement entropy arise from the bulk entanglement across the extremal surface. For this to make sense, it must be true that the extremal surface naturally decomposes the spacetime in the form \eqref{eq:bulk4d}. The bulk entanglement $S_{\homsurfA}^{bulk}  = S_\regA^{1-loop}$ can be computed on any spacelike surface foliating the entanglement wedge owing to the fact that $\EWA$ is a domain of dependence.  Since we can compute the QFT entanglement using the knowledge of $\rhoA$, we would learn that there is  non-trivial bulk entanglement entropy at
 ${\cal O}(1)$ from $S_{\homsurfA}^{bulk} $. This information being recoverable in the field theory, it must therefore be true that the bulk subregion dual to $\rhoA$ is similarly aware of the amount of entanglement between $\regA$ and $\regAc$. Heuristically, the presence of an EPR pair separated in the bulk across $\extrA$ should be detectable using the information contained in $\rhoA$ alone. This naturally pins down $\EWA$ as the bulk subregion dual to the boundary
 $\domdA$.

 On the other hand, the causal wedge for a boundary region $\CWA$ is not its own domain of dependence \cite{Hubeny:2014qwa}. To see why this is the case, realize that the causal wedge is defined starting from the boundary $\domdA$ \eqref{eq:cwedge}. We can try to pick a spacelike codimension-1 surface ${\cal Q}_{\cal A}$  within $\CWA$ that is anchored on $\regA$ and the causal information surface $\CIS$,  i.e., $\partial {\cal Q}_{\cal A}  = \CIS \cup \regA$,
 and construct its domain of dependence $\tilde{D}[{\cal Q}_\regA]$, but this bulk causal domain will differ generically from $\CWA$. The reason it does so is that $\CWA$ is obtained by following ingoing null geodesics from the boundary, which by definition are complete towards the boundary. These geodesics will caustic deep into the bulk, resulting generically in $\CIS$ being a non-smooth surface.  On the other hand, the boundary of $\tilde{D}[{\cal Q}_\regA]$ is constructed by shooting null geodesics off $\CIS$, whence they are complete towards $\CIS$, but would caustic before approaching the boundary. Generically the two loci of caustics are distinct and thus the two sets non-equivalent, as emphasized in \cite{Hubeny:2014qwa}. Moreover, the causal wedges for a region and its complement do not meet at a common codimension-2 surface. So no decomposition of the form \eqref{eq:bulk4d} is possible for them.  These observations render $\CWA$ unsuitable as a candidate dual to the boundary density matrix.

 While we have ruled out the causal wedge, note that this argument does not preclude $\tilde{D}[{\cal Q}_\regA]$. In fact,  consider any spacelike codimension-2 surface $\Psi_\regA$ lying in the causal shadow region of  $\regA$ and $\regAc$. We can always decompose the bulk spacetime into four causal domains across $\Psi_\regA$, as we did with the extremal surface in \eqref{eq:bulk4d}. In particular, we could compute the bulk entanglement entropy for a Cauchy slice bipartitioned across $\Psi_\regA$ and attempt to associate it as an
 $\mathcal{O}(c_\text{eff}^{-1})$ contribution of some observable associated with the spatial boundary region $\regA$. If consistency with boundary bipartitioning were the only constraint, then any $\Psi_\regA$ of the above kind, including $\Xi_\regA$  which  lies at the edge of acceptability vis a vis causality, should have been acceptable as the bulk dual of entanglement entropy. However, the crucial point is that no dynamical principle singles out $\Psi_\regA$ (or $\Xi_\regA$).  The special feature of the extremal surface is that it is picked out by the bulk gravitational dynamics when we implement the dual of the replica construction.

There is another minor point relating to an entanglement wedge which is worth bearing in mind. If we consider just codimension-2 extremal surfaces anchored on the boundary, then it turns out in many spacetimes one cannot foliate a bulk Cauchy surface with such surfaces, leading to the concept of  ``entanglement holes''.\footnote{ Some authors, e.g., \cite{Balasubramanian:2014sra} have taken to calling such regions the entanglement shadow, inspired by the idea of a causal shadow. We find this terminology extremely misleading. Utilizing causal constructs,  the natural regions in question are the entanglement wedges, which, as discussed in the text, are oblivious to these bulk holes.}  Most of the explicit cases in which  these have been discussed turn out to be static spacetimes.  The simplest examples are provided by the \SAdS{d+1} black hole spacetimes \cite{Hubeny:2013gta} (see \S\ref{sec:extdeter}), but as discussed in \cite{Czech:2012bh,Nogueira:2013if,Gentle:2013fma}, such behaviour can also happen in causally trivial spacetimes. More generally,  one can argue that there exist bulk codimension-0 regions which are not penetrated by any boundary-anchored extremal surfaces.   Despite this, it remains true that these spacetime regions are completely contained within the entanglement wedges $\EWA$ or $\EWAc$. Indeed this must be so, for otherwise there is no semiclassical decomposition of the  bulk Hilbert space to enable one to compute the bulk one-loop contribution. So from the perspective of bulk reconstruction, the entanglement holes are inconsequential in so far as the entanglement wedge reconstruction is concerned.

 There is an important consequence of identifying the entanglement wedge as the natural dual of the reduced density matrix. We take this statement to imply that the boundary observer restricted to $\domdA$ can learn about the bulk geometry in the entire  $\EWA$. We have hitherto argued that in order to satisfy causality of entanglement entropy, the extremal surface $\extrA$ has to lie in the causal shadow.  This set can however be quite large, and so $\extrA$ can lie very deep inside the bulk and, in general, well outside the causal wedge. Indeed, it is possible to construct examples in which the entanglement wedge contains a substantial part of the spacetime far beyond  black hole horizons, cf., \cite{Headrick:2014cta}.

 This brings forth a natural question: if the dual of the density matrix is the entanglement wedge, then should we not be able to construct local operators in the semiclassical limit all through this region, and not just in the smaller causal wedge?  This question has been surprisingly hard to answer, though there is now evidence that the subregion/subregion duality does involve a boundary reconstruction of the entanglement wedge \cite{Almheiri:2014lwa,Jafferis:2015del,Dong:2016eik}.

 To see the difficulty, let us first point out a simple argument precluding an expression along the lines of \eqref{eq:localhkll}.  Take two local bulk operators  $\phi_1(X_1)$ and $\phi_2(X_2)$ inserted at two distinct points $X_1$ and $X_2$ in the bulk. Let us assume that
$X_1$ and $X_2$ are timelike separated and furthermore $\{X_1, X_2\} \in \EWA\backslash\CWA$, i.e., they both lie outside the causal wedge but inside the entanglement wedge. Now these two operators do not necessarily have to commute with each other and we shall  arrange them not to do so $[\phi_1(X_1), \phi_2(X_2)] \neq 0$.
However, should there have been an explicit inversion of the extrapolate map along the lines of \eqref{eq:localhkll}, then we could have rewritten, say $\phi_1(X_1)$ in terms of local operators supported in $\domdA$, say
$\phi_1(X_1) = \int_{\domdA} \, d^dy \, K(y|X_1) \,{\cal O}(y)$.
But now note that $X_2$ is spacelike separated from all points in $\domdA$, as it lies outside the causal wedge. Therefore  the boundary representative, if $\phi_1(X_1)$ and $\phi_2(X_2) $ would have to commute, since $[{\cal O}(y), \phi_2(X_2)] =0 \,, \forall \,y\in \domdA$. This is a contradiction, which invalidates our assumption.

The issue is that we are assuming a local representative for the bulk operator in the entanglement wedge. The naive contradiction can be easily avoided if the boundary representative for the entanglement wedge fields was non-local. Indeed one of our motivating arguments for $\EWA$ was the fact that the knowledge of the reduced density matrix should somehow be factored in. One natural guess which will take care of this is to consider not  just boundary Heisenberg operators as in \eqref{eq:localhkll}, but also to take into account the modular evolved operators \cite{Jafferis:2015del}. One could for instance write a suggestive expression:
\begin{equation}
\phi(X) = \int \, ds \, \int_{\domdA}\, d^dy\, \tilde{K}(y,s|X)\; {\cal O}_\regA(y,s) \,, \qquad X \in \EWA
\label{eq:jlmshkll}
\end{equation}
Here $s$ is the modular time, and the boundary modular evolved operators are defined by conjugating with the modular evolution operator $\rhoA$, viz.,
\begin{equation}
{\cal O}_\regA(y,s) \equiv e^{-s\, \modA}  {\cal O}(y)\, e^{s\, \modA}\,.
\label{eq:modops}
\end{equation}
To understand this construction, we need better intuition for the modular evolved operators. Unfortunately the modular Hamiltonian is a rather complicated non-local operator in most situations. There are a handful of circumstances wherein the modular Hamiltonian simplifies: the vacuum state restricted to either the  Rindler wedge of Minkowski space for all relativistic QFTs or spherically symmetric ball-shaped domains in CFTs, whence it is given by an integral of the energy-momentum tensor, cf., \S\ref{sec:extdeter}. Unfortunately in these simple cases, the causal and entanglement wedges turn out to coincide, rendering them somewhat unsuitable for building further intuition.

% Figure
\begin{figure}[ht!]
\begin{center}
\hspace{1cm}
\includegraphics[width=3.1in]{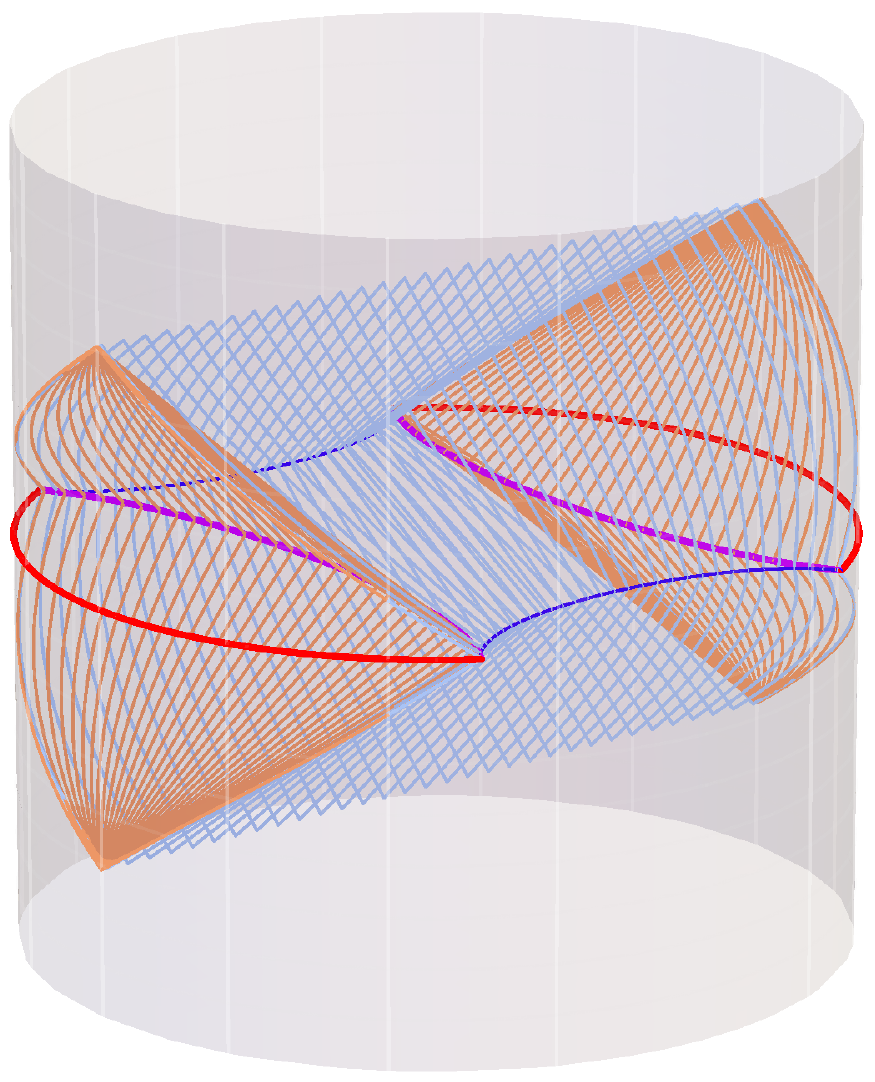}
\hspace{7mm}
\includegraphics[width=2.5in]{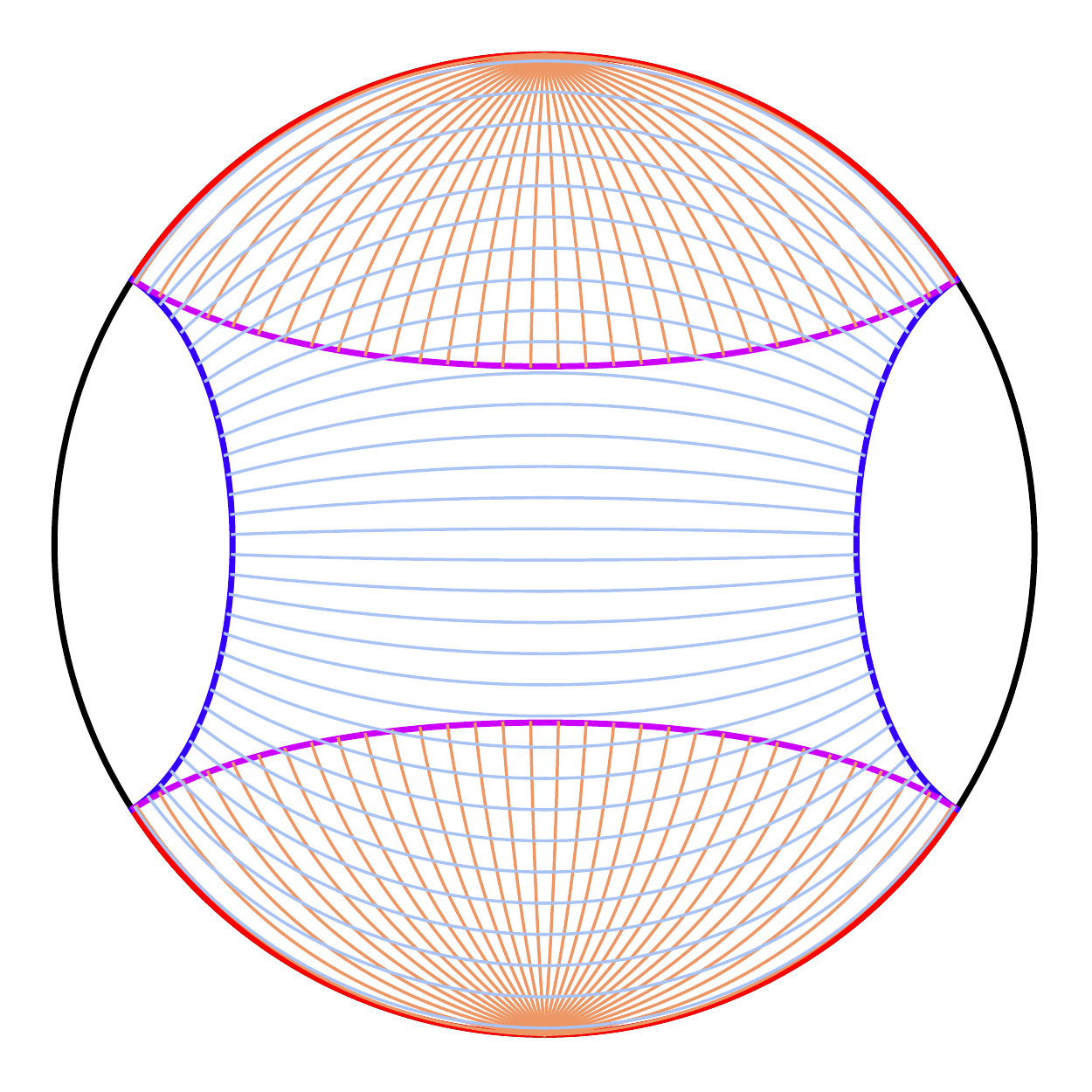}
\begin{picture}(0.3,0.4)(0,0)
\put(-90,0){\makebox(0,0){$\color{red}{\regA_1}$}}
\put(-90,70){\makebox(0,0){$\color{purple}{\Xi_{\regA_1}}$}}
\put(-90,180){\makebox(0,0){$\color{red}{\regA_2}$}}
\put(-90,110){\makebox(0,0){$\color{purple}{\Xi_{\regA_2}}$}}
\put(-30,90){\makebox(0,0){$\color{blue}{\mathcal{E}_{\regA_1}}$}}
\put(-160,90){\makebox(0,0){$\color{blue}{\mathcal{E}_{\regA_2}}$}}
\end{picture}
\caption{ The distinction between causal and entanglement wedges for a disjoint  region $\regA = \regA_1 \cup \regA_2$ in the \AdS{d+1} spacetime. We have plotted the domains for $d=2$. The left plot depicts the full wedges  $\EWA$ and $\CWA$ where we have shown the skeleton of the null geodesics that bound the domain. The right plot depicts the same projected onto the Poincar\'e wedge for ease of visualization. We have refrained from labeling the three-dimensional plot to avoid cluttering up the picture.
}
\label{f:disjointCEW}
\end{center}
\end{figure}

The simplest non-trivial case in which progress may be possible is to consider disjoint spherical ball-shaped regions for the vacuum state. The extremal surfaces and the entanglement wedge in the regime where the mutual information is non-vanishing differ significantly from the causal wedge. For the sake of visualization, we depict the situation in our familiar \AdS{3} geometry in Fig.~\ref{f:disjointCEW}.

%~~~~~~~~~~~~~~~~~~~~~~~~~~~~~~~~~~~~~~~~~~~~~~~
\section{Holography and Quantum Error Correction}
\label{sec:errors}
%~~~~~~~~~~~~~~~~~~~~~~~~~~~~~~~~~~~~~~~~~~~~~~

In the previous section, we have given arguments in favour of the entanglement wedge reconstruction for a given boundary subregion. As discussed there, the strongest evidence comes from the structure of the boundary entanglement entropy in the semiclassical limit $\ceff \gg 1$.

There are other compelling reasons to believe in the entanglement wedge reconstruction conjecture as was originally  argued using ideas borrowed from quantum error correction  by Almheiri, Dong, and Harlow (ADH) \cite{Almheiri:2014lwa} and beautifully illustrated in toy models built using tensor networks \cite{Pastawski:2015qua} (which has come to be known as the HaPPY code after the authors).\footnote{ The properties of the tensor networks built from perfect tensors as in the HaPPY code are also captured by random tensor networks \cite{Hayden:2016cfa}. We note in passing that perfect tensors can be viewed as multi-unitary matrices \cite{Goyeneche:2015aa}, which in turn are in one-to-one correspondence with absolutely maximally entangled states. } 
  Building on these ideas and the equality of the bulk and boundary relative entropies  \cite{Jafferis:2015del}, a robust argument for the entanglement wedge reconstruction was obtained in \cite{Dong:2016eik}. We will attempt to give a flavour of the ideas contained in the ADH in the discussion below.

% Figure
\begin{figure}[ht!]
\begin{center}
\includegraphics[width=4in]{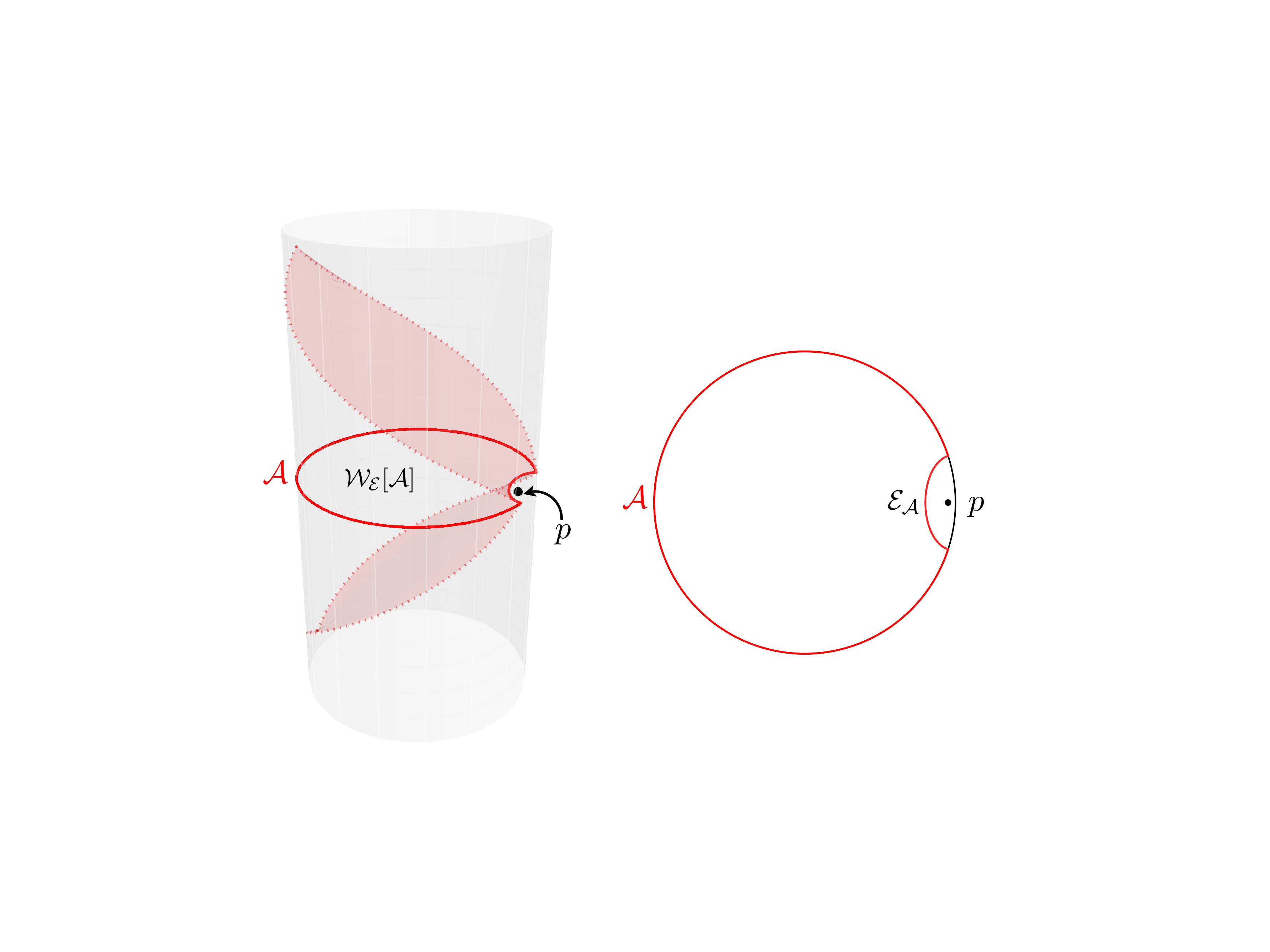}
\caption{ The bulk reconstruction of an operator close to the boundary. We have changed conventions from earlier for visualization. The boundary of the entanglement wedge is now depicted explicitly using the null generators and we also color code the region $\regA$ and its associated extremal surface $\extrA$ in the same format.  The plot on the right shows the projection onto the constant time slice $t=0$.
The operator in the bulk will have to commute with the operators in the complementary region whose entanglement wedge is shown.}
\label{f:errorC}
\end{center}
\end{figure}

We have demonstrated hitherto that there are two useful local reconstructions of the bulk: the global one which is reasonably unambiguous, and a local one that involves restricting to subregions. Now, consider a bulk operator $\phi(X)$. In the global reconstruction, we simply take the time strip related to $X$. But should we want to perform a local Rindler reconstruction, we have to face the following question: which domain on the boundary should we choose to reconstruct? We minimally want $X$ to lie on a boundary-anchored extremal surface, so that it is in the entanglement wedge of some  boundary region, which is a rather weak requirement. There are many entanglement wedges that contain the same bulk point.

The essential point we wish to make can  already be exemplified by considering the CFT vacuum, and taking the regions to all be our familiar ball-shaped regions. While here  we fail to detect the distinction between causal and entanglement wedges as $\CW{{\regAB}} = \EW{{\regAB}}$ in pure \AdS{d+1}, the distinction per se does not matter for the purposes of the argument. However, in the interest of  generality we will talk about the entanglement wedge, hopefully making it clear that our arguments apply to it more generally.

The Rindler representation leads to some a-priori counter-intuitive  properties for the local reconstructions.
Consider the following set of gedanken experiments, which illustrate the strangeness that we have to contend with.
In the first instance, let us consider a bulk point $X$ which is sufficiently close to the boundary, so that it can be represented in the entanglement wedge of a  relatively small region, say $\regA$. The bulk point is spacelike relative to the complement $\regAc$. By bulk causality, $\phi(X)$ commutes with all operators in $\regAc$. If we can localize $\regA$ arbitrarily, then we would have a problem -- the operator $\phi(X)$, and thus its boundary representative, would have to commute with all the operators on the boundary. In the limiting case, $\regAc \to \Sigma$, implying that $\phi(X)$ commutes with all operators in the boundary operator algebra. This  is in tension with the irreducibility of the operator algebra representation on the Hilbert space (and $\phi(X)$ is clearly not proportional to the identity).

% Figure
\begin{figure}[ht!]
\begin{center}
\includegraphics[width=4in]{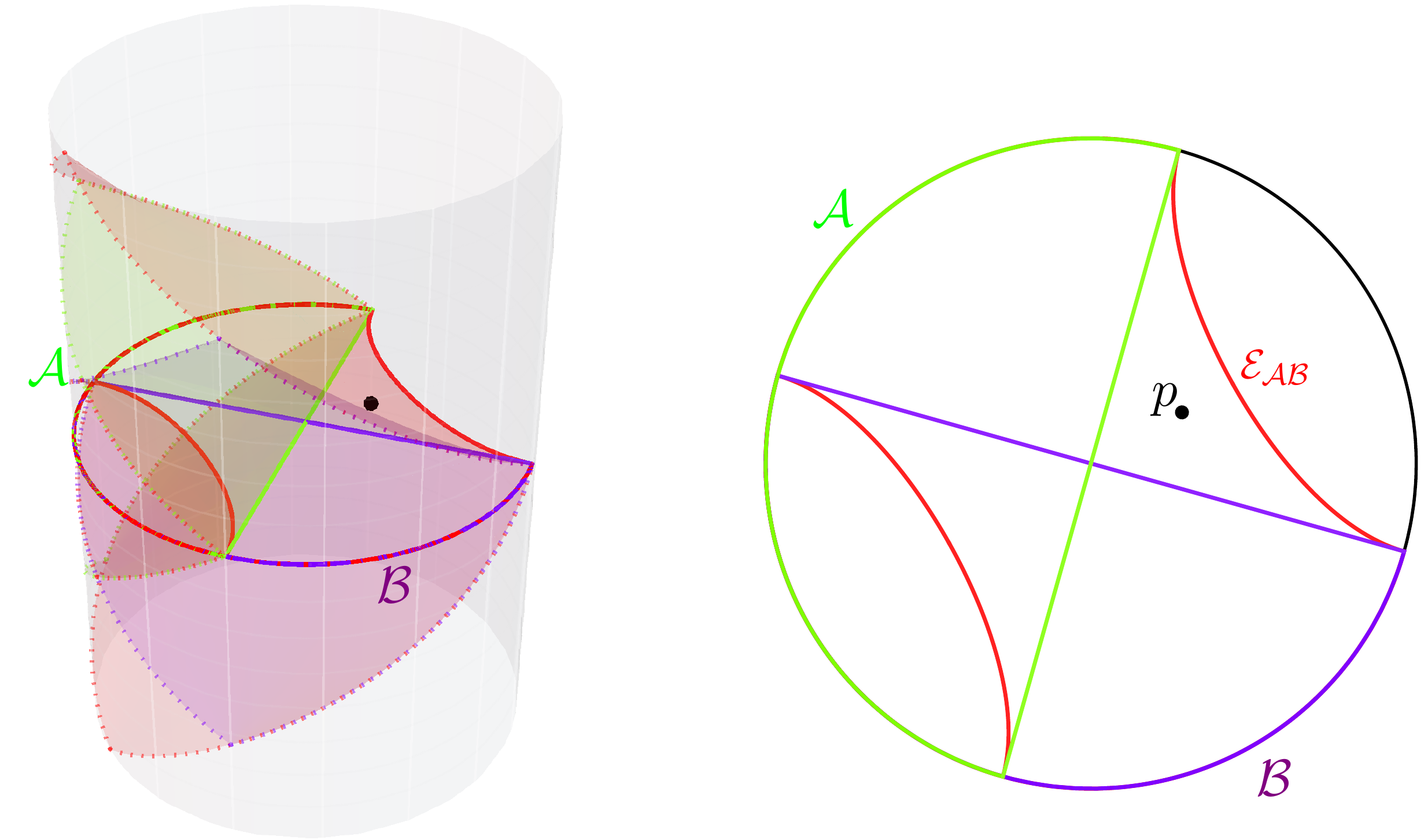}
\vspace{1cm}
\includegraphics[width=4in]{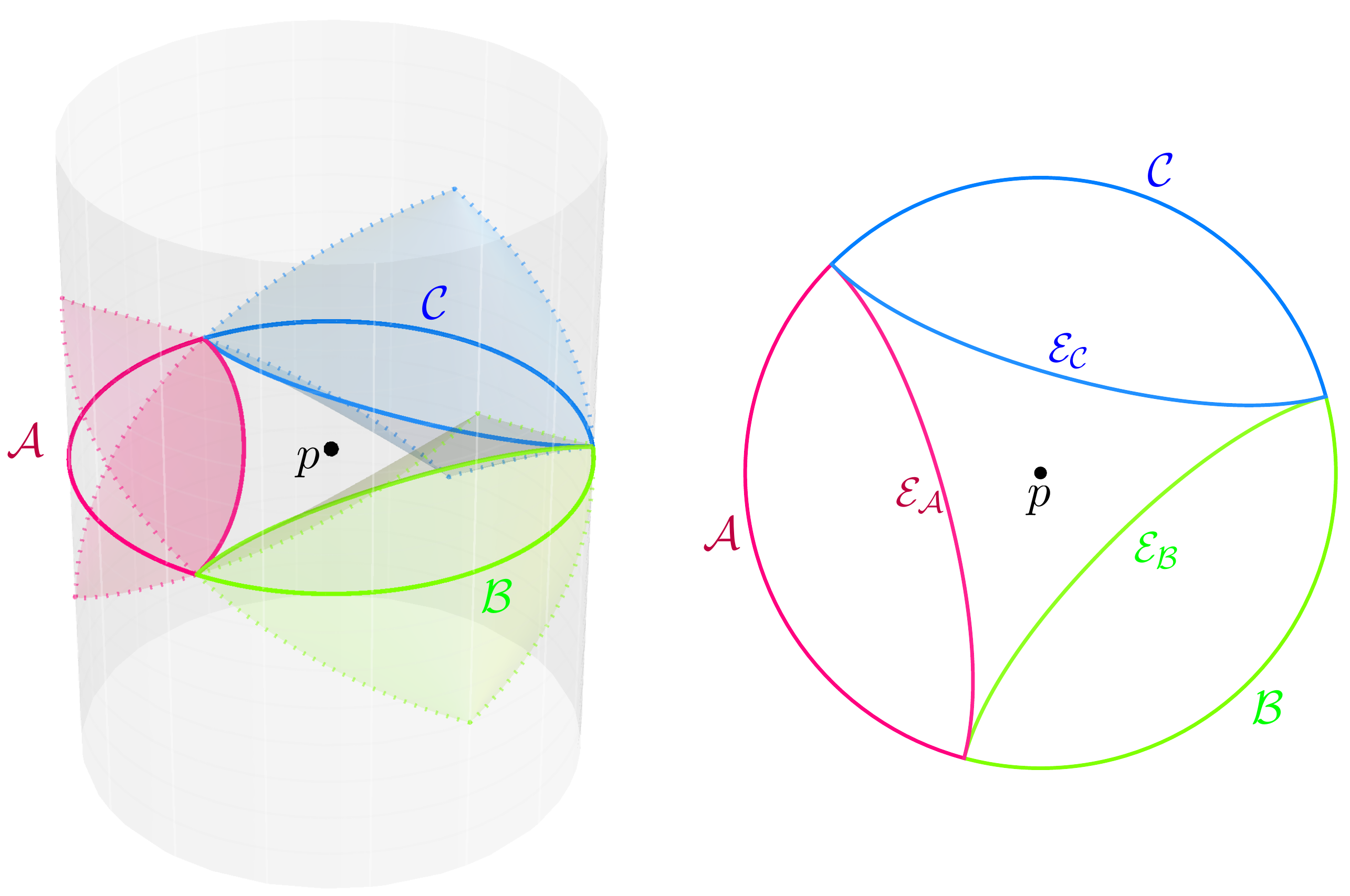}
\caption{ Two non-trivial situations involving local operator reconstruction. In the first instance, on the left, we have  a bulk point that is the  entanglement wedge of the union of two regions (red), but it cannot be reconstructed from either of the individual regions (green and purple) nor their mutual intersection (red). On the right, we illustrate the circumstance in which the bulk point is not reconstructible from any one single region, but can be represented in the union of any pair of regions.
}
\label{f:codewedgesABC}
\end{center}
\end{figure}

One may hope to do away with the above puzzle, by attributing it to regulator issues, but we can up the stakes by considering the next scenario. Say we have two regions $\regA_1$ and $\regA_2$ and a bulk point that $X \in \EW{\regA_1} \cap \EW{\regA_2}$. This bulk point now has distinct representations on the boundary in $\domd{\regA_1}$ and $\domd{\regA_2}$, respectively (among others). How are these two representations related? Clearly, they cannot correspond to the same boundary operators, since the two boundary domains may have non-overlapping elements in their respective operator algebras. Could it be that the representation only involves the common elements of the operator algebra? But this can be explicitly falsified by considering the bulk point to like outside $\EW{\regA_1\cap \regA_2}$ (see the configuration illustrated in Fig.~\ref{f:codewedgesABC}).  This means that in order to represent the
operator $\phi(X)$ in the boundary,  we need elements in the complement of $\regA_11\cap \regA_2$, or more precisely, in $\left(\domd{\regA_1} \cup \domd{\regA_2} \right)\backslash \domd{\regA_1 \cap \regA_2}$. The bulk operator `needs to know' more than the overlap of the two regions, but in slightly different form, depending on how we choose to represent it.

Things get more interesting when we realize that we can have regions $\regA_i$, which are such that no one of them contains information about the bulk operator $\phi(X)$,  lying outside each of their entanglement wedges as it does. But we can choose $X$ to lie in the common entanglement wedge of at least two of the regions. The easiest scenario to envisage here is to take three symmetric regions that are each a third of the boundary and $X$ to be at the center of AdS. No single region's entanglement wedge reaches out to the center, but the union of two adjacent regions has one large enough to contain $X$:  $\phi(X) \mapsto \domd{\regA_i \cup\regA_{i+1}}$ but $\phi(X) \nrightarrow \domd{\regA_i}$  for $i=1,2,3$ (Fig.~\ref{f:codewedgesABC}).

 We can make the situation even more interesting: consider two regions $\regA_1$ and $\regA_2$ anti-podally across from each other in global AdS. If the complementary region is large enough, then $I(S_{\regA_1} : S_{\regA_2}) =0$. In this case,  $\EW{\regA_1 \cup\regA_2} = \EW{\regA_1} \cup \EW{\regA_2}$. Let $X$ be the origin, which lies excluded from this entanglement wedge, and so $\phi(X)$ has no representation on $\domd{\regA_1} \cup \domd{\regA_2}$. Now gradually increase the sizes of the regions (still keeping them symmetrically distributed). At some point, when  $\regA_1\cup\regA_2$
 is greater than half of the total system,  we encounter a phase transition in the mutual information; the extremal surface $\extr{\regA_1 \cup \regA_2}$  (the global analog of the rainbow bridge in Fig.~\ref{f:sasurfs}) now gives the dominant contribution.   The new entanglement wedge will contain the origin, implying that we can now reconstruct $\phi(X)$ on $\domd{\regA_1} \cup \domd{\regA_2}$. On either side of the transition point, we have vastly different bulk reconstruction, though the change in the boundary algebra is infinitesimal. We can go on and concoct even more exotic behaviour by picking a collection of small regions $\regA_i$ which are initially far enough apart, and
 $\text{Vol}(\cup_i \, \regA_i) = \frac{1}{2} \, \text{Vol}(\Sigma) - \delta_V$. Deforming  the regions slightly, we again encounter a jump in the extremal surface from the disconnected to the connected one once
  $\text{Vol}(\cup_i \, \regA_i) = \frac{1}{2} \, \text{Vol}(\Sigma) + \delta_V$.  This leads to a macroscopically different domain of the spacetime being reconstructible from the subregions!

All told, these examples illustrate that the features of subregion/subregion duality are somewhat unconventional at first sight. These strange requirements are in fact a clue to the modus operandi of the local reconstruction map. What we have are different boundary realizations of the same bulk operator, but we need to ensure that the boundary avatars are inequivalent. This kind of behaviour is exactly what is necessary for quantum error correction. We won't explain in detail how quantum error correction works, but note the following salient features. One encodes the message, which we can think of as comprising for a certain  \emph{logical} $m$ number of qubits (or  degrees of freedom)  into a larger system of \emph{physical} $n$ qubits. The encoding map is described by a unitary that takes our message and scrambles it into the larger system, by suitably entangling the $m$-qubits with the remaining $(n-m)$-qubits. We may w.l.o.g.\ assume that the latter were originally in some reference state.

We want to ensure the following: should we end up losing some $l < n$ qubits, we want to be able to still read the message and protect against the loss.  Clearly, we cannot lose too many qubits; for then there would no message to read. The maximum number that we can end up missing can be estimated to be simply $l_\text{max} = \frac{1}{2}\, (n-m)$, so the message is recoverable for $l \leq l_\text{max}$. This should be intuitive: we need our message qubits to be unentangled with the lost ones. In particular, the combined message and loss subsystems of  total dimension $m+l$ must be smaller than the total number of qubits we retain  viz., $n-l$. This then leads to the requirement we impose on $l \leq l_\text{max}$  given the logical and physical qubits.
 For large systems, one can motivate this using Page's theorem  \cite{Page:1993df}, which states that random states of a bipartite system induce a maximally mixed state on the smaller factor.

The embedding of a message into a larger space is analogous to the representation of the bulk operator into the boundary Hilbert space. Not all bulk operators could be embedded this way, owing to the fact that messages are necessarily shorter than the enlarged encoding. All we care about is that any observable we could have defined on the original message can be computed accurately following the loss. This leads to the idea of the code subspace in the error correction as being the subspace of the enlarged system with sufficient entanglement to be robust against erasures. ADH argue that viewing the bulk reconstruction map in terms of an error-correction code resolves the seemingly bizarre features of the local reconstruction noted above.  The analogy with bulk reconstruction is sharply phrased in terms of the operator algebra. While  standard quantum error correction deals with the correction of states, which are passed onto a quantum channel for communication, for holography one really needs to correct for logical operation at the level of operators. One can therefore view the holographic error correction more accurately as ``operator algebra quantum error correction''.
The interested reader should refer to the original papers cited  above for a more detailed account of these developments.

%~~~~~~~~~~~~~~~~~~~~~~~~~~~~~~~~~~~~~~~~~~~~~~~
\section{Entanglement and Gravity}
\label{sec:einstein}
%~~~~~~~~~~~~~~~~~~~~~~~~~~~~~~~~~~~~~~~~~~~~~~

The connection between geometry and entanglement, as discussed in its various incarnations above, explores how the spacetime geometry may be reconstructed given the field theory entropy data.  As such this does not determine form taken by the bulk dynamics.
Nevertheless, one might argue that the fact that we use the RT/HRT prescriptions in relating entanglement entropy to the area of some surface pre-supposes that the gravitational equations of motion arise from the Einstein-Hilbert action. For other forms of gravitational dynamics, we would have to evaluate a different functional on some other surface.  Hence one might imagine that there is a way to extract the bulk dynamics directly from entanglement.

In a set of  papers \cite{Lashkari:2013koa,Faulkner:2013ica}, it was demonstrated that one can indeed obtain the  linearized gravitational equations of motion. Furthermore, \cite{Swingle:2014uza} argued that the non-linear equations should also follow. The basic idea behind these constructions is to build upon the results of \cite{Blanco:2013joa}, who explored holographic properties of relative entropy.
We need two pieces of information from field theory:
\begin{itemize}
\item The first law of entanglement \eqref{eq:e1stlaw}, which we reproduce for convenience
\begin{align}
\delta S_\regA = \vev{\modA} \,.
\label{eq:e1stlaw2}
\end{align}
Recall that this, being a statement about the linear deviations of a reference state, is upheld in any QFT.
\item The explicit expression for the modular Hamiltonian for spherical ball-shaped regions in the vacuum state of a CFT \eqref{eq:BallmodH}. We will express this result for a ball of radius $R$, centered at ${\bf x}_0 \in {\mathbb R}^{d-1}$ at $t=t_0$ in the following form:
\begin{equation}
{\cal K}_{_{\ball}} = 2\pi\, \int_{\domd{\regAB}} \, d^{d-1}x\, \frac{R^2 - ({\bf x} - {\bf x}_0)^2}{2\, R} \, \vev{T_{00}(t_0,{\bf x})} \,.
\label{eq:kball2}
\end{equation}
\end{itemize}

Focus attention on the domain of dependence of the region $\regAB $ which is conformally mapped to the hyperbolic cylinder $\mathbb{H}_{d-1} \times {\mathbb R}$; cf., \S\ref{sec:extdeter}. Being a static spacetime with metric \eqref{eq:hypc}
\begin{equation}
ds^2 = - d\tau^2 + R^2 \, \left(du^2 + \sinh^2 u \, d\Omega_{d-2}^2\right) \,,
\label{}
\end{equation}
this geometry possesses a Killing vector field $\zeta^\mu_{\mathbb{H}} = 2\pi\,R \left(\frac{\partial}{\partial \tau}\right)^\mu$. The normalization is chosen to make explicit the fact that the temperature for the theory on $\mathbb{H}_{d-1} \times \mathbb{R}$ is related to the curvature scale, $T = \frac{1}{2\pi\,R}$. The vector $\zeta^\mu_{\mathbb{H}} $ is the canonically normalized thermal vector.

We can follow this vector under the conformal map to learn that $\domd{\regAB}$ possesses a conformal Killing vector field
\begin{equation}
\zeta^\mu_{_{\ball}} = \pi\, R\, \, \left(\frac{\partial}{\partial t}\right)^\mu - \frac{\pi}{R}\, \Bigg[
\left( -(t-t_0)^2 + ({\bf x} - {\bf x}_0)^2\right) \left(\frac{\partial}{\partial t}\right)^\mu -2\, (t-t_0) (x^i -x_0^i) \, \left(\frac{\partial}{\partial x^i}\right)^\mu \Bigg] .
\label{eq:ckvball}
\end{equation}
Abstractly, the above is simply the combination of a time translation and a special conformal generator in
${\mathbb R}^{d-1,1}$; one can check $\zeta_{_{\ball}} = \frac{i\pi}{R}\, (R^2 \,P_t + K_t) $. This enables us to express the  modular Hamiltonian associated with $\rho_{_{\ball}}$, \eqref{eq:kball2},  in terms of this vector field. One can check
\begin{equation}
{\cal K}_{_{\ball}} = \int_{\domd{\regAB}}\, d\Sigma^\mu\, T_{\mu\nu}\, \zeta_{_{\ball}}^\nu \,,
\label{eq:kball3}
\end{equation}
where $d\Sigma^\mu$ is the induced volume element on a codimension-1 Cauchy slice of $\domd{\regAB}$.

The holographic dual of the CFT on the hyperbolic cylinder $\mathbb{H}_{d-1} \times \mathbb{R}$ at a temperature $T = \frac{1}{2\pi\,R}$ is the Rindler slicing of \AdS{d+1}. The geometry takes the form of a black hole spacetime with metric
\begin{equation}
ds^2 = -\frac{\varrho^2 - \lads^2}{R^2}\, d\tau^2 + \frac{d\varrho^2}{\varrho^2 - \lads^2}  + \varrho^2 \left(du^2 +\sinh^2u\, d\Omega_{d-2}^2\right)\,.
\label{eq:rads}
\end{equation}
The entanglement entropy is simply given by the entropy of the above black hole geometry, whose horizon lies at $\varrho = \lads$. This spacetime inherits the Killing field $2\pi\,R\,\partial_\tau$ whose orbits have a fixed point at the bifurcation surface of the black hole.  The  modular Hamiltonian \eqref{eq:kball3} in this presentation is a symmetry generator. In the hyperbolic black hole frame, we are simply performing a time translation which leaves the static exterior geometry invariant. The expression for the gravitational version of the modular Hamiltonian is then the energy of the black hole measured at infinity. The AdS/CFT dictionary relates this to the boundary energy-momentum tensor, so that we can write:
\begin{equation}
{\tilde {\cal K}}_{ball} = \int_{\domd{\regAB}} \,  d\Sigma^\mu\, {\tilde T}_{\mu\nu}\, \zeta_{_{\ball}}^\nu  =
\int_{\domd{\regAB}} \,  d\Sigma^\mu\,T^\text{CFT}_{\mu\nu}\, \zeta_{_{\ball}}^\nu
\,,
\label{}
\end{equation}
where we use the tilde to refer to the gravitational contribution. We can interpret the gravitational modular Hamiltonian as being the conserved charged associated with the boundary conformal Killing vector $\zeta_{_{\ball}}^\mu$.

In the original presentation of the problem, in which we have $\regAB \subset {\mathbb R}^{d-1,1}$, the dual spacetime is
Poincar\'e-\AdS{d+1} with metric \eqref{eq:pads}.  The entanglement entropy is the area of the extremal surface ${\cal E}_{\regAB}$, which is a hemisphere in this geometry \eqref{eq:rtsphere}
\begin{equation}
{\cal E}_{\regAB} = \{ t = t_0, ({\bf x} - {\bf x}_0)^2 + z^2 = R^2 \} \,.
\label{eq:rts2}
\end{equation}
 The domain of outer communication of the black hole geometry \eqref{eq:rads}, which is the region outside the horizon $\varrho \geq \lads$,   is simply the causal wedge of the ball-shaped region $\regAB$ in Poincar\'e-\AdS{d+1}. This is a special circumstance in which the causal wedge and entanglement wedge of the boundary domains coincide. Therefore we may write
\begin{equation}
\EW{\regAB} = \CW{\regAB} \xrightarrow{\text{conformal to}} \text{exterior region of hyperbolic black hole}
\end{equation}
%

%~~~~~~~~~~~~~~~~~~~~~~~~~~~~~~~~~~~~~~~~~~~~~~~
\subsection{Linearized gravity from entanglement}
\label{sec:lingrav}
%~~~~~~~~~~~~~~~~~~~~~~~~~~~~~~~~~~~~~~~~~~~~~~

We now have all the geometric ingredients in place to analyze the linearized gravitational dynamics. The idea is going to be to start with $\rho_{_{\ball}}$ and to consider perturbations atop  it. For small perturbations, we will expect the entanglement first law
\eqref{eq:e1stlaw2} to be upheld. This statement can be interpreted in the hyperbolic conformal frame as a constraint on perturbations to the black hole geometry. Physically one imagines that we start with the static black hole solution \eqref{eq:rads} and perturbs it by throwing some matter into the black hole. The linear response of the black hole respects the first law of black hole thermodynamics, which states that the changes in the gravitational energy are  compensated for by the change in entropy as long as the original configuration is on-shell, i.e., satisfies the gravitational field equations. We can relate the gravitational energy to the  gravitational modular Hamiltonian  and thus learn that:
\begin{equation}
\delta S_{bh} =  \delta E_{bh} = \delta \tilde{{\cal K}}_{ball}
\label{eq:gravrels}
\end{equation}
The first equality here is the statement of the first law of black hole dynamics \cite{Iyer:1994ys} and the second equality follows from the AdS/CFT dictionary.  This is the bulk analog of the entanglement first law and contains all the information necessary to extract the dynamics.

We will demonstrate below in \S\ref{sec:waldi} how the first law of black hole mechanics is derived. For the moment, let us take this as given and proceed to see how we obtain the gravitational equations of motion. The proof can be succinctly summarized by the following set of observations.

Recall that $S_{\regAB} = S_{bh}$, which immediately implies that $\delta S_{\regAB} = \delta S_{bh}$ holds for small variations. Using \eqref{eq:e1stlaw2}, we can further eliminate $\delta S_{\regAB}$ in favour of $\delta{\cal K}_{\regAB}$. But the latter is related to the gravitational modular Hamiltonian, which implies that $\delta S_{bh} = \delta \tilde{{\cal K}}_{ball}$, leading thence to \eqref{eq:gravrels}, upon relating the gravitational  energy to the gravitational modular Hamiltonian. What this means is that, whenever the first law of entanglement entropy holds in the boundary field theory, the gravitational first law is upheld for the dual configuration. We should note that we can interpret this statement in either conformal frame described above, and it applies to arbitrary ball-shaped regions.

Now the standard derivation of the gravitational first law \eqref{eq:gravrels} assumes the linearized field equations and thence derives $\delta S_{bh} =  \delta E_{bh} $ as an on-shell statement. We are however obtaining this relation from the properties of the CFT vacuum state interpreted holographically. Therefore we can reverse the standard derivation and infer from the fact that the CFT implies the  first law that the bulk dual must in turn satisfy the gravitational field equations: In other words, denoting the linearized Einstein's equations as $\delta \mathbf{E}_{g} =0$, we have:
\begin{equation}
\begin{split}
&\delta \mathbf{E}_g = 0
\;\; \overset{\text{\tiny{\cite{Iyer:1994ys}}}}{\Longrightarrow}\;\;
\delta \tilde{S}_{bh} =  \delta \tilde{E}_{bh}
 \\
& \delta S_{\regAB} = \vev{{\cal K}_{\regAB}}
\;\; \overset{\text{\tiny{\cite{Faulkner:2013ica}}}}{\Longrightarrow} \;\;
\delta S_{bh} =  \delta E_{bh}
\;\; \Longrightarrow \;\;
\delta \mathbf{E}_g = 0
\end{split}
\label{eq:wiee}
\end{equation}

To obtain all the components of the gravitational equations of motion, we should not just consider ball-shaped regions confined to a single time slice, but allow ourselves the freedom to work in various boundary Lorentz frames. The construction is only sensitive  to the gravitational field equations, i.e., it only picks out the dynamics of the linearized metric. If we have solutions with no (bulk) matter sources supporting the geometry, then the knowledge of $S_\regA$ is sufficient to recover the complete mapping from the field theory states in the neighbourhood of the vacuum to the linearized part of the dual spacetime.

We are assuming in this derivation that the vacuum state of the CFT is described in the gravitational picture by the pure AdS solution.  This is sufficiently innocuous,  since the symmetries preserved by the CFT vacuum state uniquely single out the \AdS{d+1} background as the putative dual. All the first law of entanglement teaches us is the dynamics of metric perturbations about this background. At no stage in our derivation have we had to say what the gravitational theory is. The discussion thus holds not just for Einstein gravity but for any general diffeomorphism-invariant gravitational dynamics with perhaps higher derivative corrections. This better be the case for the derivation of \cite{Iyer:1994ys}, which is used in \cite{Faulkner:2013ica}  and which works for any diffeomorphism-invariant theory of gravity. The precise details of the entropy functional and the form of the gravitational equations of motion change, but the set of implications described in \eqref{eq:wiee} continue to hold as stated.

We should however remind ourselves of the caveats discussed in \S\ref{sec:nhee}. It may indeed transpire that we are unable to discriminate with the data at hand the explicit gravity theory. As explained in \cite{Faulkner:2014jva,Haehl:2015rza}, it is plausible that there exists an effective Einstein-Hilbert theory with modified parameters which mimics the gravitational dynamics.

So far there isn't a very compelling argument for the non-linear Einstein's equations to be obtained from entanglement-based considerations. There have been some suggestions as to how this could be done in \cite{Swingle:2014uza}, but one would like to have a more direct argument. It is however hard to see how the discussion above generalizes, since the general constraint which replaces the first law is the positivity of relative entropy which translates into an inequality $\delta K_\regA \geq \delta S_\regA$. We describe constraints arising from the relative entropy in \S\ref{sec:cerel}.

%~~~~~~~~~~~~~~~~~~~~~~~~~~~~~~~~~~~~~~~~~~~~~~~
\subsection{The first law of black hole mechanics}
\label{sec:waldi}
%~~~~~~~~~~~~~~~~~~~~~~~~~~~~~~~~~~~~~~~~~~~~~~

To ascertain the gravitational equations of motion at the linearized level, we need to understand how the first law of black hole thermodynamics works in general. This was beautifully explained in a construction by Wald \cite{Wald:1993nt}  and subsequently elaborated upon by Wald and Iyer \cite{Iyer:1994ys} about two decades ago using standard variational calculus for a classical diffeomorphism-invariant Lagrangian.

The basic idea can be understood as follows. Say we have a diffeomorphism-invariant  action which determines the bulk dynamics. Suppose we perform a standard Euler-Lagrange variation to obtain the equation of motion. Varying the Lagrangian, we would get the equations of motion modulo some boundary terms. In the gravitational context, \cite{Iyer:1994ys} showed that these boundary terms via a Noether construction is related to the entropy of the black hole.

Let us consider a diffeomorphism-invariant gravitational Lagrangian, which we view as a $(d+1)$-form in bulk. This is to enable us to write the variational expressions without worrying explicitly about the measure factor. We denote differential forms with a bold-face font to avoid confusion. We have $S_{bulk} = \int \mathbf{L}(\phi)$, with $\phi$ being our collective label for all the fields including the bulk metric. The variational calculus for $\mathbf{L}$ is encoded in the statement:
\begin{equation}
\delta \mathbf{L} = \mathbf{E}_\phi \cdot \delta \phi   + d{\bm{\Theta}}(\phi,\delta\phi)
\label{eq:varL}
\end{equation}
in which $\mathbf{E}_\phi$ denotes the equations of motion for the field $\phi$ and $d\bm{\Theta}$ is the symplectic potential. It comprises  of boundary terms (encoded as a spacetime $d$-form) that arise upon integration by parts. These terms depend both on the fields and their first variation, as indicated. The Noether construction allows us to write down conserved charges from this basic variational statement.

We want to consider theories that are diffeomorphism-invariant. By employing the Noether construction, we can obtain the charge associated with the transformation. This involves varying the fields along the symmetry direction.  A diffeomorphism, i.e., a coordinate transformation, is implemented by Lie dragging all the fields along a vector field $\xi$ that implements the transformation.
Under such a transformation, the change of the Lagrangian is by a total derivative.

Say we consider an arbitrary vector field $\xi$ and vary the Lagrangian under a diffeomorphism generated by it. Since we vary the fields in a direction specified by $\xi^A$, the change $\delta_\xi \phi$ is obtained by taking the correct directional derivative of the field $\phi$ along the vector. This is achieved by the Lie derivative operation $\delta_\xi \phi  = \mathscr{L}_\xi \phi$. For example, metric changes by $g_{AB} \mapsto \mathscr{L}_\xi g_{AB} = \nabla_A\xi_B + \nabla_B \xi_A$.   Furthermore, denote the interior contraction of a differential form with a vector by
$\iota: \mathbf{V}_{(p)} \mapsto \mathbf{V}_{(p-1)}$. This maps $p$-forms to $(p-1)$-forms: $ \iota_\xi \mathbf{V} = \xi^A V_{A,A_1\cdots A_{p-1}}$. One useful relation to remember is that the Lie derivation along $\xi$ can be expressed as a combination of exterior derivation and interior contraction:
\begin{equation}
\mathscr{L}_\xi =\{\iota_\xi, d\} =  d\iota_\xi + \iota_\xi d
\label{eq:Ldi}
\end{equation}
The reader may find \cite{Wald:1984ai} a helpful reference for these concepts.

We can now write the change of the Lagrangian under a diffeomorphism as
\begin{equation}
\delta_\xi \mathbf{L} = \mathscr{L}_\xi \mathbf{L}  = d(\iota_\xi \mathbf{L})\,,
\label{eq:Ldiffeo}
\end{equation}
where we exploit the fact that $\mathbf{L}$ is a top-form and hence $d\mathbf{L} =0$.
The transformation \eqref{eq:Ldiffeo} must vanish owing the invariance of the theory under coordinate transformations. Now associated with any symmetry, we should be able to construct a conserved current thanks to Noether's theorem. The current associated with the diffeomorphism will be denoted as $J^A[\xi]$. Instead of writing, the current we will write an expression for its Hodge-dual $\mathbf{J}$. Define thus the
Noether  current as the spacetime $d$-form:
\begin{equation}
\mathbf{J} = \bm{\Theta}(\phi,\mathscr{L}_\xi\phi) - \iota_\xi \mathbf{L}
\label{}
\end{equation}
Note that the above is entirely analogous to the construction of a Hamiltonian from the Lagrangian through a Legendre transformation; this would indeed be the case if the symmetry was associated with time translations. The conservation $\mathbf{J}$ follows from \eqref{eq:varL}, \eqref{eq:Ldiffeo} upon using the equations of motion, for
\begin{align}
\nabla_A J^A = d\mathbf{J}[\xi] &= d\bm{\Theta}(\phi,\mathscr{L}_\xi \phi) - d(\iota_\xi \mathbf{L}) \nonumber \\
& = -\mathbf{E}_\phi \cdot \delta\phi
\label{eq:Jcons}
\end{align}
We will indicate statements that are true on-shell, i.e., upon using equations of motion with
$\overset{\textbf{\tiny{E}}}{=} $ and thus write:
\begin{equation}
d\mathbf{J}[\xi] \overset{\textbf{\tiny{E}}}{=} 0  \;\; \Longrightarrow \;\; \mathbf{J}[\xi]  \overset{\textbf{\tiny{E}}}{=}  d\mathbf{Q}[\xi]
\label{}
\end{equation}
This defines the Noether charge $(d-1)$-form $\mathbf{Q}$ on $\bulk$.

% Figure
\begin{figure}[tp]
\begin{center}
\includegraphics[width=2.25in]{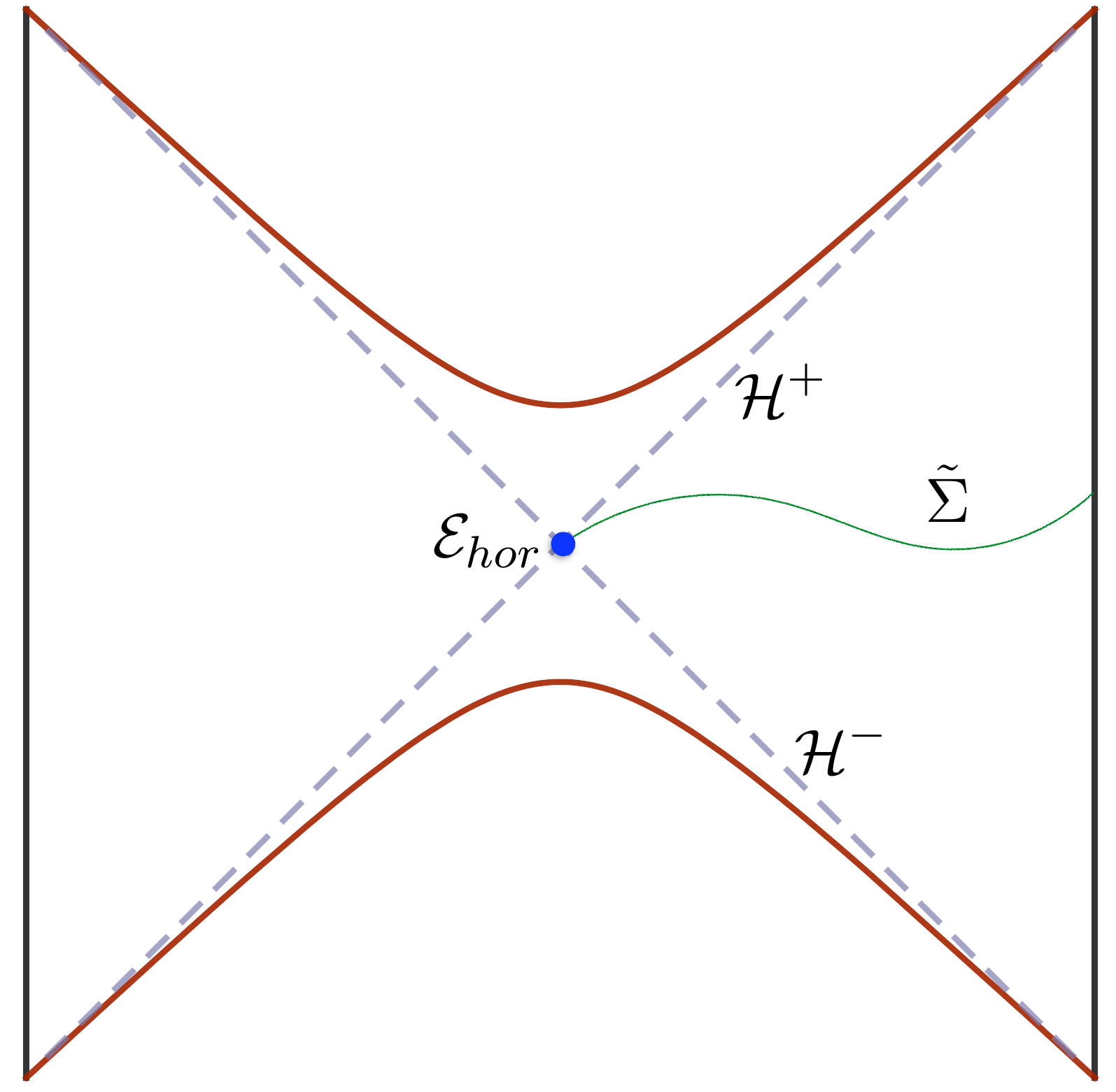}
\caption{A plot of the \SAdS{} Penrose diagram displaying the features of import in the derivation of the first law of black hole mechanics.
}
\label{f:sadsW}
\end{center}
\end{figure}

While these statements are valid for any diffeomorphism-invariant theory, we can specifically apply them to a gravitational Lagrangian  and restrict attention therein to a stationary black hole solution of the equations of motion. These geometries have a bifurcate Killing horizon, which is generated by a Killing field $\xi_{hor}^A$. The future and past horizons ${\cal H}^\pm$  are spacetime codimension-1 null surfaces ruled by the null generator $\xi_{hor}^A$ and intersect on the bifurcation surface which is fixed point of $\xi_{hor}^A$. The latter is a codimension-2 extremal surface in the spacetime -- we will call it ${\cal E}_{hor}$. Refer to  Fig.~\ref{f:sadsW} for an illustration of these features in the \SAdS{} geometry.

The horizon generator gives us the black temperature in terms of the surface gravity   $T = \frac{\kappa}{2\pi}$, with the latter being defined via $\xi^A_{hor}\nabla_A \xi^B_{hor} = \kappa\, \xi^B_{hor}$. Normalizing $\xi^A_{hor}$ such that $\kappa =2\pi$, or equivalently $T=1$, it was shown by Wald \cite{Wald:1993nt} that the black hole entropy is the Noether charge:
\begin{equation}
S_{bh} = \int_{{\cal E}_{hor}}  \, \mathbf{Q}[\xi_{hor}] \,.
\label{eq:noeW}
\end{equation}
This expression follows by interpreting the relation $\delta_\xi \mathbf{L}=0$ as the first law of black hole mechanics. One can further simplify this expression and give an explicit formula for the black hole entropy in a general theory of gravity. In general, this looks like the variation of the gravitational Lagrangian with respect to the Riemann curvature tensor (keeping the metric unchanged). We refer the reader to \cite{Iyer:1994ys} for further detail of the construction.

To obtain the first law, we also need to know what the definition of the gravitational energy is. This can be obtained from the Hamiltonian, which is encoded in the symplectic form. One considers two successive variations of the field $\delta_1 \phi$ and $\delta_2 \phi$ and takes their commutator to define the $d$-form $\bm{\omega}(\delta_1\phi, \delta_2\phi)$ through:
\begin{equation}
\bm{\omega}(\delta_1 \phi,\delta_2\phi) = \delta_2 \bm{\Theta}(\phi,\delta_1\phi) - \delta_1 \bm{\Theta}(\phi,\delta_2\phi)\,.
\label{eq:symp}
\end{equation}
To obtain the Hamiltonian that generates translations along some vector field $\xi^A$, we simply integrate $\bm{\omega}$ over a Cauchy surface $\tilde{\Sigma}$, by constraining one of the variations to be along the generator.  All told we can write:
\begin{equation}
\delta H_\xi = \int_{\tilde{\Sigma}} \, \bm{\omega}(\delta \phi, \mathscr{L}_\xi \phi)
 \label{eq:Hamgrav1}
\end{equation}
Using the variation of the Noether current
\begin{equation}
\delta \mathbf{J}[\xi]  = \delta{\bm{\Theta}}(\phi,\delta_\xi \phi) - \iota_\xi\, d\bm{\Theta}(\phi, \delta \phi) = \bm{\omega}(\delta_\xi \phi, \delta \phi ) + d\iota_\xi \bm{\Theta}(\phi,\delta \phi)
\label{eq:deltJ}
\end{equation}
we obtain upon integrating by parts:
\begin{equation}
\delta H_\xi \overset{\textbf{\tiny{E}}}{=} \int_{\partial\tilde{\Sigma}} \, \Big( \delta \mathbf{Q}[\xi] - \iota_\xi \bm{\Theta}(\phi,\delta \phi) \Big)
 \label{eq:Hamgrav2}
\end{equation}

We are now in a position to derive the statement of the first law. Consider in the  black hole spacetime a Cauchy slice $\tilde{\Sigma}$ that extends from the boundary to the bifurcation surface ${\cal E}_{hor}$. Thence $\partial \tilde{\Sigma} = {\cal E}_{hor}\cup \tilde{\Sigma}\big|_\bdy $. We define a variational $(d-2)$ form built from the Noether charge adapting the general construction to the horizon generator:
\begin{equation}
\bm{\chi}(\phi,\delta \phi, \delta_\xi \phi) = \delta \mathbf{Q}[\xi_{hor}] - \iota_{\xi_{hor}} \bm{\Theta}(\phi,\delta \phi)
\label{}
\end{equation}
The choice of the vector field to be the horizon generator ensures that the second contribution would vanish on the bifurcation surface ${\cal E}_{hor}$. Therefore we can simplify \eqref{eq:Hamgrav2} and obtain the desired expression for the black hole first law as:
\begin{equation}
\delta E_{bh} = \delta H_{\xi_{hor}} \overset{\textbf{\tiny{E}}}{=} \int_{\partial\tilde{\Sigma}} \, \delta \mathbf{Q}[\xi_{hor}]  = \delta S_{bh}\,.
\label{eq:g1law}
\end{equation}
We have made explicit that to derive the above, we need to employ the bulk equations of motion; the first law holds for linear variations about an on-shell configuration.

%~~~~~~~~~~~~~~~~~~~~~~~~~~~~~~~~~~~~~~~~~~~~~~~
\subsection{Canonical energy and relative entropy}
\label{sec:cerel}
%~~~~~~~~~~~~~~~~~~~~~~~~~~~~~~~~~~~~~~~~~~~~~~

In \S\ref{sec:critgeom}, we outlined some of the recent constraints derived in gravitational theories using relative entropy. To explain these, we need to define the idea of a canonical energy in gravitational systems \cite{Lashkari:2016idm}. We can do this directly from \eqref{eq:Hamgrav2}, which gives the variation of the gravitational Hamiltonian in terms of the Noether charge and the symplectic potential, if we could  only `integrate up' this equation in the space of variations. However, the second term, $\iota_\xi  \bm{\Theta}$, involving the explicit diffeomorphism field $\xi^A$, is not a variational total derivative, so one is not quite ready to do this in general. As described above, the strategy works for stationary black hole solutions, since they possess a Killing field which vanishes at the bifurcation surface and asymptotes to the generator of boundary time translations.

The upshot is that we need to find situations in which we can express $\iota_\xi \, \bm{\Theta} = \delta \left( \iota_\xi \,\mathbf{K}\right)$, with $\mathbf{K}$ being a spacetime codimension-2 form (i.e., a $d-1$ form in $\bulk$). This is a question of finding an appropriate vector field, since $\mathbf{K}$ is fixed by the variational principle of the Lagrangian. Moreover, the precise details of this field only matter in the vicinity of the boundary of the Cauchy surface, since in gravitational theories, diffeomorphism invariance allows us to express the Hamiltonian as a pure boundary term, as is explicit from \eqref{eq:Hamgrav2}.

Let us now consider the case in which we no longer consider the entire spacetime, but rather restrict attention to the entanglement wedge of a ball-shaped region  $\regAB$ on the boundary. We take $\tilde{\Sigma} = \homsurfAB$ so that
$\partial \tilde{\Sigma} = \regAB \cup \extr{\regAB}$.  We know that in the vacuum state, the entanglement wedge
$\EW{\regAB} = \tilde{D}[{\homsurfAB}] $ is conformal to the  domain of outer communication of a hyperbolic black hole. In this case, we clearly have a relation between the gravitational energy $H_\xi$ and the boundary energy-momentum tensor, or the modular Hamiltonian of the ball. Furthermore,$\xi^A$ is the extension of the boundary conformal Killing field $\zeta^\mu_{_{\ball}}$.

For an excited state, not necessarily in the code  subspace around the vacuum, one  has a  geometry $\bulk^{ex}$ with an extremal surface $\extr{\regAB}^{ex}$. The idea is to try to come up with a bulk vector field
$\xi^{ex}_{\regAB} =  \hat{\xi}$ defined in a neighbourhood of  $\homsurfAB^{ex}$ which resembles a Killing field with a vanishing locus around the extremal surface. The requirements can be stated as:
\begin{equation}
\begin{split}
\hat{\xi}^A\big|_{\bdy}  & = \zeta^A_{\regAB} \,, \qquad \hat{\xi}^A\big|_{\extr{\regAB}^{ex}}=  0\,,\\
\nabla_{(A} \hat{\xi}_{B)}\big|_{z\to 0}  &= {\cal O}(z^{d}) \,,  \qquad  \nabla_{[A} \hat{\xi}_{B]} = 2\pi n_{AB}
\end{split}
\label{eq:hatxidef}
\end{equation}
 where $n_{AB}$ is the unit binormal to the extremal surface. These conditions express our desire to have a vector field that serves as a Killing field in the vicinity of the homology surface, essentially giving the local neighbourhood of the extremal surface a structure of a Rindler horizon. It then follows that we can write an expression for the canonical gravitational energy as
\begin{equation}
H_{\hat{\xi}} = \int_{\regAB \cup \extr{\regAB}} \left( \mathbf{Q}[\hat{\xi}]  - \iota_{\hat{\xi}} \mathbf{K} \right)
\label{eq:canenegy}
\end{equation}

This canonical energy allows us to express the relative entropy in QFT as
\begin{equation}
S(\rho^{ex}_{\regAB}||\rho^{vac}_{\regAB}) = H_{\hat{\xi}}(\bulk) - H_{\hat{\xi}}(\text{AdS}) \,.
\label{eq:relcan1}
\end{equation}
Furthermore isolating the contribution from the extremal surface and the boundary to the canonical  energy \eqref{eq:canenegy}, we have
\begin{equation}
\begin{split}
\Delta S_{\regAB} & = \Delta \int_{\extr{\regAB} } \left( \mathbf{Q}[\hat{\xi})  - \iota_{\hat{\xi}} \mathbf{K} \right) \\
\Delta {\cal K}_{\regAB} &= \Delta \int_{\regAB} \left( \mathbf{Q}[\hat{\xi})  - \iota_{\hat{\xi}} \mathbf{K} \right) \\
\end{split}
\label{eq:relcan2}
\end{equation}
in which $\Delta$ stands for finite differences.

%~~~~~~~~~~~~~~~~~~~~~~~~~~~~~~~~~~~~~~~~~~~~~~~
\subsection{Relative entropy constraints}
\label{sec:cerel}
%~~~~~~~~~~~~~~~~~~~~~~~~~~~~~~~~~~~~~~~~~~~~~~

We can now state the various results obtained thus far from considerations of relative entropy in increasing order of generality.
\begin{itemize}
\item The first law of entanglement entropy implies the linearized Einstein's equations:
\begin{equation}
\delta S_\regA = \delta \langle \modA \rangle  \;\; \Longleftrightarrow \;\; \delta \mathbf{E} = 0
\end{equation}
\item Positivity of the quantum Fisher information implies that  the perturbative expansion of the canonical energy to quadratic order is non-negative definite. One may then write a constraint on the quadratic expansion of the
\begin{equation}
\begin{split}
\frac{1}{2}\, \frac{\partial^2}{\partial \epsilon^2}
& S( \rhoB +\epsilon\, \delta \rho || \rhoB ) =
\vev{\delta \rho\,,\, \delta \rho}_{\rhoB }  \geq 0  \\
&\qquad  \;\; \Longleftrightarrow \;\;
  \int_{\homsurfAB} \, d^dx \left( T_{AB}^\text{matter} + T_{AB}^\text{grav} \right)\xi^A \; d\Sigma^B  \Big|_{{\cal O}(\epsilon^2) }
 \geq 0
\end{split}
\label{eq:pocanquad}
\end{equation}
\item The positivity of relative entropy in general implies a positive energy theorem; the canonical energy in the entanglement wedge of a deformed spacetime is bounded from below by its value in the vacuum AdS spacetime.
\begin{equation}
S(\rho_{_{\regAB}} || \rhoB ) \geq 0  \;\; \Longleftrightarrow \;\;
 H_{\hat{\xi}}(\bulk) \geq H_{\hat{\xi}}(\text{AdS}) \,.
\label{eq:poscan}
\end{equation}
\item Finally, monotonicity of relative entropy results in a statement of  the symplectic flux computed across the homology surface is non-negative definite. A-priori it seems hard to geometrize this statement, since we have to refer to two different regions, one of whose domain of dependence  is included in the other's. The domains of integration for the canonical energy, etc., are quite distinct in the two cases.

However, we can write a simple integral inequality in a single spacetime by considering a suitable basis of regions. Let us imagine that we pick a ball-shaped region $\regAB$ and extend it into a one-parameter family
of ball-shaped regions $\regAB[\lambda]$, such that $\regAB[\lambda =0]= \regAB$ and $\regAB[\lambda=1]
= \regAB'$. The monotonicity condition should hold for any set of regions in this family as long as
$\domd{\regAB[\lambda_1]} \subset \domd{\regAB[\lambda_2]} $ for  $\lambda_1 \leq \lambda_2$. Hence we can write a variational form of the relative entropy monotonicity:
\begin{equation}
\frac{d}{d\lambda} S(\rho^{ex}_{\regAB[\lambda]} || \rho_{_{\ball}[\lambda]}) \geq 0
\label{eq:mon1}
\end{equation}
These infinitesimal constraints encode all the non-trivial relations implied by the monotonicity of relative entropy.

It is useful to eschew changes in the region, but instead allowing for the freedom to fix the region but change the state. This can be achieved by conformally mapping all the ball-shaped regions $\regAB[\lambda]$ back to $\regAB^0$. Applying this as an active transformation on the state, we can conformally map
$\rho_{\regAB[\lambda]} \mapsto \rho^\lambda_{\regAB}.$ We then write the differential statement
\begin{equation}
\frac{d}{d\lambda} S(\rho^{ex,\lambda}_{\regAB} || \rho^\lambda_{_{\ball}}) \geq 0
\label{eq:mon2}
\end{equation}

To express this statement gravitationally, we realize that  the conformal transformation  that maps a region $\regAB[\lambda]$ to $\regAB$ is achieved by a geometric action of a vector field $\mathfrak{C}^\mu[\lambda]$ on the boundary. This vector field can be extended into the bulk quite naturally to $\hat{\mathfrak{C}}^A[\lambda]$, which in turn implements a bulk diffeomorphism. We now have two independent changes associated with our configuration. On the one hand,  we have the action of the conformal transformation that rescales regions through this vector field, and on the other, we have the change associated with the vector field $\hat{\xi}^A$ capturing the excitation about the vacuum defined in \eqref{eq:hatxidef}. The gravity dual of monotonicity of relative entropy is then the statement:
\begin{equation}
\begin{split}
& \frac{d}{d\lambda} S(\rho^{ex,\lambda}_{\regAB} || \rho^\lambda_{_{\ball}})
= \delta_{\hat{\mathfrak{C}}} H_{\hat{\xi}} \geq 0
\;\; \Longleftrightarrow \; \int_{\homsurfAB} \, \bm{\omega}\left(  \mathscr{L}_{\hat{\mathfrak{C}}} g, \mathscr{L}_{\hat{\xi}} g \right) \geq 0
 \end{split}
\label{}
\end{equation}
\end{itemize}

%%%%%%%%%%%%%%%%%%%%%%%%%%%%%%%%%%%%%%%%%%%%
% \bibliography{entanglement}
% \bibliographystyle{utphys}

\providecommand{\href}[2]{#2}\begingroup\raggedright\endgroup

\end{document}